# Chapter 4

# Sulfur in hydrothermal fluids

**Gleb S. Pokrovski[1], Kalin Kouzmanov[2], and Andri Stefánsson[3]**

[1] Experimental Geosciences Team (GeoExp), Géosciences Environnement Toulouse (GET), UMR 5563 of the Centre National de la Recherche Scientifique (CNRS), Université Paul Sabatier Toulouse III, Institut de Recherche pour le Développement (IRD), Centre National d'Etudes Spatiales (CNES), Observatoire Midi-Pyrénées, 14 av. Edouard Belin, 31400 Toulouse, France; e-mails: gleb.pokrovski@get.omp.eu ; gleb.pokrovski@univ-tlse3.fr ; glebounet@gmail.com

[2] University of Geneva, Department of Earth sciences, University of Geneva, rue des Maraîchers 13, 1205 Geneva, Switzerland; email: kalin.kouzmanov@unige.ch

[3] Institute of Earth Sciences, University of Iceland, Sturlugata 7, 102 Reykjavík, Iceland; email: as@hi.is

# Glossary

*Method acronyms*

CL: cathodoluminescence
EPMA: electron probe microanalyzer
EPR: electron paramagnetic resonance
FAA: flame atomic absorption
GS: gas chromatography
GC-MS: gas chromatography-mass spectrometry
IC, IPC: ion chromatography, ion-pair chromatography
ICPAES: inductively coupled plasma atomic emission spectrometry
ICPMS: inductively coupled plasma mass spectrometry
ICP-Q-MS: inductively coupled plasma - quadrupole - mass spectrometry
ICP-SF-MS: inductively coupled plasma - sector field - mass spectrometry
HPLC: high pressure liquid chromatography
(LA-)ICPMS: (laser ablation -) inductively coupled plasma mass spectrometry
(FP)MD: (first principles) modecular dynamics
MS: mass spectrometry
(N)IR: (near) infrared (spectroscopy)
NMR: nuclear magnetic resonance
NRIXS: nonresonant inelastic X-ray scattering
PIXE: proton induced X-ray emission
SEM: scanning electron microscopy
SXRF: synchrotron X-ray fluorescence (spectroscopy)
TOF-MS: time of flight mass spectrometry
TQ-MS: triple quadrupole (or alternatively QQQ or QqQ) mass spectrometer
UV-Vis: ultraviolet-visible (spectroscopy)
XAS: X-ray absorption spectroscopy
XANES and EXAFS: X-ray absorption near edge structure and extended X-ray absorption fine structure
XES: X-ray emission spectroscopy
XPS: X-ray photoelectron spectroscopy
XRF: X-ray fluorescence (spectroscopy)

*Models, databases, software acronyms*

AD: Akinfiev Diamond (model)
CHESS: Chemical Equilibrium of Species and Surfaces, https://chess.geosciences.mines-paristech.fr/ (software)
DFT: density functional theory
EQ3/6: https://github.com/llnl/eq3_6 (software)
FDMNES: Finite Difference Method near Edge Structure, https://fdmnes.neel.cnrs.fr/ (software)
GEMS: Gibbs Energy Minimization Software, http://gems.web.psi.ch/ (software)
GW: Geochemists Workbench, https://www.gwb.com/index.php (software)
HKF: Helgeson Kirkham Flowers (model)
HCh: HydroChemistry (software)
LLNL: Lawrence Livermore National Laboratory (database)
MINTEQ: https://vminteq.lwr.kth.se/ (software)
(FP)MD: (first principles) molecular dynamics
PHREEQC: https://www.usgs.gov/software/phreeqc-version-3 (software)
RB: Ryzhenko Bryzgalin (model)
SILLS: Signal Integration for Laboratory Laser Systems (software)
SUPCRT(92): http://geopig3.la.asu.edu:8080/GEOPIG/index.html (database)
SUPCRTBL: https://models.earth.indiana.edu/applications_index.php (database)
THERMODEM: https://thermoddem.brgm.fr/ (database)

*Mineral names and their generic chemical formulas*

alunite, $KAl_3(SO_4)_2(OH)_6$
andalusite, $Al_2SiO_5$
anhydrite and gypsum, $CaSO_4$, $CaSO_4 \times 2H_2O$
apatite, $Ca_5(PO_4)_3(F,Cl,OH,S)$
argentite, $Ag_2S$
arsenopyrite, FeAsS
bastnäsite, $(La,Ce,Y)(CO_3)F$
barite, $BaSO_4$
biotite, $K(Mg,Fe)_3Si_3AlO_{10}(OH,F)_2$
bismuthinite, $Bi_2S_3$
boehmite, AlO(OH)
bornite, $Cu_5FeS_4$
braggite, (Pt,Pd,Ni)S
calcite, $CaCO_3$
chalcocite, $Cu_2S$
chalcopyrite, $CuFeS_2$
cooperite, PtS
corundum, $Al_2O_3$
covellite, CuS
digenite, $Cu_9S_5$
enargite, $Cu_3AsS_4$
epidote, $Ca_2FeAl_2Si_3O_{12}(OH)$
fayalite, $Fe_2(SiO_4)$
fluocerite, $(Ce,La)F_3$
fluorite, $CaF_2$
galena, PbS
halite, NaCl
hematite, $Fe_2O_3$
hornblende, $(Ca,Na,K)_2(Mg,Fe,Al)_5[Si_6(Al,Si)_2O_{22}](OH,F)_2$
löllengite, $FeAs_2$
magnetite, $Fe_3O_4$
microcline, $KAlSi_3O_8$
molybdenite, $MoS_2$
monazite, $(Ce,La,Nd,Th)PO_4$
muscovite, $KAl_3Si_3O_{10}(OH)_2$
orpiment, $As_2S_3$
prehnite, $Ca_2Al_2Si_3O_{10}(OH)_2$
pyrite and arsenian pyrite, $FeS_2$, $Fe(S,As)_2$
pyrrhotite, $Fe_{1-x}S$
quartz, $SiO_2$
realgar, AsS
rhenite, $ReS_2$
sanidine, $KAlSi_3O_8$
scapolite, $(Na,Ca)_4[(Al,Si)_{12}O_{24}](Cl,CO_3,S)$
sodalite, ultramarine, $(Na,Ca)_8[Al_6Si_6O_{24}](Cl,S,SO_4)_{1-2}$
sphalerite, ZnS
sperrylite, $PtAs_2$
stibnite, $Sb_2S_3$
sulfosalts, $(Ag,Pb,Cu,Fe)_x(As,Sb)_yS_z$

## Abstract:

This chapter overviews sulfur chemical speciation and behavior in different types of hydrothermal fluids across the lithosphere, spanning from active geothermal systems to basinal brines, to metamorphic and magmatic-hydrothermal fluids. The information on sulfur in these fluids stems from a wide range of recent analytical, experimental and computational methods whose advantages and limitations are discussed. A special emphasis is given to in situ approaches that have enabled unprecedented insights into the fascinating sulfur chemistry in hydrothermal fluids. These insights motivated a critical review of the chemical speciation of key base, precious, and critical metals whose behavior is intimately linked to sulfur. The results allow the role played by the different sulfur ligands on hydrothermal transport of metals and ore deposit formation to be evaluated. An outline of major challenges and emerging perspectives for sulfur-related research in geological fluids ends up this chapter.

# 4.1 Introduction

Sulfur, sulfur, sulfur… There is unlikely another volatile element that attracts so much attention and engenders so much debate among geochemists, volcanologists, magmatic petrologists, ore deposit geologists, and planetary scientists. This is because sulfur is a key element in the processes that govern the formation and distribution of ore deposits, magma evolution and degassing, volcanic hazard and isotope fractionation in mineral-melt-fluid systems. For technological applications, sulfur is an important player in chemical catalysis, nanotechnology, and materials science. Critical to all these aspects is understanding the chemical speciation, transport and partitioning of sulfur in aqueous fluids typical of those natural and industrial systems, across a wide range of temperatures ($T$), pressures ($P$) and physical-chemical environments. Nowadays, we know fairly well how to analyze total sulfur concentrations, chemical forms and isotopes in materials at ambient conditions and natural fluid, gas, or mineral samples brought to the Earth surface and accessible to collection and observation. All these samples are final products of the fluids (and silicate melts) operating at high temperature and pressure in the Earth's interior or in technological processes requiring elevated temperatures. Therefore, the fundamental question that yet remains unanswered is the following. Do all these samples and data collected therefrom correctly reflect the true sulfur speciation and partitioning in fluid(-melt-mineral) systems at elevated temperatures and pressures? The answer to this key question requires a combination of in situ experimental and analytical approaches coupled with physical-chemical modeling. Over the last 20 years, these approaches have seen significant progress that led to paradigm changes about sulfur chemical speciation in deep geological fluids and its role in metal transport and ore deposit formation. These changes will be particularly emphasized here.

In this chapter, we overview sulfur concentrations, speciation and behavior in different types of hydrothermal fluids. We obtain this information from a wide range of analytical, experimental and computational methods developed for sulfur, whose advantages and limitations will be presented. We further discuss available thermodynamic data and their reliability for predicting sulfur speciation and solubility of sulfur-bearing minerals across a wide range of $T$-$P$ parameters and fluid compositions. We overview the chemical speciation of selected key metals and the role played by the different sulfur ligands on metal transport and ore mineral precipitation. We discuss the major controls on sulfur concentration and behavior in different types of hydrothermal fluids spanning from active geothermal sources to basinal brines, to metamorphic and magmatic-hydrothermal fluids. We conclude by outlining major near-future challenges and emerging perspectives for sulfur-related research in the vast domains of hydrothermal fluids both operating in nature and used in chemistry and materials applications.

# 4.2 Analytical and spectroscopic methods for sulfur in hydrothermal fluids

### *4.2.1 Sampling techniques and wet-chemistry analyses of geothermal water*

Sulfur is among the major components in hydrothermal waters and is present both in the liquid phase like hot springs and the gas phase like in fumarole vapor. In natural liquid hydrothermal fluids, sulfate ($S^{VI}$, mainly $SO_4^{2-}$) and sulfide ($S^{-II}$, including $H_2S$ and $HS^-$, and sometimes $S^{2-}$) are commonly referred as the dominant sulfur oxidation states, but other sulfur species with intermediate oxidation states may also be present and in some cases predominate. The sulfur species of intermediate oxidation states observed in geothermal water include sulfite ($SO_3^{2-}$), thiosulfate ($S_2O_3^{2-}$), polythionates ($S_nO_6^{2-}$), polysulfides ($S_n^{2-}$), dissolved elemental sulfur ($S^0$), and aqueous metal sulfide complexes of Fe, As, Mo, Sb and W (Luther et al. 2001; Keller et al. 2014; Planer-Friedrich et al. 2007, 2020). Detailed studies on sulfur speciation in hydrothermal waters are few and typically only total sulfur or sulfate and sulfide concentrations are analyzed (e.g., Takano 1987; Xu et al. 1998, 2000; Druschel et al. 2003; Nordstrom et al. 2009; Kaasalainen and Stefánsson 2011b).

Samples of liquid geothermal water are commonly collected from hot springs and wells (Fig. 4.1). For dissolved sulfate determination, the water samples are filtered on-site through 0.45 µm (or smaller) pore size without further treatment. Alternatively, dissolved sulfide is precipitated when in solution by addition of divalent metal cation (i.e. $Zn^{2+}$ and $Cd^{2+}$) containing solutions to precipitate metal sulfide and prevent interference of dissolved sulfide upon sample storage. In the laboratory, the concentration of dissolved sulfate is analyzed, most commonly as $SO_4^{2-}$ applying ion chromatography (IC) but alternatively as total sulfur using inductively coupled plasma atomic emission spectroscopy (ICPAES) (Kaasalainen and Stefánsson 2011a). Alternatively, dissolved sulfate is determined by titration with $Ba^{2+}$ to form $BaSO_{4(s)}$ precipitate or potentiometric titrations using various types of ion selective electrodes (i.e. Ca or Pb) (APHA 2018a).

Total dissolved sulfide is commonly determined on-site on unfiltered samples by precipitation titration with Hg acetate (Arnórsson et al. 2006; Kaasalainen and Stefánsson 2011a), colorimetrically after formation of methylene blue complex (APHA 2018b) or using direct UV–Vis spectrophotometry (Giggenbach 1974a,b). Other forms of dissolved sulfur species that have been observed in hydrothermal liquid water include for example thiosulfate, polythionates and polysulfides. All these intermediate oxidation states of sulfur are unstable upon sample storage and have, therefore, been analyzed on-site when possible (Fig. 4.1a). Thiosulfate is typically analyzed using IC equipped with conductivity detector together with dissolved sulfate (Kaasalainen and Stefánsson 2011a) or collected onto ion exchange resin for sample preservation followed by elution in the laboratory and IC analysis (Druschel et al. 2003). Polythionates are also analyzed using IC as well as high pressure liquid chromatography (HPLC) and a spectrophotometric detector (Takano 1987; Kamyshny et al. 2006).

Cyclic voltammetry has also been applied successfully to determine concentrations of various dissolved sulfur species in hydrothermal waters and anoxic waters (e.g., Luther et al. 2001). Voltammetric signals are produced when redox active dissolved or nanoparticulate species interacts with the surface of a working electrode, resulting in electron flow that can be determined as a current that is proportional to concentration (Taillefert and Rozan 2002). A setup of three-electrodes is commonly applied for determination of dissolved sulfur species and consists of a silver/silver chloride reference electrode, a platinum counter electrode, and an Au-amalgam working electrode (Brendel and Luther 1995; Luther et al. 2001). Aqueous and nanoparticulate sulfur species that can be identified by cyclic voltammetry include dissolved sulfide, sulfur, polysulfide, thiosulfate and aqueous iron sulfide (e.g., Luther et al. 2001; Boyd and Druschel 2013).

Sulfur may also be present in hydrothermal vapor, for example in fumarole discharges and two-phase well discharges (liquid and vapor discharges). Hydrogen sulfide gas is the most common gaseous species, but sulfur dioxide ($SO_2$) is sometimes also detected at high temperatures (>350–400°C). Fumarole samples are collected by placing a plastic funnel over or inserting a metal tube into the vapor outlet of a fumarole emission and connecting a gas bottle using a silicone or glass tube (Fig. 4.1b). For two-phase well discharges, the liquid and vapor phase are separated at the wellhead using a Webre separator and vapor samples collected into gas bottles like for fumaroles (Fig. 4.1c). The gas bottle, most commonly of 50–250 mL total volume, is evacuated prior to sampling and contains ~10 mL of ~50% KOH or ~20% NaOH solution per 100 mL of total volume. The non-condensable gases, including hydrogen sulfide gas, are analyzed using Hg-acetate precipitation titration previously described (Arnórsson et al. 2006) or as $SO_4$ using IC after oxidation by hydrogen peroxide. Sometimes, $SO_2$ is also analyzed after precipitation of sulfide using divalent metal solution ($Cd^{2+}$, $Zn^{2+}$) followed by filtration, oxidation and analysis of dissolved sulfate using IC.

### *4.2.2 Bulk crush-leach analyses of fluid inclusions for sulfur*

Fluid inclusions are tiny volumes of single- or multiphase fluid (usually 100s to 1000s μm$^3$; Bodnar 1983) trapped in both gangue and ore minerals during their growth and posterior deformation. These microsamples provide the only direct witness of the composition of paleo hydrothermal fluids that has been recognized since 200 years (e.g., Davy 1822; Sorby 1869) and more systematically explored since 1950s (e.g., Roedder 1984; Diamond 2003; references therein). Historically, chemical analysis of the solute content (dissolved cations, anions, and volatiles, including sulfur) of fluid inclusions, termed crush-leach method, involved bulk extraction of numerous fluid inclusions from the host crystals, followed by analysis of leachates by various methods (e.g., conventional solution chemistry, spectrophotometry, ion and gas chromatography, potentiometry or mass-spectrometry). Both volatile ($H_2S$ and $SO_2$, gas-bubble hosted) and anionic (sulfate, liquid/brine-phase hosted) sulfur forms have been analyzed. For example, reduced sulfur, including $H_2S$ and COS, was analyzed by decrepitating fluid inclusions from vein quartz samples in an $N_2$ stream followed by entrapment in NaOH and fluorometric analysis, at ng S levels (Bottrell and Miller 1989). Potentiometry with $Ag/Ag_2S$ electrode was used to analyze sulfide sulfur from quartz- and fluorite-hosted fluid inclusions with detection limits of $<10^{-4}$ m S (Korsakova et al. 1991). Mass spectrometry (MS) was used to analyze $H_2S$ and $SO_2$, along with other common volatiles ($CO_2$, $N_2$, $H_2$, $CH_4$) after crushing or thermal decrepitation of fluid inclusions (e.g., Norman and Sawkins 1987). Gas-chromatography mass-spectrometry (GC-MS) analyses of volatile components of bulk fluid inclusions comprise ~30% of the total analyses published so far according to the statistical compilation of Naumov et al. (2009). Crush-leach dispositives coupled with ion chromatography were used for analyses of sulfate ion and other anions (halogens) with detection limits of less-than-ppm level in brine-type fluid inclusions in quartz, emerald, calcite, and fluorite from metamorphic and sediment-hosted deposits (e.g., Banks et al. 2000; Giuliani et al. 2003; Sośnicka et al. 2023). A remarkable suite of sulfur inorganic and organic species ($H_2S$, COS, $CS_2$, organic polysulfanes, and thiols) has been revealed by thermal decrepitation/desorption GC-MS analyses at 250 °C on fluid inclusions hosted by 3.5-billion-year-old barites from Pilbara Craton, Australia (Mißbach et al. 2021). Although in general crush-leach techniques for sulfur offer quite low detection limits for properly selected and processed samples, they are commonly prone to contamination by impurities and are limited by the availability of sufficient fluid-inclusion mass. The major general limitation of bulk methods is that such averaged data from potentially different inclusion generations in one sample provide little information about the time-space-composition evolution of fluid-mineral systems. Nowadays, these bulk methods stand as rather subordinate for sulfur (as well as most trace metals) as compared to in situ analyses using LA-ICPMS and Raman spectroscopy of individual fluid inclusions as will be shown in the following sections.

### *4.2.3 LA-ICPMS of individual fluid inclusions in hydrothermal minerals*

Individual fluid inclusions can be trapped during or after crystal growth of the host mineral and usually classified into primary, secondary and pseudo-secondary inclusions according to their origin (Roedder 1984; Goldstein 2003). Their study by microthermometry techniques can provide direct information about physical parameters and composition of the mineralizing hydrothermal fluids and their evolution in space and time in terms of temperature, pressure, fluid salinity and major volatile content. Goldstein and Reynolds (1994) introduced the use of fluid inclusion assemblages (FIAs), defined as groups of cogenetic fluid inclusions occupying particular petrographic features, such as zones of crystal growth, healed fractures, sector zoning. These features can, in most cases, be unambiguously recognized using classical petrographic or other imaging techniques (e.g., scanning electron microscopy-cathodoluminescence, SEM-CL, or optical CL; Van den Kerkhof and Hein 2001). In fluid inclusion research, the concept of FIAs is crucial to correct inclusion analysis and interpretation of the results. This concept has been largely applied to study ore-forming processes and its caveats have been recognized and thoroughly assessed. For example, Wilkinson (2001) presented a summary of fluid inclusion data from a variety of hydrothermal environments, discussing the uncertainties associated with the determination of timing relationships between fluid inclusion entrapment and ore formation, post-entrapment modifications of inclusions, diffusion of components and isotopic exchange with the host minerals. Most of the fluid inclusion studies relevant to hydrothermal ore deposits are conducted on optically transparent gangue minerals, based on the assumption that they are syngenetic with the intergrown ore minerals. However, there is growing evidence that ore mineralization in hydrothermal systems may form from episodic events of multiple fluid sources, where precipitation of ore and gangue minerals can be decoupled in time (Wilkinson et al. 2009). Near-infrared (NIR) transmitted-light microscopy has been successfully used to study fluid inclusions also in opaque ore minerals, thus providing evidence of distinct fluid inclusion compositions recorded in gangue and ore minerals which display apparent "equilibrium" textural relationships (e.g., Campbell and Robinson-Cook 1987; Campbell and Panter 1990; Kouzmanov et al. 2002, 2010).

An important caveat of microthermometry is about the determination of true entrapment temperature and pressure. Trapping pressure for fluid inclusions can only be estimated if the exact temperature of entrapment is known, or if fluid inclusions are trapped during phase separation. Fluid inclusions that have been entrapped as a single-phase fluid, can provide information, during microthermometry measurements, on the minimum temperature and pressure of entrapment only (Roedder 1984). This is because most inclusions have trapped fluids at pressures higher than their vapor pressures. Thus, to establish the true temperature or pressure of entrapment, in many cases, an important "pressure correction" is needed. This correction can be determined either using independent mineral geothermometers or from independent evidence of the depth of the system at the time of trapping. Then the pressure may be determined from *P-V-T-X* data on the inclusion fluid, which is generally approximated by a water-salt±gas ($CO_2$, $CH_4$) system (Roedder and Bodnar 1980). Such a correction may result in differences as high as several hundred degrees between temperature of homogenization and trapping temperature (see section 4.5.3 on metamorphic fluids).

Fluid inclusion microthermometry alone provides only limited information about the composition of studied inclusions and exclusively in terms of major components of the entrapped fluid. In the 1990s, two microanalytical techniques were successfully applied to quantify, for the first time, elemental concentrations in individual fluid inclusions, mainly from magmatic-hydrothermal deposits, laser ablation - inductively coupled plasma mass-spectrometry (LA-ICPMS) and proton-induced X-ray emission spectroscopy (PIXE) (e.g., Ryan et al. 1991; Heinrich et al. 1992; Rankin et al. 1992; Wilkinson et al. 1994; Audétat et al. 1998; Günther et al. 1998). Despite a destructive character of the analysis compared to PIXE (see section 4.2.5), LA-ICPMS quickly became the most powerful and efficient multi-element technique for the analysis of concentrations of a large range of major to ultra-trace elements, spanning from Li to U, in individual fluid inclusions being single-phase (liquid or gas) or polyphase (liquid+gas±solids) mixtures at the time of measurement. The unique features of the method consist of the simultaneous acquisition of the signal, via optically controlled ablation of the entire fluid inclusion volume, and accurate quantification of the elemental concentrations, using a combination of mostly matrix-independent external calibration procedures along with internal standards usually using the Na concentration determined from thermometric measurements (Pettke et al. 2012).

Early prototype analytical setups used for fluid inclusion analysis as well as their modern analogues have been described in detail by Günther et al. (1997), Günther et al. (1998), Heinrich et al. (2003), Pettke (2008) and Pettke et al. (2012). Both quadrupole and single collector sector field mass spectrometers are commonly used for element concentration analysis, allowing very fast mass jumping between the different isotopes. The high speed is an important advantage for the analysis of relatively short signals arisen from fluid inclusions, thus enabling limits of detection as low as 0.01 ppm for the heavier elements (e.g., Sr, Ba, Rb, Cs, Au, Pb). Recently, promising results using time-of-flight mass spectrometers (TOFMS) for analysis of fluid inclusions were also reported (Harlaux et al. 2015). The laser system used to drill out individual inclusions is an essential component of the setup. The use of a homogenized ArF excimer (193 nm wavelength) laser ablation systems is best suited for the controlled ablation of inclusions, thereby allowing proper ablation of quartz (at energy density of ~25 $J \cdot cm^{-2}$), which is the most common host for fluid inclusions in hydrothermal deposits. The use of Iris diaphragms, allowing either progressive or stepwise opening of the laser beam diameter (Günther et al. 1998; Guillong and Heinrich 2007), reduces unwanted common effects during ablation such as cracking of the quartz host near the crater rim that can result in explosion of the inclusion being analyzed.

As summarized by Pettke et al. (2012), fluid inclusion analysis by LA-ICPMS consists of the following steps. 1) The fluid inclusion sample (usually small chip of few mm in diameter) is placed in a small ablation cell. 2) The target fluid inclusion is selected using transmitted-light optical microscope. 3) The fluid inclusion volume is ablated using an excimer laser system with full optical control on the position of the pit during ablation. 4) The aerosol is transported out of the ablation cell using a He carrier gas mixed with Ar gas (and sometimes $H_2$) and then introduced into the ICPMS. 5) Aerosol particles are atomized and then ionized and the ions are extracted from the plasma and filtered based on their mass-to-charge ratio and measured by the mass spectrometer. 6) Transient signals of several seconds to several 10s seconds duration are produced (Fig. 4.2). 7) Quantification of the measurements is performed by analyzing external standard reference materials (SRM) bracketing the fluid inclusion analyses; additionally, the concentration of an internal standard element has to be independently known (usually Na from microthermometry determination of fluid salinity prior to LA-ICP-MS analysis). 8) Calculation of absolute element concentrations in individual fluid inclusions using available specially designed software (e.g., SILLS; Guillong et al. 2008b) or in-house developed spreadsheets and Matlab routines in various laboratories.

Easy access to several laboratories with good in-house knowledge about LA-ICPMS analysis of fluid inclusions has resulted in a large amount of data on metal concentrations in mineralizing fluids in various hydrothermal environments to be generated during the last two decades. However, sulfur concentrations were not routinely measured, mainly due to polyatomic interferences on isotope $^{32}S^+$ (95.02% abundance) from $^{16}O^{16}O^+$, and on isotope $^{34}S^+$ (4.21% abundance) from $^{16}O^{18}O^+$, originating from the ambient air or from the host mineral. Guillong et al. (2008a) reported the first quantitative analyses of sulfur in individual fluid inclusions using laser ablation inductively coupled plasma quadrupole mass spectrometry (ICP-Q-MS) and ICP sector field mass spectrometry (ICP-SF-MS) allowing the sulfur signal to be resolved from those polyatomic interferences. However, the background level of sulfur remains relatively high (around $10^5$ cps; Fig. 2a,b) in most LA-ICPMS laboratories, due essentially to contamination from the ambient air, ablation-cell environment, and previous analyses of different samples. The accuracy on developed in-house reference materials scapolite-17 and NIST glasses, reported by Guillong et al. (2008a), was within 5% for ICP-SF-MS and 7% for ICP-Q-MS. Precision (~35% RSD, relative standard deviation) and detection limits reached (~100 ppm) are limited by sometimes enigmatic and poorly reproducible sources of sulfur contamination.

Figure 4.2b shows typical transient ablation signals from four main types of natural fluid inclusions. In each case, the sulfur signal shows a distinct maximum corresponding to the fluid inclusion opening by the laser ablation system. Due to the intrinsically high background for sulfur, its signal is shorter in time and the signal-to-noise ratio is much lower than the signal of the other major and trace metals and metalloids in the inclusion fluids. Often, the maxima in the sulfur signal correspond to ablation of solids (e.g., sulfides and sulfates; Fig. 4.2a), present as daughter phases inside the inclusion vacuoles. Such solids either formed on cooling due to decrease of sulfur solubility with lowering the temperature or were accidentally trapped during host crystal growth and fluid inclusion formation.

As summarized in Table 4.2, since the work of Guillong et al. (2008a) a number of studies have been published, reporting sulfur concentrations in various fluid inclusion types from different hydrothermal environments. Total sulfur concentrations in natural fluid inclusions are highly variable, ranging from <0.001 wt% up to 12 wt% that may be reached in some hypersaline liquid inclusions from magmatic-hydrothermal systems and basinal brines. The sulfur contents in different types of ore forming fluids will be discussed in section 4.5.

### *4.2.4 Raman spectroscopy*

Raman spectroscopy holds the first place in studies of sulfur speciation in fluid inclusions from a large variety of magmatic-hydrothermal-metamorphic settings as well as in model hydrothermal fluids in laboratory experiments. Sulfur compounds have been routinely analyzed using this technique by chemists and materials scientists since 1950s (e.g., see Long 1977; Ferraro et al. 2002 for fundamentals; Singh 2002 for historical background; Lin et al. 2014 for an overview of mineral physics applications). The Raman spectroscopy method as well as its various applications in Geosciences for studying volatile elements in fluids, melts and gas bubbles (H, B, C, O, and S) have been a subject of numerous reviews over the past 20 years (e.g., Burke et al. 2001; Dubessy et al. 2012; Frezzotti et al. 2012; Neuville et al. 2014; Pasteris and Bessac 2020, to name a few); here we focus exclusively on its practical outcomes for sulfur in the fluid phase. Table 4.3 lists the major Raman characteristic wavenumbers, types of vibration and typical cross-sections (relative to sulfate) for inorganic sulfur species that can typically be analyzed in hydrothermal fluids.

*Advantages and limitations*

The main advantages of Raman spectroscopy for analyzing sulfur in dissolved forms are the following. *First*, the characteristic Raman spectral frequencies (i.e., wavenumbers) for the great majority of sulfur anions, molecules and complexes in the gas, aqueous or solid phase are well known. The three major types of bonds S-S, S-O/C and S-H display distinct wavenumber ranges for their most intense symmetric stretch vibrations, typically 350–600, 700–1400, and 2300–2600 $cm^{-1}$, respectively, making it possible to identify the different sulfur species in a single spectrum (Fig. 4.3). *Second*, these frequencies evolve little and predictably with temperature and pressure changes making it possible to perform in situ studies of sulfur in the fluid phase at elevated *T-P*. With proper corrections for Raman scattering intensity and background and with the use of standards or relative cross-sections, quantification of sulfur species in the fluid phase becomes possible both in natural fluid inclusions and experimental laboratory solutions (e.g., Pokrovski and Dubessy 2015; Barré et al. 2017; Schmidt and Seward 2017). *Third*, the resonance Raman phenomenon typical for some chromophore S species such as radical sulfur ions (e.g., Clark and Franks 1975; Clark and Cobbold 1978; Pokrovski and Dubessy 2015), which yields selective enhancement of their Raman spectral signals at certain laser wavelengths, allows detection of such species at concentrations 10 to 1000 times lower than for traditional non-resonating sulfur forms (e.g., sulfate or sulfide). *Fourth*, there exist advanced technologies of high *T-P* optical cells enabling studies of sulfur speciation in model hydrothermal fluids in the laboratory, including large-volume autoclaves with optically transparent windows (e.g., Rudolph 1996; Louvel et al. 2015; Dietrich et al. 2018; Testemale et al. 2024), fused silica capillary cells (e.g., Chou et al. 2008; Pokrovski and Dubessy 2015), synthetic fluid inclusions (e.g., Ni and Keppler 2012; Jacquemet et al. 2014), and diamond-anvil cells (e.g., Schmidt et al. 2009; Pokrovski and Dubrovinsky 2011; Schmidt and Seward 2017; Colin et al. 2020). These techniques have enabled over the past 20 years a significant advance in our understanding of sulfur high *T-P* speciation, in particular by constraining the thermodynamic stability of certain aqueous sulfur species that will be discussed further in section 4.3.4. With the development of μm-size laser beams by coupling a microscope to a Raman spectrometer, micro-Raman spectroscopy has become the method of choice since the 1980s for directly probing sulfur species in natural individual gaseous and fluid inclusions trapped in optically transparent minerals under hydrothermal conditions (e.g., see Roedder 1984, 1990; Burke et al. 2001; Frezzotti et al. 2012, for historical background and an overview). Finally, recent progress in fiber optics used to guide laser beam and collect the scattered signal allowed applications of in situ Raman probes to direct quantitative measurement of sulfur species, such as $H_2S$ and $HS^-$, in deep-ocean sediment pore waters and sea-floor hydrothermal springs (e.g., Peltzer et al. 2016; Li et al. 2023).

There are four general limitations of Raman spectroscopy for studies of dissolved species in hydrothermal fluids. *First*, intrinsic weakness of the Raman inelastic scattering signal results in relatively high detection limits for sulfur species both in fluid inclusions and high *T-P* cells, typically of 100s ppm S (except for the resonant species such as radical ions $S_3^{\bullet-}$, if combined with the right choice of the laser wavelength; e.g., Pokrovski and Dubessy 2015), as compared to much more sensitive wet chemistry analyses of geothermal fluids offering ppm-level detection limits (see section 4.2.1). *Second*, quantifying S species concentrations in dense water-rich media requires corrections for background signal, Raman cross-sections, and spectrometer response (Schmidt and Seward 2017 and references therein). This limitation is more severe for natural fluid inclusions for which systematic calibrations procedures are hard to set up. As a result, quantitative estimates of absolute sulfur species concentrations or molar ratios in natural fluid inclusions have been limited mostly to gas-phase dominated and water-poor inclusions (e.g., $CO_2$-rich; Burke et al. 2001). *Third*, Raman spectroscopy is an excellent tool for in-situ distinguishing protonated *vs* deprotonated sulfur anions (e.g., $HSO_4^-$ *vs* $SO_4^{2-}$, $H_2S$ *vs* $HS^-$), but it is not for detecting alkali ion pairing that becomes increasingly important in magmatic-hydrothermal fluids (e.g., $SO_4^{2-}$ vs $NaSO_4^-$, $HS^-$ *vs* NaHS). The Raman peak position and signal intensity of sulfur anions are generally weakly affected by complexing with $Na^+$, providing detection of the overall sulfur form (e.g., sulfate). *Forth*, the main strenching bands of some sulfur species exhibit an assymetry at the lower-wavenumber peak side that is due to both the development of hot bands with increasing temperature and hydrogen-bond dymanics (proton hopping) in aqueous solution. This is, for the example, the case for the the main stretch band of gaseous and aqueous $H_2S$ (~2610 and ~2590 $cm^{-1}$, respectively; Jiang et al. 2017; Pokrovski and Dubessy 2015; Table 4.3), and of the $HSO_4^-$ ion (Dawson et al. 1986; Schmidt and Jahn 2024). Such an asymmetry may easily be mistaken for a band of another species (e.g., $HS^-$ at 2570 $cm^{-1}$; Fansang and Zajacz 2025). *Finally*, a specific limitation, in particular for natural fluid inclusions, is laser-induced fluorescence common in the presence of organic impurities (e.g., Burke 2001), which limits data availability from fluid inclusions of sedimentary-basin fluids. This limitation may eventually be avoided by using a different laser wavelength or emerging more advanced Raman acquisition techniques (e.g., coherent anti-Stokes Raman scattering, CARS, e.g., Burruss et al. 2012; Pasteris and Bessac 2020), which have not yet been routinely applied for sulfur. In the following section, we discuss the key data on sulfur speciation obtained by Raman spectroscopy from natural fluid inclusions.

*Sulfur speciation from analysis of natural fluid inclusions*

Table 4.4 summarizes sulfur species reported so far in natural fluid inclusions using micro-Raman spectroscopy, together with selected references. Hydrogen sulfide ($H_2S$, and in some cases $HS^-$) has been the most commonly detected S species across a large variety of fluid inclusions types (gaseous, aqueous or brine) trapped mostly in quartz and more rarely in topaz, calcite, fluorite, garnet or pyroxene, from different geological environments spanning from epithermal and sedimentary basin to shallow mantle settings. Cooling down $H_2S$-rich gas-like fluid inclusions ($N_2$-$CO_2$-$CH_4$) in microthermometry analyses may produce a range of $H_2S$ clathrates and solid-phase polymorphs ($\alpha$-, $\beta$-, $\gamma$-$H_2S$; e.g., Sośnicka and Lüders 2021), which may be relevant to sulfur chemistry on cold extraterrestrial planetary systems such as Uranus and Neptune that are treated in Chapter 20 (Lodders and Figley 2025, This Book). The second, most commonly detected sulfur form is sulfate ($SO_4^{2-}$, $HSO_4^-$) in fluid inclusions from sedimentary and metamorphic settings, evaporite formations and volcanic/magmatic rocks. The unique ability of Raman to distinguish between $H_2S$ and $HS^-$ and between $SO_4^{2-}$ and $HSO_4^-$ that have distinct Raman band wavenumber positions (Table 4.3) makes it possible to estimate the pH value in moderate-temperature fluids (<350°C) by coupling the measured species ratios with the known deprotonation constants in these pairs of species (e.g., Dubessy et al. 1992; Boiron et al. 1999; see also section 4.3.4). In addition to dissolved sulfate and sulfide species, both corresponding types of minerals (e.g., anhydrite, gypsum, barite, pyrite, pyrrhotite, chalcopyrite, and some others, e.g., Frezzotti et al. 2012) have been frequently identified in fluid inclusions from sedimentary to magmatic settings. In particular, such mineral phases can be mapped in inclusions by combining Raman and X-ray fluorescence micro-tomography (e.g., Schialvi et al. 2020). Most often, these minerals are products of fluid cooling and subsequent solid phase precipitation after the inclusion entrapment at high temperatures. In some cases, however, such as in porphyry copper systems, they may also reflect post-entrapment diffusion of copper ($Cu^+$) into quartz-hosted $H_2S$-rich inclusions resulting in formation of chalcopyrite (e.g., Lerchbaumer and Audétat 2012; Seo and Heinrich 2013), or hydrogen movement in or out the inclusion that modifies the sulfate/sulfide mineral proportions and aqueous species ratios (e.g., Mavrogenes and Bodnar 1994). Even though, for the sake of commodity, most Raman analyses on natural inclusions have been performed at ambient conditions, i.e. on final products of cooled and decompressed fluids, they do attest to the common presence, and in many cases to coexistence, of sulfate and sulfide aqueous forms in geological fluids across the lithosphere.

Intermediate-valence sulfur forms are less commonly detected in fluid inclusions. The main reason is both because most of them are usually less abundant than sulfate and sulfide, in particular at low temperatures, and because their Raman signals overlap with those of the inclusion host (quartz) and other, often more abundant species (e.g., carbonates). For example, the Raman signals of $SO_2$ and $S_8$ (Table 4.3) are often obscured by more intense Raman peaks arising from quartz (e.g., Jacquemet et al. 2014). Combined with the (partial) disproportionation of $SO_2$ in aqueous solution to sulfate and sulfide on cooling (e.g., see section 4.5 and Kouzmanov and Pokrovski 2012), $SO_2$ has only been detected in rather particular fluid inclusions such as those water-poor and $CO_2$-rich inclusions from mantle derived carbonatites (e.g., Frezzotti et al. 2002). The formation of native sulfur in detectable amounts in some inclusions from evaporites and metamorphic rocks (e.g., Giuliani et al. 2003) implies high total S concentrations in the original hydrothermal fluid.

More "exotic" species, such as polysulfanes $H_2S_n$, were detected, for the first time, in $CH_4/H_2S$-rich and $S^0$-bearing inclusions in metamorphic quartzite and ruby-bearing marble (Hurai et al. 2017; Huang et al. 2017), but were likely subjected to photolysis upon exposure to the laser beam (Hurai et al. 2017). Carbonyl sulfide, COS, was also detected in the gas phase of inclusions from similar metamorphic settings (Grishina et al. 1992; Giuliani et al. 2003; Table 4.3). Finally, aqueous polysulfide ions $S_n^{2-}$ and polymeric molecular sulfur $S_n^0$ other than $S_8$, together with the trisulfur radical ion $S_3^{\bullet-}$, have recently been detected and semi-quantified in fluid inclusions in fluorite and quartz from the Carnian evaporite formation in French Alps by Barré et al. (2017). These S-rich inclusions (of total concentration up to ~3000 ppm S) contained both $SO_4^{2-}$ and $H_2S$ together with native sulfur, $S_8$(s,l), at ambient conditions, and displayed the growing formation of the polymeric sulfur species upon heating to 300°C. Their study is the first direct finding of the trisulfur radical ion in natural hydrothermal fluids, in excellent agreement with synthetic fluid inclusion Raman spectroscopy analyses (Jacquemet et al. 2014) and more systematic measurements on analogous model S-bearing solutions in laboratory experiments and derived therefrom thermodynamic properties of $S_3^{\bullet-}$ (Pokrovski and Dubessy 2015, see section 4.3.4).

In summary, Raman spectroscopy of natural fluid inclusions reveals that even though sulfate and sulfide are the apparently dominant aqueous sulfur forms, intermediate-valence sulfur species commonly form in various types of geological fluids. Their quantification at elevated *T-P* in natural fluids would require thermodynamic and kinetic constants that are derived from laboratory experiments (both in situ and ex situ) under controlled conditions. Physical-chemical insight into sulfur aqueous speciation from such data will be provided in section 4.3.4.

### *4.2.5 X-ray absorption (XAS), emission (XES and PIXE), and fluorescence (XRF) spectroscopy*

These X-ray based spectroscopies for sulfur probe electronic transitions at S K-edge (~2470 eV), *i.e.* from the core electronic level (1s) and, more rarely, at S L-edge (~163 eV), *i.e.* from 2p levels, of the sulfur atom to different higher electronic orbital levels (3p, 3s, 4d) defined by S oxidation state and chemical bonding with other atoms. These transitions may be probed by different types of spectroscopy, for determining either the total sulfur concentration by XRF and PIXE, or analyzing S redox and chemical speciation using higher energy resolution offered by XAS and XES.

*X-ray absorption spectroscopy (XAS)*

Synchrotron-based X-ray absorption spectroscopy (XAS) has been used for analyzing the chemical speciation of sulfur and other heavier elements since the mid-1980s (Teo 1986). The XAS spectrum includes the near-edge part, typically around ±10 eV of the absorption edge, called X-ray absorption near-edge structure or XANES, and the extended X-ray absorption fine structure part or EXAFS, extending to 100s eV above the edge. For light elements like sulfur, having low-energy absorption edges, the XANES part is usually used both because it is the most sensitive to S atomic environment and is relatively easy to acquire, whereas the EXAFS part is more commonly used for studying heavier absorbing elements (e.g., Teo 1986; Stöhr 1992). The shape and energy position of XANES features represent a powerful in situ method for studying the chemical and redox state of sulfur in a variety of technological and geological materials. It has been extensively applied for sulfur speciation analysis in minerals (Fleet 2005), silicate glasses (Wilke et al. 2011; Simon and Wilke 2025), organic matter and biological samples (Jalilehvand 2006; Burton et al. 2013; Qureshi et al. 2021), and meteorites (e.g., Bose et al. 2017) to name a few. Excellent reviews discuss fundamentals and practical advantages and limitations of S K-edge and L-edge XANES spectroscopy in mineralogy, environmental and life sciences (Fleet 2005; Jalilehvand 2006). Detailed applications of this type of spectroscopy to S-bearing minerals and silicate glasses are presented in other chapters of this book (Pan and Mi 2025; Simon and Wilke 2025).

Compared to Raman spectroscopy, XANES spectroscopy is a more straightforward approach because i) the energy position (extending for over >10 eV from $S^{2-}$ to $S^{6+}$; Table 4.3) and shape of the spectra are both sensitive to sulfur redox state and chemical bonding, and ii) it enables identification and potential quantification of all different sulfur forms owing to the additive nature of the X-ray absorption signal, in contrast to other types of spectroscopy (e.g., EPR, resonance Raman, UV-Vis) that do not have enough redox resolution and/or privilege only certain species (e.g., radical polysulfide ions). Combined with accurate knowledge of the absorption energy position for different S redox states (Table 4.3) and S-bearing reference mineral/organic compounds whose XANES spectra can be accessed from databases (e.g., *https://www.esrf.fr/home/UsersAndScience/Experiments/XNP/ID21/php.html*), this method offers direct quantitative information on the relative amounts of the different S redox and chemical forms in a sample. The analysis may be complemented by rapidly developing theoretical chemistry modeling tools for simulating XANES spectra (e.g., FDMNES, Joly 2022; FEFF10, Kas et al. 2021), to help relate certain observed spectral features to a particular S bonding molecular environment in organic and mineral samples such as petroleum and ultramarines (e.g., Risberg et al. 2007; Pascal et al. 2015; Rejmak 2018).

The inconvenience of XANES is both weak penetration of the incident X-ray beam into the sample at the low energies of S absorption edges (< few µm at S K-edge, ~2.5 keV) and strong absorbance of both incident and scattered radiation by the sample environment, therefore requiring He atmosphere and/or high-vacuum environment at synchrotron beamlines and weakly absorbing sample holders made of light materials (plastic films, extremely thin-walled capillaries), in particular for studies of liquid samples.

The weak penetration of soft X-rays (at energies <2.5 keV) has been the major limitation in applying this type of spectroscopy to aqueous solutions, especially at elevated temperatures and pressures. We are only aware of studies up to 70 °C. XANES spectra of aqueous solutions of common organic and inorganic sulfur compounds have been used as spectral references for comparison purposes in many studies, but fewer studies were devoted to in-depth characterization of S redox speciation and structure. Those concerned organic aliphatic and aromatic sulfur molecules in biological solutions, petroleum, and aqueous organic matter such as humic substances at ambient conditions (e.g., Qurechi et al. 2021). Inorganic aqueous sulfur species that have been studied by XAS to a significant degree of structural detail include thiosulfate in which sulfur redox state is consistent with -1 and +5 (Vairavamurthy et al. 1993), sulfate and hydrosulfate that have XANES spectra clearly distinguishable from those of their solid-phase counterparts (Pin et al. 2013), as well as sulfite, sulfonate and $SO_2$. For example, XANES spectra of hydrosulfite aqueous solutions (Risberg et al. 2007) reveal two coordination isomers, $HSO_3^-$ and $SO_2(OH)^-$ present at 25 °C in a ~1:2 ratio. This ratio slightly increases with temperature, in quantitative agreement with recent Raman measurements (Eldridge et al. 2018) that show clearly distinguishable peaks for the two isomers (Table 4.3). High-temperature low-pressure sulfur vapors, to 300°C at <0.1 bar, were also examined by S K-edge XANES in an exploratory study using a furnace design with Be windows, to show the formation of $S_{2(g)}$ and possibly other $S_n$ species, and their oxidation to $SO_2$ in the presence of $H_2O$ with increasing temperature (Engel et al. 2014). There are a few examples demonstrating that the S K-edge XANES may, in some cases, be complementary to Raman spectroscopy, revealing structural and speciation features of sulfur in fluids inaccessible by other methods. The major limitation related to the strong attenuation of X-ray signal by the cell material may be overcome in the future by recent developments of perforated diamond anvil cells for high pressure measurements. In such cells, the incoming focused X-ray beam passes through a thin diamond disk (30 µm) affixed to a fully perforated diamond anvil and X-ray fluorescence photons from the sample are collected in back-scattering geometry through the same diamond disk (Wilhelm et al. 2016). Such design holds great promise to allow us one day to study sulfur speciation by in situ XANES spectroscopy in model hydrothermal fluids as it is almost routinely being done nowadays by Raman spectroscopy (see the section above).

An alternative approach that overcomes the low penetration capacity of soft X-rays is the use of nonresonant inelastic X-ray scattering (NRIXS) or alternatively called Raman X-ray scattering (RXS) spectroscopy. In this method, the sample is excited with a hard X-ray beam (usually ~10 keV) to obtain XANES spectra at sulfur L-edges (at 160–230 eV). The technique is increasingly used at modern synchrotrons to access absorption K-edges of light elements (e.g., O, N, P) and L- and M-edges of 3d transition elements (e.g., Sahle et al. 2015). We are aware of one recent study of concentrated sulfuric acid solutions using this method (Niskanen et al. 2015), but with improvements of detector resolution and sensitivity and synchrotron X-ray beam brightness, the method is promising for studying S aqueous speciation at elevated *T-P*.

A future approach to access sulfur speciation in natural fluid phases at elevated *T-P* could be the use of zeolite-family silicate minerals belonging to the sodalite group, $(Na,Ca)_8[Al_6Si_6O_{24}](Cl,S,SO_4)_{1-2}$, which are rather common accessory minerals in a wide range of metamorphic and magmatic-hydrothermal settings (e.g., Mi and Pan 2018). These minerals are capable of hosting, in their Al-Si network structural cages, the whole range of sulfur redox forms, from sulfide to sulfate, which may potentially reflect the sulfur redox and speciation in the original fluid. In particular, the blue varieties of sodalite – lazurites or ultramarines – are capable of stabilizing the radical chromophore ions, $S_3^{\bullet-}$ and $S_2^{\bullet-}$, which confer to these natural minerals, and their synthetic analogs widely used in industry, a specific hue ranging from violet-blue to green-blue (e.g., Reinen and Lindner 1999; Farsang et al. 2023). Red ultramarines whose color is likely due to the red chromophore neutral $S_4^0$, rather than radical $S_4^{\bullet-}$, also exist (e.g., Rejmak 2020). Sulfur species have been extensively studied in both natural and synthetic ultramarines using different analytical, spectroscopic and quantum-chemistry computational methods such as Raman, UV-visible, EPR, XPS, XAS, DFT, mostly at ambient conditions (e.g., Chivers 1974; Reinen and Lindner 1999; Fleet 2005; Fleet et al. 2005; 2010; Ledé et al. 2007; Tauson et al. 2012; Chivers and Elder 2013; Ganio et al. 2018, to name a few). More detailed account of the speciation and structural features of sulfur in these minerals and the corresponding methods is provided in Chapter 3 (Pan and Mi 2025). The exceptionally rich information on S speciation and redox state offered by S-XANES spectra of lazurites is demonstrated in Figure 4.4. Coupled with appropriate sulfur reference compounds and theoretical calculations, the whole range of S oxidation states may be analyzed, including both radical ions that cannot otherwise be stabilized in quenched aqueous solution. Therefore, lazurites, together with other accessory non-sulfide minerals such as scapolite, $Na_4[Al_3S_9O_{24}](Cl,CO_3,S)$-$Ca_4[Al_6S_6O_{24}](Cl,CO_3,S)$ (Fleet et al. 2005; Qiu et al. 2021; Hamisi et al.

2023), and apatite, $Ca_5(PO_4)_3(F,Cl,OH,S)$ (Kim et al. 2017; Konecke et al. 2017; Sadove et al. 2019), which also host multiple sulfur forms and oxidation states that may reflect the redox conditions of the formation environment, emerge as promising indicators of redox conditions and S fluid-phase speciation in high *T-P* settings. However, experimental calibrations of fluid/mineral partition coefficients of each sulfur major redox form and thermodynamic and kinetics constraints on the solubility and stability of these minerals are needed to use them on a quantitative basis.

*X-ray Emission Spectroscopy (XES)*

Both S $K_{\alpha}$ and $K_{\beta}$ XES, originating from 2p-1s and 3p-1s transitions, respectively, are valuable tools, complementary to XAS, to extract information about the oxidation state and bonding structure of sulfur in various sulfur-containing compounds, including silicate glasses and minerals (e.g., Alonso-Mori et al. 2009, 2010; Wilke et al. 2011; Hettmann et al. 2012; Qureshi et al. 2021). Measurements of emitted S $K_{\alpha}$ X-rays wavelengths are performed during exposure of the solid sample to a high-energy electron beam, typically using electron probe microanalysers (EPMA). These wavelengths exhibit a systematic shift toward higher energies by about 2 eV from sulfide to sulfate. These shifts, calibrated using sulfate and sulfide standards, have been routinely used in EPMA to determine the apparent S oxidation state in glasses and minerals (e.g., see Wilke et al. 2011 for detailed examples). Aqueous and non-aqueous solutions and some organic sulfur compounds were also measured using an X-ray excitation beam instead of an electron beam to determine reduced/oxidized sulfur ratios (e.g., Holden et al. 2018), but the intrinsically low energy resolution due to relatively deep atomic electron transitions (2p), does not provide a sufficient degree of sulfur speciation details compared to XAS.

In contrast, high-resolution laboratory or synchrotron-based $K_{\beta}$ XES, probing shallower electronic levels (3p) much more sensitive to S chemical bonding, shows considerably greater chemical sensitivity, with sulfide-to-sulfate energy shifts extending over >10 eV. This large range is sometimes accompanied by spectral fine structure allowing distinction between sulfur individual chemical species, in addition to S apparent redox states (Deluigi and Riveros 2006; Kavčič et al. 2007; Alfonso-Mori et al. 2010; Qureshi et al. 2021). In general, however, S $K_{\beta}$ XES shows lesser sensitivity than S K-edge XANES, with rare exceptions such as, for example, organosulfur molecules with different types of benzoic rings (e.g., Qureshi et al 2021). Even though, in principle, XES may be made with both laboratory sources and synchrotron facilities, it requires specialized instrumentation, while XANES may be collected using fairly standard equipment available on many beamlines. At present, these limitations make the XES method generally less convenient and less sensitive than XANES for determining S speciation, but its potential contribution to S analyses in model hydrothermal fluids remains to be seen when high *T-P* optical cells, transparent enough at S K-edge X-ray energies, have been developed.

*Synchrotron X-ray Fluorescence (SXRF) and Proton Induced X-ray emission (PIXE) spectroscopies*

Fundamentally, these two types of spectroscopy probe X-rays emitted during electronic transitions to K- and L-edges of moderately heavy elements (>Na) from excited states created by either X-ray (SXRF) or proton (PIXE) beams of high energy generated by synchrotron or nuclear sources, respectively. Coupled with adequate standards and X-ray absorption matrix coefficient corrections, these methods enable quantifying bulk element concentrations without providing chemical speciation information. While both methods are quite sensitive for heavy elements having high energies of their absorption edges (e.g., Vanko et al. 2001; Ménez et al. 2002; Ryan et al. 2005; references therein), sulfur has rarely been probed by this type of spectroscopy in fluid inclusions due to both its relatively low concentrations and low K-edge energy engendering strong absorption of X-rays by the host mineral. For example, the X-ray penetration length at the S K-edge (i.e., 1/e of the transmission signal ~ 37% of total X-ray flux at the sulfur $K\alpha_1$ fluorescence line of 2.3 keV) of quartz is less than 10 μm (Vanko et al. 1993), making fluorescence detection challenging. We are only aware of a couple of reports of S qualitative detection by PIXE in magmatic brine and vapor inclusions from Mole Granite (Heinrich et al. 1992) and in metamorphic tungsten district of the French Massif Central (Gama 2000). Sulfur is very rarely mentioned in SXRF studies of host minerals and their fluid inclusions in which it is commonly below detection limit (e.g., Tang et al. 2005). With the development of the LA-ICPMS method for fluid inclusions offering greater versatility, penetration and analytical sensitivity at lower cost (see section 4.2.3), both SXRF and PIXE stand as less attractive for sulfur concentration analyses in fluid inclusions and other geological samples.

### *4.2.6 Other spectroscopic methods for sulfur speciation analysis*

*Infrared spectroscopy (IR)*

Infrared spectroscopy is complementary to Raman due to symmetry selection rules for molecules that make some chemical-bond vibrational bands active in IR, some other in Raman and some in both types of excitation. IR spectroscopy is used for remote sensing of $SO_2$ and $H_2S$ in volcanic gases (e.g., Clarisse et al. 2011), as well as in material sciences for investigation, in combination with Raman, of structural and electronic properties of sulfur species mostly in water-free solvents and mineral matrices (e.g., Sakashita et al. 2000; Song et al. 2005; Gobeltz et al. 2011). Compared to Raman spectra, IR absorption bands of the same species are generally much weaker in intensity and larger (e.g., typically with FWMH of the same band of ~100s $cm^{-1}$ for IR vs ~10 $cm^{-1}$ for Raman), rendering spectral resolution weaker. In addition, they are more affected by the spectral background and, in particular, by the presence of water that has a very strong IR absorption signal. This fundamental limitation explains the much lower amount of IR *vs* Raman studies of sulfur in aqueous solution. The strong water absorption in the IR range also poses limitations for the spectroscopic cell dimensions, with optical paths of 10s μm lengths. An advantage of IR spectroscopy (as well as UV-Vis, see below) compared to Raman is that the species concentration is linearly proportional to its absorption signal according to the Beer-Lambert law, making IR spectroscopy more quantitative than Raman in some simple aqueous systems. IR spectroscopy on aqueous solutions has been used mostly in the sulfite-bisulfite system for determining both structures and deprotonation and dimerization equilibrium constants of $S^{IV}$ species to 50 °C (e.g., Herlinger and Long 1969, Zhang and Ewing 2002; Ermatchkov et al. 2005; Risberg et al. 2007).

*UV-Visible (UV-Vis) spectroscopy*

UV-Vis spectroscopy is complementary to IR, but instead of chemical bond vibrations it probes outer-shell electronic transitions of molecules, ions and complexes as well as their interactions with the solvent. This property makes UV-Vis spectroscopy sensitive to a whole hydrated ion or complex rather than to a specific chemical bond - like Raman or IR, thereby making interpretation of UV-Vis spectra in complex systems more challenging. Like IR, UV-Vis spectroscopy obeys the Beer-Lambert law of absorption. This fundamental property makes it a quantitative method provided the identity and cross-section of the electronic transitions are known or estimated. This type of spectroscopy generally provides a more sensitive detection of aqueous species, because water is almost transparent in the UV-Vis region. This property allows quantitative application of UV-Vis spectroscopy at millimol levels of sulfur species concentrations, in contrast to Raman and IR that require far more concentrated solutions. However, the UV-Vis absorption bands are much larger than those of IR and Raman spectroscopy, and more strongly temperature-dependent. For example, the $HS^-$ ion in aqueous solution between 50 and 350 °C has an absorption band spreading over the range of 220–300 nm (Suleimenov and Seward 1997), corresponding to a FWMH value of the UV-Vis band of >6000 $cm^{-1}$, *vs* only 20–30 $cm^{-1}$ for the $HS^-$ Raman band (~2570 $cm^{-1}$, Table 4.3) in the same temperature range (Pokrovski and Dubessy 2015). Similarly, the absorption band of the blue $S_3^{\bullet-}$ ion in water or other solvents spreads over the 500–700 nm region (e.g., Steudel and Chivers 2019), which is >3000 $cm^{-1}$ in FWHM, whereas its S-S-S Raman stretch band (~530 $cm^{-1}$) has typical FWHM values of 10–50 $cm^{-1}$ (depending on the spectrometer resolution, temperature and laser wavelength; Pokrovski and Dubessy 2015; Colin et al. 2020).

The major application of UV-Vis for sulfur analysis in geosciences is ultraviolet sensing of volcanic emissions of $SO_2$ and, less frequently, $H_2S$ that both absorb below 300 nm (Oppenheimer 2010; references therein). In aqueous and non-aqueous solutions, UV-Vis spectroscopy has been used in chemistry since a longtime, in particular for colored sulfur species such as polysulfide dianions and radical sulfur ions that absorb in the visible and near-UV range (e.g., Baer and Carmack 1949; Giggenbach 1972; Chivers and Elders 2013). An interesting application of UV-Vis spectroscopic properties of sulfur is found in chemiluminescence detectors for gas and liquid chromatographic separation and analysis of a number of sulfur species in chemistry (e.g., Steudel et al. 1987; Baudouin et al. 2022). Sulfur chemiluminescence is a two-stage detection method in which the sample undergoes combustion to form sulfur monoxide (SO), which is then carried to a reaction chamber where it reacts with ozone ($O_3$) to generate sulfur dioxide ($SO_2$) and photons produced in the reaction.

The high transparency of water in the UV-Vis region allowed building, since the 1970s, of high *T-P* optical cells for studying sulfur species equilibria in model hydrothermal fluids. Those studies enabled measurements of the $H_2S$ dissociation constant to 350 °C (e.g., Tsonopoulos et al. 1976; Ellis and Giggenbach 1971; Suleimenov and Seward 1997), polysulfide ions equilibria to 240 °C (Giggenbach 1974a,b), and sulfite equilibria to 50 °C (e.g., Beyad et al. 2014; references therein). Apart from sulfur species, UV-Vis spectroscopy allowed a plethora of studies of the identity and stability of chloride and hydroxide complexes for many metals under hydrothermal conditions (e.g., As, Mo, Co, Ni, Fe, Cu, Pb). Some of those remarkable works will be reviewed in section 4.4. Continued progress in optical high *T-P* cell design and software for spectra processing makes UV-Vis spectroscopy a promising tool for experimental studies of hydrothermal fluid speciation, particularly when combined with complementary spectroscopies such as Raman (for sulfur species) and XAS (for metals).

*Nuclear magnetic resonance (NMR)*

Among the four stable sulfur isotopes (32, 33, 34, and 36), only $^{33}S$, which has a nuclear spin of 3/2, is spectroscopically active for this method. However, its nuclear spin properties coupled with the low natural abundance make this method far from ideal (Gerothanassis and Kridvin 2024; Wilke et al. 2011 for detailed discussion), even though the range of chemical shifts for the different sulfur oxidation states is very large, up to 1000 ppm. For reduced low-symmetry sulfur species the method generally provides weak and broad features, whereas for sulfates and other oxidized and more symmetrical S species the signals are somewhat better resolved (Jalilehvand 2006; references therein). Even though $^{33}S$ NMR has been applied to mineral compounds, biological samples and glasses, its ability to provide structural information is yet largely inferior to those of Raman or XAS. A comprehensive account of the method advantages and limitations and applications to silicate glasses and other geological materials with up-to-date literature can be found in Wilke et al. (2011) and Simon and Wilke (2025). We are not aware of studies of S-bearing aqueous solutions by this method. However, with the development of high-magnetic field spectrometers (>20 T) that allow increasing spectral quality both in signal-to-noise ratio and energy resolution, this method could, in the future, be used as a complementary to Raman and XAS for studying sulfur speciation in hydrothermal fluids.

*X-ray photoelectron spectroscopy (XPS)*

X-ray photoelectron spectroscopy (XPS) applied to sulfur provides information about 2p electronic levels of the element, and as such it is complementary to XANES and XES. Typically, Al Kα excitation (1487 eV) is used as the laboratory X-ray source or a monochromatic X-ray beam at < 2 keV at synchrotrons, and 2p binding energies in the range from ~160 to ~175 eV are measured for S redox states spanning from sulfide to sulfate. Because of the low X-ray energies, XPS is notoriously a surface probe and requires high-vacuum conditions (e.g., Hüfner 1995; Papp and Steinrück 2013). As a result, while XPS has been widely used in geosciences for studies of sulfur on mineral surfaces, in particular including sulfides and sulfates in weathering processes and ore processing (e.g., Chandra and Gerson 2010; Mikhlin 2020), it cannot directly be applied to aqueous solutions. As in the case of XAS, sulfur species trapped in the structural cages of lazurite minerals have been extensively studied in both natural and synthetic samples using XPS (e.g., Fleet et al. 2005; Tauson et al. 2012; Cato et al. 2018) mostly as a complement to XAS and Raman, to better constrain the relative distribution of sulfur oxidation states. Such minerals may potentially serve as proxies for fluid sulfur speciation in the future. While XPS is a relatively easily accessible laboratory technique, its strong sensitivity to the quality and state of the sample surface and corresponding sample handling requirements, combined with lower than XAS or Raman spectral resolution details as to the particular sulfur chemical species, make XPS alone potentially less attractive than XAS or Raman for studies of sulfur speciation; it should be used in combination with those techniques.

*Electron paramagnetic resonance (EPR)*

Electron Paramagnetic Resonance (EPR) or Electron Spin Resonance (ESR) spectroscopy, which probes only species with an unpaired electron, has a limited use in geosciences compared to other types of spectroscopy such as Raman, XAS or XPS. Its applications are mostly for paramagnetic transition metals in minerals and glasses (e.g., Pan and Nilges 2014) and in the petroleum domain to study organic radicals including those of sulfur (e.g., Bakr et al. 1991). In chemistry, by contrast, EPR has widely been used since the 1960s for studying organic sulfur radicals in different organic and metalorganic synthesis reactions (e.g., Sullivan 1968; Hadley and Gordy 1975). Among inorganic sulfur species, only polysulfide radical ions yield an electron resonance spectrum, which was particularly useful for measuring absolute $S_3^{\bullet-}$ concentrations and dynamic behavior in ultramarine pigments (e.g., see Chivers and Elders 2013; Steudel and Chivers 2019 for overview; and Gobeltz et al. 1998; Arieli et al. 2004; Goslar et al. 2009 for specific case studies). In addition, $S_2^{\bullet-}$, which is poorly resolved by traditional EPR, was identified in green ultramarine pigments using a specific acquisition mode (field sweep echo detected; Raulin et al. 2011). EPR spectra of polysulfides in non-aqueous solvents and lithium-sulfur batteries have been used since longtime to detect and quantify $S_3^{\bullet-}$ (Chivers 1974). An EPR signal of the third polysulfide radical, $S_4^{\bullet-}$, in low concentrations was also reported in $Li_2S_n$ solutions and ultramarine red pigments, but spectral assignment remains elusive (Steudel and Chivers 2019). We are not aware of studies of sulfur aqueous solutions, but like for XAS, EPR holds promise for its future use for direct in situ qualification of $S_3^{\bullet-}$ in hydrothermal fluids, at least from laboratory experiments.

## 4.3 Sulfur chemical speciation, kinetics and thermodynamics in hydrothermal fluids

### *4.3.1 Sulfur aqueous speciation and redox state at low temperatures*

Of all elements of the Periodic Table, sulfur likely exhibits the greatest multitude of formal oxidation states ranging from –2 to +6 that co-exist under terrestrial conditions. This remarkable redox versatility results in a plethora of corresponding species that may form at near-ambient conditions in aqueous solution, non-aqueous solvents, glasses, and solid phases known in chemistry since a long time (e.g., Cotton et al. 1999; Steudel 2003, 2020). The most important groups of these sulfur-bearing species are shown in Figure 4.5. Interestingly, the great majority of them are either molecules (i.e. of zero electrical charge) or anions (i.e. of negative charge). This property directly reflects the non-metallic chemistry of sulfur in general, even though (poly)cationic forms such as $S_8^{2+}$, $S_4^{2+}$, $H_3SO_4^+$ are known to form in specific chemical environments such as oleum (i.e. concentrated $H_2SO_4$ solutions; Cotton et al. 1999). Another fundamental property of sulfur is a strong tendency to polymerization via formation of [–S–S–] chemical bonds. This tendency is expressed in a variety of polysulfides and zero-valent sulfur molecules, as well as in oxygen bearing species, such as thiosulfate, disulfite and polythionates in which two S atoms of different oxidation state are bound one to the other. An amazing example is thiosulfate, (H)$S^-$–$S^{5+}O_3$(H), in which reduced sulfur (called sulfane sulfur) and oxidized sulfur (called sulfonate sulfur) atoms coexist (e.g., Vairavamurthy et al. 1993).

Although sulfate (+6) and sulfide (–2) are by far the major forms in most natural environments, intermediate-valence sulfur species commonly occur in aqueous solution in a variety of marine and continental sedimentary environments where they contribute to the (bio)geochemical cycle of sulfur as addressed in detail in other chapters (Turchyn et al. 2025; Bradley et al. 2025). For example, colloidal polymeric zero-valent sulfur is the third most abundant S form after sulfate and sulfide in marine sediment pore waters (Jorgensen 2021; references therein) and is quite common, together with polysulfides, in geothermal springs and fumaroles (e.g., Migdisov and Bychkov 1998; Kamyshny et al. 2008, 2014; references therein). Sulfur oxyanions, and in particular thiosulfate and polythionates, have been commonly reported both in sedimentary-basin and low-temperature (<100°C) geothermal waters since the 1970s, with concentrations sometimes reaching 20–30% of total dissolved sulfur (e.g., Boulëgue 1978a; Takano 1987; Xu et al. 1998, 2000; Kaasalainen and Stéfansson 2011; references therein). More detailed account of the formation and transformation pathways of such S species in hydrothermal fluids will be given in Section 4.5. These and other intermediate-valence S species also play an important role as reaction intermediates in the biological sulfate-sulfide redox cycle (e.g., Milucka et al. 2012; Bradley et al. 2025). Polysulfide radical ions ($S_2^{\bullet-}$, $S_3^{\bullet-}$ and, possibly, $S_4^{\bullet-}$) are important constituents of chemical engineering products (e.g., lithium-sulfur batteries, color pigments, glasses; Chivers 1974; Chivers and Elder 2013; references therein), natural zeolite minerals (e.g., Pan and Mi 2025) and S-rich magmatic-hydrothermal fluids (see below and Pokrovski and Dubessy 2015; Colin et al. 2020). To the multitude of inorganic sulfur forms shown in Figure 4.5 is added a near infinite number of organic thiol compounds (i.e., having S–C–C– bonds, hereafter '$S–C_{org}$') both molecular and anionic, along with C–S-type radical ions, all of which are generated by microbial activity in (sub)surface aqueous environments (e.g., Amend and Shock 2001; Cody 2004) and, at higher temperatures, in thermochemical sulfate reduction processes by organic matter in sedimentary rocks (Schulte and Rogers 2004; Oduro et al. 2011). Among well-known inorganic C-bearing sulfur species, thiocyanate ($SCN^-$, e.g., Patai 1977) and carbonyl sulfide (COS, e.g., Svoronos and Bruno 2002) are worth mentioning. The former is a minor product of bacterial metabolism in aquatic environments (e.g., Jiang et al. 2017), an important analytical reagent for transition metals, and a chemical reagent used in hydrometallurgy for metal recovery from ore and technological waste (e.g., Au, REE; Jha et al. 2016; Ding et al. 2019; Azizitorghabeh et al. 2021). Thiocyanate has been detected in geothermal and volcanic springs in sub-µM concentrations (e.g., Mukhin 1974; Kamyshny et al. 2014) and at 10-µM levels in Red Sea brines (Dowler and Ingmanson 1979). COS is a minor, but ubiquitous, sulfur species occurring at nM-levels in both fresh and sea water (e.g., Kamyshny et al. 2003a); furthermore, it is an important S compound in the modern terrestrial atmosphere (Brimblecombe and Norman 2025), having been probably more abundant in the Archean atmosphere (e.g., Ueno et al. 2009) where it might have mediated prebiotic synthesis of aminoacids (e.g., Leman et al. 2004). The abundance and geochemical significance of COS in hydrothermal fluids is much less known, but it has been reported a couple of times by Raman spectroscopy in natural fluid inclusions from metamorphic settings (see section 4.2.4).

According to their thermodynamic properties, the great majority of intermediate-valence S species should transform almost completely to sulfide and sulfate at equilibrium under the conditions of most terrestrial aquatic environments including both surficial and crustal fluids, at least at temperatures below 150 °C. Their equilibrium concentrations should therefore be much lower than those analyzed in these natural settings. In fact, their higher than thermodynamically allowed concentrations in such environments are due to intrinsically slow rates of sulfur redox transformations at low-to-moderate temperatures. With increasing temperature, in general these rates largely increase, yet they remain extremely variable amongst different S species types, being strongly dependent upon fluid composition and pH and controlled by different reaction intermediates. The following section discusses the fundamentals of sulfur species transformation kinetics in aqueous solution with an emphasis on hydrothermal conditions.

### *4.3.2 Kinetics of sulfur species transformation reactions in hydrothermal fluids*

The application of thermodynamic data to the interpretation of mineral assemblages in hydrothermal sulfide ore deposits and fluid-rock interaction phenomena, as well as redox conditions ($f_{O_2}$, $f_{H_2}$), acidity (pH), chemical speciation in the hydrothermal fluid phase, and the related metal transport, depends critically on the assumption that chemical equilibrium is attained. However, rates of reactions amongst sulfur species in geological fluids vary over an extremely large range, which is primarily defined by the extent of electron transfer between the reacting species.

Very fast rates are typical for reactions between sulfur species of the same or a close oxidation state, such as those involving proton transfer or ion pairing between cations and anions in aqueous solution, e.g., $H_2S = HS^- + H^+$, $HSO_4^- = SO_4^{2-} + H^+$, or $Na^+ + SO_4^{2-} = NaSO_4^-$. Such reactions typically attain equilibrium within the millisecond range even at ambient temperatures. Their reaction kinetics is controlled by the transport (i.e. diffusion) of reacting species in aqueous solution, which is very fast (Chen and Irish 1971; Weston and Schwartz 1972; Stumm and Morgan 1996; Lasaga 1998).

The rates are slower for the reactions between sulfur species of different redox states that imply an electron transfer. In general, the more electrons transferred between the reactants and products, the slower the overall reaction rate. Thus, disproportionation reactions involving $H_2S/HS^-$ and polysulfide dianions and/or radical sulfur ions both in aqueous solution and organic solvents, such as $S_n^{2-} + HS^- + OH^- = S_x^{2-} + S_y^{2-} + H_2O$ ($n + 1 = x + y$) or $S_6^{2-} = 2\ S_3^{\bullet-}$, are pretty fast, being at second-to-minute scales for 90% of reaction equilibrium attainment ($t_{90}$) at near-neutral pH conditions and millimol-levels of S species concentration (e.g., Hartler et al. 1967; Chen and Morris 1972; Kamyshny et al. 2003b; Chivers and Elder 2013). Reactions involving molecular sulfur $S^0$ and polysulfides and sulfides, such as $(1–n)\ S^0 + HS^- = S_n^{2-} + H^+$ or $2\ S_3^{\bullet-} + 2H^+ = 5\ S^0 + H_2S$, which require transfer of a small number of electrons (≤2 e per reaction) without complete breaking of S–S chemical bonds in $S^0$ and polysulfides, have rates ranging from minutes to months (in terms of $t_{90}$) for geologically relevant conditions, such as marine sediments or geothermal sulfur-bearing fluids. The exact rate would depend on the state of $S^0$, e.g., dissolved, colloidal or particulate, crystalline or amorphous (Hartler et al. 1967; Boulëgue and Michard 1977; Avetisyan et al. 2019;

references therein). Formation of $S^0$ in colloidal or particulate (i.e. solid) state is systematically observed both in natural geothermal settings and laboratory experiments during cooling of $H_2S$-$SO_4$-bearing fluids. This phenomenon directly reflects the decreasing $S_8{}^0{}_{(s,l)}$ equilibrium (thermodynamic) solubility with decreasing temperature (see below) along with fast kinetics of $H_2S$-$S_n{}^{2-}$-$S^0$ exchange reactions (Kaasalainen and Stéfansson 2011; Pokrovski and Dubessy, 2015; Kokh et al. 2020). Similarly, sulfite oxidation to sulfate, $SO_3{}^{2-} + ½\ O_2 = SO_4{}^{2-}$, with transfer of 2 electrons, is relatively fast, typically in the minute-to-hour range in seawater, the rate being also dependent on the presence of transition metals that may act as catalysts (e.g., Zhang and Millero 1991).

Reactions that require more electron transfer between the reacting S species, such as between sulfate and sulfide, with 8 electrons to exchange and to completely reconstruct S chemical bonds and coordination from reactants to products, are fundamentally much slower and involve different intermediate steps and precursors depending on fluid temperature, acidity, salinity, and sulfur speciation. The detailed discussion of their mechanisms is beyond the scope of this chapter; here we provide a brief overview of the major kinetic trends between sulfide and sulfate relevant to hydrothermal fluids. Indeed, at least at moderate temperatures of ≤350 °C, sulfate and sulfide are the major redox forms of sulfur (Fig. 4.5) and, therefore, determine fluid redox potential, the abundance of other more subordinate S redox forms as well as S stable isotope fractionations (e.g., Ohmoto and Lasaga 1982; Eldridge 2025). Experimental studies of sulfide-sulfate reaction kinetics under hydrothermal conditions were carried out since the 1930s by exploring the radioactive $^{35}S$ isotope exchange in sulfide-thiosulfate-sulfate aqueous mixtures and sulfate reduction by organic matter (e.g., Voge and Libby 1937; Voge 1939; Pryor 1960; Toland 1960; Ciuffarin and Pryor 1964). These studies recognized the complexity of the reaction mechanism, involving different intermediate-valence sulfur species such as thiosulfates and polysulfides as transition complexes, and a strong effect of temperature and pH on the reaction rates. Subsequent works explored different starting sulfur compounds, such as thiosulfate, native sulfur or mixtures of sulfide and sulfate, by analyzing the evolution versus time of the ratio of most abundant sulfur stable isotopes $^{34}S/^{32}S$ in the produced sulfate and sulfide (e.g., Grinenko et al. 1969; Robinson 1973; Igumnov 1976; Sakai and Dickson 1978). These studies have confirmed the key dependence of the reaction rates on temperature and pH inferred in earlier work. An important recognition of those studies was also that the rate of attainment of chemical equilibrium between aqueous sulfide and sulfate can be assessed from the information on the rate of sulfur isotopic exchanges between them. This is because both types of reactions, chemical and isotopic, imply breaking of sulfur bonds and exchanges of sulfur atoms in molecular structures of sulfide and sulfate.

This concept was explored by Ohmoto and Lasaga (1982) to build the first comprehensive model of sulfate-sulfide reaction rates, based on those experimental studies coupled with general knowledge of sulfate, sulfide and thiosulfate speciation as a function of pH that varies from dominant protonated to deprotonated species with increasing pH. According to those authors, the overall rate of sulfate-sulfide reaction (*R*) may adequately be described as

$$R = k_r \times [\Sigma S^{6+}] \times [\Sigma S^{2-}] \quad (4.1)$$

where $\Sigma S^{6+}$ and $\Sigma S^{2-}$ are total concentration of sulfate-type (e.g., $SO_4^{2-}$, $HSO_4^-$, $NaSO_4^-$) and sulfide-type (e.g., $H_2S$, $HS^-$, $NaHS^0$) species, respectively, and $k_r$ is the rate constant, which is a function of the solution temperature, pH, and composition. Ohmoto and Lasaga's (1982) model, in accordance with most available experimental data, indicates that, at a given temperature, the rate constant is strongly dependent on pH, as exemplified schematically in Fig. 4.6 for typical hydrothermal conditions of 300 °C. These rates are fastest in strongly acidic solutions (pH <2), decrease with increasing pH to ~4, stay almost constant between pH ~4 and ~7, and then further decrease at more basic pH, becoming extremely slow at pH >9. Such a pattern was quantitatively interpreted by Ohmoto and Lasaga (1982) by suggesting intermediate thiosulfate species depending on pH, with $H_2S_2O_3^0 + HS_2O_3^-$, $S_2O_3^{2-}$, and $NaS_2O_3^-$ acting as the rate-controlling transition complexes at acidic, neutral, and basic pH, respectively. In these thiosulfate species, intra-molecular exchange of sulfur atoms is therefore the rate determining step. Using the thiosulfate model fitted to available experimental data at different temperatures and pH values, the time necessary to attain sulfate sulfide equilibrium may be estimated at different temperatures and fluid pH as outlined in Fig. 4.7. It can be seen that in the typical pH range of hydrothermal fluids (4–7), both chemical and isotopic equilibrium between sulfate and sulfide species may be attained in most cases above 200 °C within geologically reasonable time scales of <100 years. Such time scales are generally shorter than the duration of most events relevant to the circulation of geological fluids, ore-deposit formation and the evolution of geothermal systems. However, at lower temperatures and/or more alkaline pH, significant disequilibrium in the sulfate-sulfide chemical and isotopic system may persist in the absence of bacteria-mediated processes. More discussion about the impact of sulfur redox kinetics on ore deposit formation and sulfur isotope fractionation may be found in Ohmoto and Lasaga (1982), Ohmoto and Goldhauer (1998) and Eldridge (2025).

Ohmoto and Lasaga's (1982) kinetic model was later reconsidered by Chu et al. (2004) who based their analysis on the extensive experimental data of S isotope exchange in thiosulfate-sulfide-sulfate solutions obtained by Uyama et al. (1985) coupled with earlier identification of polysulfide ions ($S_nS^{2-}$) in such solutions using UV-Vis spectroscopy by Giggenbach (1974a,b). Chu et al. (2004) suggested that the formation and disproportionation of $S_nS^{2-}$ control the exchange rates between the $S^{VI}$ and $S^{-II}$ atoms in thiosulfate through intermolecular exchange with polysulfides, rather than intramolecular exchanges within the thiosulfate molecule itself as originally proposed by Ohmoto and Lasaga (1982). However, both models were based on limited thermodynamic data available for those intermediate species and lacked unambiguous in situ evidence.

The in situ Raman spectroscopy identification of the radical $S_3^{\bullet-}$ ion in aqueous solution above 150–200 °C at the sulfide-sulfate transition (Pokrovski and Dubrovinsky 2011; Jacquemet et al. 2014; Truche et al. 2014; Pokrovski and Dubessy 2015; Barré et al. 2017) provided a fresh update to the long history of the sulfate-sulfide reaction mechanisms. This very reactive species whose maximum abundance lies within the slightly acidic to neutral pH range (pH 4–6, see below) may therefore act as the rate controlling intermediate. In contrast, no thiosulfate species or polysulfide dianions were identified by Raman spectroscopy in Pokrovski and Dubessy's (2015) high-temperature experiments using thiosulfate or mixtures of sulfate and sulfide. The $S_3^{\bullet-}$ ion was also systematically detected in the thermochemical sulfate reduction (TSR) experiments involving various organic compounds at temperatures 100–300 °C, and was hypothesized to be a key intermediate in TSR processes (Truche et al. 2014). At more acidic pH (<4) where the rates are fastest, molecular sulfur polymeric species ($S_n^0$; Tables 3 and 4) may also play a role in the sulfate-sulfide exchange reaction at moderate temperatures (<300 °C). Such species were identified by Raman spectroscopy in experimental sulfate-sulfide solutions (Pokrovski and Dubessy, 2015; Wan et al. 2024) and confirmed in natural fluid inclusions (Barré et al. 2017). In alkaline solutions, polysulfide dianions are more abundant and would likely be the rate-controlling species, in agreement with Chu et al.'s (2004) model.

Notwithstanding the exact mechanism and chemical intermediates in the sulfide-sulfate redox reactions, the key fundamental point seems to be the formation of -S-S- chemical bonds to allow electron transfer between sulfur atoms. More accurate knowledge of the abundances of such intermediate-valence S-S-type species in hydrothermal fluids would be necessary to build a comprehensive kinetic model of sulfur redox reactions that would be consistent with the whole sulfur speciation picture. The sulfur species abundances may be assessed in two ways. The most direct one is to use in situ spectroscopic methods, but they remain rare and are currently restricted to a limited *T*-*P*-compositional range. Alternatively, equilibrium thermodynamics approaches become more universal, in particular at elevated temperatures at which the kinetic limitations discussed above are minor due to the overall increase of reaction rates both within the fluid phase and between fluid and minerals. The following section overviews the major thermodynamic models and discusses thermodynamic data of the key sulfur aqueous species in hydrothermal fluids.

#### *4.3.3 Overview of thermodynamic models for sulfur in hydrothermal fluids*

Thermodynamic data for aqueous sulfur species at elevated temperatures and pressures are necessary for predicting the species equilibrium concentrations and solubility of S-bearing minerals in hydrothermal fluids. These data, together with those for other major and trace elements, can nowadays be assessed from thermodynamic databases (e.g., SUPCRT, SUPCRTBL, MINTEQ, EQ36, LLNL). Such databases are integrated in computer codes allowing thermodynamic equilibrium calculations in multicomponent fluid-mineral-gas systems at elevated temperatures and pressures (e.g., EQ3/6, CHESS, HCh, PHREEQC, GEMS, GWB to name a few). A detailed account of the existing databases, associated software, and advantages and limitations of the thermodynamic models for aqueous species and minerals in general may be found in Oelkers et al. (2009) and Pokrovski (2025a). Below we provide basic thermodynamic relationships and briefly overview major thermodynamic models used nowadays for S aqueous species and their complexes with metals. Then we discuss in more detail the origin of thermodynamic data for different classes of sulfur species and reliability of their use at the conditions of crustal fluids. Finally, using these data we quantify sulfur solubility and speciation in major types of geological fluids across a wide *T-P* range.

*Basic thermodynamic relationships and models for aqueous solution*

Thermodynamic equilibrium calculations use either the standard molal Gibbs energies for aqueous species, minerals or gases or, alternatively, reaction equilibrium constants. The Gibbs energies of individual species are defined as:

$$G_{i,T,P} = \Delta G^0_{i,T,P} + RT \ln a_i \text{ (for solid and aqueous species)} \quad (4.2)$$

$$G_{i,T,P} = \Delta G^0_{i,T,P_r} + RT \ln f_i \text{ (for gases)} \quad (4.3)$$

where $\Delta G^0_{i,T,P}$ = standard molal Gibbs energy of i-*th* species, $a_i$ = activity (aqueous or solid species) and $f_i$ = fugacity (gaseous species). The standard states commonly adopted in geochemistry of fluid-rock-gas interactions (e.g., Oelkers et al. 2009) are the pure substance ($a_i = 1$) for solid phases, pure gas substance at 1 bar pressure having ideal-gas behavior ($f_i \approx P_i$) for gases, and 1 molal (1 mole per kg of $H_2O$) solution whose behavior is ideal ($a_i \approx m_i$) for aqueous species. The standard Gibbs energy is related to the equilibrium constant of a reaction as

$$\Delta_r G^0_{T,P} = \sum_i n_i \times G^0_{i,T,P} = -2.3026RT \times \log_{10} K_{T,P} \quad (4.4)$$

where $R$ is the ideal gas constant = 8.314 J/(mol K), $T$ = temperature in Kelvin = $T$ °C + 273.15, $n_i$ is the stoichiometric coefficient of each reaction constituent, and $K_{T,P}$ is the equilibrium constant, at given $T$ and $P$, of the reaction

$$wW + xX = yY + zZ;\ \text{with}\ K = \frac{Y^y \times Z^z}{W^w \times X^x} \qquad (4.5)$$

where $W$, $X$ and $Y$, $Z$ are activities of the products and reactants, respectively, and $w$, $x$ and $y$, $z$ are their stoichiometric coefficients in the reaction. Note that both $G°$ and $K$ depend on $T$ and $P$, but do not depend on the system composition at given $T$ and $P$. At equilibrium, the total Gibbs energy of the system is minimal: $\Sigma G_{i,T,P} \rightarrow min$. Depending of the computer program, thermodynamic equilibrium calculation algorithms either search to minimize the overall Gibbs energy of the system (reactions 4.2 and 4.3) or to solve the system of reaction equations (equation 4.5), in both cases by finding a corresponding set of the species activities. Both algorithms apply mass and electrical charge balance constraints.

The major caveat of such calculations, often disregarded by many users of geochemical computer codes, is the reliability of estimations of $K$ and $G^0$ numerical values at elevated $T$-$P$. If these values at near ambient conditions are known for major sulfur species with sufficient precision, this precision worsens dramatically with increasing $T$ and $P$. This situation contrasts with that for mineral phases for which variations in $K$ and/or $G^0$ values with $T$ and $P$ are much smaller and far more regular than for aqueous species for which interactions with the solvent (hydration) represent the major contribution to the species energy. The variation of apparent standard Gibbs energy with $T$ and $P$ of any substance or species may be obtained by integrating the heat capacity and volume functions over the $T$-$P$ interval:

$$\Delta G^0_{i,T,P} = \Delta G^0_{i,T_r,P_r} - (T - T_r) \times S^0_{i,Tr,\mathrm{Pr}} + \int_{Tr}^{T} Cp_i^0\ dT - T\int_{Tr}^{T} (Cp_i^0 / T)\ dT + \int_{\mathrm{Pr}}^{P} V^0_{i,T}\ dP \qquad (4.6)$$

where $\Delta G^0_{i,T_r,P_r}$ = molal Gibbs free energy of formation from the elements at the reference temperature and pressure $T_r$ and $P_r$, which are 25°C (298.15 K) and 1 bar, respectively; $S^0_{i,T_r,P_r}$ = molal entropy of the $i$-th species at $T_r$ and $P_r$, and $Cp_i$ and $V_{i,T}$ = molal isobaric heat capacity and volume of the $i$-th species. Note that this way of expressing the Gibbs energy of a species/phase at $T$ and/or $P$ different from the reference values above ($T_r$, $P_r$), called an apparent Gibbs energy, is different from the Gibbs energy of formation from elements *sensu stricto*. In the latter case, equation (4.6), instead of the $S_i^0$, $Cp_i^0$ and $V_i^0$ terms of the species itself, would contain the terms $\Delta_r S^0$, $\Delta_r Cp^0$ and $\Delta_r V^0$ of the reaction of formation of the species from elements in their stable states (whose $\Delta G^0_{T,P}$ is zero by definition). Such values of energies were tabulated in some thermodynamic databases (e.g., Robie and Hemingway 1995) and are numerically different, at $T$ and $P$ other than the reference values, from the apparent Gibbs energies from equation (4.6), which is currently adopted in most aqueous species model formulations and databases (see below). The omission of the thermodynamic parameters of the elements in equation (4.6) greatly simplifies the $\Delta G^0$ expression and eliminates potential ambiguities related to the choice of element stable states at elevated $T$ and $P$ (e.g., solid *vs* gas). Naturally, these different numerical expressions of $G^0$ values have no effect on calculations of reaction constants (e.g., equation 4.5) or thermodynamic equilibria in multicomponent systems, provided that the same formulation is applied to all species and phases and their standard state definition is coherent (equations 4.2 and 4.3). Knowledge of $Cp^0$ and $V^0$ as a function of $T$ and $P$ is thus required for accurate calculation of the $\Delta G^0$ values (or equilibrium constants). The thermodynamic models for aqueous species mostly differ in the way they treat these functions. The most common models for solutes in aqueous fluids are briefly overviewed as follows.

*Van't Hoff equation and its extensions*

Direct information on $Cp$ and $V$ values at elevated $T$-$P$ is often absent for many species including sulfur-bearing ones. Therefore, thermodynamic equations used for describing such aqueous species assumed that the standard molal heat capacity of a reaction is zero, which allows a simple linear extrapolation of $\log_{10}K$ (hereafter $\log K$, for simplicity) as a function of the reciprocal of absolute temperature:

$$\log K_{T,P} = \frac{-\Delta_r H^0_{298}}{2.3026RT} + \frac{\Delta_r S^0_{298}}{2.3026R} \qquad (4.7)$$

where $T$ is in Kelvin and $\Delta_r H^0$ and $\Delta_r S^0$ are assumed constant, which translates to a linear equation

$$\log K = a/T + b \qquad (4.8)$$

Such a simple relationship, the so called van't Hoff approximation (e.g., Anderson 2005; Oelkers et al. 2009), is generally valid over a relatively small $T$-range (typically from ambient to ~100 °C) but may be extended to ~200–250 °C for certain isocoulombic reactions, i.e. those having similar ionic charges at both sides of the reaction. For some of such reactions, the equations above may further be simplified assuming that the entropy change is also zero:

$$\Delta_r G^0_T = \Delta_r G^0_{298K} = -2.3026RT \times \log K_T = -2.3026R \times 298.15 \times \log K_{298} \qquad (4.9),$$

which translates to

$$\log K_T = \log K_{298} \times 298.15/T(K) \qquad (4.10)$$

This simplification allows straightforward extrapolations above 25 °C (Gu et al. 1994). Equations (4.7–4.10) have been used, for example, for describing equilibria among polysulfide ions from 60 to 260 °C by Giggenbach (1971b) and, more recently, the reaction constants between polysulfide ions and $HS^-$ measured from 20 to 80°C by Kamyshny et al. (2007), and solubility of $S_{8(s)}$ in water from 4 to 80 °C by Kamyshny (2009). When the $\Delta_r Cp$ of the reaction is different from zero, but may be approximated by a constant value over the fitted *T* range, a third term is added to equation (4.8) such as:

$$\log K = a + b/T + c \times \log T \qquad (4.11)$$

When $\Delta_r Cp$ of the reaction changes with temperature, additional terms like $d \times T$, $e/T^2$ or $f \times T^2$ may be added for data regression.

$$\log K = a + b/T + c \times \log T + d \times T + e/T^2 \qquad (4.12)$$

The latter equation or similar is used, for example, to express reaction constants in the PhreeqC or LLNL databases. In the case of sulfur species, such equations have been used for describing the formation constant of the $S_3^{\bullet-}$ from sulfate and sulfide from 0 to 500 °C (see below and Pokrovski and Dubessy 2015), $HSO_4^-$ and $H_2S$ dissociation constants from 25 to 350 °C (Marshall and Jones 1966; Rudolph 1996; Suleimenov and Seward 1997), as well as a number of metal-hydrosulfide complex formation reactions (see section 4.5). Obviously, such polynomial equations should be used mostly for interpolation purposes because they lack predictive capabilities beyond the regressed *T-P* range.

*Density model*

For more extended *T* ranges, the deviation of log*K* from the linear equation (4.8) with $1/T$ at high temperatures is observed for many reactions including aqueous species (e.g., Oelkers et al. 2009), which is mostly because of the strong and non-linear variations with temperature of the key solvent properties, such as density ($\rho_{H_2O}$) and dielectric constant ($\varepsilon_{H_2O}$). The combined effect of *T* and water density on the solubility of oxide and silicate minerals in steam (e.g., Morey and Hesselgesser 1951) and on the dissociation constants of water, acids and salts in aqueous solution (e.g., Frank 1956, 1961) was observed more than 70 years ago, but it is only since the 1980s that it has been applied in a more systematic way to describe the dissociation constants of water, acids, and bases in aqueous solution and mineral solubility over a large *T*-range (e.g., Marshall and Franck 1981; Mesmer et al. 1988; Anderson et al. 1991; references therein), using equations involving *T* and $\rho_{H_2O}$:

$$\log K = A + \frac{B}{T} + \frac{C}{T^2} + \frac{D}{T^3} + (E + \frac{F}{T} + \frac{G}{T^2}) \times \log \rho_{H_2O} \qquad (4.13)$$

where *A* to *G* are *T-P*-independent constants, *T* is temperature in Kelvin, and $\rho_{H_2O}$ is the water density. Depending of the available *T-P* data range, simplified equations are also used

$$\log K = A + \frac{B}{T} + E \log \rho_{H_2O} \tag{4.14}$$

$$\log K = A + \frac{B}{T} + \frac{F}{T} \log \rho_{H_2O} \tag{4.15}$$

Such density model equations have been very efficient for describing the solubility of quartz (e.g., Manning 1994; references therein) and a few silicate minerals (Dolejs and Manning 2010) in supercritical fluid over a wide hydrothermal-magmatic *T-P* range, from hydrothermal steams to subduction-zone fluids, as well some metalloid and metal oxides (e.g., $GeO_2$, Pokrovski et al. 2005a; CuO, Palmer et al. 2004) over more limited *T-P* windows. For S-bearing species, the density model has been used to describe data on the first dissociation constant of $H_2SO_4$ in the supercritical *T-P* range (400–800°C and 500–5000 bar; Quist et al. 1965, see below). More recently, the model has been applied to the solubility reactions of Au and Pt metals and platinum sulfide, respectively in the form of $AuCl_2^-$, $PtCl_4^{2-}$ and $Pt(HS)_2^0$ or $Pt(HS)_3^-$ aqueous species, in hydrothermal fluids (Zotov et al. 2018; Tagirov et al. 2019a,b; Filimonova et al. 2021). The density model offers a straightforward description of simple solubility reactions for cases where aqueous speciation of the element is dominated by a single species, potentially in a wide *T-P* range using the best-known parameter of the fluid – water density. However, the density-model equations are rarely directly integrated in common thermodynamic databases that use either more elaborate models for $G^0$ values of aqueous species (such as HKF, see below) or polynomial least-square fits as a function of *T* for reaction constants (e.g., EQ3/6). In addition, in case of aqueous speciation strongly changing with *T-P* and/or involving multiple major species (e.g., sulfide and sulfate, multiple metal complexes), each individual reaction requires its own density model equation. Finally, the model is prone to large uncertainties when extrapolating beyond the fitted experimental data of a limited density range, for instance in liquid-like hydrothermal fluids at moderate *T-P*, where the water density changes are relatively small (e.g., in the range 25–300 °C and pressures from $P_{sat}$ to a few kbar, the change in $\rho_{H2O}$ is only from 1.0 to 0.7 g/cm$^3$).

*Ryzhenko-Bryzgalin model*

For the liquid-like type of fluids, another relatively simple equation, the so called Ryzhenko-Bryzgalin (RB) model (Ryzhenko 1981; Bryzgalin and Rafalsky 1982; Borisov and Shvarov 1992) has shown its efficiency:

$$\mathrm{p}K_{T,P} = \mathrm{p}K_{298\mathrm{K},1\mathrm{bar}} \times 298.15 / T(K) + f(T,P) \times (zz/a)_{\mathrm{eff}} \tag{4.16}$$

where p$K$ = –log$K$, $f(T,P)$ is a species-independent function as computed from the dissociation constant of water as function of *T* and *P*, and $(zz/a)_{eff}$ is a property of the complex, which is described by a linear equation as a function of the reciprocal of absolute temperature, $(zz/a)_{\text{eff}} = A + B/T(K)$, where $A$ and $B$ are $T$-independent constants intrinsic to each complexation reaction. The function $f(T,P)$ is computed from the dissociation constant of water according to Marshall and Franck (1981) using equation (4.12) with $(zz/a)_{\text{eff}} = 1.0107$.

The RB model requires only three adjustable parameters, $pK_{298K,1\text{bar}}$, $A$, and $B$ – the latter being optional and may be omitted when the data to be described are in a limited $T$-range. The RB equation is integrated in the HCh computer code (Shvarov 2008). The associated user-friendly OptimB module is offered to help the user to find the values of these adjustable parameters by regressing experimental $pK_{T,P}$ data points. More details about the model may be found in Shvarov (2015). The RB model over performs the density model particularly for liquid-like fluids where the water density variations are small (see above). The performance of the RB model with 3 adjustable parameters in some cases may even be comparable with that of the HKF model with 7 adjustable parameters (see below). However, the RB model, having been published in the Russian literature, has unfortunately received a rather limited recognition in the international geochemical community, despite its practical simplicity coupled with predictive capability. It was employed to accurately describe the stability constants of hydrosulfide complexes of Zn (Tagirov and Seward 2010) and Au (Shvarov 2015), sulfate complexes of uranyl (Kalintsev et al. 2019) or chloride complexes of Ni (Liu et al. 2012) and Y (Guan et al. 2020), derived from either experimental or molecular dynamics simulation data at ≤300 °C. Its major practical limitation is a lack of integration into widely-used thermodynamic software and databases, mostly because of the necessity for coding the $f(T,P)$ function of water. In addition, uncertainties when extrapolating the modeled stability constants beyond the $T$-$P$ range within which the regression coefficients have been obtained are hard to assess for different types of aqueous species that exhibit $T$-trends and amplitudes of dissociation different from that of water. In contrast to the density model, the RB model has not been validated on or directly applied to reactions involving solid phases and thus currently remains applicable only to dissociation of acids, ion pairs or complexes of metal cations in aqueous solution.

*HKF model*

The greatest advance in our understanding of the solubility and speciation of metals and volatiles, including sulfur-bearing species, in the hydrothermal fluid phase was with the development of the Helgeson-Kirkham-Flowers (HKF) model in early 1980s (Helgeson et al. 1981). This model describes the Gibbs energy of an aqueous species up to 1000 °C and 5 kbar by functions of the dielectric constant of water and considering two distinct contributions, the solvation (s) and non-solvation (n) terms, to the values of the partial molal volume $V^0$ and isobaric heat capacity $Cp^0$ of an aqueous species: $V^0 = V^0_s + V^0_n$ and $Cp^0 = Cp^0_s + Cp^0_n$. The model was subsequently revised by Tanger and Helgeson (1988) and Shock et al. (1992). A detailed development of the approach may be found in those original papers and an overview summary in Oelkers et al. (2009). The overall equation for the Gibbs energy is fundamentally similar to the general equation (4.6) and describes the variation of the $Cp^0$ and $V^0$ properties of an aqueous species as a function of $T$ and $P$ using 7 HKF adjustable parameters and 2 solvent (i.e. water) constants:

$$G^o_{P,T} - G^o_{P_r,T_r} = -S^o_{P_r,T_r}(T - T_r) - c_1[T\ln(\frac{T}{T_r}) - T + T_r] + a_1(P - P_r) + a_2\ln(\frac{\Psi + P}{\Psi + P_r}) -$$
$$c_2\{[(\frac{1}{T-\Theta}) - (\frac{1}{T_r-\Theta})](\frac{\Theta - T}{\Theta}) - (\frac{T}{\Theta^2})\ln(\frac{T_r(T-\Theta)}{T(T_r-\Theta)})\} + (\frac{1}{T-\Theta})[a_3(P-P_r) + a_4\ln(\frac{\Psi+P}{\Psi+P_r})] + \qquad (4.17)$$
$$\omega(\frac{1}{\varepsilon} - 1) - \omega_{P_r,T_r}(\frac{1}{\varepsilon_{P_r,T_r}} - 1) + \omega_{P_r,T_r}Y_{P_r,T_r}(T - T_r)$$

where $a_1$, $a_2$, $a_3$, $a_4$, and $c_1$, $c_2$, are the HKF coefficients for $i^{th}$ species, $\omega$ and $\omega_{P_r,T_r}$ are, respectively, the effective Born coefficients of $i^{th}$ species at the given temperature and pressure and at the reference temperature and pressure ($T_r$ = 298.15 K, and $P_r$ = 1 bar), $\Psi$ and $\Theta$ are the solvent constants, 2600 bar and 228 K, respectively, $\varepsilon$ is the water dielectric constant, and $Y$ is the derivative of $\varepsilon$ over temperature: $Y_{P_r,T_r} = \frac{1}{\varepsilon_{P_r,T_r}}[(\frac{\partial \ln\varepsilon}{\partial T})_{P_r,T_r}]$. The $a_{1-4}$ and $c_{1-2}$ coefficients describe the non-solvation $V^0_n$ and $Cp^0_n$ contributions, respectively, whereas the Born coefficient ($\omega$) accounts for the solvation contributions $V^0_s$ and $Cp^0_s$. The solvation contribution to the Gibbs energy of a species is described by the Born equation (Born 1920)

$$G^0_s = \omega(\frac{1}{\varepsilon} - 1) \qquad (4.18)$$

whose differentiation, respectively over $P$ and $T$, results in the expressions for $V^0_s$ and $Cp^0_s$ in equation (4.17). The Born parameter for neutral species is assumed to be constant at all $T$-$P$, whereas that for charged species is a function of the ion radius and the so called $g$-function, which, in turn, is a $T$-$P$ dependent solvent function retrieved from experimental data on the NaCl dissociation constant (Tanger and Helgeson 1998; Shock et al. 1992).

To reduce the number of adjustable parameters, different sets of empirical correlations among the HKF parameters and $S^0$, $Cp^0$ and $V^0$ values at 298 K and 1 bar for different types of species, such as cations and anions of different charges (Shock and Helgeson 1988; Shock et al. 1997), metal inorganic complexes (Hass et al. 1995; Shock et al. 1997; Sverjensky et al. 1997; Sassani and Shock 1998), metal organic complexes (Shock and Koretsky 1993, 1995), inorganic neutral species and volatiles (Shock et al. 1989; Plyasunov and Shock 2001; Schulte et al. 2001), aqueous organic molecules (Shock and Helgeson 1990), and thiols (Schulte and Rogers 2004), have been established and, in some cases, further revised and extended when more high *T-P* data became available. These correlations are largely based on common electrolytes and few aqueous species of different chemical types for which the range of experimental $G^0$, $Cp^0$ and $V^0$ values were available in a sufficiently wide *T-P* range to allow the full set of their HKF values to be derived directly via regressions (e.g., Tanger and Helgeson 1988). It is these correlations that provide a predictive basis for the HKF equations when little or no experimental data are available, which is the case for the majority of aqueous species.

The major advantage of the model compared to the models discussed above is its great versatility, being able to reproduce the experimental data to 1000 °C and 5 kbar and water densities above ~0.4 g/cm$^3$. Since the late 1980s, HKF parameters of a large number of aqueous solutes, including sulfur species and their complexes with metals, have been generated and incorporated in free-access thermodynamic databases such as SUPCRT92 (Johnson et al. 1992) and its updated versions including SUPCRTBL (Zimmer et al. 2016). Some equilibrium thermodynamics computer codes and associated databases for calculating mineral-fluid-gas equilibria at elevated *T-P* include the HKF model (e.g., HCh, Shvarov 2008; GEMS, Kulik et al. 2013; modified EQ3 with DEW model, Huang and Sverjensky 2019) In addition, user-friendly codes are available to allow generation and evaluation of the HKF parameters from experimental data (e.g., OptimA, Shvarov 2015; GEMFITS, Miron et al. 2015). The theoretical applicability of HKF has recently been extended to higher pressure and temperature conditions, up to 60 kbar and 1200 °C, corresponding to water densities of 1.3 g/cm$^3$, using revised and updated models for extrapolating the water dielectric constant. This HKF model extension allows equilibrium calculations of mineral solubility and element aqueous speciation under conditions of deep subduction zones and mantle metasomatism (Sverjensky et al. 2014; Huang and Sverjensky 2019; Pokrovski et al. 2022a). Furthermore, the HKF model has been successfully used for solubility and speciation calculations in supercritical mixed $H_2O$-$CO_2$ fluids ubiquitous in metamorphic settings, by using the dielectric constant of $H_2O$-$CO_2$ mixtures instead of that of pure water (e.g., Akinfiev and Zotov 1999; Kokh et al. 2017).

The model performances are very good at elevated pressures and fluid densities, but its practical application for charged species is limited to water densities in between 0.35 and 1.3 g/cm$^3$ and "exclude" the *T-P* region around the water critical point ($350<T<400$ °C at $P<500$ bar; Johnson et al. 1992; Sverjensky et al. 2014). The low-density limit reflects the corresponding limit of experimentally-based supercritical dissociation constants of NaCl from which Shock et al. (1992) retrieved the values of the *g*-function necessary for calculating the Born coefficients for ions and charged complexes. In contrast, for neutral species whose Born coefficient is *T-P* independent, the model may in principle be applied to a larger range of densities, down to 0.05 g/cm$^3$. This practical validity range refers to that of the equations used to calculate the dielectric constant and other Born functions of $H_2O$ (Johnson et al. 1992; references therein). However, given the "inaccurate" treatment of the neutral species solvation energy using the Born equations, which are strictly applicable *a priori* only to electrically charged species, the performance of the HKF model for neutral solutes in the near- and supercritical fluids has been criticized. Improvements in deriving the HKF parameters have been proposed based on more recent experimental data on vapor-liquid partitioning and direct *Cp* and *V* measurements at elevated temperatures (e.g., Plyasunov and Shock, 2001; Perfetti et al. 2008; Oelkers et al. 2009). However, for most metal complexes, including those with sulfur and chloride ligands, direct *Cp* and *V* data are not available; consequently, the accuracy of HKF predictions for the great majority of metal complexes is difficult to evaluate towards magmatic or deep metamorphic conditions and therefore requires validation by direct experimental data at elevated temperatures (>300 °C).

Nevertheless, the model performance for moderate-to-high density fluids, flexibility of parameterization and large potential for re-tuning the parameters when new experimental data become available, largely overcome these limitations. Therefore, we recommend the HKF model for most ionic S-bearing species and their complexes with metals for liquid-like geological fluids.

*Akinfiev-Diamond model*

The intrinsic limitations of the HKF model for describing nominally uncharged species using the Born electrostatic equation and its inability to properly account for the near-critical region of water make this model perform poorly for aqueous nonelectrolyte species at supercritical and near-critical conditions. A potentially powerful alternative equation of state for volatile species has been proposed by Akinfiev and Diamond (2003), hereafter the AD model. In this model, the standard Gibbs energy of an aqueous species is described using the Gibbs energy value of the equivalent ideal-gas species and three adjustable parameters accounting for the interactions of the gaseous species with water, based fundamentally on a classical virial equation of state describing the solubility of a gas in water (e.g., Prausnitz et al. 1986) and the thermodynamic properties of pure water from Hill (1990):

$$G^0_{aq,P,T} = G^0_{g,T} - RT\ln N_w + (1-\xi)RT\ln f_1^0 + RT\xi\ln\left(\frac{\tilde{R}T}{M_w}\rho_1^0\right) + RT\rho_1^0\left[a + b\left(\frac{10^3}{T}\right)^{0.5}\right] \quad (4.19)$$

where $G^0_{g,T}$ is the standard Gibbs free energy of the species in ideal gas state, $M_w$ is the molecular mass of water (18.0152 g/mol), $N_w$ is the number of moles of water in 1 kg of water (=1000/$M_w$), $f_1^o$ and $\rho_1^o$ are the fugacity (in bar) and the density (in g/cm$^3$) of pure water at given *T*-*P*, *R* and $\tilde{R}$ are the gas constants expressed in energy (1.9872 cal/(mol·K)) and volume (83.1441 cm$^3$·bar·mol/K) units, respectively, and *a*, *b* et $\xi$ are empirical *T*-*P* independent coefficients for each species. The model is fully compatible with the HKF model for aqueous electrolytes, in terms of standard and reference thermodynamic states, and therefore may be used, along with the HKF model, for describing reactions involving gases, aqueous species and minerals. The AD model performs better than HKF for describing vapor-liquid partition coefficients of volatile species and their thermodynamic properties in the near-critical region where *Cp* and *V* values of solutes exhibit drastic changes (e.g., Akinfiev and Diamond 2003; Perfetti et al. 2008; references therein). With only 3 adjustable parameters it shows an equally good predictive power as HKF that uses 7 parameters (e.g., Fig. 4.10). However, the AD model requires knowledge of the thermodynamic properties of the gaseous counterparts, which are usually not available for many poorly volatile aqueous metal complexes. The HKF model is generally expected to perform better than the AD model at densities well above that of the water critical point ($\rho_{H2O}$ > 0.33 g/cm$^3$), and is therefore more suitable for deep metamorphic fluids. For the four key volatile species, $H_2S$, $SO_2$, $O_2$ and $H_2$ in hydrothermal fluids, the use of the AD model parameters may be recommended at pressures below 1–2 kbar relevant to shallow magmatic-hydrothermal systems, while the HKF model should be privileged for the same species at higher pressures of metamorphic and subduction-zone contexts.

### *4.3.4 Thermodynamic properties and distribution of sulfur species in hydrothermal fluids*

Any thermodynamic equation of state describing the stability constants and other thermodynamic properties requires validation and/or parameterization using experimental data. Acquisition of the thermodynamic properties of sulfur species in aqueous solution from experimental measurements has a more than 100-year history. Lewis and Randall's series of studies in the early 1900s reported determination of the reaction constants and Gibbs free energies in aqueous vapor and liquid phases at elevated temperatures of some key sulfur forms such as native sulfur, $SO_2$, $H_2S$ and sulfate (e.g., Lewis et al. 1918; Lewis and Randall 1918). Since then, numerous studies have tackled the most common sulfur aqueous forms such as sulfate and sulfide in a wide range of temperatures, using a plethora of methods such as solubility, calorimetry, colorimetry, potentiometry, electrochemistry, UV-Vis and Raman spectroscopy. Intermediate valence sulfur forms have been covered to a much more limited degree. Nowadays, the extent of experimental datasets in the *T-P* space of hydrothermal fluids is highly variable depending on the given sulfur aqueous species. The amount of our knowledge of the stability of different classes of sulfur species in hydrothermal solutions and the degree of confidence we can grant to these data is evaluated.

*Sulfate species*

Sulfate-group species include sulfuric acid, $H_2SO_4$, and the anions of sulfate, $SO_4^{2-}$, and hydrosulfate (or bisulfate), $HSO_4^-$, together with their alkali and alkali-earth metal ion pairs. The sulfate ion is no doubt the most studied sulfur species in aqueous solution given its stability under terrestrial atmosphere conditions and the ease of handling and preservation of its salts. The thermodynamic properties of $SO_4^{2-}$ at near-ambient conditions stem from extensive calorimetric studies of soluble sulfate salts of Na, K, $NH_4$, and Mg and solubility measurements of some less soluble sulfate solids of Ca, Ag and Ba (see Criss and Cobble, 1964a,b; Helgeson et al. 1981; references therein). These data have been critically revised by the Committee on Data for Science and Technology (CODATA, 1978; Cox et al. 1989) to recommend the standard molal enthalpy of formation ($\Delta_f H^0$) and entropy ($S^0$) at the reference conditions of 298.15 K and 1 bar. The recommended values from these two reference compilations differ by less than 0.4 kJ/mol and 0.4 J/mol K for $\Delta_f H^0$ and $S^0$. The resulting Gibbs energy ($\Delta_f G^0$) of the sulfate ion is constrained to date to better than ±0.5 kJ/mol, making $SO_4^{2-}$ the best studied sulfur-bearing aqueous species, at least at near-ambient conditions. The $V^0$ and $Cp^0$ values of $SO_4^{2-}$ stem from heat capacity and density measurements of sulfate electrolytes up to 200 °C. Helgeson et al. (1981) analyzed these data and extracted HKF model coefficients for $SO_4^{2-}$, which were later revised by Shock and Helgeson (1988) and included in the SUPCRT92 (Johnson et al. 1992) and SUPCRTBL (Zimmer et al. 2016) databases, allowing predictions of the Gibbs energy values of $SO_4^{2-}$ at elevated *T-P*.

In saline fluids, the sulfate ion is accompanied by its ion pairs with alkali and alkali-earth cations that become increasingly predominant over free $SO_4^{2-}$, with increasing temperature and salinity and decreasing pressure according to the general reaction:

$$\text{Cation}^{n+} + SO_4^{2-} = \text{Cation}(SO_4)^{n-2} \qquad (4.20)$$

The formation constants of such ion pairs have been assessed using a variety of methods, including potentiometry, spectroscopy, conductivity, ion-exchange techniques, and densitometry, calorimetry, and solubility measurements. The major caveat is that most of these techniques do not "see" the ion pair directly. Its abundance and thermodynamic properties are derived from the difference of those of the bulk electrolyte system and those of the free cations and anions. This fundamental limitation engenders significant discrepancies in the reported stability constants depending of the analytical or experimental method used. Moreover, the definition of an ion pair itself depends on the activity coefficients model used. For example, the use of the classical Debye-Hückel model (e.g., Helgeson and Kirkham 1974) for describing anhydrite ($CaSO_4$) solubility in $H_2O$-NaCl solutions requires, in addition to the $Ca^{2+}$ and $SO_4^{2-}$ ions, consideration of all potential ion pairs and complexes with their respective stability constants (e.g., $CaSO_4^0$, $CaCl^+$, $CaCl_2^0$, $CaOH^+$, etc). In contrast, the Pitzer model (Pitzer 1991), for example, utilizes a set of binary and ternary interaction parameters between simple ions (so called virial coefficients), and only the strongest ion-ion binary interactions are accounted by ion pairs of explicit entity such as $CaSO_4^0$. Overall, the association constant values for some alkali and alkali-earth sulfate ion pairs (e.g., $CaSO_4^0$, $NaSO_4^-$ or $KSO_4^-$) typically vary over at least 0.5 log unit at ambient conditions and up to 1–2 log units at moderate temperatures (~300 °C), depending on the literature source (e.g., see Pokrovski et al. 1995; Monnin 1999; Dai et al. 2014; Kokh et al. 2020 and references therein for selected numerical examples). It is impossible to provide here a more comprehensive account of the thermodynamic data of sulfate ion pairing, given the ambiguity of an ion pair as probed by the different experimental methods or regarded by the different models. We adopt here the explicit speciation model for aqueous solution as the most versatile and flexible one for multicomponent natural fluids (Helgeson 1969 and following studies of his group). Each ion pair is regarded as an individual metal-ligand complex with its own formation constant and other thermodynamic parameters. The HKF-model parameters of sulfate ion pairs have been compiled in the SUPCRT92 and SUPCRTBL databases for $NaSO_4^-$, $KSO_4^-$, $MgSO_4^0$, $CaSO_4^0$, and $KHSO_4^0$ (Sverjensky et al. 1997). Many other aqueous species databases also use $K$ values generated from those parameters. Among them, $NaSO_4^-$ is by far the best constrained. Its original HKF predictions are well confirmed by more recent experimental studies (e.g., Pokrovski et al. 1995; Hnedkovsky et al. 2005; Scheuermann et al. 2019), with $K_{20}$ values being within <0.2 log unit at 350 °C and $P_{sat}$. Other, potentially important, ion pairs contributing to S-bearing minerals solubility at elevated $T$-$P$ (e.g., $NaHSO_4^0$ and also $NaHS^0$, see below) are absent in most commonly used databases. For such pairs, analogies with the well-known $NaCl^0$ ion pair may be used. This is because, in a first approximation, the $Cl^-$ ion may be regarded, in terms of electrostatic properties, to be similar to $HSO_4^-$ and $HS^-$. Therefore, one may assume the constant of the isocoulombic exchange reaction such as $NaCl^0$ + $HSO_4^-$ = $NaHSO_4^0$ + $Cl^-$ to be equal to one at any $T$ and $P$ (e.g., Kokh et al. 2020). For example, such analogies yielded $K$ values of $NaHSO_4^0$ very similar to those from very recently reported HKF coefficients for this ion pair (Baudry and Sverjensky 2024). For others (e.g., $SrSO_4^0$, $BaSO_4^0$, $FeSO_4^0$), simple analogies are harder to find, and thus new experimental and thermodynamic data are required to be able to explicitly use these species in modeling of fluid-rock interactions at elevated *T-P*. When approaching the salt saturation and liquid-liquid or liquid-vapor immiscibility in the alkali sulfate - water systems, more complex ion 'triples', such as , $Na_2SO_4^0$ and $K_2SO_4^0$ likely form, as indicated by recent IR and Raman spectroscopic measurements and molecular dynamics simulations (e.g., Reimer et al. 2015). The formation constants of $Na_2SO_4^0$ were first derived by Hnedkovsky et al. (2005) from electrical conductivity measurements around the temperatures of the critical point of water (350–390 °C) and fitted using the density model. These data were later refitted by Scheuermann et al. (2019) using a slightly different equation. In both cases, the limited *T*-range of the original datapoints makes extrapolations rather uncertain beyond the near-critical *T*-range.

In parallel with ion pairing, the sulfate ion undergoes increasing protonation both with decreasing pH and increasing *T*, by forming the hydrosulfate ion, $HSO_4^-$ (=$(HO)SO_3^-$) The equilibrium constant of the reaction

$$HSO_4^- = SO_4^{2-} + H^+ \quad (4.21)$$

has been measured in numerous studies since the 1930s, using a variety of methods ranging from conductivity (Hamer 1934; Davis et al. 1952; Ryzhenko et al. 1964; Hnedkovsky et al. 2005; Conrad et al. 2023) and potentiometry (Matsushima and Okuwaki 1988; Dickson et al. 1990) to calorimetry (Readnour and Cobble 1969; Oscarson et al. 1988), solubility (Lietzke et al. 1961; Marshall and Jones 1966) and to Raman spectroscopy (Rudolph 1996), covering a *T* range of 0–400 °C from $P_{sat}$ to ~300 bar. Despite the multitude of methods and studies, the values of $K_{21}$ stay in good agreement (Fig. 4.8). Even more importantly, the HKF-model predictions by Shock and Helgeson (1988) based on some of these 1950–1960s data below 250 °C are surprisingly well confirmed by more recent studies at higher temperatures (Fig. 4.8), lending credence to the HKF parameters for both sulfate and hydrosulfate ions reported in SUPCRT92. Therefore, we recommend these thermodynamic properties for both ions in high *T-P* speciation calculations. The distinguishing feature of reaction (4.21) is shifting equilibrium to the right with increasing temperature. As a result, the $HSO_4^-$ to $SO_4^{2-}$ transition is displaced from pH~2 at ambient conditions to pH~7 at 300 °C. The pressure exerts an opposit trend, favoring dissociation (Fig. 4.8). Thus $SO_4^{2-}$and its ion pairs with alkali and alkali earth cations predominate in very high-pressure (10s kbars) subduction-zone fluids.

The $HSO_4^-$ ion undergoes further protonation by forming uncharged $H_2SO_4^0$ (structurally equivalent to $(HO)_2SO_2^0$), with further increase in temperature towards water critical point and above and at relatively low pressures corresponding to vapor-like fluids, whereas at ambient conditions sulfuric acid is a very strong electrolyte, being virtually fully dissociated. As a result, published experimental measurements of its dissociation constant

$$H_2SO_4^0 = HSO_4^- + H^+ \quad (4.22)$$

are available only at temperatures >150°C (Oscarson et al., 1988), with most studies being conducted between 350 and 400 °C (Xiang et al. 1996; Sue et al. 2004; Hnedkovsky et al. 2005). These studies used calorimetry, pH-metry or conductance methods. An exceptional, and the oldest known to us, work has been that of Quist et al. (1965) who performed conductance measurements of $H_2SO_4$ aqueous solutions over an extraordinary large (for an experimental study) range of temperature, from 400 to 800 °C, and pressure, from 1 to 4 kbar. The available $K_{22}$ values and their extrapolations to lower temperatures made in some studies are summarized in Fig. 4.9a versus reciprocal temperature. For the aqueous liquid phase and supercritical fluid phases under pressures well above the water critical point (>400 bar), the dissociation of $H_2SO_4$ generally remains quite strong, so that at the pH of the water neutrality point (shown in Fig. 4.9a as a dashed blue curve at 1 kbar) $H_2SO_4^0$ species would be virtually "fully" dissociated to $HSO_4^-$. However, the dissociation is strongly pressure-dependent, becoming weaker with decreasing pressure. Therefore, $H_2SO_4^0$ is expected to become the increasingly dominant sulfate species in the magmatic vapor-like fluids (e.g., Ni and Keppler 2012). Being a neutral species and thereby much more volatile than its dissociated anionic counterparts (Pokrovski et al. 2013), $H_2SO_4^0$ may significantly contribute to sulfate degassing during magma-fluid separation at shallow depths and subsequent fluid boiling. An HKF-model parameter set of $SO_3^0{}_{(aq)}$ (HKF model convention for anhydrous $H_2SO_4$) have recently been published by Beaudry and Sverjensky (2014) who regressed the Quist et al. (1965) experimental datapoints (Fig. 4.9a). Since the pioneering work of Quist et al. (1965), it has been recognized that the $K_{22}$ value trends can be significantly linearized as a function of water density as shown in Fig. 4.9b. Nevertheless, values for similar $T$-$P$-$\rho_{H2O}$ from different studies yet remain discrepant of up to 2 log units at elevated $T$ (>350 °C), which yields to large errors, over 4 log units, when extrapolating to the reference conditions of 25 °C and 1 bar (Fig. 4.9a,b). As a result, consistent thermodynamic parameters of $H_2SO_4^0$, and even empirical equations as function of $T$ and $\rho_{H2O}$ that would cover the whole range of measured data, are currently lacking, leaving room for future thermodynamic studies.

*Sulfide species*

Sulfide-type species comprise $H_2S$, $HS^-$, $S^{2-}$ and, to a much lesser extent, their ion pairs with alkali metals.

*$H_2S_{(aq)}$.* The thermodynamic properties of $H_2S_{(aq)}$ under hydrothermal conditions in liquid-like fluids stem mostly from solubility measurements of gaseous $H_2S$ in water to 350 °C at $P_{sat}$ (Suleimenov and Krupp 1994 being the most recent; see Fig. 4.10 for complete reference list).

Another type of measurement deals with the solubility of aqueous $H_2S$ in fluids in equilibrium with common mineral buffers such as pyrite-pyrrhotite-magnetite and pyrite-magnetite-hematite performed to 500 °C and 1 kbar (Kishima 1989). All these measurements allowed derivation of the Henry constant as compiled in Fig. 4.10

$$H_2S_{(aq)} = H_2S_{(g)} \quad (4.23)$$

Coupled with more recent direct calorimetric and densimetric measurements of *Cp* and *V* of $H_2S$ (aq) (Hnedkovsky et al. 1996; Hnedkovsky and Wood 1997), these data enabled generation of robust thermodynamic parameters within the framework of the HKF (Schulte et al. 2001) and AD (Akinfiev and Diamond 2003) models. It can be seen in Fig. 4.10 that both models are in decent agreement, with the AD model performing slightly better at least for the so far experimentally covered pressure range of <1 kbar (see section 4.3.3 for general comparisons between these models). Using these models, $H_2S_{(aq)}$ concentrations in geological fluids in equilibrium with common sulfide minerals may be predicted within a plausible uncertainty of <0.5 log units to ~600 °C and ~5 kbar, thereby covering a major part of magmatic-hydrothermal and metamorphic fluids in the crust. For the vapor-like fluids with densities below 0.2 g/cm$^3$, generated by boiling processes in hydrothermal systems, $H_2O$-$H_2S$ mixing rules have also been established through *PVTX* measurements up to 400 °C (Zezin et al. 2011 and references therein). However, at higher pressures of deep subduction zone fluids, $H_2S$, similarly to $H_2O$, is expected to become increasingly dissociated thereby adding more uncertainty to the current thermodynamic predictions of sulfide-type species concentrations.

*$HS^-$ ion.* The thermodynamic properties *of the hydrosulfide ion, $HS^-$*, which is the $H_2S$ deprotonated counterpart, stem from measurements or estimations of the dissociation constant whose available studies are compiled in Fig. 4.11:

$$H_2S_{(aq)} = HS^- + H^+ \quad (4.24)$$

These data have been acquired using potentiometry, calorimetry or UV-Vis spectroscopy. Most of them were limited to temperatures of 150–200 °C, with the exception of the Ellis and Giggenbach (1971) and Suleimenov and Seward (1997) works. These experimentalists managed to acquire direct UV-Vis data on $H_2S$-NaOH aqueous solutions to, respectively, 276 and 350 °C at $P_{sat}$, thereby enabling much better constraints on $H_2S$ dissociation. Some of those data have been regressed by the HKF model in Helgeson and Shock (1988) that remains the unique set of HKF parameters for the $HS^-$ ion to date. We are not aware of more recent than 1997 experimental studies of $H_2S$ dissociation constants under hydrothermal conditions. This lack is due to notorious technical and analytical difficulties of working with toxic and aggressive sulfide solutions at elevated temperatures. The development of molecular dynamics simulation methods in the recent years made possible the first theoretical estimations, such as by Sulpizi and Sprik (2008) who reported a log$K_{24}$ value of –7±1, in perfect agreement with the numerous experimental data. However, the intrinsic uncertainties of MD methods in aqueous media are yet much greater than those of experimental measurements. The HKF and AD model calculations shown along with the whole set of experimental data in Fig. 4.11, are expected to predict the $H_2S$ dissociation constant to 400–500 °C and 1–2 kbar with a decent accuracy (i.e. within 0.5 log units). We are not aware of experimental data at pressures above $P_{sat}$ in the hydrothermal domain (<500-600 °C) except a single conductivity study of Sretenskaya (1977) to 90 °C and to 3 kbar. The HKF model that did not consider her work in its parameterization yet realistically predicts an increasing dissociation of $H_2S$ at elevated pressures (Fig. 4.11). The values of –log $K_{24}$ are comparable with those of the pH of the neutrality point of pure water (i.e. [H+] = [$OH^–$] = ½ log $K_{diss}(H_2O)$) only below 100°C where dissolved hydrogen sulfide would significantly be ionized as $HS^–$ at near-neutral pH. In contrast, at higher temperatures, the $H_2S$ ionization becomes much less significant at near-neutral pH imposed by the increasing dissociation of water. For example, only a few percent of $H_2S$ will be ionized at 500 °C and 5 kbar in an aqueous fluid of neutral pH (because p$K_{24}$ is ~2 log units lower than the pH of water neutrality point, Fig. 4.11). In high-temperature magmatic vapors (>700–800 °C, <2 kbar) $H_2S$ is predicted by all thermodynamic models to be negligibly dissociated. However, recent Raman analyses of S-bearing synthetic fluid inclusions at such conditions were interpreted by the dominant presence of $HS^–$ (Farsang and Zajacz 2025), in sharp contrast to the predictions above as well as known dissociation trends of other weak acids. This discrepancy is likely due to the growing lower-frequency asymmetry of the $H_2S$ Raman stretching peak with temperature, which was mistaken with the characteristic Raman peak of $HS^–$ situated just 20 $cm^{–1}$ to the lower wavenumbers (Table 4.3). Further experimental validation of *T*-*P*-dependent predictions would require more high-pressure and high-temperature spectroscopic and molecular modeling data. This gap is especially important to be filled for the interpretation of metal complexation because $HS^–$ is the key ligand for chalcophile metals (see section 4.5).

*$S^{2–}$ ion. The sulfide ion, $S^{2–}$,* was commonly considered in many general chemistry treatises and aquatic geochemistry textbooks (see an excellent compilation of May et al. 2018 for references). Its abundance and stability in aqueous solution is controlled by the dissociation reaction

$$HS^– = S^{2–} + H^+ \quad (4.25)$$

However, this reaction constant, both at low and high temperatures, remains a matter of debate since more than 120 years. The major data sources are compiled in Fig. 4.12 (additional references may be found in May et al. 2018). Pioneering works of Küster and Heberlein (1905) and Knox (1906) who reported log$K_{25}$ values of –13.2 and –14.9 at 25°C from, respectively, potentiometric pH and HgS solubility/Hg electrode measurements in NaHS/$Na_2S$ aqueous solutions. Other scarce studies of first half of the 20th century also reported values within this range (as summarized by Maronny 1959). The following experimental and theoretical studies after the 1950s can be divided into two distinct groups as to the $K_{25}$ value measured at near-ambient conditions.

The first, relatively earlier, group of studies that used pH-metry (e.g., Maronny 1959; Kryukov et al. 1974 and some previous studies), HgS solubility (Dickson 1966), UV spectroscopy (Ellis and Milestone 1967) and calorimetry (Stephens and Cobble 1971), report values between –15 and –13. These data were adopted in many old textbooks and used both in earlier (Helgeson 1969) and more recent (Phillips and Phillips, 2000) theoretical $T$-dependence extrapolations as well as in common databases such as LLNL that ignored subsequent data. It can be seen in Fig. 4.12 that these estimations (Helgeson 1969 and LLNL that likely use similar data) as well as some earlier experimental data (e.g., Kryukov et al. 1974) show a very strong increase of log$K_{25}$ values with temperature rise, with values reaching –7 to –8 at 300 °C, which are above those of the $H_2S$ first dissociation constant (reaction 4.24, Fig. 4.11), making $HS^-$ subordinate and $S^{2-}$ largely predominant sulfide species in near-neutral pH moderate-temperature hydrothermal fluids (e.g., Helgeson 1969). However, such predicted values at moderate temperatures, which are still being used nowadays in some thermodynamic modeling studies of S-bearing systems, are in strong contradiction with *i)* reaction (4.24) studies that are far better constrained and *ii)* the large set of gold solubility and complexation data fully consistent with the formation of $Au(HS)_2^-$ in circa-neutral (pH 6–8) hydrothermal fluids over a wide $T$-$P$ range (see Pokrovski et al. 2014 and section 4.5 for details).

The second group of datasets comprises the seminal work of Ellis and Giggenbach (1971) and Giggenbach (1971a) using improved UV-Vis spectrometry methods and of Meyer et al. (1983) by Raman spectroscopy, and later on, Licht and Manassen (1987) by potentiometry. These works reported consistently much lower log$K_{25}$ values at ambient conditions, around –17. Such values mean that the $HS^-$ ion still remains dominant even in 10 molal NaOH solutions at ambient conditions. A similar range of values has been reported by less direct methods using analogues with polysulfide ions and elemental $S^0$ surface protonation data (Schoonen and Barnes 1988; Migdisov et al. 2002). More recently, May et al. (2018) re-examined $HS^-$ in concentrated CsOH and NaOH solutions by Raman spectroscopy and found no significant changes in $HS^-$ peak intensity (note that Raman probes S-H bond vibrations and cannot directly detect the $S^{2-}$ ion). They concluded that any value of log$K_{25}$ in the range of –17 to –19 or lower would match the Raman spectra, and that there is no evidence of $S^{2-}$ formation in solutions as concentrated as 20 molal NaOH/CsOH. These authors thus proposed to exclude the $S^{2-}$ ion from any consideration in aqueous solution. Such spectacular discrepancies among the different methods and studies are due to intrinsic experimental/analytical difficulties of handling highly reactive and prone to oxidation sulfide solutions, so that general preference should be given to more direct in situ spectroscopic data. Finally, a fresh simulation study using molecular dynamics (Queizán et al. 2022) qualitatively confirmed that $S^{2-}$ rapidly protonates in aqueous solution in the presence of $H^+$ by forming $HS^-$, even though the simulation conditions were rather far from realistic aqueous solution concentrations.

If this second group of studies looks more robust at ambient conditions, the question still remains open as to the $S^{2-}$ stability at elevated temperatures and pressures. Indeed, most available studies, both experimental and theoretical, whatever the initial ambient-temperature $K_{25}$ value, indicate an increase of the reaction (4.25) constant with temperature, even though highly variable in amplitude (Fig. 4.12). For example, Migdisov et al. (2002) reported an increase by ~3 orders of magnitude of $HS^-$ dissociation degree within a very narrow *T*-range (25–70°C, Fig. 4.12). Furthermore, by analogy with the $H_2O$ dissociation, greatly increasing with pressure, it cannot be completely excluded that the $S^{2-}$ ion might be a non-negligible sulfur species in fluid of subduction zone conditions, and thereby dramatically affect the aqueous speciation of soft metals for which it has a very strong affinity. It is thus premature to say a definite goodbye to the $S^{2-}$ ion in aqueous solution, and more direct in situ spectroscopic data are urgently needed for high *T*-*P* geological fluids. In the meantime, considering the large discrepancy in the $K_{25}$ value, it is much safer to write solubility products of sulfide minerals in terms of $HS^-$ or $H_2S$, rather than $S^{2-}$ as still commonly used in chemistry (e.g., $CuS_{(s)} + H^+ = Cu^{2+} + HS^-$, or $CuS_{(s)} + 2H^+ = Cu^{2+} + H_2S$, rather than $CuS_{(s)} = Cu^{2+} + S^{2-}$).

*Sulfite species*

Sulfite-type aqueous species are those with sulfur in a formal oxidation state of IV, and include sulfur dioxide ($SO_2$), sulfite and bisulfite ($SO_3^{2-}$ and $HSO_3^-$), and their dimer ($S_2O_5^{2-}$) with its protonated counterpart ($HS_2O_5^-$). Their stabilities are constrained to date, at least at near-ambient conditions. The interest in quantifying aqueous speciation of sulfite was primarily motivated by atmospheric sciences because $SO_2$ and their products are important constituents of atmospheric aerosols (e.g., Brandt and Eldik 1995; Brimblecombe and Norman 2025). Sulfite species are important products and intermediates in microbial sulfate reduction and associated S isotope fractionation (e.g., Eldridge et al. 2016; Eldridge 2025). Moreover, $SO_2$, along with $H_2S$, has long been considered to be important constituents of magmatic fluids produced by magma degassing in the crust (e.g., Burnham and Ohmoto 1980; Wallace 2005; Wallace and Edmonds 2011). The chemical literature about sulfite species kinetics and equilibria in aqueous solution measured at laboratory conditions is too large to be overviewed in detail here, we focus here on key existing compilations that are currently used for sulfite species calculations at elevated temperatures along with some more recent data relevant to geological fluids.

*Sulfur dioxide*. The solubility and hydration structures of sulfur dioxide in aqueous solution, $SO_{2(aq)}$, have been studied in chemistry since the early 1900s. In contrast to sulfuric acid, the hypothetical sulfurous acid ($H_2SO_3$) that might serve as an intermediate in the dissolution of $SO_2$ in water, similar to carbonic acid ($H_2CO_3$), has not spectroscopically been detected in aqueous solution (e.g., Connick et al. 1982; Zhang and Ewing 2002; Wang et al. 2022). Even though some earlier theoretical DFT calculations indicated that the formation of hydrated $H_2SO_3$ might be thermodynamically favorable over hydrated $SO_2$ (Steudel and Steudel 2009a), it was not confirmed so far by more recent studies using more advanced computational tools (e.g., Wang et al. 2022). In the absence of direct data, $H_2SO_3$ will be disregarded here in favor of dissolved $SO_{2(aq)}$. The thermodynamic properties of $SO_{2(aq)}$ stem from measurements of its solubility in water described by its Henry's constant

$$SO_{2(aq)} = SO_{2(g)} \quad (4.26)$$

Figure 4.13 compiles the key available high-temperature data and major compilations that also include numerous near-ambient temperature data from older chemical literature that can be found in Goldberg and Parker (1985) and Pereda et al. (2000). In their critical compilation, Goldberg and Parker (1985) have revised the $K_{26}$ values and related thermodynamic properties of reaction (4.21) to at least 100 °C. These values are in good agreement with more recent data (e.g., Rumpf and Maurer 1992) obtained to 120 °C, which is the highest temperature at which $K_{26}$ has ever been measured, to the best of our knowledge. Similarly, these data are in good agreement with revised HKF model coefficients (Schulte et al. 2001) that include experimental data on the heat capacity of dissolved $SO_2$ to 350 °C (Sharygin et al. 1997). Another independent set of HKF coefficients (Plyasunov and Shock 2001) yields rather similar predictions to at least 350 °C (not shown). Finally, the AD model (Akinfiev and Diamond 2003) provides a good match of the available data, but deviates from the HKF model above 200 °C. These deviations are, however, rather minor (within less than 0.3 log units, Fig. 4.13) and both models may equally be used to at least 500 °C and 2 kbar for modeling shallow-crust fluids. Greater differences among these three predictions are observed at higher *T*-*P* relevant to deep subduction zones, e.g. up to 2 log units at 700 °C and 30 kbar. These differences translate to a minimal error of 2 log units for a redox or protonation reaction involving $SO_{2(aq)}$. To this minimal error, will be added uncertainties related to the other S-bearing reaction components, thereby resulting in overall uncertainties of at least 2–3 orders of magnitude for $SO_2$ predicted concentrations in the fluid at such conditions.

*Sulfite and bisulfite ions*. Sulfur dioxide in water undergoes dissociation depending on pH according to the formal reactions:

$$SO_{2(aq)} + H_2O = HSO_3^- + H^+ \quad (4.27)$$

$$HSO_3^- = SO_3^{2-} + H^+ \quad (4.28)$$

The bisulfite ion, $HSO_3^-$, in aqueous solution exists as two isomeric species, $(HS)O_3^-$ and $(HO)SO_2^-$. The first form, sometimes called sulfonate or SH isomer, has a tetrahedral geometry with the proton bound to the sulfur atom. The second form, hydrogen sulfite or OH isomer, has a pyramidal geometry in which the proton is bound to one of the oxygen atoms. Due to their different geometries and bonding, these forms have distinguishable spectral signatures (Table 4.3) as seen by Raman (Littlejohn et al., 1992; Eldridge et al. 2018), S K-edge XANES (Risberg et al. 2007) or $^{17}$O-NMR (Horner and Connick 1986) spectroscopies that allowed determination of the isomerization constant:

$$(HS)O_3^- = (HO)SO_2^- \qquad Q_{29} \quad (4.29)$$

The value of $Q_{29}$ (equilibrium quotient uncorrected for ionic strength, in contrast to $K$) at 25 °C is 2.7±0.5 (Risberg et al. 2007; Eldridge et al. 2018), implying an average ratio of 1:2.7 between $(HS)O_3^-$ and $(HO)SO_2^-$ in solution. According to the $Q_{29}$ variation with temperature, the $(HS)O_3^-$ isomer comprises from ~20 to 40% of the total bisulfite in the $T$-range 5–85 °C (Eldridge et al. 2018). The evolution of the isomerization phenomenon at higher temperatures would be important to know for interpreting S isotope signatures because S isotope fractionation between those species was calculated to be unusually large, despite the same redox state of $S^{4+}$ ($1000\times\ln^{34}\alpha$ ~25‰ at ambient conditions; Eldridge et al. 2016). The two isomers, when acting as ligands, might also have different affinities for metals in hydrothermal fluids.

In contrast to bond- and coordination-specific spectroscopy methods (Raman, XANES), most other methods such as potentiometry, calorimetry, conductivity or UV-Vis spectroscopy, which have been used to derive reaction constants (4.27) and (4.28), are not directly sensitive to isomerization reaction (4.29). Therefore, the thermodynamic properties of both isomers are treated collectively as the bisulfite ion. The properties of both bisulfite and sulfite ions are bound to reactions constants (4.27) and (4.28), which are compiled in Fig. 4.14. The thermodynamic values (enthalpy, Gibbs energy and entropy) of these reactions at ambient conditions have been reported in 1970s-1980s compilations based on critical revision of multiple experimental measurements that will not be discussed here (see Cobble et al. 1972; Goldberg and Parker 1985 and references therein). Using these data and some assumptions about the reaction (4.27) and (4.28) heat capacity changes, the constant was predicted with good confidence to at least 150–200 °C using van Hoff's types of equations (e.g., eqn. 4.11; Barbero et al. 1983; Goldberg and Parker 1985). These predictions have been confirmed by the more elaborate HKF and AD models, allowing extrapolations to higher temperatures. Both models predict weakening of dissociation with increasing $T$ that makes $SO_2$ and $HSO_3^-$ the dominant $S^{IV}$ species in hydrothermal and metamorphic fluids. Both models also predict a moderate increase of dissociation with pressure (Fig. 4.14), consistent with the rare available data (Ellis and Anderson 1961) and general trends of protonation reactions (e.g., see sulfide and sulfate above). Alkali bisulfite and sulfite ion pairs, similar to those formed by bisulfate and sulfate (see above) are expected to further contribute to $S^{4+}$ speciation at elevated temperatures in saline fluids, but their formation constants have not been thoroughly measured or estimated, and are therefore limited by the use of analogies with chloride or sulfate ion pairs.

*Disulfite ions*. Another complication of the sulfite system is formation of dimers, the pyrosulfite or disulfite ion, $S_2O_5^{2-}$, and its protonated counterpart, $HS_2O_5^-$:

$$2\ HSO_3^- = S_2O_5^{2-} + H_2O \qquad (4.30)$$

$$S_2O_5^{2-} + H^+ = HS_2O_5^- \qquad (4.31)$$

Their structure in solution likely involves S-S bonds, as shown by Raman and UV spectroscopy and is similar to the structure known in the solid phase (Connick et al. 1982; Littlejohn et al. 1992; references therein). Williamson and Rimstidt (1992), however, questioned such a structure in favor of a hypothetical S-O-S linkage that would be more coherent with Gibbs energy versus structure correlations for other oxyanion sulfur species suggested by these authors. However, their analysis was based on poorly constrained thermodynamic properties of $S_2O_5^{2-}$ from old references subjected to large uncertainties. The dimerization constant of reaction (4.30) is well constrained to date at near-ambient conditions using a variety of methods. Remarkably, 1980s-1990s data and predictions (Goldberg and Parker 1985; Connick et al. 1982; Littlejohn et al. 1992; references therein) are well confirmed by more recent spectroscopic and calorimetric measurements (Ermatchkov et al. 2005; Beyard et al. 2014) allowing reaction (4.30) thermodynamic parameters to be experimentally constrained to at least 80 °C. The value of $K_{30}$ of ~0.05 at 25 °C, implies that the dimer may only be significant at too high total sulfite concentrations to be relevant to most natural low-temperature fluids (e.g., >10% of dimer at 0.5 m of sulfite). In addition, the constant decreases with increasing temperature, making $S_2O_5^{2-}$ even less important in hydrothermal fluids. As a note, HKF parameters of $S_2O_5^{2-}$ were also published by Shock et al. (1997) using inter-parameter correlations and old Gibbs energy and entropy values from Latimer (1952), which yields $K_{30}$ values 3 orders of magnitude lower than the more recent data above, and further decreasing with the temperature rise. We therefore do not recommend using those HKF parameters for $S_2O_5^{2-}$ for modeling S-rich fluids. A protonated dimer, $HS_2O_5^-$, has been inferred to form in acidic solutions according to reaction (4.31) whose constant at 25°C was reported in a single work of Beyad et al. (2014), $\log K_{31}$ of ~3, meaning that $HS_2O_5^-$ takes over $S_2O_5^{2-}$ at pH below ~3.

*Thiosulfates*

This type of sulfur species includes the well-known thiosulfate ion, $S_2O_3^{2-}$, and its poorly known protonated counterparts, $HS_2O_3^-$ and $H_2S_2O_3$. The $S_2O_3^{2-}$ ion and its soluble alkali and alkali-earth salts are widely used chemical reagents and are kinetically stable solutes in water at neutral-to-basic pH. In geoscience-related fields, thiosulfate is used for extraction of gold and some other metals from ore because it forms very stable complexes with Au (e.g., Adams 2005; Marsden and House 2009). It is also a very convenient reagent in experimental mineralogy and petrology and isotope geochemistry, because, being nontoxic, water soluble and easy to handle at ambient conditions, it breaks down irreversibly to sulfate and sulfide, along with polysulfide species such as the radical ions (see below), when heated above 150 °C according to the formal reaction

$$S_2O_3^{2-} + H_2O = SO_4^{2-} + H_2S \qquad (4.32)$$

This overall reaction may be decomposed into the three following steps (Kokh et al. 2020)

$$S_2O_3^{2-} = S^0 + SO_3^{2-} \qquad (4.33)$$

This reaction is the most straightforward and kinetically easiest one, since it requires only breaking the S-S bond of the thiosulfate ion with minimal electron transfer between the two S atoms. With time, both $S^0$ and $SO_3^{2-}$ undergo transformation to $H_2S$ and $SO_4^{2-}$, which are thermodynamically stable at hydrothermal conditions:

$$4\ SO_3^{2-} + 2\ H^+ = H_2S + 3\ SO_4^{2-} \qquad (4.34)$$

$$4\ S^0 + 4\ H_2O = 3\ H_2S + SO_4^{2-} + 2H^+ \qquad (4.35)$$

As a result, thiosulfate provides an excellent chemical proxy for many hydrothermal fluids in which sulfate and sulfide coexist (e.g., Ohmoto and Lasaga 1982 and references therein; Pokrovski and Dubessy 2015; Colin et al. 2020; Kokh et al. 2020). In particular, its use in S-rich hydrothermal experiments allows efficient $f_{O_2}$ buffering by the sulfate-sulfide equilibrium

$$H_2S + 2\ O_2 = SO_4^{2-} + 2\ H^+ \qquad (4.36)$$

The $S_2O_3^{2-}$ ion contains 2 different sulfur atoms bound together in a S-S chemical bond, one sulfanyl (S-H) sulfur in a formal oxidation state of –1, and the other sulfonate (S-$O_3$) sulfur in +5 oxidation state which seem to be better compatible with S XANES spectra than the nominal –2 and +6 values (Vairavamurthy et al. 1993).

The thermodynamic properties of the $S_2O_3^{2-}$ ion have been determined using heat of oxidation by $Br_2$ and solubility of $Ag_2S$ and $CaSO_4$, as detailed and critically reviewed by Mel et al. (1956) and Cobble et al. (1972). The standard Gibbs energy and entropy values of the ion at reference *T-P* conditions of 25 °C and 1 bar from those studies are quite discrepant, spanning over almost 30 kJ/mol and 30 J/mol K, respectively; we are not aware of more recent experimental measurements. As a result, the reference thermodynamic properties of $S_2O_3^{2-}$ have not been updated since the 1970s and travel from one database to another with variations within those ranges. The high-temperature properties are probably best approximated by the HKF model that used in addition some limited *Cp* and *V* calorimetric data for thiosulfate salts (Shock et al. 1997). An uncertainty of >30 kJ/mol would be a realistic error when predicting the $G^0(S_2O_3^{2-})$ values at common hydrothermal conditions (200–400 °C).

Upon acidification to pH<5 at ambient conditions, thiosulfate solutions become unstable, rapidly producing $SO_2$, $H_2S$ and elemental sulfur, according to mechanisms that would require simultaneous presence of $S_2O_3^{2-}$ and its protonated forms, $HS_2O_3^-$ and $H_2S_2O_3$ (e.g., Steudel and Steudel 2009b). However, direct spectroscopic evidence of these species in solution is yet lacking, even though the hydrothiosulfate ion, $HS_2O_3^-$, has been identified in solid phase by its distinct Raman S-H and S-S frequencies (Table 4.3; Steudel and Prenzel 1989). The thiosulfuric acid, $H_2S_2O_3$, yet remains rather hypothetical from an analytical point of view. From a theoretical point of view, both protonated molecules have been extensively studied by DFT simulations that show two tautomers, $HS\text{-}SO_3^-$ and $S\text{-}SO_2(OH)^-$ (Steudel and Steudel 2009b), similar to the sulfite dimers (see above). Potentiometric measurements of the first half of 20th century performed in dilute acidic thiosulfate solutions are indicative of the reactions:

$$S_2O_3^{2-} + H^+ = HS_2O_3^- \quad (4.37)$$

$$HS_2O_3^- + H^+ = H_2S_2O_3 \quad (4.38)$$

The instability of thiosulfate upon acidification (see above) makes these reaction constants notoriously difficult to accurately measure. The work of Page (1953) reports log*K* values of 1.7 and 0.6 at 25 °C and 1 bar for reaction (37) and (38), respectively. These values mean that $S_2O_3^{2-}$ significantly protonates only below pH ~2, and hypothetical thiosulfuric acid would dominate the thiosulfate speciation only in highly acidic solutions (pH < 0.6). These *K* values were the basis for estimating high-temperature reaction constants and thermodynamic properties of the protonated forms, using approximations for $\Delta S$ and $\Delta Cp$ of these reactions (e.g., Naumov et al. 1971; Murray et Cubicciotti 1983), and HKF correlations (Shock et al. 1997). However, those theoretical estimations ignore a more recent accurate pH-metry work of Nikolaeva and Gusyakova (1989) who elegantly determined *K* values of reaction (4.38) in water and ethanol-water solutions to 90 °C. Their $\log K_{38}$ values at 25°C (1.97±0.05) agree within 0.3 log units with the Page (1953) work and a few older studies not discussed here. Their $K_{38}$ values increase by an order of magnitude to 100 °C, which is within the same magnitude as the theoretical estimations above. All estimations predict a few orders of magnitude increase in both constants to 300 °C, suggesting a significant shift in the thiosulfate speciation toward protonated species at near-neutral pH; however, the absolute *K* values among those estimations are discrepant over at least 2 log units at $T \geq 300$ °C.

To this uncertainty adds the one related to alkali ion pairs of thiosulfate anions in saline solutions whose stability has not been constrained so far. It is thus hard to accurately predict the equilibrium abundance of thiosulfate species in hydrothermal fluids above 200–300 °C. While $HS_2O_3^-$ might eventually be included with caution in sulfur speciation schemes to ~200 °C, the reported properties of its fully protonated hypothetical counterpart $H_2S_2O_3$ are not currently reliable enough to be used above ~100 °C. In situ Raman spectroscopy studies, performed on thiosulfate salts and sulfur aqueous solutions at acidic and neutral pH over a wide range of S concentrations (0.2–3.0 m) to at least 700 °C and 10 kbar, have not detected thiosulfate species within the inherently high detection limits of ~0.01–0.1 m for Raman spectroscopy at such conditions (Bondarenko and Gorbaty 1997; Yuan et al. 2013; Pokrovski and Dubessy 2015; Barré et al. 2017; Colin et al. 2020). Studies that used hydrothermal reactors, enabling fluid sampling and chemical separation of the different sulfur forms and their subsequent analyses, reported thiosulfate concentrations comparable with these detection limits. For example, Dadze and Sorokin (1993) analyzed, in acidic aqueous solutions in the presence of native sulfur at 250 to 450 °C, up to 0.01 m of thiosulfate that they tentatively attributed to the $H_2S_2O_3$ species. Kokh et al. (2020) measured, using S isotopes, the rate constants of decomposition to sulfide and sulfate of initial 0.2–0.5 m $K_2S_2O_3$ solutions at 300–450°C (see equations 4.1 and 4.32). The authors found rate values consistent with previously established ones from less concentrated solutions (Ohmoto and Lasaga 1982 and references therein; see also section 4.3.2).

*Polythionates*

<u>*Tetrathionate*</u>. A highly characteristic reaction of $S_2O_3^{2-}$ is its facile oxidation to tetrathionate, $S_4O_6^{2-}$, which – when carried with iodine ($I_2$) – is used in quantitative analysis of both reagents (e.g., Kokh et al. 2020; Steudel 2020). Tetrathionate, kinetically stable in aqueous solution at neutral and acidic pH, is representative of a long series of water soluble polythionate anions, $[O_3S\text{-}S_n\text{-}SO_3]^{2-}$, in which two $SO_3$ terminal groups are linked through -S-S- bonds with *n* from 1 to 4. Such anions may be prepared in aqueous solution and solid phase by various routes (Steudel 2020). In addition, aqueous species with chain length up to 22 S atoms were identified by liquid chromatography (e.g., Steudel et al. 1987). The rich aqueous redox chemistry and reaction kinetic mechanisms of polythionates at ambient conditions have been thoroughly addressed in the chemical literature (see Steudel 2020 for the most recent review). This interest is because, these sulfur forms, along with thiosulfate and sulfite, are of concern in microbiological and sedimentary processes as intermediates in sulfur transformation reactions (Eldridge et al. 2021; Turchyn et al., 2025 and references therein) as well as in hydrometallurgy of metals leaching from ore (e.g., Aylmore and Muir 2001; Molleman and Dreisinger 2002). Polythionates have been detected in natural near-surface aquatic environments which are out-of-equilibrium (Takano 1987; Zopfi et al. 2004; Leavitt et al. 2015; Findlay and Kamyshny 2017). However, the thermodynamic properties required for predicting polythionates abundances in hydrothermal fluids are either virtually non-existent or highly discrepant. Among them, the best thermodynamically known is the $S_4O_6^{2-}$ ion. Its Gibbs energy, entropy (and enthalpy) and heat capacity values at 25 °C/1 bar have been reported in a number of thermodynamic compilations and databases (Latimer 1952; Wagman et al. 1982; Bard et al. 1985). However, the original data from which these values were established are hardly traceable and they appear highly discrepant. A rare example how the thermodynamic values have been derived in reality is given by Cobble et al. (1972) who critically revised the rare calorimetric and solubility data and estimated $G^0_{298}$ (and $H^0$) and $S^0$ values for aqueous $S_4O_6^{2-}$. These values, compared together with those from the compilations cited above remain rather discrepant, with $G^0_{298}$ values spanning over >30 kJ/mol, and $Cp$ values over >100 J/(mol K), which corresponds to an uncertainty of at least 6 log units in the equilibrium constant value of a reaction involving $S_4O_6^{2-}$ at ambient conditions plus additional 2 to 3 log units when extrapolating the $K_i$ value to >300 °C according to equations (4.6) and (4.11) and assuming $\Delta_r Cp$ to be constant.

_Other polythionate species_. Despite such potential uncertainties, the $S_4O_6^{2-}$ ion is second "thermodynamically" best known one, after the thiosulfate ion, among the whole group of polysulfoxide species, including polythionates and dithionites, as well as peroxosulfates and associated radical oxysulfur ions (see below). Only scarce data on the thermodynamic properties have been reported for other polythionate ions and acids ($S_2O_6^{2-}$, $H_2S_2O_6^0$, $S_3O_6^{2-}$, $S_5O_6^{2-}$), which were unsystematically listed in certain widely used compilations, some of those claiming to be based on critical revisions of available data, others just referring to previous compilations (e.g., Latimer 1952; Wagman et al. 1982; Bard et al. 1985). As a result, it is hard to trace and critically evaluate the original thermodynamic measurements from those compilations, which is beyond the scope of this chapter. A typical example of the degree of discrepancy is the value of standard enthalpy of formation ($\Delta_f H^0_{298K,\ 1bar}$) of the $S_3O_6^{2-}$ ion, which is a common intermediate in the thiosulfate-tetrathionate redox transformations (e.g., Steudel 2020). This value ranges from –1200 kJ/mol as reported by Wagman et al. (1982) and Bard et al. (1985) to –1167 kJ/mol as adopted by Latimer (1952) followed by Shock et Helgeson (1988), to –1083 kJ/mol as estimated by Williamson and Rimstidt (1992), thus spanning over >117 kJ/mol. Comparable variations are observed for other sulfoxide species such as dithionite ($S_2O_4^{2-}$) for which similarly discrepant $G^0$ (and $H^0$) values can be found in the above compilations. The corresponding $S^0$ values are more rarely reported because in generally less constrained. The protonated counterparts of the dithionite and analogous dianions are yet more hypothetical than real because have never been directly observed, while being extensively modeled using theoretical quantum-chemistry methods (e.g., Drozdova et al. 1998). Curiously, protonation constants at ambient conditions (analogous to reactions 4.37 and 4.38) for the $S_2O_4^{2-}$ ion can be calculated from $G^0$ values reported for $HS_2O_4^-$ and $H_2S_2O_4^0$ in some databases (log*K* values are between 0.3 and 2.5; Wagman et al. 1982; Bard et al. 1985). These data were adopted by Shock et al. (1997) who also used inter-parameter correlations of the HKF model to report a set of HKF coefficients for the three dithionite species. They suggest that $S_2O_4^{2-}$ remains dominantly deprotonated above pH~3 at ambient conditions and above pH~4–5 at 300 °C. We were unable, however, to identify the original source of these key data for dithionite protonation, and therefore do not recommend the inclusion of the protonated species in sulfur speciation models at elevated temperatures until their direct identification is provided.

*Persulfates*

An amazing, but poorly known, group of sulfur oxidized species are persulfates, such as the peroxodisulfuric ($H_2S_2O_8^0$ and $S_2O_8^{2-}$) and peroxysulfiric ($HSO_5^-$ and $SO_5^{2-}$) acids and their anions, for which limited thermodynamic data have been reported. These species are used in industry as powerful oxidants and may be formed by oxidation of sulfuric acid in electrochemical processes or by concentrated $H_2O_2$ (Bard et al. 1985; Cotton et al. 1999). However, only $S_2O_8^{2-}$ and $HSO_5^-$ were observed in aqueous solution and are known to form alkali metal salts whose crystal structures have been investigated. Although the apparent formal oxidation state of S in these forms would be +7 and +8, respectively, the more correct structurally constrained state is +6, as in sulfate, because in persulfates the S atom is directly coordinated by 4 oxygens, one of which forming a peroxide group with a 5$^{th}$ oxygen, like [O$_3$S-O-O-SO$_3$] and [O$_3$S-O-OH]. The values of $G^0(H^0)$ and $S^0$ of $S_2O_8^{2-}$ cited in the Wagman et al. (1982) and Bard et al. (1985) books are consistent within 5 kJ/mol. They very likely stem from the same data source, but no original references could be traced to enable more critical selection. Shock et al. (1997) adopted the Wagman et al. (1982) values and reported a set of HKF coefficients for the persulfate ion. Williamson and Rimsdidt (1992) estimated an uncertainty of >30 kJ/mol for the $G^0$ value using inter-species correlations, which is of the same order of magnitude as for other oxysulfur species discussed above. Neutral $H_2S_2O_8^0$ and $H_2SO_5^0$ are very strong acids and highly unstable in aqueous solution and their dissociation constants are not known. However, one might find $G^0$, $H^0$ and $S^0$ values of $H_2S_2O_8^0$ in the common thermodynamic books cited above. Paradoxically, these values appear to be identical to those of the $S_2O_8^{2-}$ ion, which is devoid of any physical sense. A similar meaningless $G^0$ value, identical to that of the $S_2O_6^{2-}$ ion, has been attributed in those books to its protonated counterpart, $H_2S_2O_6^0$. As a result, such data for the protonated species should be disregarded. The $S_2O_8^{2-}$ ion, being stable in aqueous solution, forms ion pairs with alkalis and other common metals (e.g., Cu, Co), with stability constants similar to those with sulfate; e.g., see Smith et al. (2004) for a compilation and Chlebek and Lister (1971) and Aruga (1981) for representative original data sources. These are currently limited to near-ambient temperature solutions and cannot be used in modeling of hydrothermal fluids. The $HSO_5^-$ ion is seemingly a weak acid and would dissociate to $SO_5^{2-}$ with a reaction constant of log$K$ ~9.3 at 25 °C (i.e. at pH above 9) according to Goodman and Robson (1963), which seems to be the only accessible original source of data. However, the description of the measurement of this constant in their article is limited by only 5 (!) lines of text. This is a typical example of the low degree of reliability of thermodynamic data for S species that are unstable at ambient conditions. There is no way of any reliable extrapolation of such data to higher *T-P*, even though tentative theoretical HKF coefficients for $HSO_5^-$ as generated by Shock et al. (1988) can be found in some widely used databases (e.g., SUPCRTBL); they should be used with extreme caution.

*Radical sulfoxide species*

Persulfate ions decompose to sulfate via radical mechanisms involving $SO_4^{\bullet-}$ (Cotton et al. 1999). Other oxysulfur radical ions such as $SO_2^{\bullet-}$, $SO_3^{\bullet-}$ and $SO_5^{\bullet-}$ have also been identified as intermediates in photolysis and oxidation reactions in sulfoxide aqueous systems, relevant to industrial applications and atmospheric chemistry on Earth Venus and Io (e.g., Hayon et al. 1972; Lodders et al. 2025; Zolotov 2025). While the kinetics in such systems at near-ambient conditions have been thoroughly studied, thermodynamic properties of such radicals are yet inexistent. The concentrations of such radicals are rather low; for example in aerosol water in air $SO_4^{\bullet-}$ concentrations were estimated to be $10^{-14}$ to $10^{-12}$ M (Cope et al. 2022). However, it does not mean that a species unstable in ambient solution would definitely be unstable at elevated *T-P*. The example of the trisulfur radical ion (see below) proves the contrary. By analogy, if sulfoxide radicals turn out to be more stable at high *T-P* conditions, they might contribute to a release of highly oxidized fluids during serpentinite breakdown in subduction zones. For example, recent data based on analyses of the $As^V/As^{III}$ ratio in serpentinites from Himalayas (Pokrovski et al. 2022a), indicate $O_2$ fugacity values in the fluid close to QFM+10 while equilibrium thermodynamic modeling in the fluid-rock system involving classical S species (sulfate, sulfide and sulfite) yields $f_{O_2}$ of QFM±2. Radical oxysulfur species might, therefore, be potentially important carriers of oxidation potential during such processes whose study would require in situ spectroscopy approaches.

*Polysulfide species*

This geochemically important group of sulfur species includes polysulfide dianions, $S_n^{2-}$, and their less well-known protonated counterparts, $HS_n^-$ and $H_2S_n^0$. The structurally correct formula of polysulfides is chain-like $[S\text{-}S_n\text{-}S]^{2-}$ in which the two terminal S atoms are charged and the total chain length *n* is up to 11 atoms in aqueous solution and to 12 atoms in the solid state (e.g., Liebing et al. 2019). Comprehensive reviews of the chemistry of polysulfides, their reactions, synthesis, and detailed structural characterization and spectroscopic features in the solid state and both non-aqueous and aqueous solvents have been published (Steudel 2003, 2020; Chivers and Elders 2013; Sheldrick 2013; Steudel and Chivers 2019, to name a few). These species are important actors in sulfur-based batteries, which explains the continued interest of the chemical and material sciences communities. Polysulfide dianions have often been analyzed in sedimentary environments (e.g., Jorgensen 2021; references therein), geothermal springs (e.g., Stefánsson 2017), and more rarely in fluid inclusions (Barré et al. 2017; Hurai et al. 2017). Both in laboratory and natural systems, polysulfides commonly form by oxidation of hydrogen sulfide or reactions of native sulfur with hydrogen sulfide at pH >6 such as:

$$(n\text{–}1)/8\ S_{8(s)} + HS^- = S_n^{2-} + H^+ \quad (4.39)$$

The reaction kinetics in the $S^0$-$HS^-$-$S_n^{2-}$ system being fast at most laboratory conditions (see section 4.3.2), equilibria among the above mentioned species in aqueous solution are rapidly and reversibly attained. This property allowed acquisition of thermodynamic parameters of the polysulfide anions based on reaction (4.39) in numerous experimental studies that employed solubility, UV-Vis spectroscopy, potentiometry, conductivity or chromatography methods (e.g., Giggenbach 1972, 1974; Berndt et al. 1994; Licht et al. 1986; Licht and Davis 1997; Kamyshny et al. 2004, 2007; and references therein for earlier work). For more detailed account of the polysulfide speciation, data analysis and stability constants from <1990s works, the reader is referred to Hamilton (1991) whose theoretical analysis represents a remarkable work for an unpublished MSc thesis. Among earlier published work, Giggenbach studies using high-temperature (up to 260 °C) UV-Vis spectroscopy should particularly be acknowledged; they are also most relevant to hydrothermal fluids. These and other lower-temperature works have reported reaction (4.39) constant values to 80–150°C, which served as the basis for thermodynamic compilations in which $\Delta_f G^0$, $\Delta_f H^0$ and $S^0$ values at 25 °C and 1 bar for polysulfide dianions with S chain length up to 5 atoms were derived (Wagman et al. 1968, 1982, subsequently cited by Bard et al. 1985). These data were used to estimate the HKF-model coefficients for these ions with $n \leq 5$ (Shock and Helgeson 1988). They were included in major databases of aqueous species such as SUPCRT92 and SUPCRTBL, allowing estimations of polysulfide ions abundances in S speciation calculations at hydrothermal conditions. After those works, the acquisition of thermodynamic properties of polysulfide ions has almost stagnated for 20 years, until the early 2000s when Kamyshny et al. (2004, 2007) made a significant advance in polysulfide equilibrium speciation using a new approach, so called fast chemical derivation in a single phase. Their method is based on fast reactions of polysulfide ions with methyl triflate in aqueous solution with formation of dimethylpolysulfides and their subsequent analysis by HPLC coupled with UV detection. This approach allowed the initial distribution of inorganic polysulfide ions to be determined in experimental solutions corresponding to reaction (4.39) from 25 to 80 °C:

$$S_n^{2-} + 2\ CF_3SO_3CH_3 = (CH_3)_2S_n + 2CF_3SO_3^- \qquad (4.40)$$

Kamyshny seminal studies allowed a revision of thermodynamic properties of the existing polysulfides to $n = 5$, and generation of $\Delta_f G^0$, $\Delta_f H^0$ and $S^0$ values for longer-chain ions, up to $n = 8$. To enable the use of the new data at elevated temperatures, Barré et al. (2017) combined Kamyshny et al.'s values with the Shock and Helgeson (1988) and Sverjensky et al. (1997) correlations to generate HKF-model coefficients for the polysulfide ions from $S_2^{2-}$ to $S_8^{2-}$, as well as to integrate these data in the form of log $K$ versus $T$ polynomial equations compatible with the LLNL database of the PHREEQC code to 300 °C (Parkhurst and Appelo 1999). These data are actually the most complete set of thermodynamic properties of polysulfide ions for hydrothermal fluids. However, a very recent update was the study of Avetisian and Kamyshny (2022) that followed up the improved protocols of Kamyshny's work by specifically focusing on the disulfide ion, $S_2^{2-}$, and its protonation constants (see below) to be measured from 10 to 70 °C. The re-determined $\Delta_f G^0$ value of $S_2^{2-}$ at 25 °C and 1 bar is by 7 kJ/mol more positive and the $S^0$ value 50 J/mol K more negative than those previously reported by Kamishny et al. (2004, 2007) themselves. The new values make $S_2^{2-}$ even less stable (i.e. less abundant) in aqueous solution compared to the previous studies. However, whatever the exact values for the $S_2^{2-}$ and higher-order polysulfide ions, the majority of earlier and more recent studies remarkably agree about the predominance of tetra-, penta- and hexasulfide ions, and they yield quite similar total polysulfide concentrations in solution ($\Sigma S_n^{2-}$), whatever the exact set of data is chosen. Therefore, considering the relatively small overall differences in thermodynamic properties of the polysulfide ions among the different studies and the general robustness of the HKF model extrapolations for ionic species, these data may be used with confidence for modeling polysulfide dianions concentrations in liquid-like hydrothermal fluids.

In contrast to the dianions, the data on their protonated counterparts in aqueous solution, $HS_n^-$ and $H_2S_n^0$, described by the following reaction constants

$$S_n^{2-} + H^+ = HS_n^- \qquad K_{0\text{-}1} \qquad (4.41)$$

$$HS_n^- + H^+ = H_2S_n^0 \qquad K_{1\text{-}2} \qquad (4.42)$$

are far more scarce and discrepant. This is mostly because protonation of the dianions occurs at neutral and acidic pH where $S_8{}^0{}_{(s)}$ solubility is much lower than in alkaline solutions in which $S_n^{2-}$ has commonly been measured. As a result, acidification causes precipitation of native sulfur inducing large and time-dependent losses of polysulfide concentration from solution making it hard to perform accurate measurements. The seminal experimental study of Schwarzenbach and Fischer (1960) employed potentiometric pH titrations in 0.1 M KCl solutions at 20 °C and derived values of reaction constants (4.41) and (4.42) for tetra- and pentasulfide ions and tentative values for disulfide and trisulfide. Remarkably, all theoretical estimations and extrapolations of protonation constants in the polysulfide series that followed are based on that seminal work. Using the Schwarzenbach and Fischer (1960) data, Schoonen and Barnes (1988) applied a simple concept of the linear dependence of association constants versus reciprocal of the chain length (1/*n*) to estimate the second dissociation constant of $H_2S$ (i.e. $n = 1$; see above for detailed assessment of the $H_2S$ dissociation). With increasing *n*, the charges of the two terminal S atoms become more and more localized and independent of one another, resulting in identical protonation constants for both atoms at *n* tending to infinity (Fig. 4.15). This trend places an additional constraint on extrapolations to $n = 1$. Kamyshny et al. (2004) used the same concept and basis data for $S_4^{2-}$ and $S_5^{2-}$ to estimate the constants for higher ($n$ = 6 to 8) and lower ($n$ = 2 and 3) polysulfides. Chadwell et al. (1999), using electrochemical titrations, experimentally confirmed the $K_{0\text{-}1}$ value for pentasulfide within <0.3 log units of that of the Schwarzenbach and Fischer (1960) study. Finally, Liu et al. (2013a) used molecular dynamics approaches (FPMD) to provide a good independent confirmation of $K_{0\text{-}1}$ of sulfide and disulfide within <0.5 log unit, but the agreement with the experimental data and 1/*n* relationships for longer-chain polysulfides was less good, with differences of >1–2 log units (Fig. 4.15).

It follows from this overview that in the polysulfide group, disulfide and pentasulfide dianions and their protonated species are currently the best thermodynamically constrained ones. Their respective relative distributions as a function of pH at 25 °C is shown in Fig. 4.16e,f. There are currently no published thermodynamic coefficients for the protonated polysulfide species allowing extrapolations to elevated temperatures. Qualitatively, it would be expected that the pH range of their predominance domains shown in Fig. 4.16 would become larger with increasing *T*, compared to that of $S_n^{2-}$. The uncharged sulfanes $H_2S_n^0$, being volatile species similarly to $H_2S$ and $SO_2$, may further contribute to increasing partitioning of sulfur in the vapor phase during boiling and vapor-liquid immiscibility in hydrothermal systems.

*Molecular sulfur*

Molecular sulfur, $S_n^0$, has an extremely rich solid-, liquid- and gas-phase chemistry thoroughly discussed in comprehensive reviews (e.g., Meyer 1976 and Steudel 2003, 2020 to name a few). It is known to form a large variety of allotropes composed of ring molecules, with *n* from 6 to at least 20, isolated as individual compounds. In the molten state (typically >120 °C), it exists as polymeric chains or $S_n$ rings with *n* typically from 8 to 12, as well as less abundant higher-*n* cycles. In the gas phase, at least to 1000 °C, all $S_n$ molecules with *n* ranging from 2 to 8 form in equilibrium (Rau et al. 1973). Remarkably, among all these S allotropes and molecules, cyclooctasulfur, $S_{8(s)}$, is the thermodynamically stable solid, and $S_{8(l,g)}$ is the most abundant molecule both in the molten and gas phase over a wide *T* range, from ambient to at least 500 °C. Other shorter-chain sulfur allotropes, such as $S_2$, $S_3$ and $S_4$, are likely deposited from high-temperature magmatic plumes at the surface of Io, a famous Jupiter's satellite, yielding green and red colors, which rapidly transform to yellow $S_8$ (e.g., Davies and Vorburger 2023). Cyclooctasulfur and most other molecular sulfur allotropes exhibit an extremely large range of solubilities in different organic solvents, spanning from <0.1 wt% in ethanol, to 2–3 wt% in acetone and hexane, and to ~35 wt% S in carbon disulfide at ambient temperature for $\alpha$-$S_{8(s)}$ (see Meyer 1976 and Steudel 2020 for an extended list). Being hydrophobic, elemental sulfur easily forms colloids in water, so called sulfur or Weimarn sols, which are persistent for durations of days in aqueous solution at near ambient temperature, before being precipitated as $S_{8(s)}$ or further oxidized. This colloidal sulfur is rather common in sedimentary pore waters, geothermal springs and ocean where it forms by oxidation of $H_2S$, microbial activity or during fluid cooling (e.g., Jorgensen 2021). The solubility of $S_{8(s)}$ as the $S_8^0$ molecule in aqueous solution is very small at ambient conditions, ~$10^{-8}$ m of $S_8$, as measured in the few experimental studies (Boulëgue 1978; Kamyshny 2009; Wang and Tessier 2009). Like in most organic solvents, sulfur aqueous solubility increases strongly with temperature rise, but both speciation and thermodynamics of $S_n^0{}_{(aq)}$ remain poorly constrained in hydrothermal fluids.

Only two experimental studies, to the best of our knowledge, have reported the stability of molecular sulfur in aqueous solution in equilibrium with solid or molten sulfur above 25 °C. Kamyshny (2009) conducted accurate measurements of $\alpha$-$S_{8(s)}$ solubility in water from 4 to 80 °C and interpreted the data in terms of equilibrium constant of the reaction:

$$\alpha\text{-}S_{8(s)} = S_8^0{}_{(aq)} \quad (4.43)$$

Their data allowed derivation of $\Delta_f G^0$, $\Delta_f H^0$ and $S^0$ of $S_8^0{}_{(aq)}$ at 25 °C, 1 bar by fitting the $K_{43}$ values to a van't Hoff equation. Dadze and Sorokin (1993) analyzed $S^0$ concentrations as hexane extracts, along with other sulfur species such as sulfate, sulfide and thiosulfate, in aqueous solutions in equilibrium with molten sulfur from 200 to 440 °C and from $P_{sat}$ to 450 bar in a hydrothermal reactor. Their experiments produced 4 to 5 orders of magnitude higher $S^0$ concentrations than those of Kamyshny (2009), attesting increasing solubility of sulfur molecules whose exact stoichiometry could not be assessed from bulk solubility data. Pokrovski and Dubessy (2015) tentatively regressed the whole set of $S^0$ solubility data assuming reaction (4.43), and reported a set of HKF coefficients for a conventional $S_8^0{}_{(aq)}$. However, applying the HKF model to Kamishny (2009) low-temperature data only results in much lower $S_8^0{}_{(aq)}$ concentrations predicted above 250 °C (Fig. 4.17). This discrepancy points out to the predominance at elevated temperatures of molecular sulfur species other than $S_8^0{}_{(aq)}$. Their evidence is indirectly supported by the temperature trend of Dadze and Sorokin (1993) data points, which is opposite to that of $S_8^0{}_{(aq)}$. A more direct proof of such new $S_n^0$ species, with their maximum abundance around 300 °C, was provided by in situ Raman spectroscopic data in solutions in the presence of molten sulfur (Fig. 4.3), which requires, however, rather high dissolved sulfur concentrations (>0.1 wt% $S_{tot}$) atypical for natural fluids at such temperatures. With increasing temperature above 300 °C, these sulfur polymers break down to sulfur radical ions that are described as follows.

*Tri- and disulfur radical ions*

Sulfur has a rich radical chemistry, with a variety of polysulfide, sulfoxide and organic thiol radicals as intermediates in different sulfur transformation reactions both in aqueous and non-aqueous solution and gas phase. Among them, the trisulfur radical ion, $S_3^{\bullet-}$, is the most ubiquitous and well-studied radical species since the early 1970s. Compared to the trisulfide dianion ($S_3^{2-}$, and other polysulfide species discussed above), the trisulfur ion has an unpaired electron (that is why it is termed radical) and the resulting electrical charge of –1 (see Fig. 4.3 for its molecular structure). It should be noted that its formula is sometimes misunderstood or misspelled in modern literature including high-rating journals (e.g., Sverjensky 2025) by mistaking it for a triply charged $S^{3-}$ ion that does not exist in sulfur chemistry. The $S_3^{\bullet-}$ ion is stable in a variety of non-aqueous materials such as S-bearing organic and inorganic solvents, alkali halide melts, borosilicate glasses, ultramarine pigments, and zeolite-type minerals whose blue color is due to this chromophore ion (see Chivers and Elder 2013; Steudel and Chivers 2019 for comprehensive reviews). It is also an important intermediate in organic synthesis reactions and alkali metal-sulfur battery cycles (Song et al. 2023). The $S_3^{\bullet-}$ ion is easily detectable by a variety of spectroscopic techniques such as Raman, UV-Vis and EPR (see section 4.2), and has been the subject of theoretical molecular modeling studies (e.g., Tossell 2012; Blanchard et al. 2024). Although being much less stable at ambient conditions in water than in non-aqueous matrixes mentioned above, the blue chromophore $S_3^{\bullet-}$ ion has been known since long time to form on heating above 100 °C in aqueous solution of polysulfide ions and sulfur (e.g., Chivers 1974). However, its blue color and UV-Vis and Raman spectral patterns in aqueous solution in some earlier spectroscopic studies (Giggenbach 1971b; Uyama et al. 1985; Bondarenko and Gorbaty 1997) were erroneously attributed to $S_2^{\bullet-}$ or other species (see Pokrovski and Dubessy 2015 for discussion). Note that hydrothermal batch-reactor experiments in S-rich high-*T* aqueous solutions (Dadze and Sorokin 1993; Kokh et al. 2020) could not directly detect $S_3^{\bullet-}$, because of its rapid breakdown to sulfate, sulfide and molecular sulfur during fluid sampling or quench. The temperature-reversible formation of $S_3^{\bullet-}$ in aqueous solutions of thiosulfate, sulfur and sulfate-sulfide (e.g., Fig. 4.3) was unambiguously demonstrated by recent in situ Raman spectroscopy work over a wide range of *T-P* (Pokrovski and Dubrovinsky 2011; Jacquemet et al. 2014; Pokrovski and Dubessy 2015). Using this method Pokrovski and Dubessy (2015) were able to quantify the equilibrium concentrations of $S_3^{\bullet-}$ in sulfate-sulfide aqueous solutions to 500 °C and 2 kbar. Combined with previous spectroscopic work (Pokrovski and Dubrovinsky 2011; Giggenbach 1971b), they have generated the first set of equilibrium constants of the reaction of $S_3^{\bullet-}$ formation from sulfide and sulfate (Fig. 4.18)

$$2\ H_2S + SO_4^{2-} + H^+ = S_3^{\bullet-} + 0.75\ O_{2(g)} + 2.5\ H_2O \qquad K_{44} \qquad (4.44)$$

These independent data that arise from Raman measurements that employed 3 different laser wavelengths (623, 514, and 457 cm$^{-1}$), two different types of optical cell (HDAC and silica capillaty), and covered 2 log units of $S_3^{\bullet-}$ concentrations, along with a UV-Vis dataset (Giggenbach 1971b), are all in mutual agreement, confirming the $K_{44}$ value validity across the whole *T-P* range covered (200–500 °C, 15 bar–15 kbar), and with no evidence of potential artefacts related to Raman signal absorption in colored solutions as claimed in a later study (Schmidt and Seward 2017). Naturally, $S_3^{\bullet-}$ thermodynamic stability obeys the same laws as for any other aqueous species. Note that the formal reaction (4.44) was exclusively intented to report equilibrium constants within a conventional thermodynamic framework to show the dependence on the major fluid thermochemical parameters (pH, $f_{O2}$ and sulfide and sulfate activity). In no way it reflects any chemical mecanism of $S_3^{\bullet-}$ formation, as was misleadingly interpreted in some recent studies (e.g., Farsang and Zajacz 2025). The obtained thermodynamic data, both in the form of a van't Hoff-type equation and a set of HKF-model parameters compatible with the SUPCRT database, can be used for predicting $S_3^{\bullet-}$ concentrations in different types of geological fluids across wide *T* and *P* ranges, up to 700 °C and 30 kbar. It may sound surprising, but $S_3^{\bullet-}$ stands out now as one of the sulfur species that has been studied experimentally over the largest *T-P* range. Fig. 4.19 illustrates the distribution of $S_3^{\bullet-}$, together with other major sulfur species discussed above, in representative hydrothermal fluids, at 450 °C as the middle temperature of the hydrothermal range, as a function of fluid key parameters ($P$, $S_{tot}$ concentration, $f_{O2}$ and pH), but the pattern is similar in a wide $T$ range, at least from 300 to 600 °C. Reaction (4.44) can be re-arranged in terms of the sulfide/sulfate ratio as

$$19\ H_2S + 5\ SO_4^{2-} + 2\ H^+ = 8\ S_3^{\bullet-} + 20\ H_2O \qquad (4.45)$$

According to the stoichiometry of reaction (4.45), the maximum of $S_3^{\bullet-}$ concentration (or activity) at a given pH is reached at a sulfide/sulfate ratio of 19/5, i.e. at ~80% of the total sulfur as sulfide. Thus, the key parameter for $S_3^{\bullet-}$ is redox conditions of the sulfide-sulfate (±$SO_2$) coexistence, which are within the range of $f_{O2}$ between the hematite-magnetite (HM, at $T$≤500°C) and nickel-nickel oxide (NNO, $T$≥600°C) buffers. Additional factors favorable for $S_3^{\bullet-}$ are *i*) temperature above 250 °C in a wide $P$ range of liquid-like fluids ($\rho$ > 0.4–0.5 g/cm$^3$), *ii*) elevated $S_{tot}$ concentrations (> 0.5 wt%), *iii*) moderately acidic-to-neutral pH (4<pH<6) and, additionally *iv)* the excess of alkali or alkali-earth cations over chloride, offering a favorable electrical balance for the negatively charged trisulfur ion. Such conditions commonly occur in two crustal settings: magmatic-hydrothermal porphyry-epithermal Cu(-Au-Mo) systems associated with volcanic arcs, and metamorphic belts hosting orogenic Au deposits. The $S_3^{\bullet-}$ abundance in typical fluids from such settings is illustrated in Fig. 4.20.

The thermodynamic predictions of $S_3^{\bullet-}$ abundance in geological fluids at moderate temperatures (<350 °C) of Pokrovski and Dubessy (2015) are in very good agreement with independent studies that have identified $S_3^{\bullet-}$ in thermochemical sulfate reduction experiments in the presence of organic matter (Truche et al. 2014) as well as natural fluid inclusions (Barré et al. 2017). In constast, the Raman analyses of polysulfide solutions at high pressures (Schmidt and Seward 2017) overestimate, by at least a factor of 10, the $S_3^{\bullet-}$ concentrations predicted for their conditions using thermochemical data of Pokrovski and Dubessy (2015). Predictions above 500 °C, beyond the range so far covered by direct measurements strongly depend on the source of the thermodynamic properties of $H_2S_{(aq)}$ (see above subsection "Sulfide species" and detailed discussions in Pokrovski and Dubessy 2015). Realistic overall uncertainties for $S_3^{\bullet-}$ concentration estimations in most types of liquid-like hydrothermal fluids at pressures below 5 kbar, using currently available thermodynamic properties, are about ±0.5 log unit below 450 °C, ±1 log units around 500 °C and approaching ±1.5 log units toward 700 °C. Note that this is a fairly good precision compared to the other intermediate valence sulfur species discussed above. However, uncertainties of thermodynamic model extrapolations for all sulfur species increase dramatically above 700 and 5 kbars, but were often disregarded in recent modeling work (e.g., Baudry and Sverjensky 2024). In particular, the use of data sources for the major sulfur species, hydrogen sulfide, sulfate and its ion pairs, other that those that were employed in derivation of the $S_3^{\bullet-}$ thermodynamic properties, introduces a fundamental inconsistency that resulted in misleading interpretations of $S_3^{\bullet-}$ stability in magmatic fluids at shallow-crust conditions (>800 °C, 2 kbar; Farsang and Zajacz 2025; Pokrovski 2025b).

The discovery of sulfur radical species in hydrothermal fluids invigorated the interest to $S_3^{\bullet-}$ in geosciences, having been until recently limited mostly to the mineralogy of ultramarines, by opening the door to the domain of deep geological fluids and magma-fluid transition. The potentially important role of $S_3^{\bullet-}$ (and $S_2^{\bullet-}$) in sulfur transfers from silicate melts to aqueous fluids under subduction zone conditions has recently been demonstrated by direct in situ Raman spectroscopy measurements in a diamond-anvil cell in fluid-felsic melt systems at 700 °C and 3–15 kbar by Colin et al. (2020). In the aqueous fluid phase along with sulfate and sulfide both radicals were found to be abundant, in good agreement with the HKF model predictions of Pokrovski and Dubessy (2015), but in the coexisting hydrous silicate melt phase, they were close to detection limit. Remarkably, the radicals partition 10 to 1000 times more than sulfide and sulfate into the fluid phase relative to the melt (Fig. 4.21). The study has direct implications for sulfur degassing from magmas in subduction zone settings. This degassing would be enhanced, together with accompanying metals, in the redox window of the sulfate-sulfide transition where the radical ions have the highest abundance. Due to its capability to form very strong complexes with chalcophile metals such as Au, PGE and, potentially, Mo and Re, the $S_3^{\bullet-}$ ion could play a key role in the formation of economic metal resources in subduction zone settings (see section 4.4 below).

In contrast to its $S_3^{\bullet-}$ counterpart, the disulfur radical ion ($S_2^{\bullet-}$, which is a diatomic ion with an unpaired electron and a –1 charge, Fig. 4.3 and 4.5) remains much poorly constrained. Similarly, it is misspelled in some publications, by mistaking it for the sulfide doubly-charged ion ($S^{2-}$). Pokrovski and Dubessy (2015) tentatively estimated concentrations of $S_2^{\bullet-}$ in their Raman spectroscopy experiments at 450 and 500 °C and 700–1400 bar (e.g., Fig. 4.3), assuming that Raman scattering cross-sections of $S_2^{\bullet-}$ and $S_3^{\bullet-}$ are equal at 457 nm excitation (Ledé et al. 2007). The data at 450 °C were used by Pokrovski et al. (2019) to estimate, by thermodynamic speciation calculations, the equilibrium constant of the reaction of $S_2^{\bullet-}$ formation from sulfate and sulfide analogous to that of $S_3^{\bullet-}$ (4.45):

$$13\ H_2S + 3\ SO_4^{2-} = 8\ S_2^{\bullet-} + 2\ H^+ + 12\ H_2O \qquad (4.46)$$

To complete this small dataset, we used the same speciation calculations for the few points reported by Pokrovski and Dubessy (2015) to estimate the reaction constant value at 500 °C. Thus, the currently available $\log K_{46}$ values are –16.2±0.5 at 450 °C and 700 bar and –18.4±3.7 at 500 °C and 1000(±300) bar (±1 s.d.). These first values were used to estimate $S_2^{\bullet-}$ concentrations in the magmatic-hydrothermal fluid in Fig. 4.20a. Combining reactions (4.45) and (4.46), the relative trends in the $S_3^{\bullet-}$ *vs* $S_2^{\bullet-}$ equilibrium distribution can be identified:

$$4\ S_3^{\bullet-} + 4\ H_2O = 4\ S_2^{\bullet-} + 3\ H_2S + SO_4^{2-} + 2\ H^+ \qquad (4.47)$$

According to reaction (4.47), at given *T-P*, the disulfur ion would be favored over the trisulfur ion at *i)* more alkaline pH, *ii)* lesser total S concentrations dominated by sulfate and sulfide, and *iii)* slightly more reduced (i.e. more $H_2S$) fluids. The pH trend has been clearly confirmed by Colin et al. (2020) showing that $S_2^{\bullet-}$ was more abundant in fluids that were in equilibrium with more alkaline silicate melts. The tentative $S_2^{\bullet-}$ trend with *T* shown in Fig. 4.20a is supported by general entropy considerations that higher temperatures (i.e., more disorder) favor smaller and more compact species. Thus, $S_2^{\bullet-}$ forming at the expense of $S_3^{\bullet-}$ may become an important sulfur species at magmatic temperatures. Larger radical congeners, $S_n^{\bullet-}$ with $n$ = 4–8, have recently been envisioned by theoretical molecular modeling but await physical evidence (Chivers and Oakley 2023).

### *4.3.5 Sulfidation state concept and sulfur aqueous speciation*

A long-standing simple concept in mineral deposit research uses the notion of sulfur fugacity, $f_{S2}$, to describe sulfide mineral reactions and associations found in hydrothermal ore deposits. Although $S_2$ is not a dominant sulfur species in the natural gas phase and is non-existent in aqueous solution, it is a useful species to portray mineral equilibria in the framework of a simple thermodynamic formalism, similar to the use of oxygen fugacity ($f_{O2}$). The quantitative use of $f_{S2}$ has been facilitated by the existence of robust thermodynamic data on sulfide mineral equilibria acquired in anhydrous systems since the 1950s. These data allowed quantifying mineral stability domains and equilibria in terms of temperature, $f_{S2}$ and $f_{O2}$ (Holland 1959, 1965; Barton 1970; Barton and Skinner 1979; references therein). They constitute the basis of the sulfidation state concept reviewed and further developed in the seminal paper of Einaudi et al. (2003). According to it, the different sulfide mineral associations observed in hydrothermal ore deposits are a function of the $f_{S2}$ value, i.e., sulfidation state, as pictured in Figure 4.22. It is commonly qualified being in between "very low" and "very high" by ore deposit geologists, according to the major Fe and Cu mineral assemblages present, as exemplified by the formal equilibria (Fig. 4.22):

$$Fe_{(iron)} + S_{2(g)} = FeS_{(pyrrhotite)}\text{, very low} \qquad (4.48)$$

$$FeS_{(pyrrhotite)} + FeAs_{2(löllingite)} + ½\ S_{2(g)} = 2\ FeAsS_{(arsenopyrite)}\text{, low} \qquad (4.49)$$

$$FeS_{(pyrrhotite)} + ½\ S_{2(g)} = FeS_{2(pyrite)}\text{, intermediate} \qquad (4.50)$$

$$5\ CuFeS_{2(chalcopyrite)} + S_{2(g)} = Cu_5FeS_{4(bornite)} + 4\ FeS_{2(pyrite)}\text{, high} \qquad (4.51)$$

$$2\ S_{(liquid)} = S_{2(g)}\text{, very high} \qquad (4.52)$$

Sulfidation state goes along with oxidation state, with lower $fS_2$ generally corresponding to lower $fO_2$ or higher $fH_2$ (both being in equilibrium with $H_2O$ in the hydrothermal fluid or vapor), which led to the use of the redox parameter, RH = log ($fH_2/fH_2O$), and sulfidation parameter, RS = log ($fH_2/fH_2S$), as alternatives of $fS_2$ and $fO_2$, to interpret vapor compositions and mineral associations in active geothermal systems (Giggenbach 1987; Einaudi et al. 2003). Figure 4.22 pictures different types of hydrothermal deposits, illustrating the different fluid environments in terms of temperature and $fS_2$ parameters whose detailed evolution may be found in Einaudi et al. (2003).

However, the $fS_2$-$fO_2$ thermodynamic formalism has obvious limitations in interpreting the true sulfur chemical state and related mineral solubility in the fluid phase, which requires knowledge of major sulfur and metal aqueous species. Figure 4.23 shows the concentrations of major sulfur species and dissolved Fe and Au in the fluid as a function of temperature, along a typical pressure gradient in shallow-crust sulfide deposits, in equilibrium with iron mineral assemblages representative of low (quartz-fayalite-magnetite-pyrrhotite), intermediate (magnetite-pyrrhotite-pyrite), and high to very high (pyrite-sulfur) sulfidation states. These calculations use the thermodynamic data for sulfur discussed above and referenced in Table 4.5; those for metals will be detailed in the following section (see also Table 4.6). It can be seen in Fig. 4.23 that dissolved sulfur aqueous concentrations, dominated by $H_2S$, are similar between the low and intermediate sulfidation states as imposed by the mineral solubility. Aqueous $H_2S$ increases by more than $10^4$ times from 100 to 600 °C, along with the increase at elevated temperatures of more oxidized S species such as sulfate (representing the sum of all sulfate and hydrosulfate species and their alkali ion pairs), $SO_2$ and $S_3^{\bullet-}$, with sulfate and $SO_2$ becoming comparable to $H_2S$ at the highest temperatures at intermediate sulfidation states. For the high sulfidation states, such as represented by native sulfur(+ pyrite), sulfur solubility is much greater, attaining mol/kg fluid levels above 400 °C, along with a growing abundance, at the expense of sulfate and sulfide, of intermediate valence sulfur species, $SO_2$ and $S_3^{\bullet-}$. At all sulfidation states, Fe solubility (as chloride complexes) is comparable above 300–400 °C attaining ~1000 ppm levels, but the temperature dependence is different at high and lower sulfidation states, invoking different potential precipitation mechanisms (cooling vs fluid oxidation or sulfur scavenging by Fe-bearing rocks). Gold solubility (as hydrosulfide complexes), in contrast, is greatly favored by both temperature and sulfidation (and oxidation) state, because of the increasing $H_2S$ ligand concentrations and, at high sulfidation state, by the growing abundance of $S_3^{\bullet-}$ that is also a strong ligand for gold. Thus, full appreciation of the sulfidation state concept, based on mineral associations, may only be achieved by its combination with detailed sulfur and metal chemistry in hydrothermal fluids. An in-depth account of metal speciation and mineral solubility is given in the next section.

## 4.4 Metal speciation and sulfur-metal relationships in hydrothermal fluids

### *4.4.1 General tendencies of metal-ligand complexation in aqueous solution*

*Fundamentals of metal speciation in the fluid phase*

To appreciate the role of metal complexes with sulfur in hydrothermal fluids it is necessary to place the metal-sulfur interaction issues into a more general picture of mineral solubility and metal speciation in the fluid phase, with the simultaneous presence of other geologically ubiquitous non-sulfur ligands, in particular chloride, hydroxide and alkalis as well as some other compounds (e.g., carbonate, organic carbon, fluoride, bromide). In this section, we briefly outline the basics of metal-ligand complexation and general tendencies, before providing an in-depth account of complexes of different groups of metals with various sulfur ligands.

Chemical elements are dissolved in aqueous solution in the form of cations (electrically positively charged ions formed by most metals, e.g., $Ag^+$, $Fe^{2+}$, $Zn^{2+}$, $Al^{3+}$) and anions (negatively charged ions formed by nonmetals, usually called ligands, e.g., $Cl^-$, $SO_4^{2-}$, $HS^-$). The ions are all coordinated by water molecules to some extent that can be directly "seen" by some spectroscopic methods (e.g., X-ray absorption or nuclear magnetic resonance spectroscopy). For example, $Zn^{2+}$ has a clearly defined first-shell atomic sphere with 6 water molecules by forming Zn-$OH_2$ chemical bonds, $Zn(H_2O)_6^{2+}$, along with more loosely bound outer-sphere $H_2O$ molecules. Anions, like $Cl^-$ or $HS^-$, are much weaker solvated by $H_2O$ with no clearly defined coordination spheres. Both types of ions associate together in some extent, usually by displacing a water molecule from the metal first coordination sphere and forming a direct chemical bond (e.g., Zn-Cl):

$$Zn(H_2O)_6^{2+} + Cl^- = ZnCl\,(H_2O)_5^+ + H_2O \quad (4.53)$$

From a thermodynamic point of view, this reaction may be simplified to

$$Zn^{2+} + Cl^- = ZnCl^+ \quad (4.54)$$

by omitting water molecules in this reaction because water activity is close to one in an aqueous fluid phase.

With increasing salt content in the fluid and raising temperature, such associations become more abundant, and more ligands may be bound to the cation (e.g., forming $ZnCl_2^0$ and higher-order chloride complexes, $ZnCl_3^-$ and $ZnCl_4^{2-}$). All cations, anions and their complexes are present simultaneously in the aqueous fluid phase, but at different concentrations (or activities). This fluid propery contrasts to pure mineral phases (e.g., pyrite, sphalerite, or hematite) whose thermodynamic activity is either 1 or 0 and their formation and disappearance in *T-P*-composition space is governed by the phase rules. There is thus a fundamental physical-chemical difference between a phase and a species, often misunderstood and sometimes misinterpreted by geologists (e.g., Pokrovski 2014). The total metal content in the fluid phase is the sum of all species concentrations:

$$\Sigma Zn_{(aq)} = Zn^{2+} + ZnCl^{+} + ZnCl_2^{0} + ZnCl_3^{-} + ZnCl_4^{2-} \qquad (4.55)$$

To this equation should also be added other, non-chloride, species (e.g., such as hydrosulfides and sulfates, see sections below). However, among the multitude of possible aqueous species for a given metal, there is usually one or a couple that is dominant (i.e. most concentrated) within the given *T-P* and fluid composition range, whereas all other species are often much less abundant. Their contribution to the total metal amount transported by the fluid would be very small and may therefore be neglected. With the temperature rise, two general changes in the fluid properties occur such as *i)* the increase of thermal motion and, therefore, disorder, along with *ii)* the decrease of dielectric constant of the fluid phase. In many cases, these changes lead to a simplification of the metal speciation schemes in the high-temperature fluids compared to near-ambient aqueous solution. Thus, neutral and weakly charged complexes (e.g., $ZnCl_2^0$, $Zn(HS)_2^0$) are generally favored due to the weakening of Born solvation contribution to the species Gibbs energy when $\varepsilon$ decreases (equation 18). Similarly, the increasing thermal disorder renders more favorable species of lower entropy (i.e. with more compact and more symmetrical geometries, such as $ZnCl_2^0$ compared to $ZnCl_3^-$). However, these solvent-controlled tendencies are complicated by other factors such as specific electronic structure of the chemical bond, metal coordination and steric constraints, and relativistic electron effects, inherent to particular metals as will be selectively discussed in the following subsections. For more extended general literature on the topic concerning different types of ligands the reader is referred to Seward (1981), Brimhall and Crerar (1987), Wood and Samson (1988), and Barnes (1997).

*Soft-hard relationships*

What are the dominant complexes for a given metal in hydrothermal fluids? Their nature is governed by two factors: the metal chemical affinity for a particular ligand and the abundance of the ligand itself in the fluid. Metal-ligand affinities are fundamentally defined by the electronic structure of the complexing elements. These affinities may be predicted using the so called soft-hard classification of chemical elements (hard and soft acid and base theory, HSAB) appeared in chemistry in the 1960s (Pearson et al. 1963; Pearson 1997 and references therein). While behind this theory lie numerical functions such as ionization potential, bond dissociation energy and electronegativity, the concept in itself is simple and elegant. It is outlined below for a general reader who is also referred to Crerar et al. (1985), Brimhall and Crerar (1987), Wood and Samson (1988), and Barnes (1997) for more detail and variations of presentation. According to the HSAB classification illustrated in Fig. 4.24, metal cations (called acids) are divided in three categories. The first, shown in red, is represented by either small and strongly charged ($Be^{2+}$, $B^{3+}$, $Zr^{4+}$, $Nb^{5+}$) or big and easily ionizable ($Cs^{+}$, $Ba^{2+}$). These properties make such elements "hard", with their electronic clouds hard to deform (i.e. polarize) to share electrons with ligands in a chemical bond. Such chemical bonds are termed ionic and are ensured by dominantly electrostatic attraction between the charged cation and anion spheres. The second category, shown in blue, is constituted by large, weakly charged, and strongly polarizable cations ($Ag^{+}$, $Au^{+}$, $Pt^{2+}$, $Cu^{+}$), with electronic clouds easy to deform to enable sharing electrons with a ligand in a chemical bond. Such bonds are termed covalent. The intermediate category of metal cations lies in between these two types (e.g., $Fe^{2+}$, $Zn^{2+}$, $Co^{2+}$, $As^{3+}$) and shown in yellow in Fig. 4.24. Analogous three types are applicable to non-metals (or ligands), with $O^{2-}$, $OH^{-}$, $F^{-}$, $C^{4+}$ (as carbonate $CO_3^{2-}$) being hard, $I^{-}$, $Te^{-}$, $Br^{-}$ – soft, and $Cl^{-}$ – intermediate. The fundamental rule is the following: a soft "loves" a soft and a hard "loves" a hard. It means that hard metals will preferably complex with hard ligands, and soft metals – with soft ligands. The degree of hardness (or softness) is also dependent of the element oxidation state, with higher oxidation states being harder. Sulfur is an excellent example of such dualism because it forms both very soft sulfide ($S^{2-}$) and very hard sulfate ($S^{6+}$). Similarly, arsenic and antimony in their highest oxidation state of +5 as arsenates and antimonates form almost exclusively oxyhydroxides both in aqueous solution and solid phase at the Earth's surface. In oxidation state +3, stable in reduced sedimentary settings and geothermal systems, both oxyhydroxide and sulfide types of complexes and minerals are known for those elements. The most reduced formal states –1 to –3, as arsenides and antimonides, are soft ligands similar to sulfide. Although qualitative, this classification not only elegantly explains the predominance of the different complexes in aqueous solution, but it also accounts for the general distribution of elements on Earth, since it is applicable to solid phases as well. Remarkably, the soft-hard classification strongly resembles the geochemical classification of elements into lithophiles (i.e. $SiO_2$-loving and, therefore, hard), chalcophiles (i.e. copper-loving meaning reduced sulfur-loving such as in the main copper ore chalcopyrite, $CuFeS_2$, and, therefore, soft), and siderophiles (iron-loving and, therefore, intermediate). Thus, a hard metal cation, like $La^{3+}$, representative of the REE group, forms the most stable complexes with the hardest ligand of the halogen group, $F^{-}$, while a moderately soft metal, $Ag^{+}$, forms the strongest complexes with $I^{-}$.

However, it does not mean that fluoride and iodide complexes, respectively, would be the dominant types transporting REE and Ag by most hydrothermal fluids. Another parameter, the ligand concentration itself must be taken into account. Aqueous fluoride is a relatively minor component of most hydrothermal fluids because of poor solubility of its solid phases (e.g., fluorite, apatite), and its strong affinity for silicate minerals and melts rather than for aqueous fluids. Iodine concentrations are usually negligibly small in the fluids due to its very low natural abundance (average I concentration in the upper continental crust is only ~0.1 ppm; Rudnick and Gao 2014). Consequently, neither $F^–$ nor $I^–$ ligands are present in sufficient concentrations to complex the metals above. As a result, both types of metals are transported predominantly as chloride complexes due to the large abundance of $Cl^–$ in most types of hydrothermal fluids. However, this general picture may be complicated by additional complexing with sulfate and sulfide for REE and Ag, respectively, in some types of hydrothermal fluids as will be discussed below.

An important advantage of the hard-soft classification is also the ability to qualitatively predict tendencies in metal speciation with temperature in light of the competition among the different ligands. This is because the thermodynamics of hard and soft types of bonding are different. For hard-hard interactions dominated by electrostatic attractions, the increase in temperature and concomitant decrease in dielectric constant generally reinforce bonding according the Coulomb electrostatic law. This dielectric constant control results in the increase of stability constants with temperature for most hydroxide, chloride and sulfate metal complexes. In contrast, soft-soft interactions, being covalent and strongly steric-controlled, generally weaken with the increase of thermal disorder. Thus, compared to chloride and hydroxide complexes, hydrosulfide complexes of many intermediate and moderately soft metals (Zn, Cu, Ag, As, Sb) are generally significant only in low-temperature hydrothermal and sedimentary fluids, but become almost negligible in higher-temperature hydrothermal-magmatic fluids. However, it should not be forgotten that this general knowledge is based on hard-soft principles established using metal-ligand complexation data in aqueous solution at ambient conditions. The soft-hard properties of elements themselves may change with temperature and pressure. Recently, based on the revised and corrected scale of the electronegativity of chemical elements of Tantardini and Oganov (2021), Dong et al. (2022) theoretically evaluated the effect of very high pressures (up to 500 GPa) on the chemical hardness of elements. They found that the degree of hardness changed differently depending on the group of elements thereby leading to surprising changes in the element reactivity compared to ambient pressures. These theoretical predictions appear to be consistent with experimental studies relevant to planetary interiors (e.g., pressure induced metallization). The effect of temperature yet remains to be quantitatively assessed, which might challenge the traditional view outlined above.

*General overview of metal-transporting complexes in geological fluids*

Our current knowledge of the dominant complexes of different metals in typical liquid-type hydrothermal fluids is summarized in Table 4.6. This knowledge is the result of a large body of experimental, spectroscopic and theoretical studies since the 1950s that have established the stoichiometry of the dominant metal complexes and generated their stability constants. For details on the methods and complexation specificities depending of the element, the reader is referred to the representative papers cited in Table 4.6 as well as to more general reviews (Brimhall and Crerar 1987; Barnes 1997; Wood and Samson 1998; Kouzmanov and Pokrovski 2012; Pokrovski et al. 2014; Brugger et al. 2016; Migdisov et al. 2016). Studies of the chemical speciation of different metals in aqueous fluids at elevated temperatures and pressures benefiting of state-of-the-art approaches, such as solubility measurements, in situ spectroscopy and molecular modeling, regularly appear in the geochemical literature, and our vision of some elements hydrothermal chemistry and behavior continuously evolves as will be detailed for major ore metals in the following sections.

The current general situation is resumed below. In typical hydrothermal fluids within the *T*-*P*-density range of liquid-like to supercritical fluids, most metalloids (As, Sb, Si) dominantly form uncharged hydroxide complexes. The speciation of transition and base metals (Fe, Cu, Ag, Zn, Pb) is generally dominated by chloride species. "Hard" metals (Al, Zr, Nb, Ti, Cr) mostly form hydroxide complexes. For rare earth elements (REE), chlorides and sulfates are likely to be most important while hydroxide and fluoride are rather subordinate. "Soft" metals (Au, PGE) largely prefer the hydrosulfide ($HS^-$) and trisulfur radical ion ($S_3^{\bullet-}$) ligands. In addition to the exact complex stoichiometry cited in Table 4.6, our ability to predict metal solubilities equally depends on the robustness of the values of stability constants for all these complexes. This parameter may be expressed in terms of the uncertainty associated with thermodynamic calculations of ore and gangue mineral solubilities within the typical *T-P* domain of hydrothermal fluids (<500 °C, <5 kbar). The solubility of the major solid phases of metals and metalloids like As, Au, Si, Al, Cu, Zn, and Pb in hydrothermal fluids may reasonably be predicted within 0.5–1 log units. The equilibrium predictions are much less precise for REE, which form different complexes depending on the context, and more importantly, different silicate, oxide and fluoride solid phases and solid solutions whose thermodynamic properties are poorly known, bringing additional uncertainty. The "worst" situation is probably for Mo (and its geochemical twin, Re, not shown) whose complexes with major ligands (such as Cl and S) are yet very poorly known. Modeling trace elements that do not usually form their pure mineral phases, but rather get incorporated into major minerals in chemically bound states is virtually non-existent to date because of the lack of a robust thermodynamic framework for such mineral-hosted trace elements. The uncertainty of thermodynamic predictions generally grows with increasing *T* and *P*. Note that a typical error of 1 log unit intrinsic to best known metals and minerals may be considered as a fairly good precision in the face of much larger variability of natural metal concentrations that commonly display variations of several orders of magnitude as demonstrated by natural fluid inclusion studies (see section 4.5 and Yardley 2005; Kouzmanov and Pokrovski 2012; Pokrovski et al. 2014).

*Solubility-controlling reactions and the effect of sulfur*

Table 4.6 shows that the great majority of base metals and metalloids forming hydrothermal sulfide ore (As, Sb, Zn, Pb, Fe, Cu, Ag, Mo) do not predominantly form complexes with sulfur in the fluid phase under typical ore forming conditions in the Earth's crust, whereas sulfide minerals are omnipresent solid phases at such conditions. There is, therefore, an important divide between the hydrothermal fluid phase and the solid phase as to the global affinity of metals to sulfur. This difference is due to the presence of water as the major solvent on Earth that imposes controls on elements' state, hydration, bonding and ligand composition and activity as shown above. Thus, in addition to the intrinsic chemistry of the dominant dissolved species, the second key factor governing metal solubility in the fluid is the identity and stability of the metal-bearing solid phase as shown by the following reactions for selected ore metals:

$CuFeS_{2(s)} + 3\ H^+ + 4\ Cl^- + ½\ H_{2(g)} = CuCl_2^- + FeCl_2^0 + 2\ H_2S_{(aq)}$ (4.56)

$FeS_{2(s)} + 2\ H^+ + 2\ Cl^- + H_{2(g)} = FeCl_2^0 + 2\ H_2S_{(aq)}$ (4.57)

$½\ Ag_2S_{(s)} + H^+ + 2Cl^- = AgCl_2^- + ½\ H_2S_{(aq)}$ (4.58)

$ZnS_{(s)} + 2\ H^+ + n\ Cl^- = ZnCl_n^{2-n} + H_2S_{(aq)}$, $n = 2, 3, 4$ (4.59)

$PbS_{(s)} + 2\ H^+ + n\ Cl^- = PbCl_n^{2-n} + H_2S_{(aq)}$, $n = 2, 3, 4$ (4.60)

$MoS_{2(s)} + 4\ H_2O = HMoO_4^- + H^+ + 2\ H_2S_{(aq)} + H_{2(g)}$, (4.61)

$MoS_{2(s)} + 2\ H_2O + Na^+ = NaHMoO_2S_2^0 + H^+ + H_{2(g)}$, *magmatic fluids, tentative* (4.62)

$MoS_{2(s)} + 2\ H_2S_{(aq)} = MoS_4^{2-} + 2\ H^+ + H_{2(g)}$, *in sulfide-rich fluids below 300 °C* (4.63)

$Au_{(s)} + H_2S_{(aq)} + HS^- = Au(HS)_2^- + ½\ H_{2(g)}$, *in near-neutral and alkaline fluids* (4.64)

$Au_{(s)} + H_2S_{(aq)} + S_3^{\bullet-} = Au(HS)S_3^- + ½\ H_{2(g)}$, *in acidic-to-neutral S-rich fluids* (4.65)

$Au_{(s)} + H_2S_{(aq)} = AuHS^0 + ½\ H_{2(g)}$, *in acidic S-poor fluids* (4.66)

$Au_{(s)} + H^+ + 2\ Cl^- = AuCl_2^- + ½\ H_{2(g)}$, *in acidic, saline, $H_2S$-poor fluids above 500 °C* (4.67)

In these reactions, all species are in aqueous state (unless indicated), $H_2S_{(aq)}$ is the dominant S aqueous form and $H_{2(g)}$ is the conventional redox gaseous species related to oxygen fugacity by the chemical equilibrium with water: $H_{2(g)} + ½\ O_{2(g)} = H_2O_{(l)}$.

It can be seen from these reactions, sulfur exhibits an important effect on the mobility of most metals, including those that do not form direct complexes with sulfur in the fluid phase but do form sulfide minerals. Because dissolved $H_2S$ is directly involved in these reactions, the solubility of all metals strongly depends on $H_2S$ content of the fluid phase, with amplitudes comparable to those exerted by salt concentration, pH or redox, depending on the stoichiometric coefficients of these constituents in the solubility reaction. Figure 4.25 shows the solubility of major ore minerals as a function of four key parameters, temperature, acidity, salinity, and $H_2S$ concentration, calculated using the available thermodynamic data for minerals and aqueous species cited in Table 4.6 (note that Mo solubility curves are tentative since the effect of chloride, alkalis and other sulfur ligands is currently poorly constrained (see section 4.4.10). The differences in aqueous speciation between base metals plus Ag (chlorides), Au (predominantly sulfides), and Mo (oxyhydroxides, alkali ion pairs and possibly sulfides in sulfur-rich moderate-temperature systems) yield contrasting solubility trends versus pH, $H_2S$, and salinity for these three groups of metals (Fig. 4.25). The only feature common to all metals is an increase in solubility with temperature, for sulfur-, redox-, and pH-buffered systems, typical for rock-dominated porphyry-epithermal environments (e.g., Giggenbach 1997; Kouzmanov and Pokrovski 2012). The metal concentrations are highly variable depending on salinity, pH, and S content. For acidic-to-neutral fluids ($4 < pH < 6$) of moderate salinity (10 wt% NaCl equivalent) and S contents (<0.5 wt%), the absolute dissolved concentrations of Cu, Fe, Zn (and Pb, not shown) in equilibrium with their respective sulfide minerals are generally higher than for Au and Mo. However, this trend may be inversed in S-rich, low-salinity and alkaline pH fluid compositions. For all metals except Au, solubility increases significantly with salinity, the largest increase being for Zn and Pb (not shown), which likely form tri- and tetrachloride species (Table 4.6, section 4.4.2). Fluid pH exerts a comparably strong negative effect on base metal solubility as $H_2S$ content does. In contrast, both parameters (pH and $H_2S$) have a positive effect on Au solubility in near-neutral and S-rich fluids. Molybdenite solubility increases with increasing pH, but decreases with increasing $H_2S$; however Mo solubility predictions are affected by much larger uncertainties than the other metals because of poorly known contribution of chloride and sulfur-type complexes (see section 4.4.10).

The generalized trends shown in Fig. 4.25 illustrate the complexity of mineral solubility that is both dependent on the exact metal speciation that determines the controlling dissolution/precipitation reactions, and on the concentrations of other reaction constituents defined by the specific fluid composition. A few in-depth reviews of the identity and thermodynamic stability of complexes of metals and the solubility of their minerals in hydrothermal fluids were published in the 1970–1990s (e.g., Barnes 1979; Crerar et al. 1985; Wood and Samson 1998). The sections below provide an extended update of the speciation data for major ore metals obtained since then, in particular using cutting-edge methods and combined approaches that were unavailable in the past. The role of sulfur cannot, of course, be regarded as standalone without assessment of the role of other competing complexes, in particular chlorides and hydroxides, in the whole picture of the aqueous speciation for a metal. That is why non-sulfur aqueous complexes will also be overviewed and compared here for each selected metal.

### *4.4.2 Base metals - Zn, Pb, and Cu*

Although this group of metals is believed to form dominantly chloride complexes in typical liquid-like hydrothermal fluids (Table 4.6), the role of sulfur-bearing complexes remains a subject of great controversy for some of those metals (Zn) and appears to be more important in high-temperature fluids for others (Cu).

*Zinc*

Although zinc chloride and sulfide complexes in hydrothermal fluids have been extensively studied experimentally since the 1960s, this metal provides a rather paradoxical example that the more data are accumulated, the more discrepancy is added to the topic, in particular for sulfide complexes. Chloride complexes in hydrothermal fluids have been investigated in numerous solubility studies critically reviewed by Sverjensky et al. (1997), Wood and Samson (1998), Bazarkina (2010), and Akinfiev and Tagirov (2014). To these works, have recently been added studies that used X-ray absorption spectroscopy and molecular dynamics (e.g., Mei et al. 2015). The major common outcome of those studies is the identification of four chloride species in saline hydrothermal fluids ($ZnCl^+$, $ZnCl_2^0$, $ZnCl_3^-$, $ZnCl_4^{2-}$) and generation of their stability constants. There currently exist two HKF model (Sverjensky et al. 1997; Akinfiev and Tagirov 2014) and one RB model (Mei et al. 2015a) set of coefficients derived from different experimental studies and allowing extrapolation of these stability constants over the hydrothermal *T-P* range. The resulting distribution of Zn-Cl species is somewhat different in temperature-composition space, with the Sverjensky et al. (1997) dataset predicting the predominance of $ZnCl_2^0$ at temperatures >300-350 °C and salinities >1 wt% NaCl, whereas both Akinfiev and Tagirov (2014) and Mei (2015a) sets favor $ZnCl_4^{2-}$ at such conditions. Despite these discrepancies, the overall solubility of the major Zn-bearing minerals ($ZnS_{(s)}$ and $ZnO_{(s)}$) consistently shows a large increase with increasing temperature and salinity (Fig. 4.25), with 100s to 1000s ppm of Zn that may be transported by moderate-temperature fluids (~300 °C) and with close-to-wt% Zn concentration levels in magmatic fluids (>500 °C).

The situation with sulfide complexes is less consistent. Numerous low-temperature studies have reported a plethora of Zn-S(-OH) polymeric complexes that will not be discussed here (see

Wood and Samson 1998; Rickard and Luther 2006). In contrast, only a handful of studies is available at hydrothermal conditions, which is primarily due to intrinsic experimental difficulties for working in reduced sulfur systems and much lower ZnS solubility (<ppm level) posing analytical challenges. The most remarkable studies are those of Bourcier and Barnes (1987) and Hayashi et al. (1990) who measured $ZnS_{(s)}$ solubility to 350 and 240 °C, respectively ($P_{sat}$), in concentrated $H_2S$-NaHS solutions (to 3–4 m). They suggested $Zn(HS)_2^0$, $Zn(HS)_3^-$, and $Zn(HS)_4^{2-}$ to be the solubility-controlling complexes at acidic to neutral pH, whereas at alkaline pH and/or low temperatures, mixed Zn-HS-OH and deprotonated Zn-S species were inferred (e.g., $Zn(OH)(HS)^0$, $Zn(OH)(HS)_2^-$, and $Zn(OH)(HS)_3^{2-}$). Almost 20 years after, Tagirov et al. (2007) and Tagirov and Seward (2010) measured sphalerite solubility in less concentrated $H_2S$-NaHS solutions (<0.2 m) to 250 °C and derived the same set of three major complexes, $[Zn(HS)_{2,3,4}]$, along with a deprotonated sulfide complex, $Zn(HS)S^-$, only significant in alkaline (pH>8–9) solutions. Note that in the absence of direct spectroscopic evidence, the $Zn(HS)S^-$ species is thermodynamically equivalent to $Zn(OH)(HS)_2^-$, proposed by Hayashi et al. (1990), since they differ by one $H_2O$ molecule. Tagirov and Seward (2010) generated a RB equation coefficients for the four species derived that may be used with reasonable confidence to 300–350 °C. The absolute values of Tagirov's stability constants for the three Zn-HS species at temperatures above 100 °C are, however, up to 2–3 orders of magnitude lower than in the previous work above, clearly reflecting the difficulties of measuring and interpreting ppb-level metal solubilities. Akinfiev and Tagirov (2014) re-fitted Tagirov data within the framework of the HKF model. To allow better constraints on the HKF coefficients, they assumed that the thermodynamic functions at 25 °C and 1 bar ($\Delta_r S^0$, $\Delta_r Cp^0$, and $\Delta_r V^0$) of the isocoulombic reactions of ligand exchange with equivalent previously HKF-processed Zn-Cl complexes are equal to zero. This assumption should, however, be considered with caution because metal affinity for $Cl^-$ vs $HS^-$ ligands is very different (see below). A more recent word to the Zn-S complex issue has been added by Mei et al. (2016) who calculated, using ab-initio molecular dynamics in simulation boxes with Zn/HS ratios of 4 to 10, both geometry and energy of Zn-HS complexes with 1 to 4 $HS^-$ ligands to 600 °C. Combining these data with limited XAS measurements in 2 m NaHS solutions to 500 °C, they concluded that $Zn(HS)_3^-$ and $Zn(HS)_4^{2-}$ are the dominant species at these conditions. They further fitted their MD-generated stability constants with the RB model. Their stability constants in moderate-temperature hydrothermal fluids (<400 °C) for the two complexes above appear, however, 3 to 4 orders of magnitude higher than those of Tagirov's group. This discrepancy may be due to both intrinsically high uncertainties of the MD thermodynamic integration approach (e.g., see Laskar et al. 2022) and the limited-temperature range solubility dataset (<250 °C) which was used for HKF and RB models parameterization. Therefore, any prediction for Zn-HS types of complexes at temperatures above 300–350 °C should be taken with extreme caution.

The existing data may further be rationalized using isocoulombic ligand exchange reactions between the major chloride and hydrosulfide complexes (Fig. 4.26):

$$ZnCl_n^{2-n} + n\,HS^- = Zn(HS)_n^{2-n} + n\,Cl^-, n=2, 3, 4 \quad (4.68)$$

For consistency, the data for hydrosulfide complexes were taken from Tagirov and Seward (2010), Akinfiev and Tagirov (2014) and Mei et al. (2016), whereas the corresponding data for equivalent chloride complexes are from Akinfiev and Tagirov (2014) and Mei et al. (2015a), respectively. Three major features are apparent in Fig. 4.26. First, the chemical affinity of Zn for the $HS^-$ ligand is much greater than for the $Cl^-$ ligand (up to 4–5 log unit per ligand at ambient temperature, in terms of the exchange reaction constant whose absolute value is normalized to the number of ligands). Second, the Zn affinity for reduced sulfur decreases systematically with increasing temperature, in agreement with the general soft-hard relationships discussed above. Finally, there is large discrepancy in the ligand affinity between Tagirov and Mei groups' data. Adopting a typical $Cl^-/HS^-$ ratio of $10^3$ to $10^4$ in near-neutral pH (pH 5–6) fluids around 300 °C (i.e. typical temperature of ZnS deposit formation), the $K$ values reported in Fig. 4.26 imply that chloride complexes are thousands to millions times more abundant than their sulfide counterparts according to Tagirov's data set, whereas the two types of complexes may have comparable abundances according to Mei's data set.

Reconciling these discrepancies would require more systematic experimental data on both Zn sulfide and chloride types of complexes over a larger $T$-$P$ range. In the meantime, we would recommend using Sverjensky et al. (1997) or Akinfiev and Tagirov (2014) HKF model parameters for Zn-Cl complexes because they are based on a large number of studies and more robust and validated HKF relationships than those of the RB model. For Zn-HS complexes, the solubility data of Tagirov and Seward (2010) should be privileged at present, being most direct and analytically accurate ones, with caution about the stability constant values extrapolation above 350 °C.

*Lead*

Complexes of lead in hydrothermal fluids have received less in-depth investigations than those of its counterpart, zinc. The lower $PbO_{(s)}$ and $PbS_{(s)}$ solubility than that of the corresponding Zn solid phases might have contributed to discouraging more systematic experimental work. Like for Zn, the major Pb-bearing complexes in saline hydrothermal fluids are chlorides, investigated in the 1960s–1990s by solubility and UV-Vis spectroscopy, as reviewed in detail by Wood and Samson (1998). Sverjensky et al. (1997) have generated a consistent set of HKF coefficients for a series of Pb-Cl complexes with 1 to 4 $Cl^–$ ligands based on low-temperature studies and Seward (1984) UV-Vis work to 300 °C. The three species, $PbCl^+$, $PbCl_2^0$ and $PbCl_3^–$, showed formation constant values consistent among those works, while the tetrachloride, $PbCl_4^{2–}$, was much less certain. According to these and other limited higher-temperature solubility work (e.g., Hemley et al. 1992), $PbCl_2^0$ and $PbCl_3^–$ are the dominant species in a wide range of temperature and salinity. A more recent UV-Vis study of Luo and Millero (2007) to 40 °C demonstrated a good agreement for the mono-, di- and tri-chloride species, but did not detect $PbCl_4^{2–}$. The latter complex was also rejected by Etschmann et al. (2018) on the basis of X-ray absorption spectroscopy measurements of saline solutions to 10 m NaCl and ~500 °C, supplemented by MD simulations. Their data indicate the predominance of $PbCl_3^–$ with the possible presence of one or two $H_2O$ molecules in the Pb first coordination sphere. Compared to Zn, for which $ZnCl_4^{2–}$ is the major species, the absence of the Pb tetrachloride complex was explained by the presence of stereochemically active lone electron pair ($6s^2$) hampering additional coordination by $Cl^–$. Etschmann et al. (2018) further refitted HKF-model parameters for the first three Pb-Cl complexes based on previous work (Seward 1984; Hemley et al. 1992; Luo and Millero 2007). While the resulting stability constants for Pb-Cl complexes are very similar to those of the original HKF model of Sverjensky et al. (1997), which was partly based on the same data sources, some of the coefficients reported by Etschmann et al. (2018) appear to be unphysical. Those include, for example, large negative Born parameter values (ω) that contradict to the HKF model fundamentals that ω values for charged species must be positive (e.g., Pokrovseki et al. 2024), along with inconsistencies in $a_3$ and $a_4$ coefficients that show either zeroed values or anomalously large negative values for some complexes (e.g., $\omega = –0.83\times10^5$ cal/mol, $a_3 = –232$ cal/(mol bar), and $a_4 = 0$ cal/mol for $PbCl_2^0$). For these reasons, the Etschmann et al. (2018) HKF dataset cannot be recommended, and the HKF parameters of Sverjensky et al. (1997) for the mono-, di- and trichloride complexes should be used in Pb speciation and solubility modeling.

Much more limited data exist for lead sulfide complexes. Sverjensky et al. (1997) produced the first set of HKF coefficients for $Pb(HS)_2^0$ and $Pb(HS)_3^–$ based on the formation constants for those complexes from the solubility study of Hemley (1953) at ambient conditions. Remarkably, these HKF-model predictions are in good agreement, within 1 log unit, with the later solubility work of Giordano and Barnes (1979) who reported, from $PbS_{(s)}$ solubility measurements in NaHS-$H_2S$ solutions the formation constants for $Pb(HS)_2^0$ and $Pb(HS)_3^–$ from 30 to 200 °C ($P_{sat}$) along with a third complex of the tentative stoichiometry $Pb(HS)_2(H_2S)^0$ at 200–300 °C. The latter species is, however, less certain, even though it has also been proposed in some low-temperature $PbS_{(s)}$ solubility work (Anderson 1962; Nriagu 1971) that might have been affected by analytical issues.

Wood et al. (1987) reported tabulated values for a single $Pb(HS)_2^0$ formation constant from 200 to 350 °C on the basis of extrapolations of the Giordano and Barnes (1979) experimental values and other theoretical estimations (Naumov et al. 1971). The Wood et al. (1987) tabulated values have recently been used by Etschmann et al. (2018) to report another set of HKF coefficients for $Pb(HS)_2^0$. However, their Born parameter had an unphysical, anomalously negative, value for a non-volatile and water-hydrated metal complex whose ω is expected to be close to zero and positive in sign (Plyasunov and Shock 2001; Perfetti et al. 2008; Tooth et al. 2013; Pokrovski et al. 2024).

This overview highlights the difficulty of extracting the origin of true experimental data used in generation of thermodynamic model coefficients and of assessing their errors. We recommend using the Sverjensky et al. (1997) dataset for Pb hydrosulfide complexes in speciation calculations to 300–350 °C, whereas any higher-temperature extrapolation should be taken with extreme caution. As for zinc, the affinity of Pb to $HS^-$ is stronger than to $Cl^-$ and decreases with temperature as shown by constants of ligand exchange reactions analogous to those for Zn (Fig. 4.26):

$$PbCl_n^{2-n} + n\ HS^- = Pb(HS)_n^{2-n} + n\ Cl^-,\ n = 2 \text{ and } 3 \quad (4.69)$$

It follows from the plot in Fig. 4.26 that, like for Zn, Pb-HS complexes may only be significant in alkaline, sulfur-rich and low-temperature fluids, whereas chloride complexes are largely responsible for the formation of most hydrothermal lead deposits characterized by saline acidic-to-neutral fluids with Cl/S ratios of >1000.

*Copper*

Complexes of copper in hydrothermal fluids have been a subject of extensive studies in the 1970s–1990s, elegantly overviewed in detail by Wood and Samson (1998), and complemented by more recent work that will be briefly outlined here. Most studies employed the hydrothermal reactor solubility and synthetic fluid inclusion methods and, more recently, UV-Vis and XAS (both XANES and EXAFS) spectroscopy, complemented by quantum-chemistry and molecular dynamics (MD) simulations of complex structures and stabilities. Experimental studies span from low-temperature aquatic systems to silicate melt-fluid systems. In addition, Cu speciation in natural fluid inclusions from different magmatic-hydrothermal settings was also investigated using XAS and XRF (e.g., Mavrogenes et al. 2002; Cauzid et al. 2007; Berry et al. 2009). These and many other studies published in the chemical literature evidence a plethora of $Cu^{II}$ complexes, including hydroxides, chloride, sulfides and polysulfides that form in aqueous solution at circa-ambient conditions, whereas $Cu^{I}$ chlorides and hydrosulfides are the major Cu-transporting species in geological fluids at elevated temperatures. The two latter types will, therefore, be considered in more detail here.

*Chloride complexes of $Cu^I$* are believed to be dominant in most saline fluids (see Table 4.6). The number of specific experimental studies devoted to these complexes is too large to allow reviewing here; instead, we evaluate the major thermodynamic compilations of those data that are used by the geochemical community. There exist at least four sets of HKF-model parameters for $Cu^I$ chloride species. The first set of Sverjensky et al. (1997) for $Cu^+$, $CuCl^0$, $CuCl_2^-$ and $CuCl_3^{2-}$, included in the major thermodynamic databases of aqueous species (SUPCRT92, SUPCRTBL), has been based on 25 °C stability constants and apparently involved no other, higher-temperature, data source in the HKF-parameter derivation. Akinfiev and Zotov (2001) produced a second HKF set for those Cu species, based on $Cu_{(s)}$, $CuCl_{(s)}$, and $Cu_2O_{(s)}$ solubility data mostly from Russian work to 450 °C and 1.5 kbar (overviewed by Zotov et al. 1995) and those of Xiao et al. (1998) to 300 °C at $P_{sat}$, and using isocoulombic reactions with analogous silver complexes to better constrain thermodynamic values at 25 °C ($S^0$, $C_p^0$, $V^0$), necessary for the HKF parameter derivation. A few years later, Lui and McPhail (2005) produced another set of HKF coefficients for $CuCl^0$, $CuCl_2^-$, $CuCl_3^{2-}$, and $CuCl_4^{3-}$, based on the Lui et al. (2002) UV-Vis spectroscopy study to 250 °C. Their HKF set for $CuCl^0$ and $CuCl_2^-$ is in excellent agreement with that of Akinfiev and Zotov (2001), despite differences in the choice of experimental data regressed and calculation algorithms, yielding equilibrium constant values within <0.5 log units to at least 400–450°C and 1 kbar for the reaction

$$Cu^+ + n\ Cl^- = CuCl_n^{1-n},\ n = 1 \text{ or } 2 \qquad (4.70)$$

However, in contrast to the Akinfiev and Zotov (2001) speciation scheme that ignored higher-order complexes ($n$>2), the results of Liu and McPhail (2005) imply that $CuCl_3^{2-}$ and $CuCl_4^{3-}$ would be the dominant species in high-temperature saline fluids (>400 °C, >2 m Cl). More recently, however, Brugger et al. (2007) demonstrated by in situ X-ray absorption spectroscopy that the dominant complex in high-temperature saline fluids is dichloride, $CuCl_2^-$. Its dominant formation in high-temperature fluids was also qualitatively confirmed by other XAS and MD studies (Fulton et al. 2000; Sherman 2007). The trichloride ($CuCl_3^{2-}$) was only significant at temperatures <250 °C and salinities above 4–8 m Cl. In light of the new XAS-derived speciation, Brugger et al. (2007) re-interpreted previous UV-Vis measurements (Lui et al. 2002), rejected the tetrachloride species, and produced a third set of HKF coefficients for $CuCl_3^{2-}$. Finally, a fourth HKF dataset was recently added by the IGEM team (Trofimov et al. 2023), who selectively re-analyzed previous experimental and theoretical data in terms of $CuCl_2^-$ and $CuCl^0$ (Xiao et al. 1998; Archibald et al. 2002; Migdisov et al. 2014), complemented by their own data and Russian work cited therein. These authors also proposed an alternative set of AD model parameters for $CuCl^0$ for vapor phase and low-density supercritical fluids. This historical background demonstrates that caution should be given to speciation interpretations and thermodynamic data derivation based on a single method or limited datasets. To summarize, the present-date Cu speciation in saline hydrothermal fluids is believed to be dominated by dichloride ($CuCl_2^-$) and either Akinfiev and Zotov (2001) or Lui and McPhail (2005) or Trofimov et al. (2023) sets of HKF coefficients for this species may be used with confidence in modeling Cu solubility in magmatic-hydrothermal fluids to at least 500 °C and 1–2 kbar (e.g., Kouzmanov and Pokrovski 2012). $CuCl^0$ is a very minor species in saline liquid-type fluids, according to the above models. For lower-temperature epithermal-sedimentary brines, $CuCl_3^{2-}$ may also be added to Cu speciation models.

At higher temperatures, closer to conditions of silicate melt - aqueous fluid separation, and in S-poor vapor phase produced by supercritical fluid unmixing, the low fluid dielectric constant favors ion association and uncharged species stability thereby changing Cu speciation. Indeed, $Cu^I$ species such as $CuCl^0$, $(Na,K)CuCl_2^0$, and/or $CuCl(HCl)_n^0$ were inferred from solubility, partitioning, and molecular simulation studies at such conditions (e.g., Zajacz et al. 2011; Pokrovski et al. 2013a; Mei et al. 2014; Alex and Zajacz 2022; references therein), but quantitative assessment of their exact stoichiometries, abundances and thermodynamic properties over the magmatic *T*-*P*-density range is yet lacking. The role of sulfur in Cu speciation and transport, in particularly in high-temperature fluid systems, also remains poorly quantified, as will be overviewed below.

*Sulfide complexes*. Extensive geo-biochemical literature attests for the formation of a variety of $Cu^I$ complexes, both mono- and polymeric, with reduced sulfur ligands (sulfide, polysulfides, thiosulfate) in near-surface aquatic and biological systems (Rickard and Luther 2006; Pluth and Tonzetich 2020, for review). By contrast, only a handful of studies is available for hydrothermally relevant conditions. Among before-2000s studies (see Wood and Samson 1998 for review), the work of Crerar and Barnes (1976) is the rare one that reported thermodynamic values for Cu sulfide complexes. The authors measured the solubility of the chalcopyrite-pyrite-bornite assemblage in $NaHS-H_2S$ fluids from 200 to 350 °C and reported the first values of stability constants for $Cu(HS)_2^-$ and $Cu(HS)_2(H_2S)^-$. Mountain and Seward (1999) measured the solubility of chalcocite ($Cu_2S_{(s)}$) at 22 °C in $NaHS-H_2S$ solutions over a wide pH range (4.0–11.5) using flow-through reactors, and produced an accurate set of stability constants for $CuHS^0$, $Cu(HS)_2^-$ and $Cu_2S(HS)_2^-$, dominant in sulfide-bearing solutions at acidic, neutral, and alkaline pH, respectively. Mountain and Seward (1999) also re-assessed the high-temperature data of Crerar and Barnes (1976) in light of the speciation scheme above, and reported van't Hoff-type equations (e.g., equation 4.8) describing the temperature dependence (up to 350 °C) of *K* values of the formation reaction for the two monomeric complexes (Fig. 4.27a,b):

$$Cu^+ + n\ HS^- = Cu(HS)_n^{1-n},\ n = 1, 2 \quad (4.71)$$

Mountain and Seward (2003) further confirmed the data for $Cu(HS)_2^-$ by extending $Cu_2S_{(s)}$ solubility measurements to 95 °C at near-neutral pH. Their reported set of log$K_{71}$ for $Cu(HS)_2^-$ that also incorporated the data from the three studies above remains the most up-to-date one from 25 to 300 °C at $P_{sat}$ (Fig. 4.27b). However, because their log$K$ *vs* $T$ functions did not offer a possibility of extrapolating to supercritical conditions, two attempts have been made to place their data into the framework of the HKF equation of state. Akinfiev and Zotov (2001) used only the ambient-temperature data of Mountain and Seward (1999) and isocoulombic reactions with analogous $Ag^I$ complexes to produce a first set of HKF parameters for $CuHS^0$ and $Cu(HS)_2^-$. Their resulting log$K_{71}$ values are within 0.5 log units of those from Mountain and Seward (2003) empirical equations, lending credence to the HKF-model approach which was based exclusively on low-temperature data. Akinfiev and Zotov (2010) further refined the HKF parameters for $Cu(HS)_2^-$ (notably the entropy and the related $a$ and $c$ parameters) by incorporating solubility data (<100 °C) of Mountain and Seward (2003). They rejected the higher-temperature data of Crerar and Barnes (1976) claiming that they were inconsistent. Paradoxically, the revised HKF parameters yield $K_{71}$ values 1 to 2 log units lower than the van't Hoff estimations above 200 °C Mountain and Seward (2003) (Fig. 4.27b).

The first in situ verification of Cu hydrosulfide speciation at supercritical conditions was made by Etschmann et al. (2010). They measured $Cu_2S_{(s)}$ solubility and local structure in concentrated NaHS solutions to 600 °C and 600 bar using in situ X-ray absorption spectroscopy. Their structural data clearly demonstrate the predominance of complexes with a linear S-Cu-S geometry, which would be consistent with $Cu(HS)_2^-$ (note that XAS cannot "see" hydrogen atoms in most cases). Remarkably, their measured solubilities to 450 °C are in excellent agreement with the thermodynamic predictions using the HKF parameters of $Cu(HS)_2^-$ from Akinfiev and Zotov (2001) further lending credence on their data, and confirming $Cu(HS)_2^-$ to be the dominant $Cu^I$ hydrosulfide complex in liquid-like fluids over a very large range of fluid acidity and sulfur concentration. This conclusion has recently been re-confirmed by new solubility data of $Cu_2S_{(s)}$ obtained in $H_2S$ solutions at 350 and 450 °C at 500 and 1000 bar by a Russian team (Trofimov et al. 2023 and Russian references therein). The authors combined their few solubility datapoints with those of Mountain and Seward (1999, 2003), while rejecting Crerar and Barnes (1976) values, to propose another set of HKF parameters for $CuHS^0$ and $Cu(HS)_2^-$. Their set yields, however, virtually identical $K$ values (within <1 log units, Fig. 4.27b) to those of Akinfiev and Zotov (2001) to at least 600 °C and 2 kbar.

Theoretical molecular simulations complemented the collection of Cu complexes. Mei et al. (2013a), using thermodynamic integration approach in MD simulations, were able to predict the Cu-Cl and Cu-HS complex stability constants between 22 and 327 °C to be within 1–4 log units of the key experimental values (depending on the species and temperature). However, the MD uncertainties are yet too high compared with those of experimental measurements and corresponding thermodynamic models, even though they may provide qualitatively similar trends (Fig. 4.27a,b). In addition, these MD simulations indicate favorable formation of a mixed-ligand complex, $Cu(HS)Cl^{-}$, that might potentially be dominant in hydrothermal fluids at Cl/S ratios of $10^3$ to $10^4$ at near-neutral pH around 300 °C. Complexes of Cu with sulfur of yet unknown stoichiometry have been hypothesized from in situ XRF-XANES spectra acquired on re-heated to 500 °C natural fluid inclusions from the Yankee Lode porphyry deposit (Cauzid et al. 2007). Mixed HS-Cl complexes for Cu and Au have also been inferred from experiments at magmatic temperatures and quantum-chemistry simulations (Zajacz et al. 2010, 2011), attesting a significant affinity of $Cu^{I}$ for reduced sulfur even at very high temperatures.

This affinity is further expressed in the S-rich hydrothermal-magmatic vapor phase in which Cu forms relatively volatile complexes with reduced sulfur. Such complexes promote Cu partitioning into the vapor compared to other transition metals such as Fe, Zn or Pb, as has been thoroughly demonstrated by laboratory measurements of vapor-liquid partition coefficients and natural fluid inclusion analyses in porphyry Cu systems. Note, however that natural fluid inclusions may be affected by post-entrapment enrichment through $Cu^{+}$ diffusion into the inclusion from the external fluid, leading to an overestimation of Cu vapor-brine partitioning (see Kouzmanov and Pokrovski 2012; Pokrovski et al. 2013a, for extensive reviews on vapor-like geological fluids).

### *4.4.3 Silver*

Like for copper, complexes with chloride and hydrosulfide ligands control silver transport by hydrothermal fluids. A set of four *chloride complexes* of $Ag^{I}$, $AgCl^0$, $AgCl_2^{-}$, $AgCl_3^{2-}$ and $AgCl_4^{3-}$, forms over a wide salinity range at low-to-moderate temperatures (<200 °C). With increasing temperature to 300–350 °C, the predominance domain of the dichloride species, $AgCl_2^{-}$, widens significantly, covering salinities from ~0.01 m to at least 2 m NaCl. However, at supercritical temperatures and higher salinities, there is more discrepancy. Based on available data from $AgCl_{(s)}$ solubility experiments to 350 °C (Seward 1976), Sverjensky et al. (1997) generated a set of HKF parameters for the four chloride species. Their model implies a growing abundance of $AgCl_4^{3-}$ at supercritical temperatures and salinities >2 m NaCl. However, their set is in disagreement with that of Akinfiev and Zotov (2001) who based their HKF-model for $AgCl^0$ and $AgCl_2^{-}$ on more extensive solubility studies (Zotov et al. 1995; Tagirov et al. 1997; Akinfiev and Zotov 1999) that do not show evidence of tri- and tetrachloride above 300–350 °C. Their data imply that $AgCl_2^{-}$ is the major chloride species above 300°C and >2 m NaCl (>10 wt% salt). Thus, depending on the chosen thermodynamic data set, Sverjensky et al. (1997) *vs* Akinfiev and Zotov (2001), solubility predictions of $Ag_2S_{(s)}$ in moderate-salinity fluids buffered by iron sulfide/oxide and alkali-aluminosilicate assemblages above 400 °C differ by a factor of 10–100 (see Pokrovski et al. 2013b for detailed discussion). This discrepancy is a consequence of both a lack of validation of a thermodynamic model (HKF) beyond the range of data covered by measurements and difficulties intrinsic to deriving species stoichiometry from regressions of bulk-solubility data.

In the 2010s, with the advent of spectroscopic and molecular modeling approaches, important constraints have been brought. Liu et al. (2012) showed by MD simulations that the only stable chloride species in hydrothermal range would be $AgCl^0$ and $AgCl_2^-$, while higher-order chloride complexes are unstable even at low temperatures. Pokrovski et al. (2013b) combined in situ X-ray absorption spectroscopy in supercritical fluids with MD simulations to demonstrate that $AgCl_2^-$, of quasi-linear Cl-Ag-Cl geometry and without $H_2O$ molecules in the first Ag atomic shell, is the only spectroscopically detectable species above >350 °C and salinities to at least 40 wt% NaCl. In addition, triangular $AgCl_3^{2-}$ may account for a moderate part of total dissolved Ag (<30 %) below 300 °C and salinities above 10–15 wt% NaCl. These results are in very good agreement with the thermodynamic properties of $AgCl_2^-$ of Akinfiev and Zotov (2001), and of $AgCl_3^{2-}$ of Zotov et al. (1995). Their thermodynamic dataset, as compiled in Pokrovski et al. (2013b), is thus recommended for modeling Ag transport in saline fluids over the hydrothermal-magmatic range. Further support to this speciation scheme has recently been added by Zotov et al (2020) from $Ag_{(s)}$ solubility measurements showing that the $AgCl_2^-$ complex dominates up to remarkably high salinities at supercritical temperatures (425–475 °C, 0.1–18 m NaCl). These authors demonstrated that the traditional Debye-Hückel equation for activity coefficients (Helgeson et al. 1981) can be applied with confidence to very high salinities (at least for 1:1 electrolytes).

*The hydrosulfide complexes of silver* in hydrothermal fluids are dominatly $AgHS^0$ and $Ag(HS)_2^-$, at acidic and neutral pH, respectively, whereas at alkaline pH and low temperatures other, likely polynuclear, species exist (e.g., $Ag_2S(HS)_2^{2-}$). In addition, complexes with polysulfides, thiosulfate and sulfite may form in more oxidizing environments (Wood and Samson 1998; Rickard and Luther 2007; references therein). The mono- and dihydrosulfide complexes (reaction 4.72 below) are the most relevant to hydrothermal fluids. Their stability has been derived from $Ag_2S_{(s)}$ solubility studies in $H_2S$-NaHS solutions, both at ambient and elevated temperatures (Schwarzenbach et al. 1958; Schwarzenbach and Widmer 1966; Renders and Seward 1989; Melent'yev et al. 1969; Sugaki et al. 1987; Gammons and Barnes 1989; Stefánsson and Seward 2003a), and more recently from theoretical MD simulations (e.g., He et al. 2016). Among the experimental studies that generated species stability constants, the solubility measurements of Stefánsson and Seward (2003a) using flow-through reactors cover the widest *T-P* range, from ambient conditions to 400 °C and 500 bar. These data are in good agreement with previous work and significantly strengthen the robustness of the available experimental dataset. These data were regressed by the authors as polynomial equations of log*K* vs *T*, which are strictly valid within the experimental *T-P* range (see section 4.3.3). Consequently, there have been several attempts to place some of those experimental data into the HKF-model framework to enable predictions to higher temperatures and pressures. However, the data selection procedures were not sufficiently explicit or coherent in those attempts. Sverjensky et al. (1997) produced the first set of HKF parameters for $Ag(HS)_2^-$ based of inter-parameter correlations and using the stability constant at 25 °C from Gammons and Barnes (1989). Akinfiev and Zotov (2001) regressed a more extended solubility dataset that included $Ag_2S_{(s)}$ solubility works at elevated temperatures and an improved description of $H_2S_{(aq)}$ thermodynamic properties (AD model, see section 4.3.3) to produce HKF-model parameters for $Ag(HS)_2^-$ and $AgHS^0$. Akinfiev and Zotov (2010) further revised the $AgHS^0$ parameters based on the low-temperature solubility studies of Schwarzenbach et al. (1958) and Schwarzenbach and Widmer (1966). Those two studies reported the lowest $Ag_2S_{(s)}$ solubilities, which likely correspond to the well-crystalline sulfide phase, and were therefore judged to be most accurate. Fig. 4.27c,d compares the formation constants of both complexes for the reaction

$$Ag^+ + n\ HS^- = Ag(HS)_n^{1-n},\ n = 1 \text{ or } 2 \qquad (4.72)$$

Despite the differences in data choice, the values from both HKF datasets, as well as the most recent solubility work of Stefánsson and Seward (2003a) not considered by those models, are all in decent agreement, with differences ≤1 log unit to 300 °C at $P_{sat}$. It should be noted that this discrepancy is in part due to different choices of the thermodynamic properties of the other reaction constituents ($Ag^+$, $H_2S$ and $HS^-$) adopted in data processing in different works. The agreement may be improved if similar basis data are used. For example, regressing Stefánsson and Seward (2003a) log$K_{72}$ values, using the RB equation and the $HS^-$ and $Ag^+$ thermodynamic properties used by Akinfiev and Zotov (2010), improves their agreement with the independent HKF predictions at elevated temperatures. The stability constants of the Ag hydrosulfide complexes extrapolated to 400–600 °C are likely to be uncertain within <2 log units. We recommend using Akinfiev and Zotov's (2001, 2010) HKF dataset for silver species for consistency, in particular if comparisons with Cu and Au complexes are to be made, because the three elements have been processed within the same thermodynamic framework. According to the available data overviewed above, silver hydrosulfide complexes dominate in sulfide solutions below 100 °C and in dilute hydrothermal fluids (<300 °C) of meteoric origin typically found in active geothermal systems (e.g., Otaki and Broadlands, New Zealand). The silver dichloride species usually dominate in high-temperature hydrothermal fluids of sea-water or higher salinity (see Stefánsson and Seward 2003a for details of Ag species distribution at different solution compositions).

At magmatic temperatures, similarly to $Cu^I$, the ion pair between $AgCl_2^-$ and $Na^+$, $NaAgCl_2^0$, becomes dominant in S-poor fluids. In S-bearing fluids over a typical $f_{O2}$ range of arc magmatism (NNO±2), another mixed-ligand ion pair, $NaAg(HS)Cl^0$, predominantly forms, as has recently been inferred from elegant solubility measurements of Cu-Au-Ag alloys of a given metal activity using synthetic fluid-inclusion techniques for trapping the fluid phase at 900 °C and 2 kbar (Yin and Zajacz 2018; Alex and Zajacz 2022). Further qualitative evidence for another mixed-ligand complex, $Ag(HS)Cl^-$, and its ion pair with $Na^+$ has been provided by MD simulations (Lai et al. 2022) suggesting that these species may control $Ag^I$ speciation in hydrothermal fluids with $H_2S$ concentrations >0.01 m and temperatures >300 °C. However, the realistic uncertainties on the stability constants of these and other chloride and hydrosulfide species that may now be calculated by MD methods yet remain rather high, and do not systematically reproduce the more robust experiment-based data trends (e.g., Fig. 4.27c). Nevertheless, in light of the recent work, mixed HS-Cl complexes could be important for chalcophile metals (Cu, Ag, Au). More data on their stability constants over wider *T-P* ranges are required to place such species, along with their more traditional chloride and hydrosulfide congeners, in a unified thermodynamic framework for hydrothermal-magmatic fluids.

### *4.4.4 Gold*

Among all metals, gold is definitely the "king" of hydrothermal ore-deposit research. As a result, it is the most studied metal in hydrothermal fluids as to its chemical speciation and solubility. Being notoriously soft, aurous gold, $Au^I$, has a much stronger chemical affinity to reduced sulfur in aqueous solution than $Cu^I$ and $Ag^I$ but, due to relativistic effects affecting its electronic structure, its affinity to the $OH^-$ and $Cl^-$ ligands is also much stronger. As a result, there is competition between those three types of ligands for $Au^I$. The speciation is further complicated by the existence of auric gold, $Au^{III}$, which has a strong affinity for both $OH^-$ and $Cl^-$. Tetrachloride, $AuCl_4^-$, is the most popular $Au^{III}$ species in chemistry and may easily be prepared by gold metal dissolution in aqua regia. Auric gold species are only stable at strongly oxidizing conditions of the Earth's surface. Aurous gold can also be stabilized at ambient conditions as very strong complexes with cyanide, $Au(CN)_2^-$, famous for its use in heap leaching operations for gold recovery. The formation of strong $Au^I$ complexes with intermediate-valence sulfur ligands ($S_2O_3^{2-}$, $S_n^{2-}$, $SCN^-$, organic thiols) occurs at conditions of sulfide ore oxidation in mine drainage sites (e.g., Vlassopoulos and Wood 1990), and is used in chemical engineering applications such as gold ore processing (Adams 2005) or gold nanoparticle synthesis (Häkkinen 2012). These ligands are, however, generally present in much smaller concentrations than $H_2S$ and $HS^-$. As a result, the fate of gold in crustal fluids and silicate melts has been believed to be controlled by $Au^I$-OH-Cl-HS types of complexes. They have been the subject of a large number of experimental and thermodynamic studies, reviewed in detail in Wood and Samson (1998) and Pokrovski et al. (2014). More recently, tri- and disulfur radical ions have emerged as strong ligands for $Au^I$ in hydrothermal-magmatic fluids. Below we overview the available data on $Au^I$ hydroxide, chloride, sulfide and polysulfide species with an emphasis on >2010s work. Unfortunately, many thermodynamic databases of gold species, widely used in geochemical modeling of fluid-rock interactions and ore deposit formation, contain sometimes obsolete and confusing information on some of those species and do not take into account the more recent data. Another caveat of the thermodynamic properties of reaction constants of Au solubility the reader can face in the literature is the use of different thermodynamic data sources for the reaction constituents such as $H_2S/HS^-$, $Cl^-/HCl^0$, $H_{2(aq)}/H_{2(g)}$ that renders different datasets thermodynamically inconsistent and may lead to large uncertainties on gold solubility predictions at elevated temperatures.

*$Au^I$ aqua ion and its hydroxide complexes*. Aurous gold ion, $Au^+$, often used in expressions of formation constants of Au complexes and is traditionally present in common databases (e.g., SUPCRT92, SUPCRTBL), is in reality virtually non-existent in aqueous solution at ambient conditions at any physical pH value because it undergoes hydrolysis to $AuOH^0$.

$$Au^+ + H_2O = AuOH^0 + 0.5\ H^+ \qquad (4.73)$$

The reaction constant value, $\log K_{73}$, is between 6 and 11 at ambient conditions according to available estimations (e.g., Akinfiev and Zotov 2001; Stefánsson and Seward 2003b; Rauchenstein-Martinek et al. 2014). This range of values means that $Au^+$ can only form in significant fractions at pH < –6 in the best case, which is simply an unphysical condition in aqueous solution. Note that the instability of $Au^+$ strongly contrasts with that of the $Cu^+$ and $Ag^+$ ions that are dominant in aqueous solution at pH <9–11 in the absence of complexing ligands (S, Cl, see above). Paradoxically, $Au^+$ thermodynamic properties and HKF-model coefficients have been estimated and revised with some minor modifications by the HKF people in several papers over almost 20 years (Helgeson et al. 1981; Shock and Helgeson 1988; Shock et al. 1997) and used in some more recent work (e.g., Akinfiev and Zotov 2001; Stefánsson and Seward 2003c). However, it is hard to track the true original data source, and in particular the Gibbs energy of formation value of the $Au^+$ aqua ion, since all the papers mentioned above just refer one to another for this value. A perseverant reader finally finds that it stemmed from the compilation of Latimer (1952) in which it has been tentatively estimated from the solubility product of $Au^I$ iodide ($AuI_{(s)}$), which, in turn, has been very roughly estimated using an analogy with $CuI_{(s)}$ and $AgI_{(s)}$. This example shows how unreliable thermodynamic data can be for some aqueous species, in particular those poorly stable within the range of experimentally realizable conditions. Needless to say that extrapolation of such data to elevated temperatures suffers from large uncertainties that may lead to unphysical results in speciation calculations for hydrothermal fluids. For example, the value of log$K_{73}$ that may be calculated at 600 °C and 2 kbar using the HKF parameters of Rauchenstein-Martinek et al. (2014) is –2.2, which implies that $Au^+$ would be the dominant $Au^I$ species in solution at realistic pH of 2–3. This result is unphysical, however, since it contradicts everything we know about the large increase of cation hydrolysis with temperature making most metal cations negligible in hydrothermal fluids in favor of their hydroxide counterparts (e.g., Wood and Samson 1998). As a result, no set of $Au^+$ thermodynamic properties published so far could be recommended, and the use of $Au^+$, for example in expressing cumulative complexation constants as commonly done for much more stable $Ag^+$ and $Cu^+$ (equations 4.70–4.72) should be avoided for Au species. The thermodynamic properties of $Au^I$ hydroxide, chloride, and sulfur-ligand complexes have been derived from solubility measurements of gold metal. Therefore, $Au_{(s)}$ solubility reactions, such as equations (4.64–4.67) and others given below, should be used to avoid the misleading $Au^+$ issue.

The stability of monohydroxide complex, $AuOH^0$, dominant over a geologically realistic range of pH (pH <10–12) in dilute S- and Cl-poor solutions, has been investigated by gold solubility measurements from 25 to 600 °C (Vlassopoulos and Wood 1990; Zotov et al. 1995 and references therein; Stefánsson and Seward 2003b) for the reaction:

$$Au_{(s)} + H_2O = AuOH^0 + 0.5\ H_{2(g)} \qquad (4.74)$$

These data, are in good agreement, in particular at temperatures above 200 °C where Au solubilities approach the 100-ppb level and, therefore, become more easily measureable. They were regressed using the HKF model based on different approaches and solubility data sources by Akinfiev and Zotov (2001) and Rauchenstein-Martinek et al. (2014). Both sets of HKF parameters predict similar $K_{74}$ values, within <0.5 log units between 200 and 600 °C and up to several kbars (Fig. 4.28a). Therefore, both datasets may be used in thermodynamic modeling studies. The $AuOH^0$ species is only significant for gold transport in Cl- and S-poor fluids in which its concentrations may attain 100s ppb Au. The $Au(OH)_2^-$ complex, invoked in few studies in highly alkaline solutions at near-ambient temperature, is unlikely to be important in geological fluids. The tentative HKF parameters proposed so far for this species (Pokrovski et al. 2014 and references therein; Rauchenstein-Martinek et al. 2014) are not recommended for extrapolation to hydrothermal conditions.

*Au^I chloride complexes*. Chloride complexes of $Au^I$ have received considerable attention given the common presence of salt in geological fluids. To the best of our knowledge, the first quantitative estimations of Au-Cl complex stability and concentrations in hydrothermal fluids have been made by Helgeson and Garrels (1968). They combined rare available experimental data and thermodynamic extrapolations with estimations of gold budget in quartz veins based on geological observations to conclude that gold is transported as aurous chloride complexes and precipitated mostly in response to cooling. Their conclusions were, however, criticized by Boyle (1969) who, based on field observations and geochemistry of gold ore, argued that Au is primarily transported as complexes with S, As and Sb and deposited in response to boiling and fluid-rock interactions rather than cooling. These first debates have been followed by a substantial number of experimental solubility studies that quantified the equilibrium constant of reaction (4.67). Among the available experimental studies, those of Nikolaeva et al. (1972) to 90 °C, Zotov et al. (1991) to 500 °C and 1.5 kbar, Gammons and Williams-Jones (1995) to 300 °C and $P_{sat}$, and Stefánsson and Seward (2003c) to 600 °C and 1.8 kbar, are the most consistent ones (see Wood and Samson 1998; Pokrovski et al. 2014 for detailed assessment). They have been used in derivation of HKF-model parameter sets of variable degree of confidence for $AuCl_2^-$, considered as the main complex. The Sverjensky et al. (1997) set, included in the SUPCRT92 and SUPCRTBL databases, was based on low-temperature data (Nikolaeva et al. 1972) and HKF correlations, whereas the Akinfiev and Zotov (2001) set was derived by optimization of more extended solubility data mostly from Russian teams (Nikolaeva et al. 1972; Zotov et al. 1995 and references therein). In the range 400–600 °C and 0.5–2 kbar, both sets predict Au solubility to be 1 to >3 log units higher than Stefánsson and Seward's (2003c) extensive solubility measurements. The latter experimental work was apparently used in a third HKF dataset (Rauchenstein-Martinek et al. 2014) that, however, did not report explicit information on the parameter derivation procedure. Recently, Zotov et al. (2018) have performed additional gold solubility measurements at 450 °C and 0.5–1.5 kbar over a large $f_{H2}$ range (>6 log units). They combined these data with those cited above, as well as more recent gold solubility measurements at magmatic temperatures (to 1000 °C and 2 kbar; Zajacz et al. 2010; Guo et al. 2018) reinterpreted in terms of a single species, $AuCl_2^-$, to propose a density-model equation for the thermodynamic constant of the following solubility reaction that was modified from reaction (4.67) to render it isocoulombic and allow a better linearization vs $T$:

$$Au_{(s)} + HCl^0_{(aq)} + Cl^- = AuCl_2^- + 0.5\ H_{2(aq)} \qquad (4.75)$$

Their simple model allows the stability of $AuCl_2^-$ to be predicted with an uncertainty of <1 log units at hydrothermal temperatures (<600 °C), with $K_{75}$ values being in between Akinfiev and Zotov's (2010) and Stefansson and Seward's (2003c) results, thereby partly reconciling the two datasets (Fig. 4.28b). The data of Zotov et al. (2018) await to be incorporated in common thermodynamic databases, e.g., in the form of an HKF-parameter set, to enable its wider use in hydrothermal-magmatic fluid models. The large *T*-*P*-salinity range of the $AuCl_2^-$ stability has independently been confirmed by direct in situ XAS measurements at hydrothermal conditions that evidence a quasi-linear Au coordination with two Cl atoms (Pokrovski et al. 2009a; Tagirov et al. 2019a).

These in situ data, along with most solubility studies discussed above, do not provide any robust evidence for mono- and trichloride species, $AuCl^0$ and $AuCl_3^{2-}$, whose significance was also doubted in some previous studies (e.g., Gammons and Williams-Jones 1995). Despite the lack of evidence for both species, two theoretical HKF parameter sets for them have been suggested (Sverjensky et al. 1997; Akinfiev and Zotov 2001). Their values should be, however, regarded as highly unconstrained and their use in gold speciation modeling in hydrothermal fluids is not recommended.

*$Au^I$ hydrosulfide complexes*. This most important type of $Au^I$ species in hydrothermal fluids has been the subject of extensive solubility measurements since the pioneering work of Ogryzlo (1935) who reported elevated Au solubility in alkaline sulfide solutions. Since then, more than 40 studies have attempted to constrain Au (hydro)sulfide species stoichiometry and stability constants via the solubility method (see Pokrovski et al. 2014 for detailed review). Among early studies, those of Seward (1973), Renders and Seward (1989) and Shenberger and Barnes (1989) are probably the most constrained ones. It follows from these and other studies that $Au^I$ dihydrosulfide, $Au(HS)_2^-$, is likely the major complex in the near-neutral pH range of $H_2S$-bearing geological fluids. In addition, other Au-S complexes of various stoichiometry have also been proposed in different studies dealing with acidic and alkaline pH regions and more oxidizing conditions (see below). Despite the common agreement on the $Au(HS)_2^-$ stoichiometry, the absolute values of the solubility constants of reaction (4.64) diverged significantly among before-2000s studies (up to 3 log units at typical hydrothermal temperatures of 300 °C). Such discrepancies may be explained by difficulties in controlling or buffering the three key parameters, $H_2S$, $H_2$ and pH, on which Au solubility depends, as well as different choices of the thermodynamic datasets for $H_2S_{(aq)}$ and its dissociation constant, and $H_{2(aq)}/H_{2(g)}$ equilibria, as well as the presence of other sulfur ligands (polysulfides, thiosulfates) which also form strong complexes with $Au^I$. Some data from these early studies have been used to produce first sets of HKF parameters of $Au(HS)_2^-$ by Sverjensky et al. (1997) and Akinfiev and Zotov (2001) still being used in modeling studies. While the most available experimental data and these predictions are in good agreement at moderate temperatures (200–350 °C), the HKF predictions at supercritical temperatures likely overestimate Au solubility values by a few orders of magnitude, as was shown by 1990–2000s studies (e.g., Benning and Seward 1996; Gibert et al. 1998). Among them, gold solubility measurements of Stefansson and Seward (2004) and Tagirov et al. (2005, 2006) are the most extended and analytically elaborated ones, covering a large range of $T$ (25–500 °C) and $P$ ($P_{sat}$–1 kbar). These works are in good mutual agreement (Fig. 4.28c,d) and provide important constraints on Au speciation in supercritical liquid-like fluids. The data of these three studies have been processed by Akinfiev and Zotov (2010) (as overviewed and corrected in Pokrovski et al. 2014) to produce an improved set of HKF model coefficients. Another set of $Au(HS)_2^-$ HKF parameters has been proposed by Rauchenstein-Martinek et al. (2014) based on the solubility data of Stefansson and Seward (2004). The first direct molecular evidence for $Au(HS)_2^-$ in near-neutral hydrothermal fluids was provided by in situ X-ray absorption spectroscopy combined with molecular modeling (Pokrovski et al. 2009b), showing that Au is linearly coordinated by two $[(H)S)]^-$ ligands. Gold bulk solubility measured from the spectrum absorption edge amplitude was in excellent agreement with gold solubility predictions using the thermodynamic properties of $Au(HS)_2^-$ (Akinfiev and Zotov 2010; Pokrovski et al. 2014). As a result, at present, the stability of $Au(HS)_2^-$ is well constrained in a wide $T$ (20–500 °C) and $P$ (<1.5 kbar) range of hydrothermal fluids.

The second "popular" Au sulfide species in hydrothermal fluids is *gold monohydrosulfide*, $AuHS^0$, which is believed to dominate gold speciation at weak-to-moderate $H_2S$ concentrations at acidic pH. Being much less soluble than $Au(HS)_2^-$, it has been constrained less thoroughly experimentally based exclusively on solubility data. The low $AuHS^0$ concentrations at most experimental conditions (typically below the ppm level) have not yet enabled direct spectroscopic proof of its existence. Moreover, MD calculations showed $AuHS^0$ to be structurally a quasi-linear complex HS-Au-$OH_2$, coordinated by a water molecule that may easily be exchanged with a stronger ligand such as $HS^-$ or $H_2S$ (Liu et al. 2011). Solubility-derived reaction constants of $AuHS^0$ (equation 4.66) from before-2000s works (e.g., Hayashi and Ohmoto 1991; Benning and Seward 1996; Gibert et al. 1998) were rather discrepant (e.g., >2 log units at 300 °C; see more detailed analyses in Stefánsson and Seward 2004, and Tagirov et al. 2005), due to numerous technical issues coupled with the complexity of experimental system compositions employed promoting the $AuHS^0$ competition with Au chloride, hydroxide and other S-ligand complexes forming at acidic pH. The works of Stefánsson and Seward (2004) and Tagirov et al. (2005) prevail amongst the most accurate solubility studies to date for $AuHS^0$, which are in mutual agreement (within ~0.5 unit of $\log K_{66}$; Fig. 4.28d). HKF-model parameter sets derived by Akinfiev and Zotov (2010) and Rauchenstein-Martinek et al. (2014), based on those works, are in good agreement and may be used for quantitative modeling of Au solubility in sulfide-poor acidic fluids. The former HKF set of Akinfiev and Zotov (2001), yet used in some databases, should be considered as obsolete because was based on experiments  in which both Au and S speciation was insufficiently constrained (e.g., Hayashi and Ohmoto 1991).

*Other Au (hydro)sulfide and polysulfide species* have also been invoked in some solubility and more recent X-ray spectroscopy studies in acidic-to-neutral sulfur-rich solutions to account for much higher measured Au solubility than predicted using the $AuHS^0$ and $Au(HS)_2^-$ complexes above. The additional complexes include $AuHS(H_2S)^0$ (Hayashi and Ohmoto 1991), $AuHS(H_2S)_3^0$ (Loucks and Mavrogenes 1999), or $AuHS(H_2S)^0$ or $AuHS(SO_2)^0$ (Pokrovski et al. 2009b). However, none of these species has been confirmed by more recent work. For example, the very popular in the 1990s $AuHS(H_2S)_3^0$ complex in high-temperature metamorphic fluids, inferred from limited synthetic fluid inclusion analyses of Au concentrations in saline fluids at 550–725 °C and 1–4 kbar buffered by iron oxide/sulfide and alkali-aluminosilicate assemblages (Loucks and Mavrogenes 1999), has later been rejected for a number of reasons. First, these solubility data could equally be interpreted by traditional $AuHS^0$ + $Au(HS)_2^-$ + $AuCl_2^-$ species (Stefánsson and Seward 2004). Moreover, its stability constants calculated using the HKF set proposed in the original study, significantly overestimate the in situ measured Au solubilities at 300–400 °C (Pokrovski et al. 2009b). Direct evidence against the tetra-hydrosulfide complex has been brought by in situ XAS measurements and MD simulations. Such species stoichiometry contradicts the linear S-$Au^I$-S geometry inherent to $Au^I$ coordination chemistry as directly shown by XAS in hydrothermal fluids (Pokrovski et al. 2009b, 2014). Recent MD simulations have demonstrated $AuHS(H_2S)_3^0$ to be unstable, breaking down to the linear [L-Au-(SH)] (L = $H_2O$ or $H_2S$/HS) and 2 $H_2S$ molecules (Liu et al. 2011). The $[HS\text{-}Au\text{-}SH_2)]^0$ species, which is more consistent with $Au^I$ coordination chemistry, has been regarded as a possible alternative for acidic solutions as well as for S-rich hydrothermal-magmatic vapor phases (Pokrovski et al. 2008a, 2009b). However, there are rather disparate thermodynamic data for this species, and the direct determination of its stoichiometry faced the limitations of traditional XAS methods that are poorly sensitive to light atoms like hydrogen. Alkali ion pairs (e.g., $NaAu(HS)_2^0$) have been proposed in magmatic vapors (e.g., Zajacz et al. 2010). Such ion pairs may easily account for Au transport in magmatic vapor-like fluids in contrast to recent claims that the traditional Au sulfide species, $AuHS^0$ and $Au(HS)_2^-$, dominant at lower-temperature hydrothermal conditions should be revised (Farsang and Zajacz 2025). In alkaline NaHS solutions, a binuclear species, $Au_2(HS)_2S^{2-}$ in which two [(HS)Au-] moieties are linked together via an [-S-] bridge, has been inferred from analyses of pH and ligand concentration trends in $Au_{(s)}$ solubility data (Seward 1973). However, direct spectroscopic or molecular dynamics evidence for such bulky species in hydrothermal fluid is lacking. Polysulfide-ligand ($(H)S_nS^{2-}$) $Au^I$ complexes of unconstrained stoichiometry and stability have equally been invoked in some Au solubility studies in $S^0$/$H_2S$-bearing hydrothermal fluids both at relatively low and high temperatures (100–150 °C, Berndt et al. 1994; 600–800 °C, Hu et al. 2022) to explain the elevated measured gold concentrations (100s ppm Au); however, their interpretation was complicated by poorly constrained sulfur speciation in such experiments with unknown or neglected contributions of the thiosulfate, sulfite or di- and trisulfur ion ligands to Au complexing.

*Complexes with sulfur radical ions*. From the dawn of ore-deposit research, the interpretation of Au speciation in S-rich fluids both at hydrothermal and magmatic conditions has been based on the assumption that the major sulfur forms are hydrogen sulfide and sulfate, the former being the ligand of choice and the latter having a negligible affinity for $Au^I$. This view was challenged in the 2010s by the discovery of the $S_3^{\bullet-}$ ion (Pokrovski and Dubrovinsky 2011; Pokrovski and Dubessy 2015), which turned out to be abundant in acidic-to-neutral solutions in the redox range of the sulfate-sulfide coexistence (see section 4.3.4 and Figs. 18–20). This discovery has first inspired theoretical molecular modeling studies of $S_3^{\bullet-}$ interactions with chalcophile metals. These works indicated that $S_3^{\bullet-}$, similarly to $HS^-$, might be able to form stable complexes with $Cu^I$ and $Au^I$ (Tossell 2012; Mei et al. 2013b). Direct quantitative experimental evidence has been brought by Pokrovski et al. (2015) who combined in situ solubility measurements with EXAFS spectroscopy and FPMD modeling in hydrothermal solutions in which sulfate and sulfide coexist, to demonstrate the formation of a mixed complex, $Au(HS)S_3^-$, across a large *T-P* range of crustal fluids (equation 4.65). Its exact stoichiometry and structure have further been confirmed by Pokrovski et al. (2022b) using high-resolution XANES spectroscopy coupled with quantum-chemistry and molecular modeling that enabled its unambiguous spectroscopic distinction from its "rival" $Au(HS)_2^-$ (Fig. 4.29) that was virtually unresolvable using classical XAS techniques (e.g., Pokrovski et al. 2009b, Trigub et al. 2017).

Based on the robust thermodynamic properties of $Au(HS)_2^-$ and $S_3^{\bullet-}$, Pokrovski et al. (2015) demonstrated that the equilibrium constant is equal to 1 over the entire *T-P* range investigated (300–500 °C, 0.1–1 kbar) for the ligand exchange reaction

$$Au(HS)_2^- + S_3^{\bullet-} = Au(HS)S_3^- + HS^- \quad (4.76)$$

These data open new perspectives for our understanding of Au speciation in crustal fluids, allowing a more accurate assessment of gold transport and precipitation phenomena. Fig. 4.30 illustrates the effect of trisulfur ion complexing on gold solubility in a porphyry Cu-Au setting, predicted using the thermodynamic data for S and Au species discussed above. The $Au(HS)S_3^-$ abundance closely follows that of $S_3^{\bullet-}$ itself (e.g., Fig. 4.19, 4.20) and accounts for >90–95% of the total dissolved gold at temperatures ≥350 °C and pressures $\geq P_{sat}$ in S-rich (>0.5 wt% S) acidic-to-neutral (pH 4–6) fluids under redox conditions of the $H_2S$-$SO_2$ or sulfide-sulfate coexistence (Pokrovski et al. 2015). This gold complex therefore stands as a major gold carrier in porphyry environments, capable of extracting from magma and transporting 10s to 100s ppm of dissolved Au (Fig. 4.30), thus accounting for natural Au concentrations analyzed in fluid inclusions from such settings (Kouzmanov and Pokrovski 2012). In can be seen in Fig. 4.30 that the other traditional Au species with $Cl^-$, $OH^-$ and $HS^-$ ligands considered above represent only a minor fraction of dissolved gold in S-rich fluids at magmatic-hydrothermal temperatures. It is only upon fluid cooling below 300–350 °C and $S_3^{\bullet-}$ decrease that Au hydrosulfide complexes, such as $Au(HS)_2^-$ (and eventually $AuHS^0$, at lower S concentrations and more acidic pH), take the lead in gold transport to epithermal deposits. In metamorphic fluids relevant to orogenic gold deposits (section 4.5.3), the generally lower sulfur concentrations and lower $f_{O_2}$ render $Au(HS)S_3^-$ less abundant compared to $AuHS^0$ and $Au(HS)_2^-$. It is only at hot stages of metamorphism corresponding to amphibolite facies (>550–600 °C) potentially favorable for large S release in the fluid due to breakdown of pyrite to pyrrhotite in some lithologies (e.g., Tomkins 2010; Schwarzenbach and Evans, 2025) that the gold-trisulfur complex might become important. However, thermodynamic predictions at temperatures above 500–600 °C are exclusively based on extrapolations and therefore become highly dependent on the choice of thermodynamic data sources for metal and sulfur species as well as other fluid constituents. The use of consistent thermodynamic datasets for the key species such as $H_2S$ and $Au(HS)_2^-$ (Pokrovski et al. 2014), based on which the reaction (4.76) constants have been derived, allows significantly reducing uncertainties of extrapolations above 500 °C. As a result, current realistic uncertainties of Au solubility predictions at 550–700 °C and 1–5 kbar for a given fluid composition are probably within 1 log unit (see Pokrovski et al. 2015 for detailed discussion). Note that the estimated Au solubilities above 500 °C in Fig. 4.30 are pessimistic in light of the possible existence of Au complexes with the disulfur radical ion that becomes more abundant at closer-to-magmatic temperatures (Fig. 4.20a; 4.21), as well as of mixed alkali-Cl-HS complexes whose stability remains to be quantified at the conditions of the magmatic-hydrothermal transition.

The findings of stable Au-trisulfur ion complexes help better constrain gold deposit formation models (e.g., He et al. 2024) and, in the future, may open perspectives for improving methods of gold recovery from ore or hydrothermal synthesis of gold nanomaterials (Pokrovski et al. 2015, 2022b). Assuming similar affinities of $S_3^{\bullet-}$ and $HS^-$ for base metals ($Cu^I$, $Zn^{II}$) and $Ag^I$, with equilibrium constants equal to one for exchange reactions of their dihydrosulfide complexes ($Cu(HS)_2^-$, $Zn(HS)_2^0$, $Ag(HS)_2^-$) analogous to reaction (4.76), the predicted effect of $S_3^{\bullet-}$ on these metals solubility was found to be almost negligible (Pokrovski et al. 2015). This result is in line with the much lower affinity to $HS^-$ of those metals than of $Au^I$ (Fig. 4.31b) and their dominantly chloride-type speciation in most hydrothermal fluids. By contrast, the effect of $S_3^{\bullet-}$ on Pt, Pd and, potentially, Mo speciation appears to be much stronger as discussed in the following sections.

#### *4.4.5. Platinum group elements (PGE)*

This group of metals includes Pd, Pt, Rh, Ir, Ru and Os, which are notoriously chalcophile being concentrated in (ultra)mafic magmatic settings along with Fe-Ni-Cu sulfide minerals and melts, such as in the famous Bushveld, Norilsk and Sudbury deposits from which >95% of the current PGE georessources are mined (e.g., Mungall and Naldrett 2008; Gunn 2014). The mechanisms of PGE concentration in magmatic deposits are believed to be controlled by different processes of fractional crystallization, fractionation and segregation of silicate and sulfide melts amply discussed in the literature (e.g., Kerr and Leitch 2005; Brenan 2008; Lorand et al. 2008; Barnes and Liu 2012; Mudd and Jowitt 2014; Barnes and Ripley 2016; Kiseeva et al. 2025; Simon and Wilke 2025). However, the potential role of aqueous magmatic or post-magmatic fluids in PGE ore formation processes has long been debated (e.g., Ballhaus and Stumpfl 1986; Kanitpanyacharoen and Boudreau 2013; Le Vaillant et al., 2016a,b; Boudreau, 2019; Sullivan et al., 2022a,b, to name a few). Furthermore, there are numerous instances of PGE (re)mobilization and concentration in various hydrothermal settings such as porphyry Cu-Au-Mo deposits, black-shale hosted gold deposits, and serpentinites from subduction zones (e.g., Ballhaus and Stumpfl 1986; Rehkämper et al. 1997; Wood 2002; Hanley 2005; Mungall 2005; Luguet et al. 2008; Economou-Eliopoulos 2010; Bazarkina et al. 2014; Arai and Miura 2016; Barnes and Ripley 2016; Holwell et al. 2017; Boudreau 2019), with Pd and Pt concentrations locally reaching 1000 times their average crustal abundance, thereby making such settings potentially exploitable for these metals. The interest in aqueous, including hydrothermal, chemistry of PGE has been rapidly growing in the last 20 years, pushed by the increasing use of these critical metals in many high-technology fields spanning from the automotive and petrochemical industries to pharmaceuticals and nanomaterials (Brenan 2008; Yam 2010; Gunn 2014). However, compared to the base and precious metals discussed above that all form nominally hydrothermal deposits (Zn, Pb, Cu, Ag, Au), only a small number of studies is available in hydrothermal fluids for Pd and Pt whose aqueous speciation and thermodynamic properties yet remain highly uncertain. Compared to the rest of the platinoids group, Pd and Pt are more abundant in the Earth's continental crust (0.8 ppb *vs* 0.04–0.08 ppb for the other PGE; Chen et al. 2016), more soluble and mobile in aqueous fluids, and more widely used by the industry. These properties make Pd and Pt more available, cheaper and therefore more amenable to scientific research. Their hydrothermal speciation and the role of sulfur in their fluid-phase transport will be overviewed here. The other PGE remain virtually unknown in hydrothermal fluids and only very scarce experimental data and thermodynamic estimations are available for some of them at temperatures above ambient (e.g., see Sassani and Shock 1998; Wood 2002).

*Palladium*

This element exists almost exclusively in $Pd^{II}$ oxidation state with an electronic configuration enabling square planar complexes with most ligands. Among them, water, chloride and sulfur are most relevant to hydrothermal conditions. The $Pd(H_2O)_4^{2+}$ ion, of square-planar coordination with 4 $H_2O$, undergoes hydrolysis at ambient conditions, by substitution of $H_2O$ by $OH^-$, by forming $PdOH^+$ at pH~1, and $Pd(OH)_2^0$ and $Pd(OH)_3^-$ at higher pH (the coordinating water molecules are omitted in these complexes for more fluency). However, the available estimations of the HKF parameters of $Pd^{2+}$ and high-temperature extrapolations of its hydrolysis constants (Sassani and Shock 1998; Tagirov et al. 2013, 2015) must be regarded with caution because the data are both very scattered and do not predict the expected increase in hydrolysis at elevated temperatures common for metal cations. Sulfate complexes of Pd were reported at ambient conditions; these data were used by Sassani and Shock (1998) to propose a set of HKF parameters for $PdSO_4^0$, $Pd(SO_4)_2^{2-}$, $Pd(SO_4)_3^{4-}$. However, both hydroxide and sulfate complexes appear to be largely negligible at hydrothermal conditions dominated by chloride and sulfide ligands for which Pd has a much greater chemical affinity.

Chloride complexes. In acidic chloride solutions below 100 °C, stepwise chloride complex formation, with four square-planar complexes, from $PdCl(H_2O)_3^+$ to $PdCl_4^{2-}$, is evidenced by a plethora of solubility, spectroscopy, potentiometry and quantum-chemistry studies, with $PdCl_4^{2-}$ being dominant at salinities above 0.1–0.3 m NaCl (see the reviews of Wood et al. 1992; Sassani and Shock 1998; Bazarkina et al. 2014). At typical hydrothermal conditions (<500 °C), the agreement on the dominant species and its thermodynamic stability is less good. First comprehensive theoretical estimations, based on low-temperature data and using the HKF model, predicted $PdCl_2^0$ and $PdCl_3^-$ as the dominant species (Sassani and Shock 1998), in contrast to earlier theoretical estimations (Wood et al. 1992) that privileged $PdCl_3^-$ and $PdCl_4^{2-}$. The latter speciation scheme, while being independent, appears in better agreement with pioneering solubility measurements of Gammons' group who used different Pd solid phases (Pd, $Pd_4S$, $Pd_{16}S_7$, PdS, and Ag-Pd alloy) to constrain the formation of $PdCl_3^-$ and $PdCl_4^{2-}$ to 300 °C (Gammons et al. 1992; Gammons and Bloom 1993a, Gammons 1995). More recent solubility measurements to higher temperatures (500 °C, 1 kbar) supported $PdCl_4^{2-}$ as the dominant species at salinities above 0.1 m NaCl; they were coupled with an analysis of selected existing data to generate a second set of HKF parameters for Pd chloride complexes (Tagirov et al. 2013). First systematic in situ XAS measurements at hydrothermal conditions, combined with $PdO_{(s)}$ and $PdS_{(s)}$ solubility determinations in saline solutions to 450 °C and 600 bar (Bazarkina et al. 2014), demonstrate that $PdCl_4^{2-}$ forms at the expense of $PdCl_3(H_2O)^-$ at salinities above 0.1–0.3m NaCl in a wide *T* range, from ~100 to 450 °C. Molecular dynamics simulations at 300 °C, using a thermodynamic integration method, imply $PdCl_4^{2-}$ to be the most stable (i.e. most abundant) among the four chloride complexes (Mei et al. 2015b). Solubility measurements of $Pd_{(s)}$ at magmatic temperatures in saline fluids (900 °C, 2 kbar, to 60 wt% NaCl), however, point to lower-order chloride complex stoichiometries, $PdCl_2^0$ and $PtCl_3^-$ (Sullivan et al. 2022a).

More importantly, the discrepancies in the thermodynamic stability of Pd chloride species among those studies are remarkably large for the same species, for example attaining 7 orders of magnitude for the $PdS_{(s)}$ solubility reaction as $PdCl_4^{2-}$ at temperatures 300–450 °C (see Bazarkina et al. 2014 for detailed comparisons). This difference translates to an equivalent uncertainty of 7 orders of magnitude (!) in aqueous Pd concentrations in equilibrium with a Pd-bearing solid phase in typical hydrothermal fluids. For example, those of Sassani and Shock (1998) yield the lowest Pd concentration values (<1 ppt) and those of Tagirov et al. (2013) – the highest (~1 ppm) at salinities of 10 wt% NaCl at acidic (pH~4) and oxidizing (HM buffer) conditions. The discrepancy becomes even more spectacular at magmatic temperatures, up to 12 orders of magnitude between those two HKF-model predictions, with predicted solubilities of $Pd_{(s)}$ of <10 ppt (Sassani and Shock 1998) and >$10^6$ ppm (Tagirov et al. 2013) at 900 °C, 2 kbar, 10 wt% NaCl and $f_{O2}$ of the NNO buffer. Both predicted values seem unlikely to be realistic, compared to direct experimental measurements that report ~10 ppm Pd at such conditions (Sullivan et al. 2022a).

We are thus unable to recommend any of the thermodynamic datasets published so far for estimating the concentrations of Pd chloride complexes in hydrothermal-magmatic fluids. Based on the data from direct $PdS_{(s)}$ and $Pd_{(s)}$ solubility measurements as such conditions (Bazarkina et al. 2014; Sullivan et al. 2022a) as well as those extrapolated from lower temperatures (e.g., Gammons and Bloom 1993a), Pd transport as chloride complexes may only be significant (i.e. above 1–10 ppb) in acidic, sulfur-poor, and highly-saline fluids at temperatures above 400–450 °C. Therefore, to account for Pd mobilization by most types of sulfur-bearing near-neutral pH hydrothermal fluids, other types of complexes, in particular those with sulfur, are likely to be required.

*Sulfide complexes*. This type of Pd complexes has received far more limited experimental quantification than chlorides. This limitation is primarily due to intrinsically lower Pd dissolved concentrations that could be obtained in equilibrium with Pd sulfide solids, which are the stable phases in $H_2S$-bearing systems, in contrast to $PdO_{(s)}$ and $Pd_{(s)}$ commonly used in solubility studies in S-free Cl-bearing fluids discussed above. Furthermore, the Pd-S system is characterized by the existence of several sulfide phases (e.g., PdS, $Pd_4S$, $Pd_{16}S_7$) whose thermodynamic properties and the resulting stability regions are yet poorly quantified. For example, the value of the standard Gibbs energy of formation, $\Delta_f G^0$ at 25 °C and 1 bar, of stoichiometric $PdS_{(s)}$ varies over >15 kJ/mol in different studies (see Bazarkina et al. 2014 for a recent compilation). This range of values corresponds to a $PdS_{(s)}$ solubility variation of >1.5 log units, bringing additional uncertainty to the stability constants of Pd complexes derived in different studies. However, this uncertainty is yet much smaller than that associated with the derivation of aqueous Pd-S complexes from solubility studies.

First systematic investigations of such complexes have been made by C. Gammons's and S. Wood's groups. Gammons and Bloom (1993) measured $PdS_{(s)}$ and $PtS_{(s)}$ solubility in reduced $H_2S/HS^-$ solutions (pH = 0.5–8.4, 0.4–1.0 m S) from 200 to 300° C at $P_{sat}$ using fused silica tubes quenched at the end of experiment. Their interpretation suggests three Pd ($Pd(HS)_2^0$, $Pd(HS)_3^-$ and $Pd(HS)_4^{2-}$) and Pt ($Pt(HS)_2^0$, $Pt(HS)_3^-$ and $Pt(HS)_4^{2-}$) complexes, among which di- and trihydrosulfide would be the most abundant for both metals. However, their solubility-pH-$H_2S$ trends were too scattered to allow robust constraints on the exact $HS^-$ ligation numbers. Their study contrasts with that of Pan and Wood (1994) and Wood et al. (1994) who used $H_2S/SO_4^{2-}$ solutions of comparable pH and sulfide concentration ranges (pH 5.9–9.4; 0.3–2.2 m S) to simultaneously measure $PtS_{(s)}/PtS_{2(s)}$ and $PdS_{(s)}$ solubility from 25 to 90 °C and from 200 to 350 °C at $P_{sat}$ using hydrothermal reactors allowing fluid sampling. The authors tentatively assumed $Pd(HS)_2^0$ and $Pt(HS)_2^0$ to be the major species. However, they failed to provide enough constraints on the stoichiometry and proposed to the reader a large choice of alternative complexes of different ligation numbers (from $Pd(HS)^+$ to $Pd(HS)_4^{2-}$ and from $Pt(HS)^+$ to $Pd(HS)_4^{2-}$) with their corresponding formation constants. Some of their tabulated constants were later identified to be in error (Pokrovski et al. 2021b). While the measured Pt and Pd concentrations in these two studies were comparable, the predicted absolute $PdS_{(s)}$ and $PtS_{(s)}$ solubility values using corrected Pan and Wood's (1994) stability constants are, however, 3 to 4 orders of magnitude higher than those from Gammons and Bloom (1993) at the same temperature and solution composition. This discrepancy has remained unresolved for more than 15 years. Tagirov and Baranova (2009) measured $PdS_{(s)}$ solubility in $H_2S$ (up to 0.6 m)-NaOH-HCl solutions (2<pH<10) from 5 to 200 °C using ampoule- or batch-reactor quench-based methods. Despite a large solubility data scatter (over ~3 log units), the authors tentatively derived from the analyses of solubility trends 3 complexes, $Pd(HS)_2^0$, $Pd(HS)_3^-$, and $PdS(HS)_2^{2-}$, which were dominant at acidic, neutral, and alkaline pH, respectively. Their data and extrapolations to higher temperatures for $Pd(HS)_2^0$ appear to be strongly inconsistent with both Gammons' and Wood's works, with stability constants being from 4 to 8 log units lower, while for $Pd(HS)_3^-$ discrepancies are somewhat smaller. Tagirov et al. (2013) further used their data and ligand exchange reactions with the analogous Pd-Cl species to generate a set of HKF parameters for those Pd-S complexes. However, in light of the discrepancies discussed above, their use cannot currently be recommended. Additional caution as to the Pd-S complex stoichiometry was added by MD calculations (Mei et al. 2015b), which predicted much lower constants (by 3 to 6 log units) for $Pd(HS)_2^0$ and $Pd(HS)_3^-$ and, instead, the dominant formation of $Pd(HS)_4^{2-}$, rejected by Tagirov's team. Very recently, Laskar (2022) and Laskar et al. (2022, 2024) reported $PdS_{(s)}$ and $PtS_{(s)}$ solubility measurements in reduced $H_2S$-NaOH solutions at slightly acidic to alkaline pH with the pyrrhotite-pyrite buffer from 50 to 300 °C and 100–500 bar. They used a flexible-cell reactor equipped with an accurate sampling device and analytical setups allowing metal concentration analyses down to ppb levels. The authors demonstrated the dominant formation of $Pd(HS)_4^{2-}$ and $Pt(HS)_4^{2-}$ (see below), and reported a first set of HKF parameters for both complexes. The stability constant of the $Pd(HS)_4^{2-}$ complex derived in their work is in agreement with Mei et al.'s (2015b) predictions, lending more credence to the high stability of the symmetrical tetra-hydrosulfide species, rather than mixed $[Pd(HS)_n(H_2O)_{4-n}]^{2-n}$ species that have chemically contrasting ligands (S and O) in the metal coordination sphere. Such mixed-ligand complexes are generally expected to be less stable, in agreement with the fundamental hard-soft relationships. Laskar (2022) further measured $PtS_{(s)}$ and $PdS_{(s)}$ solubility in sulfate-sulfide solutions at 300 °C, 300–500 bar, and revealed large solubility enhancements for both Pt and Pd (from ppb level in reduced $H_2S$ solutions to 1–10 ppm level), thereby confirming the formation of recently discovered stable complexes with the $S_3^{\bullet-}$ ion (Pokrovski et al. 2021b), as will be discussed below for the case of platinum.

*Platinum*

Platinum primarily distinguishes itself from palladium by the existence along with $Pt^{II}$ of another oxidation state, $Pt^{IV}$. Both $Pt^{II}$ and $Pt^{IV}$ form oxide, chloride and sulfide solid phases and corresponding aqueous species within the range of terrestrial redox conditions. Divalent $Pt^{II}$ and its aqueous complexes and solid phases (e.g., $PtS_{(s)}$) are more common in hydrothermal environments, even though $Pt^{IV}S_{2(s)}$ also occurs, along with stable $Pt^{IV}$ complexes with hydrosulfide and polysulfide ligands, whereas $Pt^{IV}$-O and $Pt^{IV}$-Cl types of compounds and aqueous species are only stable in surficial oxygenated environments (e.g., see Reith et al. 2014 for a recent review). Similarly to $Pd^{II}$, the great majority of $Pt^{II}$ aqueous complexes are square-planar, whereas those of $Pt^{VI}$ are octahedral (Pokrovski et al. 2021b).

The divalent $Pt^{2+}$ cation is strongly hydrolyzed in water and may only be significant at extremely acidic negative pH (<–2.5) above which the $PtOH^+$ and $Pt(OH)_2^0$ complexes likely to form (Baes and Mesmer 1976; Wood 1991; Sassani and Shock, 1998). Because of the instability of $Pt^{2+}$ in aqueous solution, published estimations of its Gibbs energy vary over more than 30 kJ/mol at ambient conditions and up to 50 kJ/mol at hydrothermal temperatures (Sassani and Shock 1998;

Tagirov et al. 2015). Like the $Au^{+}$ ion, $Pt^{2+}$ should be avoided in thermodynamic modeling in aqueous solution. Caution should be taken when deriving and reporting absolute values of stability constants of $Pt^{II}$ complexes from cumulative reactions of complex formation involving $Pt^{2+}$ – the approach commonly used in MD theoretical studies; such data will be affected by at least the same uncertainty (see Laskar et al. 2022 for extended discussion). According to the HKF-model parameters reported by Sassani and Shock (1998), hydroxide and sulfate complexes of $Pt^{II}$ would represent a negligible contribution to Pt mobility in typical hydrothermal fluids, with less than 1 Pt atom per kg of water. Like for Pd and other metals overviewed above, the two major ligands for Pt considered so far in hydrothermal research are chloride and (hydro)sulfide.

*Chloride complexes*. Because of the much stronger $Pt^{2+}$ (and $Pt^{4+}$) hydrolysis, $Pt^{II}$ and $Pt^{IV}$ chloride complexes in low-temperature aqueous solution (<100 °C) are less constrained than those of $Pd^{II}$. A series of $PtCl_n^{2-n}$ complexes with $n$ from 1 to 4 has usually been considered in the chemical literature. The estimations of their formation constants from $Pt^{2+}$ at ambient conditions, based on meagre existing data, most of them being measured at high ionic strength, are uncertain by at least 3 to 8 log units, with even greater uncertainties when extrapolated above 100 °C (see the review of Sassani and Shock 1998). Direct experimental data at hydrothermal temperatures are limited to a handful of studies. Gammons' group conducted first systematic solubility measurements of $Pt_{(s)}$ and $AgCl_{(s)}$ and in HCl-NaCl-$SO_4$ saline solutions to 300 °C at $P_{sat}$ (Gammons et al. 1992; Gammons 1995, 1996). They identified $PtCl_3^-$ and $PtCl_4^{2-}$ for $Pt^{II}$ and $PtCl_5^-$ and $PtCl_6^{2-}$ for $Pt^{IV}$ as the dominant complexes. Their results indicate that $Pt^{IV}$ chlorides are restricted only to very acidic and oxidizing conditions corresponding to atmospheric $O_2$ pressures. Under hydrothermal conditions, $Pt^{II}$ chloride complexes may only be significant (> 1 ppb Pt) in unusually acidic (pH<4), highly saline (>15 wt% NaCl) and oxidizing (hematite stable) fluids. The interest for Cl complexes of Pt has therefore faded until a new generation of researchers coming with improved experimental and analytical methods. Tagirov et al. (2015) extended Gammons' data to supercritical conditions by measuring $Pt_{(s)}$ solubility in a narrow $T$ range of 400–475 °C at 1 kbar, and interpreted the data as the $PtCl_3^-$ complex. Based on these and some previous data the authors generated a new set of HKF parameters for $Pt^{2+}$ and the four $Pt^{II}$-Cl complexes. However, in their following study of $Pt_{(s)}$ solubility aided by in situ XAS measurements to 575 °C and 5 kbar, Tagirov et al. (2019b) reconsidered their data in terms of the dominant $PtCl_4^{2-}$ and proposed a density-model equation to describe, up to magmatic temperatures, the solubility constant of the reaction

$$Pt(s) + 2\ HCl^0_{(aq)} + 2\ Cl^- = PtCl_4^{2-} + H_{2(aq)} \quad (4.77)$$

Their equation reasonably matches the $Pt_{(s)}$ solubility data in the acidic NaCl brine phase at 800 °C in the system vapor-brine-rhyolite melt (Simon and Pettke 2009), whereas in the corresponding vapor phase it significantly underestimates the measured solubilities, which were instead tentatively interpreted by the uncharged complex $PtCl_2^0$. The tendency to form less chlorinated and less charged complexes with increasing temperature and decreasing salinity seems to be confirmed by the recent study of Sullivan et al. (2022b) who studied $Pt_{(s)}$ solubility as a function of NaCl (to ~30 m) and HCl (to ~3 m) concentrations in aqueous fluids at 800–1000 °C and 2 kbar and redox conditions of NNO±1. Their solubility dependences would be more consistent with $PtCl_2^0$ and $PtCl_3^-$ than with $PtCl_4^{2-}$, but the absolute solubility values (10–100 ppm Pt) are also equally matched by $PtCl_4^{2-}$ of Tagirov et al. (2019b). For comparison, older HKF predictions of $Pt_{(s)}$ solubility of Sassani and Shock (1998) and Tagirov et al. (2015), respectively, underestimate by 7 log units and overestimate by 4 log units the measured solubilities at the Sullivan et al. (2022b) experimental conditions. Thus, at present, Tagirov et al.'s (2019b) thermodynamic data for $PtCl_4^{2-}$ are recommended for modeling Pt transport in saline fluids over a range of hydrothermal-magmatic conditions. The contribution of Pt-Cl complexes is very weak in the hydrothermal domain (<500 °C) being limited to oxidized, highly-saline, S-poor, and acidic conditions, but grows towards magmatic temperatures at which $PtCl_2^0$ and/or $PtCl_3^-$ may also contribute. Their thermodynamic quantification awaits more systematic data coupled with in situ approaches. Like for Pd, accounting for Pt mobility in sulfide-bearing, near-neutral pH and reduced hydrothermal-magmatic fluids requires the assessment of complexes with reduced sulfur ligands.

*Sulfide complexes*. Platinum sulfide ($PtS_{(s)}$ and $PtS_{2(s)}$) solubilities were investigated together with $PdS_{(s)}$ in $H_2S$-bearing fluids under hydrothermal conditions in early work of Gammons' and Wood's groups, as detailed above for the case of palladium. Whereas the measured Pt concentrations were similar in those works, the reported solubility constants for the same reactions (e.g., 4.78) presented similarly large discrepancies as for Pd-hydrosulfide complexes. It is only by the 2020s that some advance has been made towards resolution of these discrepancies. Kokh et al. (2017) used a flexible-cell reactor to measure $PtS_{(s)}$ solubility in supercritical $H_2O$-$CO_2$ fluids with pH and redox buffered by potassium aluminosilicate and sulfide/oxide iron mineral assemblages at 450 °C and ~700 bar (pH = 4–6, 0.02–0.06 m $H_2S$). The constancy of the solubility as a function of $CO_2$ content (up to 50 wt%) was consistent with the uncharged $Pt(HS)_2^0$ complex with equilibrium concentrations of 1–2 ppb:

$$PtS_{(s)} + H_2S_{(aq)} = Pt(HS)_2^0 \quad (4.78)$$

Such measured concentrations were, however, at least 100 times lower than those predicted by extrapolating the solubility constants for $Pt(HS)_2^0$ of Pan and Wood (1994). Filimonova et al. (2021) measured $PtS_{(s)}$ solubility in $H_2S/HS^-$ solutions (pH = 3–8, 0.01–0.2 m S) at 50–75 °C, 1 bar, and at 450 °C, 1000 bar, using batch reactors quenched at the run end. They tentatively interpreted their data by $Pt(HS)_2^0$ as the dominant species, with $PtS_{(s)}$ solubility magnitudes matching those of Gammons and Bloom (1993) and Kokh et al. (2017). However, the ppb-level Pt concentrations found might have been subjected to significant uncertainties by the use of a quench technique that often suffers from elevated analytical backgrounds. Thus, the bulk-solubility approach alone for PGE in hydrothermal sulfur-bearing systems does not provide sufficient resolution that would allow unambiguous constraints on sulfide complex stoichiometry and stability. Very recently, Laskar et al. (2022) combined in a single study three complementary approaches – solubility measurements, in situ XAS experiments, and MD simulations – to demonstrate that the tetrahydrosulfide complex, $Pt(HS)_4^{2-}$, is the dominant species, to at least 300 °C in $H_2S$-rich (0.2–2.0 m S) near-neutral to alkaline ($5 \leq pH \leq 9$) solutions. They derived a coherent set of constants of the dissolution reaction

$$PtS_{(s)} + 3\ H_2S_{(aq)} = Pt(HS)_4^{2-} + 2\ H^+ \qquad (4.79)$$

over the temperature range 25–300 °C and pressure range $P_{sat}$–600 bar, consistent with the results of the three independent methods employed. Furthermore, their MD estimations yield the solubility constants of lower-ligation number complexes, $Pt(HS)_2^0$ and $Pt(HS)_3^-$ (analogous to reaction 4.79), to be 7 to 15 orders of magnitude lower than suggested in the previous solubility studies above. Such a difference is well beyond the errors associated with the determination of stability constants from MD, supporting the overwhelming predominance of $Pt(HS)_4^{2-}$ over those complexes. Laskar et al. (2022) further re-analyzed Gammons' and Wood's solubility data in terms of $Pt(HS)_4^{2-}$ and obtained stability constant values better matching those of $\log K_{79}$ above thereby, at least partly, reconciling those previous data. Thus, it turns out that, similarly to Pd, the symmetrical, fully S-coordinated, tetrahydrosulfide species would be the dominant carrier of Pt at moderate temperatures in reduced sulfide-bearing hydrothermal fluids of near-neutral pH. However, in sulfur-poor fluids and at higher temperatures, lower $HS^-$ ligation number complexes, such as $Pt(HS)_2^0$, may over compete $Pt(HS)_4^{2-}$, as indicated by rare solubility data existing for supercritical fluids (Kokh et al. 2017; Filimonova et al. 2021). This tendency would be in line with the water dielectric constant decrease favoring ion association and uncharged or weakly charged species stability. From the other hand, the increasing ion pairing with alkalis, similar to Au and Cu chloride and hydrosulfide complexes (see above), coupled with the high species symmetry, might maintain the high stability of Pt tetrahydrosulfide up to magmatic temperatures.

It follows that $PtCl_4^{2-}$ and $Pt(HS)_4^{2-}$, which are the only Pt species directly evidenced by in situ spectroscopy, would be the major Pt-bearing complexes in most moderate-temperature hydrothermal fluids (<400 °C). At higher temperatures, $PtCl_2^0$, $PtCl_3^-$, and $Pt(HS)_2^0$ may over compete those species, but their direct spectroscopic proof is yet lacking and more robust constraints are required on their stability constants over the large *T-P* range. The solubility of $PtS_{(s)}$ in moderately-saline sulfur-bearing fluids of ~10 wt% NaCl equivalent and 0.1 m $H_2S$ under redox conditions corresponding to those of the conventional Ni-NiO mineral oxygen buffer (NNO) is illustrated in Fig. 4.31a as a function of pH at 300 °C and 500 bar. The solubility is dominated by $Pt(HS)_4^{2-}$ (thermodynamic data from Laskar et al. 2022) at moderately-acidic to alkaline pH, whereas the domain of predominance of $PtCl_4^{2-}$ is limited to very acidic pH (pH<3). The contribution of hydroxide and sulfate complexes (Sassani and Shock 1998; not shown) is virtually negligible being <1 Pt atom per kg of fluid. The maximum solubility is attained at near-neutral pH (6–7) as $Pt(HS)_4^{2-}$, but the total Pt dissolved concentrations remain too small (<0.1–1 ppb Pt at pH 5–7) to account for significant Pt transport in most epithermal environments. For example, such Pt dissolved concentrations would have required an amount of fluid equivalent to the global flux of water through all subduction zones operating for 1000 years (e.g., Wallace 2005) to form a modest deposit of 10 tons of Pt provided much focused fluid flow and 100% precipitation efficiency. Such a scenario would be geologically implausible. Therefore, it is unlikely that Pt-Cl or Pt-HS types of complexes may explain the multiple instances of PGE remobilization and concentration commonly observed in sulfide-dominated hydrothermal environments.

*Comparison with gold*. The behavior of PGE in hydrothermal fluids contrasts with that of gold, which is far more soluble and mobile than Pt and Pd in most epithermal systems (<300–350 °C) that host large gold deposits (10s to 100s tons). In epithermal fluids, the speciation of $Au^{I}$ is dominated by similar hydrosulfide complexes ($Au(HS)_2^-$) easily attaining ppm-level concentrations (e.g., Fig. 4.30; Kouzmanov and Pokrovski 2012 and references therein), which is ~1000 times more than Pt. The chemical affinities of $Pt^{II}$ and $Au^{I}$ to the aqueous $HS^-$ ligand compared with $Cl^-$ are astonishingly similar as can be seen from the ligand exchange reactions:

$$PtCl_4^{2-} + 4\ HS^- = Pt(HS)_4^{2-} + 4\ Cl^- \quad (4.80)$$

$$AuCl_2^- + 2\ HS^- = Au(HS)_2^- + 2\ Cl^- \quad (4.81)$$

Their equilibrium constant values at 300 °C and 500 bar ($\log K_{80}$ and $\log K_{81}$ of 23.2 and 11.4, respectively), derived by combining the thermodynamic properties of Pt and Au complexes recommended in this study (Table 4.6), correspond, respectively, to 5.8 and 5.7 per $HS^-$ ligand when normalized to the number of ligands in the Pt (/4) and Au (/2) complexes. Thus, the key difference between Au from one hand and Pt and Pd from the other hand in hydrothermal systems is the different solubility-controlling solid phases. The formation of poorly soluble Pt and Pd sulfide minerals is the major limiting factor for PGE mobility compared to Au whose solubility is largely controlled by the native metal. Had Pt (and Pd) metal been the stable phase in S-bearing environments, its solubility would have been comparable with that of Au metal (Fig. 4.31a). This difference in the solid-phase control of sulfur on the noble metal transport is the main cause of the general scarcity of Pt and Pd (and by inference other PGE) compared to Au in epithermal deposits, despite their similar Clarke values (~1 ppb) and almost identical chemical affinities for the hydrosulfide ligand in aqueous solution.

*Comparison with base metals*. The gold and platinum affinity for $HS^-$ versus $Cl^-$ ligands may further be compared with that of base metals (Fe, Cu, Zn, Pb) and silver as illustrated in Fig. 4.31b in terms of symmetrical ligand exchange reactions (such as 4.68, 4.69, 4.80 and 4.81) for di-chloride and di-hydrosulfide complexes, which are the best constrained for all those metals. Two major tendencies emerge from this figure. First, the affinity of $Au^I$ and $Pt^{II}$ for $HS^-$ *vs* $Cl^-$ is almost 10 orders of magnitude greater (in terms of log*K* value) than that of the four base metals and silver at ambient temperature with differences slightly diminishing with temperature increase. Second, for all metals shown (note that there is only a single data point for Fe as will be discussed in the next section) the affinity for $HS^-$ vs $Cl^-$ systematically decreases with increasing temperature. The temperature trend directly reflects the general soft-hard metal-ligand relationships such as i) stronger soft metal (Au, Pt) - soft ligand ($HS^-$) complexes and ii) weakening the stability of any soft ligand complexes in favor of harder ligand complexes with increasing temperature and thermal disorder and decreasing the water dielectric constant. As discussed in sections above, the uncertainties on the log*K* values are strongly temperature-dependent, with the best constrained data being around 300±50 °C. At such temperatures, the order of increasing metal affinity for reduced sulfur is the following: $Cu^I \leq Ag^I \approx Zn^{II} \leq Pb^{II} \approx Fe^{II} << Au^I \approx Pt^{II}$. At higher, supercritical, temperatures, the order among the base metals is more uncertain, both because of significant uncertainties on their log*K* values and the potential presence of higher-order chloride as well as mixed Cl-HS species as discussed above. Finally, an interesting trend with pressure can be seen for Au, which is the best studied metal as to its $HS^-$ and $Cl^-$ complexes. The affinity for reduced sulfur seems to slightly increase with decreasing pressure, which is also in line with greater volatility of sulfur-bearing Au and Pt complexes in the vapor phase (Pokrovski et al. 2013a), which is beyond the scope of this chapter. Among the considered metals, data for $Pt^{II}$ hydrosulfide complexes remain the least constrained to date; however, even within their large uncertainties, PGE sulfide solubilities as sulfide and chloride complexes appear to be negligibly small in typical hydrothermal fluids.

*Complexes with the trisulfur radical ion*. To account for Pt and Pd mobility in hydrothermal fluids, complexes with ligands other than $HS^-$ or $Cl^-$ would be required. The most plausible one is the trisulfur radical ion, stable in a wide range of temperature and capable of forming stable complexes with $Au^I$ (see section 4.4.4.) The first recognition of the role of the $S_3^{\bullet -}$ ion in PGE transport came from a recent study of Pokrovski et al. (2021b) who measured $PtS_{(s)}$ solubility in sulfate-sulfide solutions containing significant concentrations of $S_3^{\bullet -}$ within a narrow *T-P* and S concentration range (275–300 °C, 500–700 bar, 1–2 m S), using different types of hydrothermal reactors (both quench- and sampling-based) and in situ X-ray absorption spectroscopy, coupled with quantum chemical modeling of the species structures and XANES spectra. Their data show that the presence of $S_3^{\bullet-}$ enhances Pt solubility by 3 to 5 log units compared to $H_2S$-dominated solutions of similar sulfur concentrations (Laskar et al. 2022), due to the formation of mixed $HS^-/S_3^{\bullet-}$ complexes with both $Pt^{II}$ and $Pt^{IV}$ according to the reactions

$$PtS_{(s)} + H_2S_{(aq)} + 2\ S_3^{\bullet-} = Pt^{II}(HS)_2(S_3)_2^{2-} \quad (4.82)$$

$$PtS_{(s)} + 2\ H_2S_{(aq)} + 2\ S_3^{\bullet-} + H^+ + 0.5\ O_{2(g)} = Pt^{IV}(HS)_3(H_2O)(S_3)_2^{-} \quad (4.83)$$

Laskar (2022) further confirmed the reaction (4.82) stoichiometry and equilibrium constant by additional solubility measurements at 300 °C. He reported an analogous complex for Pd with a similar stability constant value. The stability constants of reactions (4.82) and (4.83) enable predictions of Pt solubility in epithermal fluids as illustrated in Fig. 4.32. The solubility enhancement due to the formation of these complexes, apparent in this figure at sulfur concentrations above 0.1 wt% (Fig. 4.32a), $f_{O_2}$ around the sulfate-sulfide transition (≈ HM±1; Fig. 4.32b) in the acidic-to-neutral pH range (Fig. 4.32c), appears to be much stronger than for $Au^I$ (Fig. 4.32d). Indeed, the ligand exchange reaction

$$Pt(HS)_4^{2-} + 2\ S_3^{\bullet-} = Pt(HS)_2(S_3)_2^{2-} + 2\ HS^- \quad (4.84)$$

has a $K$ value of ~10,000 as calculated by combining the data of Pokrovski et al. (2021b) and Laskar et al. (2022), whereas the corresponding value for an analogous $Au^I$ exchange reaction (4.76) is 1. An equally high value was obtained by Laskar (2022) for the $Pd^{II}$ exchange reaction analogous to reaction (4.84). The much higher affinity of $S_3^{\bullet-}$ than $HS^-$ both for $Pd^{II}$ and $Pt^{II}$ is likely due to a combination of the PGE unfilled valence electron shells ($4d^8$ and $5d^8$, respectively) and 4-coordinated $Pd^{II}$ and $Pt^{II}$ (as well as 6-coordinated for $Pt^{IV}$) geometries, both offering more flexibility for binding with $S_3^{\bullet-}$, compared with $Au^I$ that has its 5d shell filled ($5d^{10}$) and almost exclusively linear coordination geometries.

The discovery of stable Pd and Pt complexes with the trisulfur ion might provide a plausible explanation for PGE enrichment cases observed in sulfide-dominated hydrothermal environments. It further emphasizes the potential role of aqueous S-bearing fluids in PGE magmatic deposit formation. This role would be consistent with the growing abundance of both $S_3^{\bullet-}$ and $S_2^{\bullet-}$ in magmatic-derived fluids (Figs. 20, 21). Selective complexing by the radical ions of Pt and Pd, compared to the other more refractory and less chalcophile PGE (Ru, Rh, Ir, Os), may induce fluid-mineral fractionations affecting PGE elemental and isotope signatures in mantle-derived rocks widely used as tracers of core-mantle evolution (Kiseeva et al. 2025 and references therein). The highly soluble $S_3^{\bullet-}$ complexes with both Pd and Pt may open new routes for hydrothermal synthesis of PGE-based nanomaterials as well as PGE ore processing. There is thus an urgent need to extend these first findings over the wider hydrothermal-magmatic *T-P* range via more systematic studies of PGE interactions with the sulfur radical ions.

### *4.4.6 Iron*

Iron holds a particular place in metal-sulfur interactions as pyrite ($FeS_2$) is by far the most abundant metal sulfide mineral in the Earth's crust. However, Fe-S relationships in the fluid phase are quite different. Both $Fe^{II}$ and $Fe^{III}$ oxidation states coexist in natural fluids and minerals, with $Fe^{III}$ being thermodynamically stable in oxygenated Earth's surface environments. Increasing temperature and depth generally makes $Fe^{II}$ more abundant both in solid and fluid phases, but $Fe^{III}$ still remains an important redox form in a wide range of oxygen fugacity, as manifested by ubiquitous coexistence of both redox states in mineral assemblages (e.g., magnetite-hematite-pyrite, with equilibrium $f_{O2}$ ~$10^{-30}$ bar at 300 °C), and by the presence of both $Fe^{II}$ and $Fe^{III}$ in major silicate minerals (e.g., serpentinite, garnet) and silicate glasses over a wide $T$-$P$-$f_{O2}$ range of metamorphic and magmatic conditions. In sulfide-dominated aquatic environments, both near-surface and hydrothermal, pyrite is notoriously stable, being transformed to pyrrhotite when going towards magmatic and high-metamorphic (amphibolite facies) conditions. In aqueous solution, the $Fe^{2+}$ cation is dominant to pH~9, whereas $Fe^{3+}$, being chemically "harder", is hydrolyzed at pH>2 (Baes and Mesmer 1976). Both cations form hydroxide and/or organic ligand complexes and sulfide clusters in near surface aquatic environments, but in hydrothermal fluids their speciation is very different. Both $Fe^{II}$ and $Fe^{III}$ hydroxide species domains greatly widen with increasing temperature at the expense of the cations. These species are reasonably well constrained to at least 350 °C by solubility and spectroscopic experiments (e.g., Sweeton and Baes 1970; Tremaine and Leblanc 1980; Diakonov et al. 1999; Stefánsson and Seward 2008; references therein) and thermodynamic predictions (Shock et al. 1997). Discrepancies among the stability constant values for the given $[Fe(OH)_n]$ complex in different studies typically amount 1–3 log units.

Irrespective to the exact $K$ values, the solubilities of iron oxides as hydroxide complexes are quite low in typical hydrothermal fluids (≤ ppm levels of Fe). The major Fe carrier is believed to be chloride. In near-ambient aqueous solution, both $Fe^{II}$ and $Fe^{III}$ are known to form at least 4 chloride complexes, from mono- to tetrachloride. Yet, there are large discrepancies as to their stability and composition in hydrothermal fluids, despite a large amount of work on $Fe^{II}$-chloride complexation. Similarly to other base metals such as Zn and Pb (see above), most available solubility work tends to converge to the dominant presence of dichloride species stoichiometries, $FeCl_2^0$, over a wide salinity range (see Wood and Samson 1998; Sverjensky et al. 1997; Simon et al. 2004; Testemale et al. 2009; Saunier et al. 2011; references therein). In contrast, spectroscopy-based approaches (UV-Vis, XAS and Raman) tend to favor higher-order, $FeCl_4^{2-}$ (and/or $FeCl_3^-$) species at similar fluid salinities (Liu et al. 2006; Testemale et al. 2009; Scholten et al. 2019). Recent hematite and pyrite solubility measurements at supercritical temperatures (e.g., Simon et al. 2004; Saunier et al. 2011; Syverson et al. 2013; Pokrovski et al. 2021c) are well accounted for by the predictions using the thermodynamic properties of $FeCl_2^0$ available within the HKF model (Sverjensky et al. 1997). In contrast, those of $FeCl_4^{2-}$ (Testemale et al. 2009) lead to significant overestimations of the solubility results. Other recent solubility experiments also show $Fe^{II}$ dichloride to be the major complex in low-density hydrothermal vapors and supercritical fluids (400–500 °C, ρ<0.3 g/cm$^3$) produced by seawater boiling and fluid unmixing in active geothermal systems of mid-ocean ridges (Xing et al. 2021). Therefore, higher-order $Fe^{II}$ chloride complexes still await more in-depth experimental and thermodynamic efforts to be considered in quantitative modeling of iron transport by hydrothermal-magmatic fluids.

Data on $Fe^{III}$ chloride species are less abundant under hydrothermal conditions. A handful of solubility and spectroscopic studies (Liu et al. 2006; Saunier et al. 2011; Stefánsson et al. 2019; Gammons and Allin 2022) demonstrates the dominant formation of $Fe^{III}Cl_3^0$ and $Fe^{III}Cl_4^-$ at oxidizing conditions (hematite stable). However, the numerical thermodynamic data for those complexes available so far are too disparate or available within too narrow temperature ranges to allow reliable predictions of Fe(III) abundance at hydrothermal-magmatic conditions. For example, the reported solubility constants of hematite with formation of the $FeCl_4^-$ complex differ as much as 10 log units at 300 °C (e.g., Stefánsson et al. 2019 vs Gammons and Allin 2022). Given the growing recognition that subduction-zone fluids may be quite oxidized, with $f_{O_2}$ values sometimes above that of the HM buffer (e.g., Debret et al. 2016; Debret and Sverjensky 2017; Pokrovski et al. 2022a; Schwarzenbach and Evans 2025; references therein), it is likely that $Fe^{III}$ chloride, and possibly sulfate, complexes might be the major carriers of Fe in deep subduction zones.

Stable $Fe^{II}$-(hydro)sulfide complexes are known at near-ambient conditions in the presence of reduced sulfur, such as $FeHS^+$, along with a number of polynuclear hydrosulfide (e.g., $Fe_3(HS)_5^+$), polysulfide (e.g., $FeS_5^0$, $Fe_2(S_5)_2^{2+}$), and sulfide ($Fe_xS_x$, conventionally expressed as $FeS^0$) clusters, for which tentative stability constant values at ambient conditions and high ionic strength have been reported in few studies (see Rickard and Luther 2006, 2007 for extensive reviews). Precursors of $[Fe_xS_y]$-like stoichiometries, with short-range structures resembling those of mackinawite ($FeS_m$) or troilite (FeS), have traditionally been assumed to operate over a wide temperature range, from ambient up to at least 300 °C during pyrite and marcasite precipitation (e.g., Murowchick and Barnes 1986; Schoonen and Barnes 1991; Graham and Ohmoto 1994). At moderate temperature conditions (<100 °C), the formation of pyrite has commonly been regarded to occur via two possible mechanisms, by reaction of soluble FeS precursors either with $H_2S$ or with polysulfide dianions, $S_n^{2-}$ (Rickard and Luther 2007 for details). However, few constraints on the mechanisms of pyrite formation are available at hydrothermal temperatures, since very few data exist on Fe-S aqueous speciation at such conditions. By analogy with other metals overviewed above, monomeric $Fe(HS)_n^{2-n}$ complexes would be expected to predominate at elevated temperatures both because of the increasing thermal disorder destabilizing polymeric species more common at ambient conditions, and very low equilibrium concentrations of polysulfide dianions themselves (e.g., Fig. 4.29). Equilibrium constants for $FeHS^+$ have only been reported at ambient conditions in seawater (as compiled in Rickard and Luther 2006, 2007) with values of log$K$ ≈ 4–6 of the reaction

$$Fe^{2+} + HS^- = FeHS^+ \quad (4.85)$$

These values compare favorably with those of the analogous $CoHS^+$ species deduced from $CoS_{(s)}$ solubility experiments by Migdisov et al. (2011), with log$K$ ≈ 5.8–6.5 in the temperature range 25–300 °C for the reaction analogous to (4.85). Higher-order Fe-HS species, such as $Fe(HS)_2^0$ and $Fe(HS)_3^-$, were invoked in some low-temperature studies, but their stoichiometry remains tentative. Recently, Laskar (2022) conducted exploratory pyrite solubility experiments in NaHS-HCl-NaOH solutions (0.6 m S) at 300 °C and 500 bar and found that $Fe(HS)_2^0$, with a log$K$ value of 11±1 for the reaction

$$Fe^{2+} + 2\ HS^- = Fe(HS)_2^0 \quad (4.86)$$

would account for the measured $FeS_{2(s)}$ solubility in a wide pH range (4–8), with Fe concentrations ($10^{-5}$ to $10^{-4}$ m) being 2 to 6 orders of magnitude higher at the experimental solution compositions than those of the known Fe chloride and hydroxide complexes discussed above. Reaction (4.86) constant, combined with the thermodynamic data of $FeCl_2^0$ from Sverjensky et al. (1997), translates to log$K$ of ~7.5 for the $HS^-$ *vs* $Cl^-$ ligand exchange reaction

$$FeCl_2^0 + 2\ HS^- = Fe(HS)_2^0 + 2\ Cl^- \quad (4.87)$$

This value is comparable with those for analogous exchange reactions (4.68) and (4.69) for Zn and Pb (log$K_{Zn,Pb}$ ≈ 6–8 at 300 °C). Therefore the Fe hydrosulfide complexes may be comparable in abundance with Fe chlorides in moderate-temperature and moderate-salinity hydrothermal fluids with Cl/S ratios of 1000.

Experiments of Laskar (2022) in sulfide-sulfate solutions revealed further enhancement of pyrite solubility (up to $10^{-3}$–$10^{-2}$ m Fe) compared to sulfide solutions of similar S concentration and pH. This solubility increase was tentatively interpreted by a complex with the $S_3^{\bullet-}$ ion, $Fe(HS)S_3^0$, with a log$K$ value of ~2.7 at 300 °C and 500 bar for the ligand exchange reaction:

$$Fe(HS)_2^0 + S_3^{\bullet-} = Fe(HS)S_3^0 + HS^- \quad (4.88)$$

These, yet exploratory, data indicate that the affinity of $Fe^{II}$ for $S_3^{\bullet-}$ may be greater than for $HS^-$, in line with the other transition metals such as Pt and Pd as was discussed above.

This first experimental evidence for [Fe-HS-$S_3$] complexes implies that such species may play a role in pyrite formation mechanisms and control Fe isotope exchange kinetics between pyrite and hydrothermal fluid. Pokrovski et al. (2021c) performed in situ pyrite precipitation experiments at 400–450 °C in saline hydrothermal fluids, along with analyses of Fe isotopes in both fluid and mineral combined with DFT quantum-chemistry calculations of electronic and molecular structures of [Fe-HS-$S_3$]-type complexes of different stoichiometry and their corresponding β-factors (see Eldridge et al. 2025 for isotopic definitions). Such transient complexes (e.g., $Fe(HS)_2(H_2O)_2^0$, $Fe(HS)_2(S_3)_2(H_2O)_2^{2-}$) may account for the observed kinetic isotope fractionation experiments between the fluid and precipitating pyrite. Using their data along with those from recent Fe isotope fractionation studies during pyrite precipitation at lower temperatures (300–350 °C, Syverson et al. 2013, 2015), Pokrovski et al. (2021c) developed a two-stage kinetic model (Fig. 4.33) accounting for non-equilibrium Fe isotope signatures commonly observed in pyrite *vs* fluid from submarine hydrothermal vents. The model may further enable estimating, using Fe isotope signatures, the dynamic of relatively short (<year-scale) mineralizing events in ancient hydrothermal metal sulfide deposits that cannot be assessed using more traditional dating approaches. Thus, (poly)sulfide complexes may open new doors to our quantitative understanding of both transfer and precipitation of a number of metals in hydrothermal fluids.

Finally, another interesting facet of the Fe-S relationships recently emerged for magmatic conditions (they are covered in detail by Simon and Wilke 2025). In addition to chemical transport of sulfur and metals as dissolved complexes, recent experimental and modeling work showed that iron sulfide melt micro particles (FeS), which are primary hosts of chalcophile metals (Cu, Au, Mo, PGE) in arc magmas, may physically be transported by magmas. This transport may be possible particularly for small particles whose sinking velocity (governed by Stokes' law) is slower than the magma ascent rate (Heinrich and Connolly 2022). Vapor nucleation in magmas at shallow depth may further promote sulfide upward transport and concentration, by flotation of sulfide melt on vapor bubbles (Mungall et al. 2020; Iacomo-Marziano et al. 2022). This mechanism provides additional potential for metal mobilization and concentration prior or during fluid degassing and subsequent metal transport as chemically dissolved complexes to magmatic-hydrothermal ore deposits.

#### *4.4.7 Arsenic*

Arsenic is distinguished from the base and noble metals discussed above by *i)* a much larger range of oxidation states under natural conditions, spanning from arsenide (–III to 0) to arsenite (III) and arsenate (V) in the solid phase, and *ii*) distinct aqueous speciation largely dominated by $As^{III}$ and $As^{V}$ oxyhydroxide species.

For $As^{III}$, arsenious acid, $As(OH)_3^0$, and its first deprotonated counterpart ($AsO(OH)_2^–$, at pH >8–9) are the major forms in aqueous solution, but oxy-hydrosulfide and other thionate species may also be abundant at certain conditions (see below). In its species, $As^{III}$ is almost exclusively tri-pyramidal coordinated with O(H) or S(H) ligands, with a non-bonding lone electron pair ($4s^2$) occupying a fourth summit of the distorted tetrahedron, as imposed by $4s^24p^3$ hybridization of outer electron orbitals. There is an apparent ambiguity in some thermodynamic databases, by expressing the stoichiometry of $As^{III}$ oxyhydroxide species as $HAsO_2^0$ and $AsO_2^–$ (e.g., Shock et al. 1997) vs $As(OH)_4^–$ (Baes and Mesmer 1976). They differ from the more structurally correct formula above by one $H_2O$ molecule and are based on the convention that the thermodynamic properties of hydration reaction such as $HAsO_2^0 + H_2O = As(OH)_3^0$, are equal to zero at any temperature and pressure. Therefore, this choice has no impact on the complexation or solubility constants involving such species in water-dominated fluids in which $H_2O$ activity is close to 1. These stoichiometry differences might, however, become meaningful in fluids of water activity distinctly less than 1 (e.g., in carbonic fluids) or in low-density water vapors. However, a number of Raman and X-ray absorption spectroscopy and solubility studies has convincingly shown that $As^{III}$ keeps a $As(OH)_3^0$ stoichiometry across a very wide range of water activity and fluid density (e.g., Pokrovski et al. 1999, 2002a, 2013a). Neither cationic $As^{III}$ hydroxide species, nor $As^{III}$ chloride complexes are known in aqueous solution of pH>0 (Baes and Mesmer 1976; Nordstrom et al. 2014). The dissociation constant of $As(OH)_3^0$ to $AsO(OH)_2^–$ is well constrained at <100 °C. In contrast, at higher temperatures, unexplained discrepancies yet persist between, from one hand, potentiometric measurements (Pokrovski, 1996) supported by independent theoretical predictions (Shock et al. 1997; Smith et al. 1998), and, from the other hand, UV-spectroscopy measurements (Zakaznova-Herzog et al. 2006a). The differences attain 1.5–2.0 log units at 300 °C, meaning that the arsenite anion would be dominant at pH>7 according to the latter work, instead of pH>8–9 as per the earlier studies.

Pentavalent $As^{V}$, stable in surficial oxygenated environments, displays a stronger affinity to oxygen or hydroxyl ligands than $As^{III}$, with a series of tetrahedrally coordinated oxyhydroxide species of arsenic acid, $AsO(OH)_3^0$, $AsO_2(OH)_2^–$, $AsO_3(OH)^{2–}$, and $AsO_4^{3–}$, subsequently dominating from pH<2 to pH>11 at ambient conditions – a dissociation pattern very similar to that of phosphoric acid ($H_3PO_4$=$(HO)_3PO$). Their dissociation constants were measured only up to 95

°C (Zhu et al. 2016 and references therein). These data may be extrapolated with reasonable confidence to temperatures of ~300 °C using the HKF model (Shock and Helgeson 1988 as revised by Shock et al. 1997) that proved itself to be robust for dissociation constants (see section 4.3.3).

Further extension of the HKF model prediction reliability for both $As^{III}$ and $As^{V}$ species has been enabled by direct in situ measurements of heat capacity and volume of arsenic and arsenious acids to 350 °C and 300 bar (Perfetti et al. 2008). These data provided further constraints on the HKF-model coefficients for both $As(OH)_3^0$ and $AsO(OH)_3^0$. Coupled with solubility measurements of arsenopyrite to 450 °C (Pokrovski et al. 2002b) and available calorimetric data for iron sulfarsenide phases, Perfetti et al. (2008) generated a consistent set of thermodynamic properties in the system $FeAs_{(s)}$-$FeAs_{2(s)}$-$FeAsS_{(s)}$-$As(OH)_3^0{}_{(aq)}$. These properties were used by Xing et al. (2019) to propose a first quantitative thermodynamic model of arsenic solid solution in arsenian pyrite, $Fe(As,S)_2$, which is a common host of arsenic in sediment-hosted, epithermal, and orogenic gold deposits. These data allow predictions of arsenic transport in high-temperature hydrothermal fluids (300–500 °C) that is mainly controlled by arsenian pyrite and arsenopyrite. The predicted $As(OH)_3^0$ concentrations (10–100 ppm As) are in good agreement with analyses of natural fluid inclusions from Cu-W-Sn-As magmatic-hydrothermal deposits (Pokrovski et al. 2002b). In lower temperature epithermal environments (<200 °C), arsenic sulfides, such as orpiment ($As_2S_3$) and realgar (AsS), become the As solubility controlling phases (Pokrovski et al. 1996). Less certainty as to the exact As speciation and redox exists at present for higher *T-P* fluids (>500 °C and 2 kbar). Extrapolations of the $As^{V}$ oxyhydroxide species thermodynamic data to magmatic and deep-metamorphic conditions indicate that $As^{V}$ oxyhydroxide complexes might over compete $As^{III}$ complexes only in extremely oxidizing fluids of $f_{O2}$~QFM+10. Such fluids may be released during out-of-equilibrium serpentinite breakdown in deep subduction-zone settings (Pokrovski et al. 2022a; references therein).

The major role of sulfur in arsenic fate in high-temperature hydrothermal fluids is thus mostly to control the solubility of arsenic sulfide and iron sulfarsenide minerals rather than to directly complex arsenic, which stays as $As(OH)_3^0$ both in the fluid and vapor phases and hydrous silicate melts (Pokrovski et al. 2005b, 2008a; Borisova et al. 2010). In contrast, at lower temperatures (<200–300 °C), sulfide complexation becomes increasingly important. There exists a plethora of experimental solubility, potentiometric and spectroscopic studies at temperatures <150 °C, thoroughly reviewed by Wood and Samson (1998) and Rickard and Luther (2006). These data show extremely variable [$As^{III}$-S]-type complexes stoichiometries, spanning from trimeric species with As-S-As bonds (e.g., $H_2As_3S_6^-$ or more structurally correct $As_3S_4(SH)_2^-$, e.g., Spycher and Reed 1989; Helz et al. 1995) to a series of monomeric complexes (e.g., $As(SH)_3^0$, $AsS(SH)_2^-$, $AsS_2(SH)^{2-}$ and $AsS_3^{3-}$, see Zakaznova-Herzog and Seward 2012 for a recent compilation of available references), along with mixed [As-(O(H),S)3]-type species (e.g., $As(OH)_2(SH)^0$, $As(OH)S_2^-$, e.g., Wilkin et al. 2003; Bostick et al. 2005; Beak et al. 2008), depending of As concentration, As/S ratio, pH, and saturation degree with As sulfide phases ($As_2S_3$). These experimental findings have inspired both static DFT and MD computational studies of structures and energetics of selected As-S species (e.g., Tossell and Zimmermann 2008; He et al. 2017). However, until now, no consensus has been reached on the species nature and stability and there is no reliable estimation of their stability constants at hydrothermal conditions. Recent analyses of thioarsenates and thioarsenites in natural geothermal waters (see section 4.4.12) point to transient out-of-equilibrium nature of such species. They may play a role of precursors during the formation of arsenic sulfide minerals, as well as of potential ligands for selectively complexing some chalcophile metals (e.g., Au, Ag, Cu) – the hypotheses that still require experimental and spectroscopic confirmation for hydrothermal fluid conditions.

### *4.4.8 Antimony*

Antimony is quite similar to arsenic, both in redox and speciation, with a range of antimonide, sulfide and oxide minerals with Sb nominal oxidation states from –III to V, and major aqueous species as $Sb^{III}$ and $Sb^{V}$ oxyhydroxides in dilute aqueous solution. However, the major differences with As are *i)* generally lower solubility of the analogous Sb solid phases in aqueous solution, *ii)* different coordination and electronic structure of $Sb^{III}$ and $Sb^{V}$ species, and *iii)* greater affinity of $Sb^{III}$ to chloride.

Pentavalent $Sb^{V}$ is stable in oxygenated near-surface waters as octahedrally-coordinated species $Sb(H_2O)(OH)_5^0$ dominant at acidic pH and deprotonating to $Sb(OH)_6^-$ at pH>3 (Tella and Pokrovski 2012 and references therein). The octahedral coordination, enabled by the empty 4d shell of Sb, is sterically favorable for the formation of chelate complexes of 5- and 6-membered rings with oxygen-bearing organic ligands such as fulvic and humic acids in surficial environments. This specific coordination control makes $Sb^{V}$ different from $As^{V}$ whose tetrahedral [$AsO_4$] coordination does not allow formation of stable complexes with di-functional organic ligands (Pokrovski and Schott 1998; Tella and Pokrovski 2012). Similarly to $As^{III}$, trivalent $Sb^{III}$, which is stable at hydrothermal conditions, forms oxyhydroxide complexes, with trigonal-pyramidal $Sb(OH)_3^0$ (sometimes expressed as $HSbO_2^0$) being dominant in the pH range from ~2 to ~12 at ambient temperatures (Baes and Mesmer 1976; Zakaznova-Herzog and Seward 2006b). In more acidic and more alkaline solutions, respectively, cationic ($Sb(OH)_2^+$ or $H_4SbO_3^+$) and anionic counterparts, with variable expressions of their stoichiometry by addition or subtraction of water molecules ($Sb(OH)_4^-$, $H_2SbO_3^-$ or $SbO_2^-$), become dominant; however, more recent in situ XAS measurements demonstrated that the true $Sb^{III}$ coordinations in solution are distorted fourfold configuration with O(H) ligands, $Sb(OH)_2(H_2O)_2^+$ and $Sb(OH)_4^-$, respectively (Tella and Pokrovski 2009). In contrast with $As^{III}$, which always remains trifold, such fourfold coordinations have much less steric constraints for the formation of organic chelate species in natural waters (Tella and Pokrovski 2009). At near-ambient conditions, $Sb^{III}$ is known to form complexes with chloride in strongly acidic and with (hydro)sulfide in neutral-to-alkaline solutions, but no consensus about the dominant stoichiometry and stability constant values has ever been reached, as reviewed in detail by Filella and May (2003) and Rickard and Luther (2006).

Under hydrothermal conditions, $Sb^{III}$ appears to be more speciation-versatile than $As^{III}$, by forming oxyhydroxide, sulfide and chloride complexes. Trihydroxide, $Sb(OH)_3^0$, stands out as the major species at elevated temperatures. Paradoxically, its domain of predominance in pH space narrows down with increasing temperature according to rare available data (e.g., $2<pH<12$ at 25 °C *vs* $3<pH<9$ at 300 °C; Zakaznova-Herzog and Seward 2006b). This behavior is, however, in apparent disagreement with the general tendency of strengthening hydrolysis and favoring the stability of uncharged species; it was not supported by other earlier data, showing $Sb(OH)_2^+$ predominance field to be retracted towards more acidic pH with increasing temperature (Popova et al. 1975). More in situ data are required to resolve these discrepancies. It may not be excluded that changes in the $Sb^{III}$ coordination in acidic and alkaline solutions with temperature might partly be responsible for such hydrolysis pattern.

The thermodynamic properties of $Sb(OH)_3^0$, dominant within the typical hydrothermal fluid pH range, have thoroughly been constrained by extensive $Sb_2O_{3(s)}$, $Sb_{(s)}$ and $Sb_2S_{3(s)}$ solubility works to supercritical conditions (450 °C, 1 kbar) by Zotov et al. (2003). The HKF-model dataset derived using their own data and those from some previous studies (see Wood and Samson 1998), is recommended for modeling antimony solubility and transport at elevated temperatures, which is mainly controlled by the reaction

$$0.5\ Sb_2S_{3(s)} + 3\ H_2O = Sb(OH)_3^0 + 1.5\ H_2S_{(aq)} \qquad (4.89)$$

The HKF model predictions of Zotov et al. (2003) were further supported by recent solubility studies of $Sb_2S_{3(s)}$ in sulfide-bearing solutions that reported similar $K_{89}$ values between 200 and 350 °C (within <0.5 log units; Olsen et al. 2018, 2019). By contrast, the old HKF values of $HSbO_2^0$ (Shock et al. 1997) based on low-temperature data underestimate $Sb_2S_{3(s)}$ solubilities up to 4 orders of magnitude above 300 °C (Fig. 4.34); they thus require to be updated in commonly used databases (e.g., SUPCRTBL).

Unlike arsenic, $Sb^{III}$, having a more metallic character, forms different chloride and oxychloride complexes in acidic saline solutions both at ambient (e.g., Filella and May 2003 for a review) and hydrothermal conditions. The latter have been explored by solubility, XAS and Raman spectroscopy studies (e.g., Oelkers et al. 1998; Ovchinnikov et al. 1982; Pokrovski et al. 2006; Persaud 2022). They reveal $SbCl_3^0$, $SbCl_4^-$, and $SbCl_6^{3-}$ in concentrated HCl solutions (>1 m HCl), and $Sb(OH)_2Cl^0$ and $Sb(OH)_3Cl^-$ in less acidic, and thus more geologically relevant, fluids (pH>2). The formation constants of those species were reported within rather narrow *T* ranges or were limited in some studies to exchange reactions between different Sb-Cl complexes with no systematic referencing to the key species $Sb(OH)_3^0$ or Sb solid phases to be applicable to natural fluids (e.g., Persaud 2022). The rare geologically relevant available data indicate that $Sb(OH)_3Cl^-$ may compete with $Sb(OH)_3^0$ in saline near-neutral pH solutions above 300 °C and 5–10 wt% NaCl (Pokrovski et al. 2006), but more data over larger *T* and salinity ranges are required. The uncharged oxychloride $Sb(OH)_2Cl^0$ and trichloride $SbCl_3^0$ have been shown to be more volatile than the other $Sb^{III}$ species discussed above, and therefore responsible for enhanced Sb partitioning into the vapor phase in acidic vapor-brine systems (Pokrovski et al. 2008b), in contrast to $As^{III}$ partitioning, which is unaffected by the presence of HCl (Pokrovski et al. 2005b). This contrasting vapor-liquid fractionation pattern between $As^{III}$ and $Sb^{III}$ may help tracing the acidity conditions operating in magmatic-hydrothermal systems using the As/Sb ratios in coexisting vapor and brine fluid inclusions.

Like for arsenic, the role of sulfur in Sb behavior in geological fluids remains the least constrained, mainly due to the notorious difficulties of experimenting and measuring in sulfur-bearing aqueous systems in general, along with the low solubility of the thermodynamically stable $Sb_2S_{3(s)}$ solid thereby posing analytical challenges. A plethora of different Sb-S-OH complexes, both monomeric and dimeric having Sb-S-Sb bonds, has tentatively been proposed at low temperatures, with caution that more polymerized species may equally exist due to a strong tendency of $Sb^{III}$, similar to $As^{III}$, of forming colloidal solutions in sulfide systems (Krupp 1988; Rickard and Luther 2006). At hydrothermal conditions, only a few theoretical and experimental studies are available that reported both stoichiometry and stability constants (see Wood and Samson 1998 and Olsen et al. 2019 for more detailed comparisons). A theoretical attempt has been made by Spycher and Reed (1989) who critically re-evaluated available experimental data on Sb and As sulfide systems and suggested a first systematic set of solubility constants of $Sb_2S_{3(s)}$ to 300 °C, with the dominant formation of $H_2Sb_2S_4^0$, $HSb_2S_4^-$ and $Sb_2S_4^{2-}$ complexes in acidic (pH<3), near-neutral (4<pH<8) and alkaline (pH>9) solutions, respectively

$$Sb_2S_{3(s)} + HS^- + n{-}1\ H^+ = H_nSb_2S_4^{n-2},\ n = 0,\ 1,\ \text{and}\ 2 \qquad (4.90)$$

An analogous pH-dependent scheme with three trimeric complexes was suggested for $As^{III}$ (section 4.4.7). The Spycher and Reed (1989) evaluation of existing experimental studies was, however, criticized by Krupp concerning different aspects in the tough business of extracting species stoichiometries and equilibrium constants. Such a debate that has seen two comments-replies (Krupp 1990a,b; Spycher and Reed 1990a,b) remains a rare example in sulfur hydrothermal research. Krupp (1988) $Sb_2S_{3(s)}$ solubility measurements from 25 to 350 °C represent a remarkable experimental study of the last century in the metal-sulfur speciation topic. His work reported the same three dimeric sulfide species as above, but with stability constants of reaction (4.90) being lower as much as 2 log units for the most geologically relevant $HSb_2S_4^-$ complex. In addition to those sulfide complexes, Krupp (1988) suggested the formation of a hydroxysulfide species, $Sb_2S_2(OH)_2^0$, that became increasingly dominant over the sulfide species in acidic-to-neutral pH above 200 °C. More discrepancy has been added by another theoretical evaluation of some selected available studies by Russian teams (Akinfiev et al. 1993, as cited in Zotov et al. 1995) that reported a set of HKF coefficients for the three Sb sulfide complexes above. Their study ignores, however, Krupp's mixed species and yields $K_{90}$ values for $HSb_2S_4^-$ of 1 to 3 log units lower than those of Krupp (1988). The issue of Sb-S complexation in hydrothermal fluids has remained without any advance for more than 25 years, until recent solubility works of a New Zealand team (Olsen et al. 2018, 2019). They conducted systematic $Sb_2S_{3(s)}$ solubility measurements in NaOH-$H_2S$ solutions over the 30–400 °C range from $P_{sat}$ to 300 bar using a flow-through reactor along with Sb K-edge XAS measurements at 30 °C. Their proposed speciation scheme at 30 °C remains quite complex, with at least 5 species, $H_3SbS_2O^0$, $HSb_2S_4^-$, $H_2Sb_2S_5^{2-}$, $Sb(OH)_3^0$ and $Sb_2S_4^{2-}$, from pH 2 to 12 in the order of their respective maximum abundance. This speciation scheme significantly simplifies at hydrothermal conditions, with $HSb_2S_4^-$ and $Sb_2S_4^{2-}$ being predominant below 150–200 °C and with $Sb(OH)_3^0$ at higher temperatures. Their reaction (4.89) and (4.90) constants are within 0.5 log units of those of Zotov et al. (2003) and Krupp (1988), respectively. The Olsen et al. (2019) dataset for $HSb_2S_4^-$ and $Sb_2S_4^{2-}$ is recommended in Sb speciation modeling. According to these and most previous studies, $Sb(OH)_3^0$ is responsible for Sb transport in most hydrothermal fluids above 300 °C. These fluids are often undersaturated with stibnite because of increasingly high $Sb_2S_{3(s)}$ solubility with the temperature rise (>100s to 1000s ppm Sb at $H_2S$ concentrations 0.001–0.01 m at 300–400 °C) compared to natural Sb fluid concentrations (typically <1–10 ppm, Kouzmanov and Pokrovski 2012). Sulfide complexes may only be important for Sb transport in geothermal waters and epithermal fluids at temperatures below 150–200 °C as well as in sulfide-rich sedimentary fluids, if mineral-fluid equilibrium is reached; however, out-of-equilibrium Sb (and As) transport as colloidal sulfide particles and/or dissolved thionate species was also evidenced in rapidly cooling and decompressing geothermal waters (see section 4.4.12).

### *4.4.9 Bismuth*

Bismuth pursues the $As^{III}$-$Sb^{III}$ trend in aqueous speciation, with its major oxidation state of $Bi^{III}$ in most chemical compounds, with more metallic character, and growing affinity to chloride and sulfide both in solid phase and aqueous solution. At ambient conditions, the $Bi^{3+}$ cation is only stable in very acidic solutions (pH<1); with increasing pH, even in very dilute solutions (<$10^{-5}$ m Bi), it hydrolyses to a variety of polynuclear hydroxide cations (e.g., $Bi_6(OH)_{12}^{6+}$, $Bi_9(OH)_{22}^{5+}$) that are converted at pH>8 to $Bi(OH)_3^0$ and possibly to $Bi(OH)_4^-$ at pH>13, similar to $Sb^{III}$ (Baes and Mesmer 1976). Bismuth trioxide, $Bi_2O_{3(s)}$, is about 1 order of magnitude less soluble in water than $Sb_2O_{3(s)}$, with solubility increasing from $10^{-5}$ to 0.01 m between 25 and 300 °C. With increasing temperature, the pH space of the stability domain of $Bi(OH)_3^0$ further widens similar to that of $Sb^{III}$. Its stoichiometry in hydrothermal solutions was confirmed by in situ XAS (Tooth et al. 2013), and its thermodynamic properties were derived from solubility measurements to 600 °C and 800 bar (Kolonin and Laptev 1982; Tooth et al. 2013). The HKF model parameters of $Bi(OH)_3^0$ reported by Tooth et al. (2013) are recommended for thermodynamic modeling, whereas an older HKF data set for dehydrated $BiO_2^0$ of Shock et al. (1997) based on lower-temperature data underestimates $Bi_2O_{3(s)}$ solubility by >2 orders of magnitude above 300 °C. Bismuthinite, $Bi_2S_{3(s)}$, a typical hydrothermal Bi-bearing mineral, is much less soluble as $Bi(OH)_3^0$ than stibnite or orpiment (Fig. 4.34), with equilibrium Bi concentrations less than ppb at $H_2S$ concentrations of 0.001–0.01 m, typical of moderate-temperature hydrothermal fluids. These solubility predictions largely underestimate Bi concentrations, of 10s to 100s ppm, measured by LA-ICPMS in some natural fluid inclusions (e.g., Kouzmanov and Pokrovski 2012 and references therein; Schirra et al. 2022; Galdos et al. 2024), thereby calling for complexes other than with hydroxide.

Indeed, being more metallic than As and Sb, $Bi^{III}$ has a much greater affinity for the $Cl^-$ ligand. Chloride complexes, $BiCl_n^{3-n}$ with $n$ from 1 to 6, have been widely recognized at ambient conditions in saline acidic solutions. However, more discrepancy persists at elevated temperatures between two major sources of data. Kolonin and Laptev (1982) suggested $BiCl_6^{3-}$ as the major complex based on solubility and UV-Vis measurements to 300 °C at $P_{sat}$. In contrast, Etschmann et al. (2016) employed in situ XAS methods to show that $BiCl_3^0$ largely dominates over $BiCl_6^{3-}$ at typical pH and salinities of hydrothermal fluids. Etschmann et al. (2016) derived the stability constants of Bi chloride complexes based on the variation of the number of Cl atoms in the Bi first coordination shell from EXAFS and using the RB model. Their thermodynamic constants for the reaction

$$BiCl_3^0 + 3\ H_2O = Bi(OH)_3^0 + 3\ Cl^- + 3\ H^+ \quad (4.91)$$

imply that in moderately saline hydrothermal fluids ($a_{Cl^-} = 1$), the chloride complex would dominate over the hydroxide at pH<5 at 350 °C and $P_{sat}$, which may explain the elevated Bi concentrations analyzed in saline fluid inclusions.

The major unknown in Bi hydrothermal chemistry is the effect of sulfur. Complexes with thiol-organic ligands (e.g., thiocarbamide and its derivatives) may stabilize dissolved $Bi^{III}$ in solution in the absence of sulfide. Reduced sulfur ($H_2S$) would play a negative effect on Bi sulfide minerals solubilities if Bi formed dominantly hydroxide or chloride species in the fluid (see section 4.4.1 and Fig. 4.25; e.g., like Mo or Zn, Fe, Cu). The low $Bi_2S_{3(s)}$ solubility compared to analogous Sb and As sulfides (Fig. 4.34) makes the measurements of sulfide complexes analytically challenging and/or requires quite elevated temperatures to increase solubility (>400 °C). The only information, we are aware of, about such complexes under hydrothermal conditions are theoretical estimations of Skirrow and Walshe (2002). These authors adopted an analogy with sulfide complexes of arsenic and antimony and tentatively assumed $Bi^{III}$ complex stoichiometry similar to that of $Sb^{III}$ ($Bi_2S_4^{2-}$, $HBi_2S_4^-$ and $Bi_2S_2(OH)_2^0$). However, given the large differences in the affinity of these elements to the different ligands (see above), such estimations should be regarded with caution, thus awaiting experimental evidence. Similarly, the role of polysulfide anions and radical ions cannot be excluded. Interestingly, the only structurally characterized metal complex of a polysulfide radical ion is that of $S_4^{\bullet-}$ with $Bi^{III}$ in metalorganic compounds (Schwamm et al. 2017). Thus, given the elevated Bi affinity to reduced sulfur, $S_3^{\bullet-}$ might be a potentially important ligand for $Bi^{III}$ in S-rich fluids, and in lesser extend for $As^{III}$ and $Sb^{III}$.

In contrast, in sulfur-poor, reducing (pyrrhotite stable), high-temperature (>400 °C) environments, bismuth metal melts may be stable at the expense of Bi sulfides or sulfosalts and coexist with a hydrothermal fluid, as found in skarn, intrusion-related, orogenic and volcanogenic massive sulfide gold deposits. Such melts have been suggested to be very efficient scavengers of gold from the hydrothermal fluid, thereby providing an additional mechanism of concentrating the metal (e.g., Tooth et al. 2008, 2011; Du et al. 2023 and references therein). The presence of reduced $As^{-I}$ in arsenian pyrite, arsenopyrite and löllingite stable in some of such S-poor reducing systems will be an additional competing mechanism of Au concentration, via coupled redox reactions between $Au^I$ and $As^{III}$ in hydrothermal fluid leading to the formation of $[Au^{II}As^I_nS_{6-n}]$ complexes in the $Fe^{II}$ crystallographic sites of the sulfoarsenide mineral structure (Pokrovski et al. 2021a).

### *4.4.10 Molybdenum*

Among metals and metalloids forming sulfide ore in the Earth's crust, molybdenum is the most chemically variable one, exhibiting redox states from –II to VI and coordinations from 4 to 9, in compounds with different O, Cl and S ligands known in chemistry and material sciences. However, at naturally relevant conditions, the most common redox states are IV and VI in which Mo can be both tetrahedrally and octahedrally coordinated with $O^{2–}/OH^–$, $Cl^–$ and $S^{2–}$ ligands. The most common stable solid phase in hydrothermal-magmatic systems is molybdenite, $Mo^{IV}S_{2(s)}$, whereas oxides, $Mo^{VI}O_{3(s)}$ and $Mo^{IV}O_{2(s)}$, and molybdates (e.g., $CaMo^{VI}O_{4(s)}$), being far more soluble, occur only occasionally. In contrast with the solid state, Mo aqueous speciation is more variable and remains very poorly quantified at hydrothermal conditions. Most available work currently converges to the conclusion that hexavalent $Mo^{VI}$ is the major redox state in aqueous solution over a wide $T$-$P$-$f_{O2}$ range (Wood and Samson 1998; Zotov et al. 1995; references therein). Even though the $Mo^{VI}$ primary ligand is oxygen or hydroxyl in molybdate species, this extremely versatile element is equally capable of forming a plethora of oxy-chloride-sulfide complexes.

*Oxyhydroxide complexes*. $Mo^{VI}$ is a hard cation strongly hydrolyzed in dilute aqueous solution (<$10^{–4}$ m Mo) at ambient conditions, by forming at pH<4 molybdic acid conventionally expressed as $H_2MoO_4$ and, at higher pH, its dissociation products, the tetrahedral (hydro)molybdate ions, $HMoO_4^–$ and $MoO_4^{2–}$:

$$H_2MoO_4 = HMoO_4^– + H^+ \quad (4.92)$$

$$HMoO_4^– = MoO_4^{2–} + H^+ \quad (4.93)$$

In contrast to its deprotonated counterparts, molybdic acid likely has a structural formula of $MoO_2(OH)_2(H_2O)_2$ in which Mo is in a distorted octahedral coordination made up by O, OH and $OH_2$ ligands, rather than $Mo(OH)_6$, $MoO_3(H_2O)_3$ or $H_2MoO_4$ formulations commonly encountered in the literature (Liu et al. 2013b; Zhang et al. 2019; references therein). Cationic monomer species, such as $MoO(OH)_2^+$ and $MoO_2^{2+}$, with $Mo^{VI}$ in octahedral coordination complemented by water molecules, have also been invoked at very low Mo concentrations in strongly acid (pH<1) aqueous solutions (Gruywagen 1999; Zhang et al. 2019). However, they are unlikely to be significant for natural fluids and were not considered in hydrothermal research. A remarkable feature of $Mo^{VI}$ (and its chemical congener, tungsten, $W^{VI}$) aqueous chemistry, often ignored by geoscientists, is a very strong tendency to form polymeric anions at acidic pH at concentrations as low as $10^{–3}$ m (~100 ppm Mo). Such anions are composed of clusters with 7 or more $Mo(O,OH)_6$ octahedra linked by their edges, such as $Mo_7O_{24}^{6–}$ and $Mo_8O_{26}^{4–}$ and their partly protonated counterparts. Their formation constants from $MoO_4^{2–}$ have been determined at ambient conditions (e.g., Baes and Mesmer 1976; Gruywagen 1999). In addition, there is a plethora of heteropoly molybdate anions in which the central cavity of the cluster is occupied by a different element in tetrahedral $XO_4$ coordination (where X = Si, Ge, P, As). Among them is the famous $[X^{n+}Mo_{12}O_{40}^{n-8}]$ often called the Keggin ion after its discoverer (see Cotton et al. 1999). These remarkable configurations can undergo reduction to blue mixed-valence species ($Mo^{IV,V,VI}$) without loss of structure. The blue color of such partly reduced Mo species is used since a long time in quantitative colorimetry of monomeric silica in aqueous solution. Silicic acid is trapped as $SiO_4$ in the Mo Keggin structure, which is then partly reduced using oxalic acid thereby developing a stable blue color (e.g., Charlot 1966). Interestingly, a blue color due to reduced Mo species has been observed in the sampled or quenched fluids from Mo hydrothermal solubility experiments (e.g., Kudrin 1989; Kokh et al. 2016), indicating that reduced Mo forms might also be present in hydrothermal fluids. Polymeric molybdates have commonly been ignored in the interpretation of hydrothermal experiments even at elevated Mo concentrations (>0.01 m), arguing that growing thermal disorder would be unfavorable for their stability (Ivanova et al. 1975; Dadze et al. 2017a). Interestingly, $W^{VI}$ has recently been demonstrated by in situ Raman spectroscopy to form stable polymeric oxy-hydroxides at hydrothermal conditions (Carocci et al. 2022). By analogy, a similar behavior might be expected for $Mo^{VI}$, awaiting direct spectroscopic evidence. Monomeric molybdic acid dissociation reaction constants to $HMoO_4^-$ ($K_{92}$) and $MoO_4^{2-}$ ($K_{93}$) have currently been constrained to 350 °C by recent UV-Vis spectroscopy and $MoO_{3(s)}$ and $CaWO_{4(s)}$ solubility measurements (Minubaeva and Seward 2010; Dadze et al. 2018a). These $K$ values are in good mutual agreement (within 1 log unit), but differ, especially for $K_{92}$, from those reported in older solubility studies of $MoO_{2(s)}$ (e.g., Kudrin 1989). In the absence of direct evidence for Mo polymeric species at elevated temperatures, we recommend using the Dadze et al. (2018a) set of $K_{92}$ and $K_{93}$ values in hydrothermal fluid modeling to at least 350 °C and <1 kbar. The $K_{92}$ value shows a particular temperature behavior of increasing dissociation in contrast to analogous acids of similar strength ($H_3PO_4$, $H_2SO_4$, $H_3AsO_4$). Such a difference is likely due to Mo coordination change from 6 to 4 upon molybdic acid dissociation (reaction 4.92) and may therefore be explained by the entropy principle postulating that more compact species of lower coordination are generally favored at elevated temperatures. In this regard, $Mo^{VI}$ contrasts with $P^V$, $S^{VI}$ and $As^V$ that always keep their tetrahedral coordination with oxygen.

*Alkali ion pairs*. In saline (NaCl and KCl) solutions, a moderate increase of $MoO_{2(s)}$ solubility at temperatures 300–450 °C has originally been interpreted by the formation of the $NaHMoO_4^0$ and $KHMoO_4^0$ ion pairs (Kudrin 1989) such as:

$$Na^+ + HMoO_4^- = NaHMoO_4^0{}_{(aq)} \quad (4.94)$$

The same Na ion pair was recently suggested as the dominant species in $MoO_{3(s)}$ solubility measurements in NaCl and $NaCF_3SO_3$ (triflate) solutions at 250, 300 and 350 °C (Shang et al. 2020).

These data show a moderate increase in $MoO_{3(s)}$ solubility (<1.5 log unit) in the presence of high sodium contents (>1 m Na) as opposed to more dilute solutions (<0.1 m Na). However, their data were criticized by Plyasunov (2020) demonstrating that Shang et al.'s (2020) $MoO_{3(s)}$ solubilities in Na-poor solutions appear to be systematically lower, by a similar magnitude, than most existing data (Ivanova et al. 1975; Dadze et al. 2017a,b). Dadze et al. (2017b) interpreted a similar increase in $MoO_{3(s)}$ solubility in $NaClO_4$ solutions (up to 2.2 m Na) at 300 °C by an activity coefficient effect rather than ion pairing. In light of the relatively high $MoO_{3(s)}$ solubility in water at elevated temperatures (~0.01m at 300 °C), along with the potential tendency to form $Mo^{VI}$ polymeric oxyhydroxide species (see above), the addition of an electrolyte in such systems might favor disruption of the polymeric structures thereby increasing solubility. Dadze et al. (2018a) interpreted their $CaMoO_{4(s)}$ solubility measurements ($<5\times10^{-4}$ m Mo) in highly saline solutions (up to 3.8 m NaCl) at 300 °C as monomeric molybdic acid and molybdate ions, with no evidence for statistically significant contributions from ion pairs or chloride complexes. Kokh et al. (2017) tentatively suggested the $KHMoO_4^0$ complex to describe the solubility of $MoS_{2(s)}$ in supercritical $H_2O$-$CO_2$ fluids of low salinity (450 °C, ~700 bar, <0.1 m KCl) in the presence of the pyrite-magnetite-pyrrhotite and potassic feldspar-quartz-muscovite assemblages, but other alternative $Cl^–$ and $HS^–$/$S^{2–}$ complexes could not be tested. More recently, Guan et al. (2023) added further confusion to the debate about the Na ion pairs, by publishing a set of HKF coefficients for $NaHMoO_4^0$ derived from the three temperature datapoints, 250, 300 and 350 °C, of Shang et al. (2020). The reported thermodynamic functions predict an unrealistically strong stability for $NaHMoO_4^0$ at temperatures <300 °C, with an ion pair formation constant value of reaction (4.94) at 25 °C of 8 log units higher than any known-so-far ion pair between $Na^+$ or $K^+$ and –1 and –2 charged anions (e.g., $Cl^–$, $HSO_4^–$, $SO_4^{2–}$, $B(OH)_4^–$, $HCO_3^–$ and $MoO_4^{2–}$; see Pokrovski et al. 2024). In addition, some of the reported HKF parameters, such as the Born parameter value (ω) of $NaHMoO_4^0$ is several times more negative that those of notoriously gas-like aqueous species (e.g., dissolved noble gases). Such values are unphysical for hydrated ion pairs and neutral complexes that are known to have positive ω (e.g., Plyasunov and Shock 2001; Perfetti et al. 2008; Pokrovski et al. 2024). Both stability constants and HKF parameters of sodium hydromolybdate ion pair are, therefore, artifacts of the HKF parameter derivation procedures used in Guan et al. (2023) and should be rejected. Thus, the effect of salt in general, and of alkalis in particular, on Mo aqueous speciation in hydrothermal fluids awaits more constrained studies before it could be considered in quantitative speciation models and used as the basis for interpreting the effect of chloride and sulfur discussed below.

*The effect of chloride*. Oxychloride complexes, of general stoichiometry $Mo^{VI}O_2Cl_n(H_2O)_{4–n}^{2–n}$, with growing number of Cl ligands when increasing temperature from 25 to 340 °C, have been reported using in situ X-ray absorption spectroscopy only in strongly acid solutions (2–6 m HCl) (Borg et al. 2012). In contrast, only the spectral signal from molybdate anions was observed in the same study for highly saline (to 6 m NaCl) near-neutral pH solutions in the same *T* range. At higher temperatures, in a synthetic fluid inclusion study by Ulrich and Mavrogenes (2008), Mo aqueous concentrations in the system $Mo_{(s)}$-$H_2O$-KCl at 500–800 °C to 3 kbar were tentatively interpreted by potassium ion pairs or oxychloride species. However, Ulrich and Mavrogenes's interpretation of in situ XANES spectra acquired from a few individual fluid inclusions was later criticized by Borg et al. (2012) showing that these spectra closely resembled that of molybdenum oxide solid ($MoO_{2(s)}$) rather than of an aqueous species. Kudrin (1989) tentatively interpreted his $MoO_{2(s)}$ solubility data (see above) in terms of $KHMoO_4^0$ rather than chloride complexes. We note that the effect of alkalis *vs* chloride is virtually impossible to separate in NaCl or KCl solutions (Wood and Samson 1998). A potentially pertinent Cl species, $MoO_2(OH)_2Cl^-$, has recently been proposed based on $MoO_{3(s)}$ solubility measurements in moderately saline fluids at 300 °C (Dadze et al. 2018b). This complex was, however, ignored by the same authors in an accompanying study of $CaMoO_{4(s)}$ solubility at 300 °C and 100 bar in more concentrated NaCl solutions (to 3.5 m NaCl, Dadze et al. 2018a), pointing to a lack of sufficient resolution in detecting such weak complexes by the solubility method at moderate temperatures. Similarly, at much higher *T* (600–800 °C) and *P* (26 kbar), a synthetic fluid inclusion study of Bali et al. (2012) reported only a fairly moderate increase in $MoO_{2(s)}$ solubility, by a factor of 5 with a 40-fold increase in salinity, pointing either to potential weak chloride and/or alkali complexes or an activity coefficient effect. Thus, the effect of chloride on Mo mobility in hydrothermal fluids yet remains both inconsistent and controversial. Its quantification would require more concerted studies using combined approaches.

*The effect of sulfur*. In ambient aqueous solution, hexavalent $Mo^{VI}$ has been known since a long time to form a series of colored monomeric oxysulfide complexes with increasing sulfide concentration, by stepwise replacement of the $O^{2-}$ by $S^{2-}$ ligand in the molybdate ion: $MoO_4^{2-}$, $MoO_3S^{2-}$, $MoO_2S_2^{2-}$, $MoOS_3^{2-}$, and $MoS_4^{2-}$, with the color changing from transparent to yellow, orange, orange-red, and red, respectively. Both thermodynamics and kinetics of these complexes formation have been thoroughly studied using UV-Vis spectroscopy (Erickson and Heltz 2000; Rickard and Luther 2006; references therein). The oxysulfide species, among which tetrasulfide is major (Fig. 4.35a), dominate the $Mo^{VI}$ speciation in sulfidic pore waters and euxinic marine and lake water columns at near-neutral pH and sulfide concentrations above 0.01 millimol (Helz 2021; references therein). A paradoxical feature of these complexes is the association of Mo in its highest oxidation state (VI) with S in its lowest oxidation state (–II), which is likely due to electron delocalization via π-bonding. The remarkably strong bonding affinity between molybdenum and sulfur in contrasting oxidation states suggests that even greater Mo-S affinities might be expected for lower-valence Mo states (e.g., IV) and with intermediate-valence S species such as polysulfides in hydrothermal fluids – an hypothesis requiring thorough investigation.

The effect of sulfur on Mo transport in sulfur-bearing hydrothermal systems, where molybdenite is the stable phase remains extremely poorly documented. The few early $MoS_{2(s)}$ solubility studies in $H_2S$-bearing fluids, overviewed by Wood and Samson (1998) and Zotov et al. (1995), showed a negative effect of $H_2S$ on the solubility, likely consistent with molybdate-type dominant species (e.g., reactions 4.92 and 4.93). In contrast, Kokh et al. (2016) observed an increase in both $MoS_{2(s)}$ liquid-phase solubility and Mo vapor-liquid partition coefficient with increasing $H_2S$ concentration at 350 °C in a multicomponent minerals-aqueous solution-vapor system ($FeS_2$-$MoS_2$-$CuFeS_2$-$SnO_2$-$PtS_2$-Au-KCl-HCl-$K_2S_2O_3$-$CO_2$-$H_2O$), pointing to the formation of Mo-S type of complexes both in liquid and vapor phases. Zhang et al. (2012) measured, using the synthetic fluid inclusion method, $MoS_{2(s)}$ solubility in saline fluids with $f_{S_2}$ and $f_{O_2}$ buffered by the pyrite(py)-pyrrhotite(pyh)-magnetite(mag), py-pyh-Ni-NiO, pyh-QFM, and py-pyh-Co-CoO assemblages at 600–800 °C and 2 kbar. Their Mo concentration trends were formally consistent with a tentative complex stoichiometry of $NaHMoO_2S_2^0$. No thermodynamic constant has been proposed by the authors who fitted their solubility trends by an empirical equation as a function of temperature, salinity, and sulfur and oxygen fugacity, therefore restricted within the parameters ranges of their experiments. More recently, in an analogous synthetic fluid inclusions study (at 600 °C, 2 kbar), Li et al. (2021) obtained similar $MoS_{2(s)}$ solubilities (10s ppm Mo) in $H_2O$-$CO_2$ fluids in the presence of the py-pyh-mag assembladge, but the species identity could not be constrained. Liu et al. (2020) applied the XAS method to alkaline NaHS solutions to 340 °C coupled with MD calculations of species structures and their XANES spectra. Their spectroscopic data are consistent with formation of mixed-ligand complexes, $MoO_nS_{4-n}^{2-}$ ($n$ = 0 to 4). The sulfide ligand number of the dominant complex decreased both with increasing temperature for a given solution composition and with decreasing NaHS concentration at a given temperature. Alternatively, their data were equally consistent with a mixture of the endmember complexes, $MoO_4^{2-}$ and $MoS_4^{2-}$, with a decrease in the sulfide complex fraction with temperature and NaHS concentration. Based on the only available 25 °C data of Erickson and Helz (2000) and using an isocoulombic extrapolation (equation 4.10), Liu et al. (2020) further estimated the stability constants of the following exchange reaction to 300 °C

$$MoO_4^{2-} + n\ H_2S_{(aq)} = MoO_{4-n}S_n^{2-} + n\ H_2O_{(liq)};\ n = 1 \text{ to } 4 \quad (4.95)$$

The predicted reaction (4.95) constant values systematically decrease with increasing temperature for all sulfide species. For example, the $\log K_{95}$ value for the major $MoS_4^{2-}$ complex ranges from 19.9 at 25 °C to 10.3 at 300 °C. These values mean that $MoS_4^{2-}$ will over compete $MoO_4^{2-}$ at >10⁻ [5] and >0.003 m $H_2S$ in the fluid, respectively at 25 °C and 300 °C. The other intermediate Mo-O-S complexes show up at hydrothermal conditions only at rather alkaline pH (Fig. 4.35b). The theoretical predictions of the $MoS_4^{2-}$ stability were confirmed by Laskar (2022) and Kokh et al. (2024) by direct $MoS_{2(s)}$ solubility measurements coupled with in-situ XAS at 300 °C in NaHS-HCl solutions of near-neutral and alkaline pH (pH>6–7). In more acidic fluids (pH<5–6) and in the presence of both sulfide and sulfate, $MoS_{2(s)}$ solubility values were too high to be explained by $MoS_4^{2-}$ alone, thereby implying additional complexes, likely with the trisulfur radical ion. These data were interpreted by a mixed sulfide-trusulfur complex, $MoS_3(S_3)^-$, according to the reaction

$$MoS_{2(s)} + H_2S_{(aq)} + S_3^{\bullet-} + 0.5\ O_{2(aq)} = MoS_3(S_3)^- + H_2O_{(liq)} \qquad (4.96)$$

with an equilibrium $\log K_{96}$ value of 14.3±0.3 at 300°C and 500 bar. These first results point to polysulfide ions as potentially important carriers of Mo in porphyry-type hydrothermal fluids (Kokh et al. 2024).

As a conclusion, at present, only oxyhydroxide (Dadze et al. 2018a) and oxysulfide (Liu et al. 2020; Kokh et al. 2024) complexes of Mo may be modeled with a good accuracy in moderate-temperature hydrothermal fluids (<350 °C), but they await to be placed in a more widely used thermodynamic framework (e.g., HKF model). More systematic solubility, spectroscopy and modeling work is needed to quantify the effect of chloride, alkalis and different sulfur ligands, including sulfur radical ions, on Mo speciation at hydrothermal-magmatic conditions.

### *4.4.11 Sulfate complexes of metals*

In light of the overwhelming abundance of chloride in most hydrothermal environments along with their generally reducing conditions favoring sulfide and polysulfide ligands, sulfate complexes of metals in hydrothermal fluids have received rather modest attention. Because the $SO_4^{2-}$ ion is rather rigid and its charge is high, monodentate coordination is generally most common and the order of the complexes is typically limited to 1:1 and 1:2, at least at near-ambient conditions. However, examples of bidentate coordination and metal complexes with three sulfate ligands exist in the literature. Sulfate is a chemically hard anion. Therefore, the stability of its metal complexes generally increases from soft to intermediate and to hard cations, as shown in Fig. 4.36. This figure compiles available formation constants for the 1:1 metal sulfate complexes, which represent the largest set amongst the known stoichiometries, plotted versus their corresponding first hydrolysis constants at 25 °C and 1 bar for the three groups of cations – hard, intermediate and soft. Sulfate ion pairs with alkali and alkali-earth metals and the solubility of their sulfate minerals were discussed in more detail in section 4.3.4; here we focus on other metals.

*Soft metals* (shown in red in Fig. 4.36) generally form the weakest complexes with sulfate. As a result, few data are available at ambient conditions, and virtually no data exist at hydrothermal temperatures. High-temperature thermodynamic properties of 1:1, 1:2, and 1:3 sulfate complexes of $Pt^{II}$, $Pd^{II}$, $Rh^{II,III}$, and $Ru^{II,III}$ have been theoretically estimated by Sassani and Shock (1998) using HKF model correlations on the basis of meagre data at ambient conditions. Such complexes have negligibly low abundances in virtually any natural hydrothermal environments thereby being unable to compete with the $Cl^-$, $HS^-$ and $S_3^{\bullet-}$ ligands (e.g., Pokrovski et al., 2021b; Laskar et al. 2022).

*Intermediate or borderline cations* (shown in blue in Fig. 4.36) form stronger sulfate complexes whose stability is linearly correlated with their hydroxide counterparts. This group of cations, as well as the soft group above, generally has four-fold (tetrahedral or square-planar) and six-fold (octahedral) coordinations in solution forming single-bond linkages with sulfate via an oxygen atom, [M-O-$SO_3$]. Such coordinations are poorly compatible with a bidentate linkage with a sulfate anion via two oxygen atoms due to steric strain. Bidentate (i.e. edge-sharing $MO_n$ and $SO_4$ polyhedra) are equally uncommon in these metal sulfate solid structures (Pan and Mi 2025). A very limited amount of experimental work exists on sulfate complexes of base metals (Fe, Mn, Zn, Pb, Cu, Ni) in aqueous solution at elevated temperatures. Most stability constants are theoretical estimations based on near ambient-temperature data. For example, $Fe^{II}$ and $Fe^{III}$ sulfate complexes such as $Fe^{II}HSO_4^+$, $Fe^{II}SO_4^0$, $Fe^{III}SO_4^+$, and $Fe^{III}(SO_4)_2^{2-}$ have been considered in the chemical literature, among which $FeSO_4^0$ appears to be the best constrained one. Their stability constants and basic thermodynamic functions ($\Delta_f H^0$, $S^0$) at 25 °C were thoroughly revised by Lemire et al. (2013). On the basis of this and earlier compilations, van't Hoff-type isocoulombic extrapolations to 200–300 °C of the formation constants of these and some other metal sulfate complexes (e.g., $ZnSO_4^0$, $NiSO_4^0$) can be found in the LLNL or Thermoddem databases. We are unaware of any set of HKF-model parameters for base-metal sulfate complexes (except $MnSO_4^0$, Sverjensky et al. 1997) that would allow reliable extrapolations to supercritical temperatures. Many sulfate minerals, including alkali, alkali-earth and base metals, have a retrograde solubility in dilute aqueous solution with temperature, at least at pressures close to $P_{sat}$. Those of heavy metals are weakly soluble in water even at ambient conditions (e.g., congruent solubility, in molality units, of $BaSO_4 \approx 10^{-5}$, $PbSO_4 \approx 10^{-4}$, $CdSO_4$ and $SrSO_4 \approx 10^{-3}$, and $Ag_2SO_4 \approx 10^{-2}$ m), rendering it difficult measurements of aqueous sulfate complexation with such metals. Because complexes with hard ligands generally become stronger with increasing temperature, sulfate might compete with chloride complexes of $Fe^{II}$ and $Zn^{II}$ in oxidized and salt-poor fluids from subduction zones. For example, formation of sulfate complexes was hypothesized to account for relatively "light" $\delta^{56}Fe$ and $\delta^{66}Zn$ signatures inferred for oxidized fluids released by prograde metamorphism of the subduction slab (Debret et al. 2016, 2020; references therein). Thermodynamic data on sulfate complex stability at such conditions are required to allow quantitative interpretations of metal isotope signatures in subduction zones.

*Hard cations*, such as alkali-earth and rare earth metals, $Al^{3+}$, and actinides are prone to forming much stronger complexes with the hard $SO_4^{2-}$ ligand than those of soft and borderline metals mentioned above, as can be seen in Fig. 4.36 (shown in green). However, all these metals are also strongly hydrolyzed in aqueous solution at elevated temperatures, so that there is competition between the $SO_4^{2-}$ (and $HSO_4^-$) and the $OH^-$ ligands for these metal cations in aqueous solution. In contrast to the soft and borderline groups of cations, there is much more scatter from the average linear trend of $SO_4^{2-}$ vs $OH^-$ complex stability for the hard group. This scatter is due to *i)* larger uncertainties related to the determination of sulfate and hydroxide complex stability for strongly hydrolyzed and poorly soluble cations of this group, and *ii)* larger variety and versatility of the metal coordination for which large and distorted $MO_n$ polyhedral are typical, with $n$ from 7 to 9. The latter property allows for easier formation of bidentate sulfate complexes with less steric strain compared to lower-coordination cations of the intermediate and soft groups. For example, bidentate $YO_n$-$SO_4$ linkages in hydrothermal fluids were inferred from spectroscopic and molecular simulation approaches (e.g., Guan et al. 2020, 2022). The potential tendency to form stronger bidentate complexes for this group of metals in aqueous solution is in line with the numerous examples of such cation-sulfate geometries, which are commonly present along with monodentate in alkaline earth and rare earth sulfate solids (Pan and Mi 2025).

Because of the competition with hydrolysis, sulfate complexes (and eventually hydrosulfate, $HSO_4^-$, which is a weaker ligand) are expected to be more significant at mildly acidic conditions (typically $2<pH<4$) at which the metal cation is not yet fully hydrolyzed and the sulfate anion is yet not fully protonated (e.g., Fig. 4.16b). For example, Tagirov and Schott (2001) generated an HKF-coefficients dataset for $AlSO_4^+$, which eventually may be dominant in acidic, sulfate-rich, but fluoride- and salt-poor fluids. Whereas a substantial dataset exists at near-ambient conditions on the uranyl ion ($U^{VI}O_2^{2+}$) sulfate complexes (see the review of Grente et al. 2020), it is only recently that the stability constants of $UO_2SO_4^0$ and $UO_2(SO_4)_2^{2-}$ were measured to 200 °C using UV-Vis spectroscopy by Kalintsev et al. (2019), and RB-model equations of log$K$ generated by the authors may reliably be used to ~300 °C. Their results suggest that sulfate may be the dominant carrier of $U^{VI}$ in oxidized fluids of acidic pH (<4) with sulfate concentrations above ~1000 ppm S. In the recent years, there has been a surge of interest for REE sulfate complexes in deep fluids given the multiple instances of REE economic concentration in carbonatites and similar rocks whose mineralogy and fluid/melt inclusions provide clear evidence for the presence of alkali metal sulfate-carbonate-chloride brines (e.g., Walter et al. 2020a,b, 2021; references therein). These findings motivated recent experimental studies of REE complexes with sulfate as well as the solubility of sulfate minerals of REE and major alkali and alkali earth metals. A significant quantitative advance, compared has been made by Migdisov's team through systematic generation of key experimental data for different REE complexes ($Cl^-$, $OH^-$, $F^-$, $PO_4^{3-}$) including those with sulfate at hydrothermal temperatures (see the review of Migdisov et al. 2016), thereby bringing reliable constraints to previous theoretical predictions almost exclusively based on low-temperature data (Wood et al. 1990; Haas et al. 1995). More specifically, based on their own experimental data, Migdisov et al. (2016) have generated a set of RB-model parameters for the 1:1 and 1:2 sulfate complexes of the 14 rare earth elements. Their data suggest that sulfate complexes may efficiently over compete $Cl^-$ and $F^-$ in sulfur-rich (1–2 wt% sulfate) acidic (2<pH<5) fluids at 200–400 °C and relatively elevated pressures (>1 kbar) at which the solubility of Na and K sulfates becomes prograde (e.g., Wan et al. 2023; references therein). Thus, drop in pressure and the resulting precipitation of sulfate minerals as well as unmixing to sulfate-rich brine and vapor may be the major controls on REE distribution and precipitation (e.g., Wan et al. 2021; Guan et al. 2022).

### *4.4.12 Speciation of thioanions in natural hydrothermal waters*

In addition to our ability to thermodynamically predict metals speciation and the effect of sulfur of their solubility in hydrothermal fluids as overviewed above, recent progress in analytical techniques has also allowed some metal and metalloids speciation to be directly measured in modern geothermal fluids. Many natural hydrothermal waters contain substantial concentrations of dissolved sulfide (see Table 4.1) and are enriched in various elements having high affinity for reduced sulfur; for example, arsenic, antimony, molybdenum, and tungsten (W) (e.g., Xu et al. 1998; McCleskey et al. 2010; Kaasalainen and Stefánsson 2011a,b, 2012). These trace elements are known to form oxyanions as well as thioanions in aqueous solution (see above).

Aqueous As, Sb, Mo and W thioanions have been encountered in hot spring waters by in situ analyses involving chromatographic separation followed by inductively coupled plasma mass spectrometry (IC-ICPMS and IPC-ICPMS) or hydride generation and atomic fluorescence spectrometry (IC-HG-FAA) (Vassileva et al. 2001; Keller et al. 2014a,b; Lohmayer et al. 2015; Planer-Friedrich and Scheinost 2011; Planer-Friedrich et al. 2020). The results reveal arsenite ($H_nAs^{III}O_4^{n-3}$), arsenate ($H_nAs^VO_4^{n-3}$), thioarsenite ($H_nAs^{III}S_3^{n-3}$) and thioarsenate ($H_nAs^VS_xO_{4-x}^{n-3}$, with $x \leq 4$) to coexist in natural thermal waters (Planer-Friedich et al. 2007; Keller et al. 2014a). Similar species distribution is observed for aqueous Sb, Mo and W, with coexisting antimonite ($H_nSb^{III}O_3^{n-3}$) and antimonate ($H_nSb^VO_4^{n-3}$) together with tri- and tetra-thioantimonate ($H_nSb^VS_3O^{n-}$

$^{3}$ and $H_nSb^{V}S_4O^{n-3}$), as well as molybdates ($H_nMo^{VI}O_4{}^{n-2}$) and thiomolybdates ($H_nMo^{VI}S_xO_{4-x}{}^{n-2}$) and tungstate ($H_nW^{VI}O_4{}^{n-2}$) and thiotungstates ($H_nW^{VI}S_xO_{4-x}{}^{n-2}$) (e.g., Planer-Friedrich and Scheinost 2011; Ullrich et al. 2013; Planer-Friedrich et al. 2020). At low sulfide-to-metal ratios, the aqueous oxyanions predominate, whereas with increasing sulfur concentrations, the thioanions become more important accounting, for example, for up to 90, 80, 75 and 50% of the total dissolved As, Mo, W and Sb, respectively, in thermal waters with of 0.005 to 0.5 mmol/kg of $H_2S$, pH > 8 and temperatures of 50–100°C (Planer-Friedrich et al. 2020).

The thioanions form by nucleophilic attachment of $HS^-$ to the metal(loid)-oxygen bond (Erickson and Helz 2000). Such reactions are favored by the increased abundance of the $HS^-$ ion upon pH increase and high sulfide to metal(loid) ratios. At acidic pH, formation of thioanions is limited due to precipitation of sulfides whereas at alkaline pH, excess of $OH^-$ over $HS^-$ favors oxyanion formation. Moreover, thermodynamic species distribution is not attained (e.g., Keller et al. 2014a), indicating that the formation and relative abundance of thioanions are kinetically controlled, which is clearly seen by the fact that oxy- and thioanions with As and Sb of variable oxidation state (III *vs* V) coexist in the same water samples.

## 4.5 Sulfur concentration, speciation and reactions in geological fluids

Hydrothermal fluids in general may be subdivided into modern and ancient systems. The former can be observed in situ and directly sampled and analyzed. The latter are not accessible for direct observation, and the only straightforward information on their composition stems from fluid inclusions. Using these two major sources of information, coupled with mineral paragenesis compositions and physical-chemical modeling for both types of fluids, we outline the key controls on sulfur behavior in modern geothermal systems and in three major types of paleo ore-forming environments controlled by magmatic-hydrothermal, metamorphic, and sedimentary-basin fluids.

### *4.5.1 Active geothermal on-land and deep-sea sources*

Geothermal systems are typically classified into volcanic and non-volcanic systems (Goff and Janik 2000). Volcanic geothermal systems occur in areas of active volcanism where permeability and geothermal gradient are high. Hydrothermal fluid temperatures of such systems are usually between 200 and 350°C but may be as high as 450°C. Non-volcanic geothermal systems are commonly observed within permeable rock formations or within tectonically active areas with hydrothermal fluid temperatures ranging from above ambient to ~150°C. The hydrothermal fluids are sourced from meteoric water or seawater or a mixture thereof. Components of connate, magmatic and metamorphic fluids may also be present in hydrothermal fluids (Chambefort and Stefánsson 2020).

Hydrothermal fluids have been divided into reservoir (i.e. so called primary) fluids and surface (i.e. secondary) fluids. Reservoir fluids are those observed at depth within geothermal systems. The main types of reservoir hydrothermal fluids are NaCl waters, acid-sulfate waters and high-salinity brines. Surface hydrothermal fluids have chemically evolved upon ascent to surface through processes such as *i)* boiling to form liquid water and vapor, *ii)* phase separation of saline fluids into a hypersaline brine and a more dilute vapor, *iii)* vapor condensation in shallow ground water or surface water to produce steam-heated waters, and *iv)* mixing of hydrothermal fluids with shallower non-thermal ground and surface water (Arnórsson et al. 2007).

The composition of reservoir hydrothermal fluids is typically obtained by combining liquid and vapor phase well discharge concentrations (m) with the vapor fraction (x), i.e. $m = x^{vap} \times m^{vap} + (1 - x^{vap}) \times m^{liq}$. The concentrations of sulfate and $H_2S$ for the reservoir fluid are variable, typically ranging from a few ppm up to 100s ppm in some cases (Table 4.1). Other intermediate-valence sulfur species in reservoir hydrothermal fluids are considered to be insignficant and usually not analyzed. The origin of sulfur in these fluids is the source water, together with volcanic gas and rock leaching. The concentration of $SO_4$ is controlled by the source water composition and the degree of leaching from rocks and sediments, but can also be influenced by anhydrite saturation in some cases (Fig. 4.37a; Stefánsson and Arnórsson 2002) through the reaction

$$\text{Anhydrite} = Ca^{2+} + SO_4^{2-} \quad (4.97)$$

Regarding aqueous sulfide, it is commonly assumed to be controlled by equilibrium with sulfur containing mineral buffers (Fig. 4.37), for example reactions involving pyrite, pyrrhotite and magnetite (PPM buffer):

$$\text{Pyrite} + 2\ \text{Pyrrhotite} + 4\ H_2O = \text{Magnetite} + 4\ H_2S \quad (4.98)$$

or more complex reactions among minerals like epidote and prehnite (PPPE) (Stefánsson and Arnórsson 2002):

$$\text{Pyrite} + \text{Pyrrhotite} + 2\ \text{Prehnite} + 2\ H_2O = 2\ \text{Epidote} + 3\ H_2S \quad (4.99)$$

The concentrations of sulfur species in surface hydrothermal fluids are very variable. Hot springs with alkaline pH values that are formed upon boiling of reservoir fluids have similar dissolved sulfate but lower dissolved sulfide concentrations relative to parent reservoir fluid. Other intermediate species, particularly thiosulfate, have been observed in such waters, with concentrations ranging on a scale of ppb to few ppm. These intermediate sulfur species are considered to be produced upon oxidation of dissolved sulfide at alkaline pH values. Thiosulfate may also form upon oxidation and hydrolysis of native sulfur and dissolved polysulfides, the latter being formed upon oxidation of dissolved sulfide at acid to mildly alkaline pH values (Xu et al. 1998; Kaasalainen and Stefánsson 2011b). Lastly, thiosulfate may also form upon hydrolysis of $SO_2$ of magmatic origin (Giggenbach 1987). These oxidation and disproportionation reactions may also involve other intermediate sulfur oxidation states, for example dissolved sulfite and polythionates (Xu et al. 2000; Kaasalainen and Stefánsson 2011b).

In steam-heated acid hot springs, the concentration of sulfate can be very high, up to a few thousand ppm. Dissolved sulfide in reservoir fluids partitions into the vapor formed upon depressurization boiling when fluids ascend to surface resulting in $H_2S$ enriched vapor. At surface, condensation of such vapor and its mixing with shallow groundwater and surface water followed by oxidation result in enrichment in dissolved sulfate (Fig. 4.38). The low pH further enhances dissolution of many rock-forming elements resulting in high metal content of many of these acid waters (Giggenbach 1987, 1997; Hedenquist et al. 1993, 1994; Markússon and Stefánsson 2011; Kaasalainen and Stefánsson 2012).

The concentration of $H_2S$ in the vapor phase at surface discharged from boreholes and fumaroles depends on the reservoir concentrations and the nature and degree of boiling, generally being in the range of a few ppm to 100s ppm. Typically, $H_2S$ quantitatively (>90%) partitions into the vapor phase upon initial boiling resulting in high concentrations. Prolonged boiling, however, results in a decrease in $H_2S$ concentrations as the vapor fraction increases.

In deep-sea geothermal systems, circulating seawater results in hydrothermal vent fluids being emitted into cold seawater. Hydrothermal vent fluids contain elevated chloride concentration, similar to or higher than seawater. Dissolved sulfide is the dominant form of sulfur in such fluids with low to insignificant amounts of sulfate. Sulfur sources in hydrothermal vent fluids are similar as for terrestrial geothermal systems and include leaching of sulfide present in crustal host rocks and sediments, reduction of seawater-derived sulfate to hydrogen sulfide, and magmatic sulfur volatiles. Mixing of vent fluids with seawater usually results in metal sulfide precipitation and deposition. Moreover, the heating up of seawater sulfate in the recharge zone of the systems, together with retrograde solubility of anhydrite, may further result in deposition of calcium sulfate.

#### *4.5.2 Magmatic-hydrothermal fluids*

Magmatic-hydrothermal systems include deposits from the porphyry family (porphyry Cu-Mo-Au deposits and associated skarn, Cordilleran, high- and intermediate-sulfidation epithermal deposits), vein-type and Sn-W skarn deposits, as well as iron-oxide copper gold (IOCG) deposits. Various fluid types are involved in mineralization processes in these systems (Table 4.2; Kouzmanov and Pokrovski 2012 and references therein). They may be classified into four types according to their origin (Table 4.2): *i)* supercritical fluids that separate from hydrous magma, *ii)* and *iii)* high to low-density vapors and hypersaline liquids (brines), both generated by phase separation of the supercritical fluid upon its ascend, typical for proximal high-temperature parts of the systems, and, finally, *iv)* lower-temperature and lower-salinity aqueous fluids, typical for more distal parts of these systems.

A detailed interpretation of metal concentrations in fluid inclusions from magmatic-hydrothermal systems has been made by Kouzmanov and Pokrovski (2012), based on a large LA-ICPMS dataset, along with very limited data for sulfur. Over the last decade, the sulfur dataset has been largely extended owing to recent publications (Table 4.2, Supplementary Table 4.S1). Figure 39a,b compiles sulfur concentrations in the four major types of fluid inclusions from magmatic-hydrothermal systems as a function of temperature and salinity, as measured by LA-ICPMS. Sulfur concentrations vary within more than three orders of magnitude – from 100 ppm (its typical detection limit) to ~100,000 ppm. Sulfur concentrations from high-temperature fluid inclusions (supercritical, brine and vapor types) are independent of inclusion homogenization temperature and salinity. Low-temperature and low-salinity fluid inclusions (aqueous type) show an overall tendency of decreasing sulfur concentrations with decreasing temperature (Fig. 4.39a). Thus, the temperature trend mostly reflects the cooling of the magmatic system. Such cooling results in a large decrease of metal sulfide solubility, which is the major phenomenon in the magmatic-hydrothermal environment. Thus, the 2–3 orders of magnitude decrease in sulfur concentration, parallel with up to a 4–5 orders of magnitude decrease of metal concentration in the fluids in individual systems, is due mainly to direct sulfide mineral precipitation from the fluid upon cooling (e.g., Fig. 4.25a; Kouzmanov and Pokrovski, 2012 and references therein). This sulfide formation mechanism is accompanied by that of sulfidation of Fe(II)-bearing mafic minerals (e.g., biotite; hornblende) in the host intrusions and wall rock. Cooling is generally accompanied by a salinity decrease due to fluid mixing with external waters, resulting in a further decrease of metal solubility. Sulfur often remains in excess of metals (see below and Fig. 4.40). The apparent sulfur decrease with increasing salinity observed in some aqueous-type inclusions (Fig. 4.39b) is likely to reflect the origin of this type of inclusions. Many of them are the product of condensation of S-rich supercritical fluid or vapor of relatively low salinity (<5–10 wt% NaCl) at the closer-to-surface parts of the magmatic-hydrothermal systems (e.g., Heinrich 2005; Pokrovski et al. 2013a). The total sulfur budget of a magmatic-derived fluid thus controls the degree of sulfide precipitation on cooling (Heinrich 2006; Seo et al. 2009). Comparing the amount of total sulfur and water, necessary to precipitate all major metal sulfides in a polymetallic vein within a porphyry-centered district, with the measured sulfur concentrations in fluid inclusion assemblages from the same vein, Catchpole et al. (2011) concluded that ore-forming fluids were enriched in sulfur compared to metals. Their findings thus demonstrate a largely sufficient amount of sulfur to precipitate all of the available metals from the fluid as sulfide minerals, despite possible loss of part of sulfur due to $H_2S$ degassing and sulfidation of wall rock.

Figure 4.40 summarizes the relationships between sulfur and selected metal concentrations of the four fluid inclusion types in magmatic-hydrothermal systems. As mentioned above, most of the inclusions from the supercritical, vapor and aqueous types of fluids are significantly enriched in sulfur compared to most metals (plotting below the 1:1 mole ratio line). In contrast, brine inclusions systematically show an excess of iron over sulfur (Fig. 4.40a), in agreement with the increasing stability of Fe chloride complexes and therefore iron minerals solubility in highly saline fluids (see section 4.4.6 and Fig. 4.25b). Interestingly, a significant part of the data on the Cu vs. S plot for brine and vapor inclusions (Fig. 4.40b) cluster around the 2:1 line (when converted to mole ratio), corresponding to the chalcopyrite stoichiometry, $CuFeS_2$. This characteristic feature of the Cu-S relationships is due to a post-entrapment rapid inward diffusion of $Cu^+$ through the quartz host structure from the external fluid. This diffusion is compensated by outward diffusion of $H^+$ and/or $Na^+$ from the inclusion and is mainly controlled by the initial sulfur concentration in the fluid inclusion (e.g., Seo et al. 2011; Lerchbaumer and Audétat 2012; Pokrovski et al. 2013a; Seo and Heinrich 2013 for details of the mechanism). In contrast, due to lower diffusion coefficients, other base metals (Fe, Zn, Pb) and gold are much less affected by such a diffision process. The post-entrapment enrichment of Cu is more pronounced in vapor-type inclusions resulting in apparent vapor-liquid partitioning coefficients of Cu up to 100, as was inferred from cogenetic vapor-brine inclusion assemblages (see Kouzmanov and Pokrovski 2012 for a compilation). Such vapor-brine inclusion data resulted in overestimation of the role of vapor on Cu transport in porphyry systems in some earlier studies (e.g., Heinrich et al. 1992, 1999; Audétat et al. 1998; Cauzid et al. 2007; Seo et al. 2009). However, extensive experimental measurements of vapor-liquid distribution of metals at magmatic-hydrothermal conditions in laboratory water-salt-sulfur systems (Pokrovski et al. 2013a; references therein) clearly demonstrated that Cu partitions always in favor of the brine phase (typical $D_{vap/liq}$ (Cu) values of 0.01–0.1), along with all other base metals (Fe, Zn, Pb, Ag) and metalloids (As, B, Sb, Si). The only exception is $Au^{I}$ and $Pt^{II}$ that are able to be enriched in the vapor phase compared to the coexisting brine, due to formation of volatile hydrosulfide complexes (see Pokrovski et al. 2013a for a detailed overview of the metal vapor-phase transport). In general, sulfur concentrations in fluids from magmatic-hydrothermal systems do not show clear correlations with other metals and metalloids. A lack of correlation does not mean, however, the absence of sulfur control on those metals. Given the multiplicity of parameters simultaneously affecting metal solubility in the magmatic-hydrothermal systems such as temperature, pressure, redox, salinity, acidity and sulfur speciation and concentration (see section 4.4.1 and Fig. 4.25) along with limited inclusion datasets, the effect of sulfur may be obscured in such systems.

The key feature of supercritical, vapor and aqueous types of fluids in magmatic-hydrothermal systems is significant sulfur excess over metals forming sulfide minerals, by a factor of 10 on average (Supplementary Table S1). This excess is mostly conditioned by the fluid source, originated from magma and carrying together metals and sulfur (mainly as $H_2S$ and $SO_2$), along with its subsequent evolution upon its ascent through the crust, which is mainly controlled by cooling and decompression (e.g., Figs. 20, 25 and 30; Hedenquist and Lowenstern 1994; Hedenquist 1995; Sillitoe 2010; Kouzmanov and Pokrovski 2012; Fontboté et al. 2017, for an overview). As to the fate of sulfur, these combined processes result in a number of phenomena *i)* precipitation of part of sulfur with Fe, Cu, Ag, Au, Zn, Pb, As as sulfide ore minerals (pyrite, chalcopyrite, bornite, and, at lower temperature, sphalerite, galena, sulfosalts, e.g., reactions 4.56–4.60); *ii)* enrichment in volatile sulfur species ($H_2S$ and $SO_2$) of the vapor phase upon vapor-brine unmixing (reactions 4.23 and 4.26), whereas base metals remain more enriched in the brine; *iii)* transfer or sulfur-rich vapor phase to epithermal parts of the system followed by condensation (resulting in aqueous-type inclusions) and, sometimes, boiling closer to the surface, depending on the temperature-pressure gradient; *iv)* hydrolysis of magmatic $SO_2$ in aqueous solution upon cooling, according to the disproportionation reaction:

$$4\ SO_2 + 4\ H_2O = 3\ SO_4^{2-} + 6\ H^+ + H_2S \qquad (4.100)$$

This important reaction provides additional $S^{II-}$ necessary to form metal sulfides, the $S^{VI}$ necessary to form sulfates such as alunite and anhydrite common in these systems, and significant acidity ($H^+$) that may greatly facilitate base metal transport in the form of chloride complexes (e.g., reactions 4.56–4.60), in particular through altered rocks of low pH buffering capacity (e.g., Hedenquist et al. 2000; Heinrich 2005; Kouzmanov and Pokrovski 2012; Reed et al. 2013).

In contrast to surface geothermal fluids whose sulfur concentrations may be reasonably approximated by the solubility of anhydrite (for sulfate sulfur) and of the pyrite-pyrrhotite(-magnetite/epidote/phengite) assemblages (Fig. 4.37), sulfur concentrations analyzed in fluid inclusions of magmatic-hydrothermal fluids as well as in metamorphic and sedimentary fluids (see below and Fig. 4.41 and 4.42), appear in significant excess of those in equilibrium with the most common sulfur and oxygen mineral buffers in the upper crust such as pyrite-pyrrhotite-magnetite (PPM) and pyrite-magnetite-hematite (PMH) (Fig. 4.43). It is only at the highest temperatures recorded by the fluid inclusions (>500 °C) or pressures below 500 bar that such buffers may approximate the sulfur concentrations recorded in natural fluid inclusions. Note that sulfur concentrations in equilibrium with PPM and PMH are almost insensitive to variations in salinity and acidity, in contrast to most metal concentrations (Fig. 4.25). The significantly higher sulfur concentrations from magmatic-hydrothermal systems than those of the PPM and PMH buffers are thus mainly a result of the excess of sulfur over iron in the original fluids, and the lack of significant fluid-rock interactions that may occur in some settings where the porphyry system is overprinted by an epithermal system on already altered rocks due to significant acidity generated by reaction (4.100) on cooling (e.g., Heinrich 2005; Kouzmanov and Pokrovski 2012; Simon and Wilke 2025).

### *4.5.3 Metamorphic fluids*

Metamorphic fluids, formed by extraction of small amounts of volatiles from large crustal terrains, have rock-buffered major element and isotopic compositions because they are controlled by equilibria with the minerals of their source rocks (Yardley 1983; Cox et al. 1987; Ridley and Diamond 2000). Orogenic gold deposits which occur in metamorphic terrains are thought to involve ore fluids produced by metamorphic devolatilization (e.g., Goldfarb et al. 2005). The behavior of sulfur in metamorphic processes is discussed in more detail in this Book by Schwarzenbach and Evans (2025) and Harlov (2025). Here we focus on direct evidence of metamorphic fluid compositions from fluid inclusions. These data are provided by recently published LA-ICPMS analyses of total sulfur concentration in fluid inclusions from gold mineralization associated with late metamorphic quartz veins and Alpine fissure veins (Marsala et al., 2013; Miron et al., 2013; Rauchenstein-Martinek et al., 2014, 2016) and will be compared here with with thermodynamic predictions of sulfur solubility and speciation.

Fluid inclusions from metamorphic systems vary in composition from dominantly aqueous – in sub-greenschist and greenschist facies host rocks, to aqueous-carbonic inclusions – in amphibolite facies rocks. Volatile contents ($CO_2$, $H_2S$) in the fluid inclusions systematically increase as a function of increasing metamorphic grade, indicating that the fluids have been produced by prograde devolatilization reactions. Reported temperatures of homogenization range from 80 to 300 °C, and salinity varies between 0.4 and 10.3 wt% NaCl eq. (Table 4.2; Fig. 4.39c,d). However, the true fluid entrapment temperatures may be several 100 °C higher. Sulfur concentrations vary within two orders of magnitude, from 10s to 1000s ppm, and do not show correlations with temperature, but tend to have generally higher values in lower-salinity fluid inclusions (Fig. 4.39c,d). Metals are strongly depleted in the fluids compared to sulfur (in general, one to two orders of magnitude; two to four orders of magnitude for Au). Metal concentrations in metamorphic fluids are several orders of magnitude lower than those in fluids from magmatic-hydrothermal systems (Fig. 4.40, 4.41, Table 4.2). Positive correlations between S and As concentrations are often recorded, arsenic representing the largest data set among metals and metalloids (Fig. 4.41c). The As-S trends imply a common solubility control on both elements by dissolution-precipitation of As-bearing pyrite and arsenopyrite, which are the most common sulfide minerals in such systems. Gold concentrations do not clearly correlate with S in the fluids, and show a large scatter, between 0.001 and 1 ppm Au for sulfur concentrations ranging from <10 to 1000s ppm (Fig. 4.41d). Gold concentrations are correlated neither with As nor Ag concentrations (e.g., Rauchenstein-Martinek et al. 2014). Although gold hydrosulfide complexes such as $AuHS^0$ and $Au(HS)_2^-$ (and eventually $Au(HS)S_3^-$ at high temperatures and S concentrations) are the gold transporting species in low salinity and near-neutral pH typical of metamorphic fluids (see sections 4.4.1 and 4.4.4), the increase in sulfur concentrations with metamorphic grade does not result in respectively higher Au concentrations as would have been expected. This observation, together with thermodynamic calculations showing fluids to be under-saturated with respect to the native gold phase, indicates that Au concentrations in the generated fluid should be limited by the availability of gold in the source rocks from which the fluids were produced (Rauchenstein-Martinek et al. 2014) rather than gold solubility-controlling reactions (such as 4.64–4.66).

The main processes affecting sulfur during the ore-bearing fluid evolution in orogenic deposits are fluid-rock interactions, dominated by so called sulfidation reactions between aqueous $H_2S$ and an $Fe^{II}$-rich rock acting as a sink for sulfur via precipitation of pyrite (and pyrrhotite). Efficient $H_2S$ consumption from the fluid shifts reactions (4.64) and (4.65) to the left, leading to gold precipitation. Mafic rocks rich in ferrous iron are, thus, an excellent sink for sulfur. Wall-rock sulfidation is the primary control on metal deposition in deep and high-temperature orogenic gold systems; in addition, with decreasing depth of formation, phase separation favored by significant $CO_2$ content may also contribute to Au precipitation (e.g., Mikucki 1998; references therein). During progressive metamorphism at depth, a large amount of sulfur released into the fluid was expected at the breakdown of sedimentary pyrite to pyrrhotite occurring at temperatures >500 °C (e.g., Tomkins 2010; Large et al. 2011; Pokrovski et al. 2015). However, because $Fe^{II}$ rock content in metamorphic terrains is usually superior to that of $S^{II-}$, sulfur release may be over-competed by backward sulfide sulfur scavenging from the fluid by the excess of Fe oxide and silicate in the host rock (e.g., Zhong et al. 2015). Therefore, the solubility of the PPM buffer assemblage may be regarded as a reasonable proxy for sulfur concentrations observed in metamorphic fluid inclusions. These inclusions must, however, be corrected for the entrapment temperature typical of greenstone-amphibolite facies, which should typically be 200–300 °C higher than the homogenization temperature (section 4.2.3). It can be seen in Fig. 4.43 that such corrected fluid-inclusion data would match well the calculated PPM buffer solubility.

### *4.5.4 Sedimentary brine fluids*

Basinal brine fluids (also called basement brines; Walter et al. 2020b) are involved in mineralization processes in a number of hydrothermal environments such as sediment-hosted metal deposits (e.g., sedimentary exhalative – SEDEX, Mississippi Valley type – MVT, Kupferschiefer, Irish-type Pb-Zn, unconformity-related U deposits), various rift-related or basement-hosted polymetallic, fluorite-barite or uranium deposits, as well as IOCG deposits (e.g., Leach et al. 2005; Williams et al. 2005; Wilkinson 2010; Richard et al. 2016). Hot, oxidized, chloride-rich basinal brines are able to transport high concentrations of base metals as chloride complexes (up to 10,000s of ppm of Cu, Pb, Zn – Table 4.2; Wilkinson et al. 2009), via substantial fluid flow focusing, from the crystalline basement reservoirs to shallow crustal levels. However, mixing with other types of oxidized fluids, such as shallower sediment-derived fluids or meteoric waters alone commonly produces carbonate-barite-fluorite veins poor in sulfide. Precipitation of metal sulfide ore mineralization requires a mechanism that efficiently supplies reduced sulfur or reducing agents (e.g., $CH_4$ or organic matter) capable of reducing sulfate-bearing brines (Goldhaber et al. 1995; Fusswinkel et al. 2013; Burisch et al. 2017; Walter et al. 2020a).

LA-ICPMS data on metal concentrations in basinal brine fluid inclusions are abundant in the literature; however, sulfur data were reported only in a few studies. Recently, total sulfur concentrations in fluid inclusions of basinal brine fluids from a number of barren and Pb-Zn veins from Schwartzwald, Germany, have been published (Fusswinkel et al. 2013; Burisch et al. 2016; Walter et al. 2018, 2019). Temperatures of homogenization are relatively low (100–180 °C), and salinity varies between ~2 and ~30 wt% (NaCl + $CaCl_2$), with most values being restricted in the 20–26 wt% range (Fig. 4.39e,f; Table 4.2). Sulfur concentrations vary over four orders of magnitude, from ~10 to almost 100,000 ppm. Such concentrations are comparable to the ones recorded by supercritical, hypersaline liquid and vapor inclusions from magmatic-hydrothermal systems (see section 4.5.2; Table 4.2). Metals (Fe, Cu, Zb, Pb) and metalloids (As, Sb) are generally depleted compared to sulfur in this fluid type, with some exceptions for Fe and, more rarely, for other base metals, as this is also the case for brine inclusions in magmatic-hydrothermal systems (Fig. 40). In some case examples (not identified in the figure), negative correlations are observed between sulfur and metal (Fe, Cu, Zn, As) concentrations that vary within comparable magnitudes, ~1 to 10,000 ppm (Fig. 4.42). Such correlations are likely due to the negative effect of the fluid $H_2S$ content on base metal sulfide solubility (Fig. 4.25d). These natural cases indicate fluid mixing as the likely cause of precipitation, as inferred from halogen systematics in these systems (Fusswinkel et al. 2013; Walter et al. 2020a). However, $H_2S$ was not directly detected in those fluid inclusions (e.g., by using Raman spectroscopy, section 4.2.4), but sulfate minerals (e.g., gypsum) were observed, for example in fluid inclusions from Schwartzwald. Their crush-leach analyses show that sulfur is dominantly present as $SO_4^{2-}$ in the mineralizing fluid, at least at ambient conditions of those analyses (Walter et al. 2018, 2019), attaining high sulfate concentrations approaching those of chloride ($SO_4^{2-}/Cl^-$ mole ratios of 0.2–0.6; Walter et al. 2017). The abundance of sulfide mineralization in the district is conditioned by the availability of $H_2S$ as controlled by reactions (4.56–4.66) or similar. Generation of $H_2S$ in such environments is often driven by reducing agents (like $CH_4$, hydrocarbons, or graphite) that may react with the abundant sulfate present in the metal-transporting saline fluid. This reaction commonly occurs in sedimentary environments in which dissolved sulfate (in basinal brines and other formation waters) encounters hydrocarbons between 100 and ~160°C or, at higher temperatures, in the presence of bitumen or graphite. Upon this encounter, sulfate is reduced in a reaction called thermochemical sulfate reduction (TSR) represented in a simplified form as

$$2\ H^+ + SO_4^{2-} + CH_4 = CO_2 + H_2S + 2\ H_2O \qquad (4.101)$$

The extent and kinetics of TSR reactions and the associated S, C and O isotope fractionation have been extensively studied in numerous natural settings as well as laboratory experiments since >50 years (e.g., Orr 1974; Cai et al. 2022; Meshoulam et al. 2023; references therein). The observation of native sulfur in fluid inclusions from sedimentary-basin settings (e.g., Barré et al. 2017) is direct evidence of the existence of hydrothermal fluids containing both sulfate and sulfide from which $S^0$ precipitates on cooling. Reaction (1.101) not only leads to more reducing conditions, but also generates more $H_2S$ and consumes $H^+$. The concomitant increase in both $H_2S$ concentration and pH strongly reduces base metal solubility in the form of chloride complexes (reactions 4.56–4.60, Fig. 4.25c,d) but, at the same time, greatly favors gold solubility as hydrosulfide and trisulfur ion complexes (reactions 4.64 and 4.65). This sulfur control may be key for gold transport from magmatic bodies emplaced into a sedimentary basin and its economic accumulation in structural traps as, for example, occurred in the Shahuindo and Algamarca epithermal gold deposits in northern Peru and similar deposits worldwide (Galdos et al. 2024; Vallance et al. 2024). Hydrogen sulfide generation due to TSR, as well as lower-temperature biogenic sulfate reduction (BSR) and simultaneous or subsequent fluid mixing, are also the main processes leading to sulfide ore mineralization in MVT and five-element vein deposits (Burisch et al. 2017; Walter et al. 2019; Sośniska et al. 2023; references therein). Sulfur excess over metals with a large portion of sulfate sulfur in basinal fluids from such systems is also demonstrated by much lower sulfur (dominantly $H_2S$) concentrations predicted in equilibrium with the PPM buffer than those analyzed in fluid inclusions even if the latter are corrected to true entrapment temperature of 150–250 °C (Fig. 4.43).

This overview shows the large variability of sulfur transformation reactions and solubility controls and the complex effect of sulfur on metal fate in different types of hydrothermal fluids. This complexity will be better comprehended when a number of challenges have been addressed in our understanding of sulfur in geological fluids. Some of these future challenges are summarized in the following section.

## 4.6 Perspectives and challenges

This overview highlighted the fascinating complexity of sulfur aqueous speciation in hydrothermal fluids, its transformation reactions and its relationships with metals. Despite unprecedented advances gained over the last 20 years on these topics, many aspects of the role played by sulfur in hydrothermal fluids remain to be understood and quantified.

Our state of knowledge of the speciation of sulfur and metals in hydrothermal fluids is strongly temperature- and pressure-dependent. This knowledge weakens significantly with increasing *T* and *P* even for the most common and abundant sulfur forms such as sulfate and sulfide. For example, their degree of protonation, which is the fundamental parameter affecting sulfur mineral solubility, degassing and complexation with metals, is subject to large uncertainties at *T-P* close to and above the water critical point. The thermodynamic stability of the molecular $H_2SO_4^0$ and of the sulfide ion $S^{2-}$ at magmatic-hydrothermal conditions is poorly known. Protonation and alkali ion pairing of most sulfur anions become increasingly important with temperature rise, but very meagre direct data are available above 300 °C. The last 10-year advances of in situ approaches applied to hydrothermal fluids have revealed that many intermediate-valence sulfur species, such as $S_3^{\bullet-}$, $S_2^{\bullet-}$, $SO_2^0$, $S_n^0$, and protonated polysulfide dianions are virtually unquenchable in aqueous solution. As a result, they cannot be properly detected by traditional ex situ techniques involving fluid sampling and cooling in laboratory or from studies of natural samples which are final products of the high *T-P* processes. The quantification of these important sulfur players thus requires in situ spectroscopic approaches coupled with molecular modeling.

In-depth understanding of sulfur relationships with metals in hydrothermal fluids is another outstanding issue. If the relative affinities of metals to the S-, Cl- and O-type ligands are qualitatively known, we are far from a quantitative picture of the metal interactions with the key sulfur ligands (sulfide, sulfate and polysulfides) in front of their major rivals, $Cl^-$ and $OH^-$. There is yet great disparity in our knowledge for different metals. For example, if complexes of precious (Au, Ag) and some base (Cu) metals with the hydrosulfide and chloride ligands are relatively well quantified at present, those of Mo, Re, and many PGE are virtually unknown in hydrothermal fluids. There is growing experimental and theoretical evidence that mixed ligand (H)S-Cl-O(H)-Na species may play an increasing role for many metals (e.g., Cu, Ag, Au, Mo, PGE) in magmatic-hydrothermal fluids; however, the robust evidence of those species and their quantification across the geologically pertinent *T-P* range are lacking. Likewise, polysulfide dianions ($S_n^{2-}$) and their protonated counterparts may play a significant role in chalcophile metal transport at least at moderate temperatures (<300 °C). The di- and trisulfur radical ions have recently emerged as other potentially important sulfur ligands for chalcophile metals making complexes of high stability (e.g., with Au, Pt, Pd, Mo). These findings open perspectives for in-depth investigations of their impact on the transport and fractionation of different metals across the lithosphere.

Metal complexes with intermediate-valence sulfur ligands may also play a crucial role in the kinetics of sulfide mineral (e.g., pyrite) precipitation processes in hydrothermal systems. The mechanisms behind these processes drive trace metal incorporation into major sulfide minerals that are economic hosts for many critical and precious metals in hydrothermal ore. Their quantification requires elaborate micro- and nano-scale spectroscopic and imaging techniques coupled with novel physical-chemical models of metal and sulfur speciation at the fluid-mineral interface. Quantifying the partitioning of different sulfur species between minerals/melts and the fluid phase is another major challenge. If the distribution of major volatile sulfur forms ($H_2S$, $SO_2$) between fluids and magmas at relatively shallow depths is thoroughly constrained to date, much remains to be done for deeper subduction-zone and metamorphic environments where ionic and intermediate-valence sulfur species become increasingly abundant. Some of those environments may preserve silicate and phosphate minerals (zeolites, apatite, scapolite, cancrinite, serpentinite, and zircon) capable of accommodating in their structure a large range of sulfur redox states, which may reflect those that existed in the parental fluid. Data on fluid-mineral partition coefficients of the different sulfur species are required for using such minerals as novel tracers of the redox state and composition of deep fluids.

The challenges above can only be met through a judicious combination of analytical, spectroscopic, experimental, and modeling approaches. Among the analytical issues, improvement of sulfur preservation and separation procedures for sampled geothermal fluids (e.g., by coupled chromatography and mass spectrometry techniques) would provide a far better knowledge of sulfur intermediates in sulfur redox reactions and of metal complexes with different sulfur ligands. Lowering the detection limits for sulfur in LA-ICPMS techniques (e.g., by using a new generation of triple quadrupole, usually abreviated as TQ, QQQ or QqQ, mass spectrometers and cleaner ablation procedures) would significantly extend the fluid-inclusion database for sulfur, which is currently limited by a few studies and relatively high analytical detection limits. Extending in situ spectroscopy on sulfur in aqueous fluids at elevated temperatures and pressures to methods, such as XANES, NRIXS, EPR and UV-Vis, will be a near-future perspective for both geoscience and chemistry communities. These methods are particularly sensitive to sulfur redox and electronic states thereby providing an excellent complement to Raman spectroscopy, which is currently the only technique routinely used for sulfur in hydrothermal fluids both in nature (fluid inclusions) and experiment (spectroscopic cells). More generally, development of in situ spectroscopy for geologically pertinent fluid-mineral-melt systems should be a major advance for experimental geochemistry and petrology.

A formidable challenge of the modeling of sulfur and metals in geological fluids across the range of terrestrial *T-P* conditions is developing a common framework of physical-chemical equations of state enabling predictions of thermodynamic properties of aqueous species from "classical" aqueous solution to hypersaline brines and to low-density vapor phases. Popular thermodynamic models for aqueous fluids discussed in this chapter (e.g., HKF, density, AD, RB models) cover restricted *T-P*-density ranges and/or a limited choice of species, and are subject to increasing uncertainties in predictions beyond those ranges. Another important type of sulfur-bearing geological and planetary fluids, the mixed $H_2O$-$CO_2$(-$CH_4$) ones typical of deep metamorphic and magmatic contexts, has received only a scarce coverage by thermodynamic modeling. Therefore, experimental data on the speciation of both sulfur and economic metals at such conditions are needed to validate the model predictions and assess their uncertainties. Finally, linking thermodynamic equilibrium models of silicate melts with those of aqueous fluids in a single framework is another great perspective.

As emphasized in this overview, in the 2000s, molecular modeling approaches, in particular FPMD, have started providing new insights into the molecular structure and energetics of sulfur species and ore metal complexes thereby helping to interpret spectroscopic and solubility data. An important outcome of these approaches is the possibility to calculate, at virtually any temperature and pressure, the values of the Gibbs energies of complexation reactions for relatively simple metal complexes and sulfur species. Realistic uncertainties associated with such calculations are yet much greater than those from more direct experimental measurements, but the situation is improving with rapid advances in computer codes. In the meantime, acquiring new experimental data for sulfur and metals in hydrothermal fluids remains the primary necessity for parameterizing models of fluid-rock interactions.

Another advance of the 21$^{th}$ century is the growing application of physical hydrology approaches based on heat distribution, fluid flow models and permeability changes. Such computational methods have allowed integrated reactive transport modeling of fluid paths and three-dimensional ore distribution and shape. However, the chemistry of fluid-rock interactions, mineral solubility, and chemical element speciation are not yet quantitatively accounted for in such models, in particular for sulfur and economic metals. Integration of both chemistry and physics in the same conceptual model would thus be another major computational challenge and would contribute to our fundamental understanding of deep geochemical cycles, assessing the role of sulfur in metal deposit formation, as well as improving mineral resources management.

***Acknowledgments***

This chapter was motivated by the results of >25 years of research of our teams on sulfur and related metals in hydrothermal fluids supported by funding from Agence Nationale de la Recherche (grants SOUMET ANR-2011-Blanc-SIMI-5-6/009-01, RadicalS ANR-16-CE31-0017), Institut des Sciences de l’Univers of the CNRS (INSU-CNRS), Mission pour les initiatives transverses et interdisciplinaires (MITI) interdisciplinary programs of the CNRS “MetalloMix 2021” (grant PtS3) and “ConditionsExtremes 2024” (grant ExtremeS), Institut Carnot ISIFoR (grants OrPet and AsCOCrit), Swiss National Science Foundation (grants 200020_134872, 200021_127123, and 200021_165752), Université de Toulouse, and French Ambassy. We thank Dan Harlov for organizing and pushing forward this book, and reviewers Chris Gammons and Tris Chivers for their thoughtful and insightful comments that greatly improved the paper.

# 4.7 References

Adams MD, ed. (2005) *Advances in Gold Ore Processing*. Elsevier

Aguda R (1981) Desolvation processes in the association of peroxodisulfate and sulfate with bivalent metals in aqueous solution. A calorimetric study. J Inorg Nucl Chem 43:1859–1861

Akinfiev NN, Zotov AV (1999) Thermodynamic description of equilibria in mixed fluids ($H_2O$-non-polar gas) over a wide range of temperature (25–700 °C) and pressure (1–5000 bars). Geochim Cosmochim Acta 63:2025–2041

Akinfiev NN, Zotov AV (2001) Thermodynamic description of chloride, hydrosulphide, and hydroxide complexes of Ag(I), Cu(I), and Au(I) at temperatures of 25–500°C and pressures of 1–2000 bar. Geochem Intl 39(10):990–1006

Akinfiev NN, Zotov AV (2010) Thermodynamic description of aqueous species in the system Cu–Ag–Au–S–O–H at temperatures of 0–600°C and pressures of 1–3000 bar. Geochem Intl 48(7):714–720

Akinfiev N, Zotov A (1999) Thermodynamic description of equilibria in mixed fluids ($H_2O$-non polar gas) over a wide range of temperature (25 to 700 °C) and pressure (1 to 5000 bars). Geochim Cosmochim Acta 63:2025–2041

Akinfiev NN, Diamond LW (2003) Thermodynamic description of aqueous nonelectrolytes over a wide range of state parameters. Geochim Cosmochim Acta 67:613–627

Akinfiev NN, Tagirov BR (2014) Zn in hydrothermal systems: Thermodynamic description of hydroxide, chloride, and hydrosulfide complexes. Geochem Intl 52(3):197–214

Akinfiev NN, Zotov AV, Shikina ND (1993) Experimental investigations and self-consistent thermodynamic data in the Sb(III)-S(II)-O-H system. Geochimia 1993(12):1709–1723 (translated in Geochem Int 1994, 31(7):27−40).

Alex A, Zajacz Z (2022) The solubility of Cu, Ag and Au in magmatic sulfur-bearing fluids as a function of oxygen fugacity. Geochim Cosmochim Acta 330:93–115

Alonso-Mori R, Paris E, Giuli G, Eeckhout SG, Kavčič M, Žitnik M, Bučar K, Pettersson LGM, Glatzel P (2009) Electronic structure of sulfur studied by X-ray absorption and emission spectroscopy. Anal Chem 81(15):6516–6525

Alonso Mori R, Paris E, Giuli G, Eeckhout SG, Kavčič M, Žitnik M, Bučar K, Pettersson LGM, Glatzel P (2010) Sulfur-metal orbital hybridization in sulfur-bearing compounds studied by X-ray emission spectroscopy. Inorg Chem 49:6468–6473

Amend JP, Shock EL (2001) Energetics of overall metabolic reactions of thermophilic and hiperthermophilic Archaea and Bacteria. FEMS Microbiol Rev 25:175–243

APHA - American Public Health Association - Standard Methods Committee (2018a) American Water Works Association, and Water Environment Federation. 4500-$SO_4^{2-}$ sulfate. In: *Standard Methods For the Examination of Water and Wastewater*. Lipps WC, Baxter TE, Braun-Howland E, eds. Washington DC, APHA Press

APHA - American Public Health Association - Standard Methods Committee (2018b) 4500-$S^{2-}$ sulfide. In: *Standard Methods For the Examination of Water and Wastewater*. Lipps WC, Baxter TE, Braun-Howland E, eds. Washington DC, APHA Press

Anderson GM (1962) The solubility of PbS in $H_2S$-water solutions. Econ Geol 70:809–828

Anderson GM (2005) *Thermodynamics of Natural Systems*. Cambridge University Press

Anderson GM, Castet S, Schott J, Mesmer RE (1991) The density model for estimation of thermodynamic parameters of reactions at high temperature and pressure. Geochim Cosmochim Acta 55:1769–1779

Arai S, Miura M (2016) Formation and modification of chromitites in the mantle. Lithos 264:277–295

Archibald SM, Migdisov AA, Williams-Jones AE (2002) An experimental study of the stability of copper chloride complexes in water vapor at elevated temperatures and pressures. Geochim Cosmochim Acta 66:1611–1619

Arieli D, Vaughan DEW, Goldfarb D (2004) New synthesis and insight into the structure of blue ultramarine pigments. J Am Chem Soc 126: 5776–5788

Arnórsson S, Bjarnarson JÖ, Giroud N, Gunnarsson I, Stefánsson A (2006) Sampling and analysis of geothermal fluids. Geofluids 6:203–216

Arnórsson S, Stefánsson A, Bjarnason JÖ (2007) Fluid-fluid equilibria in geothermal systems. Rev Miner Geochem. 65:259–312

Aruga R (1981) Desolvation processes in the association of peroxodisulfate and sulfate with bivalent metals in aqueous solution. A calorimetric study. J Inorg Nucl Chem 43:1859–1861

Avetisyan K, Kamyshny A, Jr (2022) Thermodynamic constants of formation of disulfide anion in aqueous solutions. Geochim Cosmochim Acta 325:205–213

Avetisyan K, Buchshtav T, Kamyshny A, Jr (2019) Kinetics and mechanism of polysulfides formation by a reaction between hydrogen sulfide and orthorhombic cyclooctasulfur. Geochim Cosmochim Acta 247:96–105

Avrahami M, Golding RM (1968) The oxidation of the sulphide ion at very low concentrations in aqueous solutions. J Chem Soc A:647–651

Audétat A (2015) Compositional evolution and formation conditions of magmas and fluids related to porphyry Mo mineralization at Climax, Colorado. J Petrol 56:1519–1546

Audétat A, Günther D, Heinrich CA (1998) Formation of a magmatic–hydrothermal ore deposit: insights with LA-ICP-MS analysis of fluid inclusions. Science 279:2091–2094

Aylmore M, Muir D (2001) Thiosulfate leaching of gold: a review. Miner Eng 14(2):135–174

Azizitorghabeh A, Wang J, Ramsay JA, Ghahreman A (2021) A review of thiocyanate gold leaching - Chemistry, thermodynamics, kinetics and processing. Miner Eng 160:106689.

Baer JE, Carmack M (1949) The ultraviolet absorption spectra of aliphatic sulfides and polysulfides. J Amer Chem Soc 71:1215–1218

Baes CF, Mesmer RE (1976) *The Hydrolysis of Cations*. Wiley, New York, USA.

Bakr M, Akiyama M, Sanada Y (1991) In situ high temperature ESR measurements for kerogen maturation. Org Geochem 17:321–328

Bali E, Keppler H, Audetat A (2012) The mobility of W and Mo in subduction zone fluids and the Mo–W–Th–U systematics of island arc magmas. Earth Planet Sci Lett 351-352:195–207

Ballhaus CG, Stumpfl EF (1986) Sulfide and platinum mineralization in the Merensky Reef: evidence from hydrous silicates and fluid inclusions. Contrib Mineral Petrol 94:193–204

Banks DA, Giuliani G, Yardley BWD, Cheilletz A (2000) Emerald mineralisation in Colombia: fluid chemistry and the role of brine mixing. Miner. Deposita 35:699–713

Barbero JA, Hepler LG, McCurdy KG, Tremaine P (1982) Apparent molal heat capacities and volumes of aqueous hydrogen sulfide and sodium hydrogen sulfide near 25°C: the temperature dependence of $H_2S$ ionization. Can J Chem 60:1872–1880

Barbero JA, McCurdy KG, Tremaine PR (1983) Thermodynamics of aqueous carbon dioxide and sulfur dioxide: heat capacities, volumes, and the temperature dependence of ionization. Can J Chem 61:2509–2519

Bard AJ, Parsons R, Jordan J (1985) *Standard Potentials in Aqueous Solution*. CRC Press, New York

Barnes HL (ed.) (1976) *Geochemistry of Hydrothermal Ore Deposits*. Second Edition, Wiley, New York

Barnes HL (ed.) (1997) *Geochemistry of Hydrothermal Ore Deposits*. Third Edition, Wiley, New York

Barnes SJ, Liu W (2012) Pt and Pd mobility in hydrothermal fluids: Evidence from komatiites and from thermodynamic modelling. Ore Geol Rev 44:49–58

Barnes SJ, Ripley EM (2016) Highly siderophile and strongly chalcophile elements in magmatic ore deposits. Rev Mineral Geochem 81:725–774

Barré G, Truche L, Bazarkina EF, Michels R, Dubessy J (2017) First evidence of the trisulfur radical ion $S_3^-$ and other sulfur polymers in natural fluid inclusions. Chem Geol 462:1–14

Barton PB Jr (1970) Sulfide petrology. Min Soc Amer Spec Paper 3:187–198

Barton PB Jr, Skinner BJ (1979) Sulfide mineral stabilities. In: Geochemistry of Hydrothermal Ore Deposits (Barnes HL, ed.). NY, Holt, Rinehart and Wibston, p. 236–333

Baudouin D, Xiang H, Vogel F (2022) On the selective desulphurization of biomass derivatives in supercritical water. Biomass Bioenergy 164:106529

Bazarkina EF, Pokrovski GS, Zotov AV, Hazemann JL (2010) Structure and stability of cadmium chloride complexes in hydrothermal fluids. Chem Geol 276:1–17

Bazarkina EF, Pokrovski GS, Hazemann J-L (2014) Structure, stability and geochemical role of palladium chloride complexes in hydrothermal fluids. Geochim Cosmochim Acta 146:107–131

Beaudry P, Sverjensky DA (2024) Oxidized sulfur species in slab fluids as a source of enriched sulfur isotope signatures in arcs. Geochem Geophys Geosyst 25:e2024GC011542

Beak DG, Wilkin RT, Ford RG, Kelly SD (2008) Examination of arsenic speciation in sulfidic solutions using X-ray absorption spectroscopy. Environ Sci Technol 42:1643–1650

Benning LG, Seward TM (1996) Hydrosulphide complexing of Au(I) in hydrothermal solutions from 150–400 °C and 500–1500 bar. Geochim Cosmochim Acta 60:1849–1871

Bény C, Guilhaumou N, Touray J-C (1982) Native-sulphur-bearing fluid inclusions in the $CO_2$-$H_2S$-$H_2O$-S system. Microthermometry and Raman microprobe (MOLE) analysis. Thermochemical interpretations. Chem Geol 37:113–127

Beuschlein WL, Simenson LO (1940) Solubility of sulfur dioxide in water. J Am Chem Soc 62:610–612

Benning LG, Seward TM (1996) Hydrosulphide complexing of Au (I) in hydrothermal solutions from 150–400 °C and 500–1500 bar. Geochim Cosmochim Acta 60:1849–1871

Berndt ME, Buttram T, Earley III D, Seyfried Jr WE (1994) The stability of gold polysulfide complexes in aqueous sulfide solutions: 100 to 150°C and 100 bars. Geochim Cosmochim Acta 58:587–594

Berni GV, Wagner T, Fusswinkel T (2020) From a F-rich granite to a NYF pegmatite: Magmatic-hydrothermal fluid evolution of the Kymi topaz granite stock, SE Finland. Lithos 364–365:105538

Berry AJ, Harris AC, Kamenetsky VS, Newville M, Sutton SR (2009) The speciation of copper in natural fluid inclusions at temperatures up to 700 °C. Chem Geol 259:2–7

Beyad Y, Burns R, Puxty G, Maeder M (2014) A speciation study of sulfur(IV) in aqueous solution. Dalton Trans 43:2147–2152

Björke JK, Stefánsson A, Arnórsson S (2015) Surface water chemistry at Torfajokull, Iceland-Quantification of boiling, mixing, oxidation and water-rock interaction and reconstruction of reservoir fluid composition. Geothermics 58:75–86

Blanchard M, Desmaele E., Pokrovski GS, Pinilla C, Merlin Méheut M, Vuilleumier R (2024) Direct prediction of isotopic properties from molecular dynamics trajectories: Application to sulfur radical ions. Chem Geol 661:122202

Bodnar RJ (1983) A method of calculating fluid inclusion volumes based on vapor bubble diameters and P-V-T-X properties of inclusion fluids. Econ. Geol 78:535–542

Boiron M-C, Moissette A, Cathelineau M, Banks D, Monnin C, Dubessy J (1999) Detailed determination of palaeofluid chemistry: an integrated study of sulphate-volatile rich brines and aquo-carbonic fluids in quartz veins from Ouro Fino (Brazil). Chem Geol 154:179–192

Bondarenko GV, Gorbaty YE (1997) In situ Raman spectroscopic study of sulfur-saturated water at 1000 bar between 200 and 500°C. Geochim Cosmochim Acta 61:1413–1420

Borg S, Liu W, Etschmann B, Tian Y, Brugger J (2012) An XAS study of molybdenum speciation in hydrothermal chloride solutions from 25–385 °C and 600 bar. Geochim Cosmochim Acta 92:292–307.

Borisov MV, Shvarov YV (1992) *Thermodynamics of Geochemical Processes*. Moscow State University Publishing House, Moscow, 254 p. (in Russian)

Borisova AY, Pokrovski GS, Pichavant M, Freydier R, Candaudap F (2010) Arsenic enrichment in hydrous peraluminous melts: insights from femtosecond laser ablation – inductively coupled plasma – quadrupole mass spectrometry, and in situ X-ray absorption fine structure spectroscopy. Am Mineral 95:1095–1104

Borisova AY, Toutain J-P, Dubessy J, Pallister J, Zwick A, Salvi S (2014) $H_2O$–$CO_2$–S fluid triggering the 1991 Mount Pinatubo climactic eruption (Philippines). Bull Volcanol 76:800

Born M (1920) Volumen und Hydratationswarme der Ionen. Zeit Phys 1:45–48

Bose M, Root RA, Pizzarello S (2017) A XANES and Raman investigation of sulfur speciation and structural order in Murchison and Allende meteorites. Meteorit Planet Sci 52(3): 546–559

Bostick BC, Fendorf S, Brown Jr GE (2005) In situ analysis of thioarsenite complexes in neutral to alkaline sulphide solutions. Min Mag 69:781–795

Bottrell SH, Miller MF (1989) Analysis of reduced sulfur species in inclusion fluids. Econ Geol 84:940–945.

Boudreau A (2019) *Hydromagmatic Processes and Platinum-Group Element Deposits in Layered Intrusions*. Cambridge Univ Press, p. 275

Boulègue J (1978a) Metastable sulfur species and trace metals (Mn, Fe, Cu, Zn, Cd, Pb) in hot brines from French Dogger. Amer J Sci 278:1394–1411

Boulègue J (1978b) Solubility of elemental sulfur in water at 298 K. Phosphorus and Sulfur Related Elements 5:127–128

Boulègue J, Michard G (1978) Constantes de formation des ions polysulfures $S_6^{2-}$, $S_5^{2-}$ et $S_4^{2-}$ en phase aqueuese. J. Français d'Hydrologie 9:27–33

Bourcier WL, Barnes HL (1987) Ore solution chemistry: VII. Stabilities of chloride and bisulfide complexes of zinc to 350 °C. Econ Geol 82:1839–1863

Boyd ES, Druschel GK (2013) Involvement of intermediate sulfur species in biological reduction of elemental sulfur under acidic, hydrothermal conditions. Appl Env Microb 79:2061–2068

Boyle RW (1969) Hydrothermal transport and deposition of gold. Econ Geol 64:112–115

Bradley A (2025) Sulfur biochemistry. This Book.

Brandt C, Eldik van R (1995) Transition metal-catalyzed oxidation of sulfur(IV) oxides. Atmospheric-relevant processes and mechanisms. Chem Rev 95:119–190

Brenan JM (2008) The platinum-group elements: ''admirably adapted" for science and industry. Elements 4:227–232

Brendel PJ, Luther GWI (1995) Development of a gold amalgam voltammetric microelectrode for the determination of dissolved Fe, Mn, $O_2$, and S(-II) in porewaters of marine and fresh-water sediments. Environ Sci Technol 29:751–761

Brimblecombe P, Norman A-L (2025) This Book.

Brimhall GH, Crerar DA (1987) Ore fluids: Magmatic to supergene. Rev Miner 17:235–321

Brugger J, Etschmann B, Liu W, Testemale D, Hazemann J-L, Emerich H, Van Beek W, Proux O (2007) An XAS study of the structure and thermodynamics of Cu(I) chloride complexes in brines up to high temperature (400°C, 600 bar). Geochim Cosmochim Acta 71:4920–4941

Brugger J, Liu W, Etschmann B, Mei Y, Sherman DM, Testemale D (2016) A review of the coordination chemistry of hydrothermal systems, or do coordination changes make ore deposits? Chem Geol 447:219–253

Bryzgalin OV, Rafalsky RP (1982). Approximate estimation of instability constants of complexes of ore elements under high-temperatures. Geokhimiya 6:839–849 (in Russian)

Burisch M, Walter BF, Wälle M, Markl G (2016) Tracing fluid migration pathways in the root zone below unconformity-related hydrothermal veins: Insights from trace element systematics of individual fluid inclusions. Chem. Geol 429:44–50

Burisch M, Gerdes A, Walter BF, Neumann U, Fettel M, Markl G (2017) Methane and the origin of five-element veins: mineralogy, age, fluid inclusion chemistry and ore forming processes in the Odenwald, SW Germany. Ore Geol Rev 81:42–61

Burke EAJ (2001) Raman microspectrometry of fluid inclusions. Lithos 55:139–158

Burnham CW, Ohmoto H (1980) Late-stage processes of felsic magmatism. Mining Geol Spec Issue 8:1–13

Burruss RC, Slepkov AD, Pegoraro AF, Stolow A (2012) Unraveling the complexity of deep gas accumulations with three-dimensional multimodal CARS microscopy. Geology 40:1063–1066

Cai C, Li H, Li K, Wang D (2022) Thermochemical sulfate reduction in sedimentary basins and beyond: A review. Chem Geol 607:121018

Campbell AR, Panter KS (1990) Comparison of fluid inclusions in coexisting (cogenetic?) wolframite, cassiterite, and quartz from St. Michael's Mount and Cligga Head, Cornwall, England. Geochim. Cosmochim. Acta 54(3):673–681

Campbell AR, and Robinson-Cook S (1987) Infrared fluid inclusion microthermometry on coexisting wolframite and quartz. Econ Geol 82:1640–1645

Campbell WB, Maass O (1930) Equilibria in sulphur dioxide solutions. Can J Res 2: 42–64

Carocci E, Truche L, Cathelineau M, Caumon M-C, Bazarkina EF (2022) Tungsten (VI) speciation in hydrothermal solutions up to 400 °C as revealed by in situ Raman spectroscopy. Geochim Cosmochim Acta 317:306–324

Catchpole H, Kouzmanov K, Fontbote L, Guillong M, Heinrich CA (2011) Fluid evolution in zoned Cordilleran polymetallic veins - insights from microthermometry and LA-ICP-MS of fluid inclusions: Chem Geol 281:293–304

Catchpole H, Kouzmanov K, Putlitz B, Seo JH, Fontbote L (2015) Fluid evolution in zoned Cordilleran polymetallic veins - insights from microthermometry and LA-ICP-MS of fluid inclusions. Econ. Geol 110:39–71

Cato E, Rossi A, Scherrer NC, Ferreira ESB (2018) An XPS study into sulphur speciation in blue and green ultramarines. J Cult Herit 29:30–35

Cauzid J, Philippot P, Martinez-Criado G, Ménez B, Labouré S (2007) Constasting Cu-complexing behaviour in vapour and liquid fluid inclusions from the Yankee Lode tin deposit, Mole Granite, Australia. Chem Geol 246:39–54

Chadwell SJ, Rickard D, Luther GW (1999) Electrochemical evidence for pentasulfide complexes with $Mn^{2+}$, $Fe^{2+}$, $Co^{2+}$, $Ni^{2+}$, $Cu^{2+}$ and $Zn^{2+}$. Aquatic Geochem 5:29–57

Chambefort I, Stefánsson A (2020) Fluids in geothermal systems. Elements 16:407–411

Chandra AP, Gerson AR (2010) The mechanisms of pyrite oxidation and leaching: A fundamental perspective. Surface Sci Rep 65:293–315

Chang J, Li JW, Audétat A (2018) Formation and evolution of multistage magmatic-hydrothermal fluids at the Yulong porphyry Cu-Mo deposit, eastern Tibet: Insights from LA-ICP-MS analysis of fluid inclusions. Geochim Cosmochim Acta 232:181–205

Charlot G (1966) *Les Méthodes de la Chimie Analytique*. Masson & C$^{ie}$, Paris

Charlou JL, Fouquet Y, Bougault H, Donval JP, Etoubleau J, Jean-Baptiste P, Dapoigny A, Appriou P, Rona PA (1998) Intense plumes generated by serpentinization of ultra-mafic rocks at the intersection of the 15°20'N fracture zone and the Mid-Atlantic Ridge. Geochim Cosmochim Acta 62:2323–2333

Charlou JL, Donval JP, Konn C, Ondréas H, Fouquet Y (2010) High production and fluxes of $H_2$ and $CH_4$ and evidence of abiotic hydrocarbon synthesis by serpentinization in ultramafic-hosted hydrothermal systems on the Mid-Atlantic Ridge. in: *Diversity of Hydrothermal Systems on Slow Spreading Ocean Ridges Geophysical Monograph Series* 188:265–296

Chase MW Jr (1998) NIST-JANAF thermochemical Tables, 4$^{th}$ edn. J Phys Chem Ref Data Monographs, No. 9, American Institute of Physics, New York

Chen H, Irish DE (1971) A Raman spectral study of bisulfate-sulfate systems. II. Constitution, equilibria, and ultrafast proton transfer in sulfuric acid. J Phys Chem 75(17): 2672–2681

Chen KY, Morris JC (1972) Kinetics of oxidation of aqueous sulfide by $O_2$. Environ Sci Technol 6:529–537

Chen K, Walker RJ, Rudnick RL, Gao S, Gaschnig RM, Puchtel IS, Tang M, Hu Z-C (2016) Platinum-group element abundances and Re–Os isotopic systematics of the upper continental crust through time: Evidence from glacial diamictites. Geochim Cosmochim Acta 191:1–16

Chiodini G, Caliro S, Lowenstern JB, Evans WC, Bergfeld D, Tassi F, Tedesco D. (2012) Insights from fumarole gas geochemistry on the origin of hydrothermal fluids on the Yellowstone Plateau. Geochim Cosmochim Acta 89:265–278

Chivers T (1974) Ubiquitous trisulfur radical ion $S_3^-$. Nature 252:32–33

Chivers T, Elder PJW (2013) Ubiquitous trisulfur radical ion: fundamentals and applications in materials science, electrochemistry, analytical chemistry and geochemistry. Chem Soc Rev 42:5996–6005

Chivers T, Oakley RT (2023) Structures and spectroscopic properties of polysulfide radical anions: A theoretical perspective. Molecules 28, 5654

Chlebec RW, Lister MW (1971) Ion pairing and the reaction of alkali metal ferrocyanides and persulfates. Can J Chem 49:2943–2947

Chou I-M, Song Y, Burruss RC (2008) A new method for synthesizing fluid inclusions in fused silica capillaries containing organic and inorganic material. Geochim Cosmochim Acta 72:5217–5231

Chu X, Ohmoto H, Cole DR (2004) Kinetics of sulfur isotope exchange between aqueous sulfide and thiosulfate involving intra- and intermolecular reactions at hydrothermal conditions. Chem Geol 211:217–235

Ciuffarin E, Pryor WA (1964) The kinetics of the exchange of sulfur-35 between thiosulfate and sulfide. J Am Chem Soc 86:3621–3626

Clarisse L, Coheur P-F, Chefdeville S, Lacour J-L, Hurtmans D, Clerbaux C. (2011) Infrared satellite observations of hydrogen sulfide in the volcanic plume of the August 2008 Kasatochi eruption. Geophys Res Lett 38:L10804

Clark RJH, Franks ML (1975) The resonance Raman spectrum of ultramarine blue. Chem Phys Lett 34:69–72

Clark, R.J.H, Cobbold, D.G., 1978. Characterization of sulfur radical anions in solutions of alkali polysulfides in dimethylformamide and hexamethylphosphoramide and in the solid state in ultramarine blue, green, and red. Inorg Chem 17, 3169–3174

Clarke ECW, Glew DN (1971) Aqueous nonelectrolyte solutions. Part VIII. Deuterium and hydrogen sulfide solubilities in deuterium oxide and water. Can J Chem 49:691–698

Cobble JW, Stephens HP, McKinnon IR, Westrum EF, Jr (1972) Thermodynamic properties of oxygenated sulfur complex ions. Heat capacity from 5 to 300°K for $K_2S_4O_6$(c) and from 273 to 373°K $S_4O_6^{2-}$(aq). Revised thermodynamic functions for $HSO_3^-$(aq), $SO_3^{2-}$(aq), $S_2O_3^{2-}$(aq), and $S_4O_6^{2-}$(aq) at 298°K. Revised potential of the thiosulfate-tetrathionate electrode. Inorg Chem 11:1669–1674

CODATA (1978) CODATA recommended key values for thermodynamics, 1977. J. Chem Thermodyn 10:903–906

Cody GD (2004) Transition metal sulfides and origins of metabolism. Annu Rev Earth Sci 32:569–599

Colin A, Schmidt C, Pokrovski GS, Wilke M, Borisova AY, Toplis M (2020) *In situ* determination of sulfur speciation and partitioning in aqueous fluid-silicate melt systems. Geochem Persp Let 14:31–35

Connick RE, Tam TM, von Deuster E (1982) Equilibrium constant for the dimerization of bisulfite ion to form disulfite (2–) ion. Inorg Chem 21:103–107

Connolly JAD, Galvez ME (2018) Electrolytic fluid speciation by Gibbs energy minimization and implications for subduction zone mass transfer. Earth Planet Sci Lett 501:90–102

Conrad JK, Arcis H, Ferguson JP, Tremaine PR (2023) Second ionization constant of sulfuric acid in $H_2O$ and $D_2O$ from 150 to 300 °C at p = 11.5 MPa using flow AC conductivity. Phys Chem Chem Phys 25:1659–1676

Cope JD, Bates KH, Tran LN, Abellar KA, Nguyen TB (2022) Sulfur radical formation from the tropospheric radiation of aqueous sulfate aerosols. Proc Nat Acad Sci USA 119(36):e2202857119

Cotton FA, Wilkinson G, Murillo CA, Bochmann, M (1999) *Advanced Inorganic Chemistry*, 6th Edition Wiley, Chichester, 1355 p.

Cox JD, Wagman DD, Medvedev VA (1989) *CODATA Key Values for Thermodynamics*. Hemisphere Publishing Corp., New York

Cox SF, Etheridge MA, Wall VJ (1987) The role of fluids in syntectonic mass transport, and the localization of metamorphic vein-type ore deposits. Ore Geol Rev 2:65–86

Crerar DA, Barnes HL (1976) Ore solution chemistry V. Solubilities of chalcopyrite and chalcocite assemblages in hydrothermal solution at 200–350 °C. Econ Geol 71:772–794

Crerar DA, Wood SA, Brantley S, Bocarsly A (1985) Chemical controls on solubility of ore forming minerals in hydrothermal solutions. Can Mineral 23:333–352

Criss CM, Cobble JW (1964a) The Thermodynamic properties of high temperature aqueous solutions. IV. Entropies of the ions up to 200° and the correspondence principle. J Amer Chem Soc 86:5385–5390

Criss CM, Cobble JW (1964b) The thermodynamic properties of high temperature aqueous solutions. V. The calculation of ionic heat capacities up to 200°. Entropies and heat capacities above 200°. J Amer Chem Soc 86:5390–5393

Dadze TP, Sorokin VI (1993) Experimental determination of the concentrations of $H_2S$, $HSO_4^-$, $SO_{2aq}$, $H_2S_2O_3$, $S^0_{aq}$, and $S_{tot}$ in the aqueous phase in the S-$H_2O$ system at elevated temperatures. Geochem Intl 30:38–53

Dadze TP, Kashirtseva GA, Novikov MP, Plyasunov AV (2017a) Solubility of $MoO_3$ in acid solutions and vapor-liquid distribution of molybdic acid. Fluid Phase Equil 440:64–76

Dadze TP, Kashirtseva GA, Novikov MP, Plyasunov AV (2017b) Solubility of $MoO_3$ in $NaClO_4$ solutions at 573 K. J Chem Eng Data 62:3848–3853

Dadze TP, Kashirtseva GA, Novikov MP, Plyasunov AV (2018a) Solubility of calcium molybdate in aqueous solutions at 573 K and thermodynamics of monomer hydrolysis of Mo(VI) at elevated temperatures. Monatsh Chem 149:261–282

Dadze TP, Kashirtseva GA, Novikov MP, Plyasunov AV (2018b) Solubility of $MoO_3$ in aqueous acid chloride-bearing solutions at 573 K. J Chem Eng Data 63:1827–1832

Dai Z, Kan A, Zhang F, Tomson M (2014) A thermodynamic model for the solubility prediction of barite, calcite, gypsum, and anhydrite, and the association constant estimation of $CaSO_4^{(0)}$ ion pair up to 250 °C and 22000 psi. J Chem Eng Data 60:766–774

Davis CW, Jones HW, Monk CB (1952) E.M.F. studies of electrolytic dissociation. Part I. Sulphuric acid in water. Trans Faraday Soc 48:921–928

Davis AG, Vorburger AH (2023) Io's volcanic activity and atmosphere. Elements 18:379–384

Davy H (1822) On the state of water and aeriform matter in cavities found in certain crystals. Phil Trans Royal Soc London 112:367–376. https://www.jstor.org/stable/107692

Dawson BSW, Irish DE, Toogood GE (1986) Vibrational spectral studies of solutions at elevated temperatures and pressures. 8. A Raman spectral study of ammonium hydrogen sulfate solutions and the hydrogen sulfate-sulfate equilibrium. J Phys Chem 90(2):334–341

Debret B, Sverjensky DA (2017) Highly oxidising fluids generated during serpentinite breakdown in subduction zones. Sci Rep 7:10351

Debret B, Millet M-A, Pons M-L, Bouilhol P, Inglis E, Williams H (2016) Isotopic evidence of iron mobility during subduction. Geology 44:215–218

Debret B, Reekie CDJ, Mattielli N, Beunon H, Ménez B, Savov IP, Williams HM (2020) Redox transfer at subduction zones: insights from Fe isotopes in the Mariana forearc. Geochem Persp Let 12:46-51

Deluigi MT, Riveros JA (2006) Chemical effects on the satellite lines of sulfur $K_\beta$ X-ray emission spectra. Chem Phys 325:472–476

Diakonov II, Schott J, Martin F, Harrichoury J-C, Escalier J (1999) Iron(III) solubility and speciation in aqueous solutions. Experimental study and modelling: Part 1. Hematite solubility from 60 to 300 °C in NaOH-NaCl solutions and thermodynamic properties of $Fe(OH)_4^-$ (aq). Geochim Cosmochim Acta 63:2247–2261

Diamond LW (2003) Glossary: Terms and symbols used in fluid inclusion studies. Miner Assoc Canada Short Course Series 32:363–372

Dickson AG (1966) Solubilities of metallic sulfides and quartz in hydrothermal sulfide solutions. Bull Volcanol 29:605–628

Dickson AG, Wesolowski DJ, Palmer DA, Mesmer RE (1990) Dissociation constant of bisulfate ion in aqueous sodium chloride solutions to 250 °C. J Phys Chem 94:7978–7985

Dietrich M, Behrens H, Wilke M (2018) A new optical cell for in situ Raman spectroscopy, and its application to study sulfur-bearing fluids at elevated pressures and temperatures. Am Miner 103:418–429

Ding Y, Zhang S, Liu B, Zheng H, Chang C, Ekberg C (2019) Recovery of precious metals from electronic waste and spent catalysts: A review. Res Conserv Recycl 141:284–298

Dolejs D, Manning CE (2010) Thermodynamic model for mineral solubility in aqueous fluids: theory, calibration and application to model fluid-flow systems. Geofluids 10:20–40

Dong X, Oganov AR, Cui H, Zhou X-F, Wang H-T (2022) Electronegativity and chemical hardness of elements under pressure. Proc Nat Acad Sci USA 119(10):e2117416119

Dowler MJ, Ingmanson DE (1979) Thiocyanate in Red sea brine and its implications. Nature 279:51–52

Drozdova Y, Steudel R, Hertwig RH, Koch W, Steiger T (1998) Structures and energies of various isomers of dithionous acid, $H_2S_2O_4$, and of its anion $[HS_2O_4]^-$. J Phys Chem 102A:990–996

Druschel GK, Schoonen MAA, Nordstrom DK, Ball JW, Xu Y, Cohn CA (2003) Sulfur geochemistry of hydrothermal waters in Yellowstone National Park, Wyoming, USA. III. An anion-exchange resin technique for sampling and preservation of sulfoxyanions in natural waters. Geochem Trans 4:12–19

Drummond SE (1981) *Boiling and Mixing of Hydrothermal Fluids: Chemical Effects of Mineral Precipitation*. Ph.D. thesis, Pennsylvania State University, 400 p.

Du B, Wang Z, Santosh M, Shen Y, Liu S, Liu J, Xu K, Deng J (2023) Role of metasomatized mantle lithosphere in the formation of giant lode gold deposits: Insights from sulfur isotope and geochemistry of sulfides. Geosci Frontiers 14:101587

Dubessy J, Geisler D, Kosztolanyi C, Vernet M (1983) The determination of sulphate in fluid inclusions using the MOLE Raman microprobe. Application to a Keuper halite and geochemical consequences. Geochim. Cosmochim. Acta 47:1–10

Dubessy J, Guilhaumou N, Mullis J, Pagel M (1984) Reconnaissance par microspectrométrie Raman, dans les inclusions fluides, de $H_2S$ et $CO_2$ solides à domaine de fusion comparable. Bull Mineral 107:189–192

Dubessy J, Poty B, Ramboz C (1989) Advances in C-O-H-N-S fluid geochemistry based on micro-Raman spectrometric analysis of fluid inclusions. Eur J Mineral 1:517–534

Dubessy J, Boiron M-C, Moissette A, Monnin C, Sretenskaya N (1992) Determinations of water, hydrates and pH in fluid inclusions by micro-Raman spectrometry. Eur J Mineral 4: 885–894

Dubessy J, Caumon M-C, Rull F (eds) (2012) Raman spectroscopy applied to Earth sciences and cultural heritage. EMU Notes in Mineralogy, vol 12, London

Eckert W (1998) Electrochemical identification of the hydrogen sulfide system using a $pH_2S$ (glass/Ag°, $Ag_2S$) electrode. J Electrochem Soc 145:77–79

Eckert B, Steudel R (2003) Molecular spectra of sulfur molecules and solid sulfur allotropes. Topics Curr Chem 231:31–98

Economou-Eliopoulos M (2010) Platinum-group elements (PGE) in various geotectonic settings: opportunities and risks. Hell J Geosci 45:65–82

Einaudi MT, Hedenquist JW, Inan EE (2003) Sulfidation state of fluids in active and extinct hydrothermal systems: Transitions from porphyry to epithermal environments. Soc Econ Geol Spec Publ 10:285–313

Eldridge DL (2025) Sulfur isotope fractionations in aqueous and gaseous systems: $^{32}S$, $^{33}S$, $^{34}S$, $^{36}S$, and $^{34}S^{18}O$. This Book

Eldridge DL, Guo W, Farquhar J (2016) Theoretical estimates of equilibrium sulfur isotope effects in aqueous sulfur systems: highlighting the role of isomers in the sulfite and sulfoxylate systems. Geochim Cosmochim Acta 195:171–200

Eldridge DL, Mysen BO, Cody GD (2018) Experimental estimation of the bisulfite isomer quotient as a function of temperature: Implications for sulfur isotope fractionations in aqueous sulfite solutions. Geochim Cosmochim Acta 220:309–328

Eldridge DL, Kamyshny A Jr, Farquhar J (2021) Theoretical estimates of equilibrium sulfur isotope effects among aqueous polysulfur and associated compounds with applications to authigenic pyrite formation and hydrothermal disproportionation reactions. Geochim Cosmochim Acta 310:281–319

Ellis AJ, Anderson DW (1961) The effect of pressure on the first acid dissociation constants of "sulphurous" and phosphoric acid. J Chem Soc 1765–1767

Ellis AJ, Giggenbach W (1971) Hydrogen sulphide ionization and sulphur hydrolysis in high temperature solution. Geochim Cosmochim Acta 35:247–260

Ellis AJ, Millestone NB (1967) The ionization constants of hydrogen sulphide from 20 to 90°C. Geochim Cosmochim Acta 31:615–620

Engel JS, Bovenkamp GL, Prange A, Hormes J (2014) In situ speciation of sulfur vapors by X-ray absorption near edge structure spectroscopy. Chem Geol 380:1–6

Erickson BE, Helz GR (2000) Molybdenum(VI) speciation in sulfidic waters: stability and lability of thiomolybdates. Geochim. Cosmochim. Acta 64:1149–1158

Ermatchkov V, Kamps ÁPS, Maurer G (2005) The chemical reaction equilibrium constant and standard molar enthalpy change for the reaction {$2HSO_3^-(aq) \rightleftharpoons S_2O_5^{2-}(aq) + H_2O(1)$}: a spectroscopic and calorimetric investigation. J Chem Thermodyn 37:187–199

Etschmann BE, Liu W, Testemale D, Müller H, Rae NA, Proux O, Hazemann J-L, Brugger J (2010) An in situ XAS study of copper (I) transport as hydrosulfide complexes in hydrothermal solutions (25–592 °C, 180–600 bar): Speciation and solubility in vapor and liquid phases. Geochim Cosmochim Acta 74:4723–4739

Etschmann BE, Liu W, Pring A, Grundler PV, Tooth B, Borg S, Testemale D, Brewe D, Brugger J (2016) The role of Te(IV) and Bi(III) chloride complexes in hydrothermal mass transfer: An X-ray absorption spectroscopic study. Chem Geol 425:37–51

Etschmann BE, Mei Y, Liu W, Sherman D, Testemale D, Müller H, Rae N, Kappen P, Brugger J (2018) The role of Pb(II) complexes in hydrothermal mass transfer: An X-ray absorption spectroscopic study. Chem Geol 502:88–106

Farsang S, Caracas R, Adachi TBM, Schnyder C, Zajacz Z (2023) $S_2^-$ and $S_3^-$ radicals and the $S_4^{2-}$ polusulfide ion in lazurite, haüyne and synthetic ultramarine blue revealed by resonance Raman spectroscopy. Am Mineral 108:2234–2243

Farsang S, Zajacz Z. (2025) Sulfur species and gold transport in arc magmatic fluids. Nat Geosci 18: 98–104

Ferraro JR, Nakamoto K, Brown CW (2003) *Introductory Raman Spectroscopy*. 2nd ed. Elsevier, 434 p.

Filimonova ON, Tagirov BR, Zotov AV, Baranova NN, Bychkova YV, Tyurin DA, Chareev DA, Nickolsky MS (2021) The solubility of cooperite PtS(cr) at 25 – 450 °C, Psat – 1000 bar and hydrosulfide complexing of platinum in hydrothermal fluids. Chem Geol 559:119968

Filella M, May PM (2003) Computer simulation of the low-molecular-weight inorganic species distribution of antimony(III) and antimony(V) in natural waters. Geochim Cosmochim Acta 67(21):4013–4031

Findlay AJ, Kamyshny A (2017) Turnover rates of intermediate sulfur species ($S_x^{2-}$, $S^0$, $S_2O_3^{2-}$, $S_4O_6^{2-}$, $SO_3^{2-}$) in anoxic freshwater and sediments. Front Microbiol 8:1–15

Fleet ME (2005) XANES spectroscopy of sulfur in earth materials. Can Min 43:1811–1838

Fleet ME, Liu X (2010) X-ray absorption spectroscopy of ultramarine pigments: A new analytical method for the polysulfide radical anion $S_3^-$ chromophore. Spectrochim Acta Part B 65:75–79

Fleet, ME, Liu X, Harmer SL, Nesbitt HW (2005) Chemical state of sulfur in natural and synthetic lazurite by S K-edge XANES and X-ray photoelectron spectroscopy. Can Min 43(5):1589–1603

Fontboté L, Kouzmanov K, Chiaradia M, Pokrovski GS (2017) Sulfide minerals in hydrothermal deposits. Elements 13:97–103

Frank EU (1956) Hochverdichteter Wasserdampf II. Ionendissociation von KCl in $H_2O$ bis 750 °C. Z Phys Chem 8:107–126

Frank EU (1961) Uberkritisches Wasser als electrolytisches Losungsmittel. Angew Chem 73:309–322

Frezzotti ML, Andersen T, Else-Ragnhild Neumann E-R, Simonsen SL (2002) Carbonatite melt-$CO_2$ fluid inclusions in mantle xenoliths from Tenerife, Canary Islands: a story of trapping, immiscibility and fluid–rock interaction in the upper mantle. Lithos 64:77–96

Frezzotti ML, Tecce F, Casagli A (2012) Raman spectroscopy for fluid inclusion analysis. J Geochem Explor 112:1–20

Fulton JL, Hoffmann MM, Darab JG (2000) An X-ray absorption fine structure study of copper(I) chloride coordination structure in water up to 325 °C. Chem Phys Lett 330:300–308

Fusswinkel T, Wagner T, Wälle M, Wenzel T, Heinrich CA, Markl G (2013) Fluid mixing forms basement-hosted Pb-Zn deposits: Insight from metal and halogen geochemistry of individual fluid inclusions. Geology 41:679–682

Galdos R, Vallance J, Baby P, Salvi S, Schirra M, Velasquez G, Viveen W, Soto R, Pokrovski GS (2024) Origin and evolution of gold-bearing fluids in a carbon-rich sedimentary basin: a case study of the Algamarca epithermal gold-silver-copper deposit, northern Peru. Ore Geol Rev 166:105857

Gama S (2000) Evénements métallogéniques à W-Bi (Au) à 305 Ma en Châtaigneraie du Cantal: apport d'une analyse multi-spectrométrique (micro PIXE-PIGE et Raman) des minéraux et des fluides occlus à l'identification des sources de fluides hydrothermaux. PhD thesis, Université d'Orléans (in French), https://theses.hal.science/tel-00002404

Gamsjäger H, Bugajski J, Gajda T, Lemire RJ, Preis W (2005) Chemical thermodynamics of nickel. Nuclear Energy Agency Data Bank, vol. 6, Elsevier, Amsterdam

Gammons CH (1995) Experimental investigations of the hydrothermal geochemistry of platinum and palladium: IV. The stoichiometry of Pt(IV) and Pd(II) chloride complexes at 100 to 300 °C. Geochim Cosmochim Acta 59:1655–1667

Gammons CH (1996) Experimental investigations of the hydrothermal geochemistry of platinum and palladium: V. Equilibria between platinum metal, Pt(II), and Pt (IV) chloride complexes at 25 to 300°C. Geochim Cosmochim Acta 60:1683–1694

Gammons CH, Barnes HL (1989) The solubility of $Ag_2S$ in near-neutral aqueous sulfide solutions at 25 to 300°C. Geochim Cosmochim Acta 53:279–290

Gammons CH, Bloom MS (1993) Experimental investigation of the hydrothermal geochemistry of platinum and palladium: II. The solubility of PtS and PdS in aqueous sulfide solutions to 300 °C. Geochim Cosmochim Acta 57:2451–2467

Gammons CH, Williams-Jones AE (1995) The solubility of Au-Ag alloy + AgCl in HCl/NaCl solutions at 300°C: new data on the stability of Au(I) chloride complexes in hydrothermal fluids. Geochim Cosmochim Acta 59:3453–3468

Gammons CH, Allin NC (2022) Stability of aqueous Fe(III) chloride complexes and the solubility of hematite between 150 and 300 °C. Geochim Cosmochim Acta 330:148–164

Gammons CH, Bloom MS, Yu Y (1992) Experimental investigation of the hydrothermal geochemistry of platinum and palladium: I. Solubility of platinum and palladium sulfide minerals in NaCl/$H_2SO_4$ solutions at 300 °C. Geochim Cosmochim Acta 56:3881–3894

Gammons CH, Yu Y, Bloom MS (1993) Experimental investigation of the hydrothermal geochemistry of platinum and palladium: III. The solubility of Ag–Pd alloy + AgCl in NaCl/HCl solutions at 300 °C. Geochim Cosmochim Acta 57:2469–2479

Ganio M, Pouyet ES, Webb SM, Schmidt Patterson CM, Walton MS (2018) From lapis lazuli to ultramarine blue: investigating Cennino Cennini's recipe using sulfur K-edge XANES. Pure Appl Chem 90(3):463–475

Gerothanassis IP, Kridvin LB (2024) $^{33}S$ NMR: recent advances and applications. Molecules 29:3301

Gibert F, Pascal M-L, Pichavant M (1998) Gold solubility and speciation in hydrothermal solutions: Experimental study of the stability of hydrosulfide complex of gold ($AuHS^0$) at 350 to 450°C and 500 bars. Geochim Cosmochim Acta 62:2931–2947

Giggenbach WF (1971a) Optical spectra of highly alkaline sulfide solutions and the second dissociation constant of hydrogen sulfide. Inorg Chem 10:1333–1338

Giggenbach WF (1971b) The blue solutions of sulfur in water at elevated temperatures. Inorg Chem 10:1306–1308.

Giggenbach WF (1972) Optical spectra and equilibrium distribution of polysulfide ions in aqueous solution at 20°C. Inorg Chem 11:1201–1207

Giggenbach WF (1974a) Equilibria involving polysulfide ions in aqueous sulfide solutions up to 240°C. Inorg Chem 13:1724–1730

Giggenbach WF (1974b) Kinetics of polysulfide–thiosulfate disproportionation up to 240°C. Inorg Chem 13:1730–1733

Giggenbach WF (1987) Redox processes governing the chemistry of fumarolic gas discharges from White Island, New Zealand. Appl Geochem 2:143-161

Giggenbach WF (1997) The origin and evolution of fluids in magmatic-hydrothermal systems. In: Barnes, H. L. (ed.), *Geochemistry of Hydrothermal Ore Deposits*, 3rd ed., p. 737–796. New York, John Wiley & Sons

Giordano TH, Barnes HL (1979) Ore solution chemistry. 6. PbS solubility in bisulfide solutions to 300 °C. Econ Geol 74:1637–1646

Giuliani G, Dubessy J, Banks D, Vinh HQ, Lhomme T, Pironon J, Garnier V, Trinh VT, Long PV, Ohnenstetter D, Schwarz D (2003) $CO_2$-$H_2S$-COS-$S_8$-AlO(OH)-bearing fluid inclusions in ruby from marble-hosted deposits in Luc Yen area, North Vietnam. Chem Geol 194:167–185

Gobeltz N, Demortier A, Lelieur JP, Duhayon C (1998) Correlation between EPR, Raman and colorimetric characteristics of the blue ultramarine pigments. J Chem Soc Faraday Trans 94(5):677–681

Gobeltz N, Ledé B, Raulin K, Demortier A, Lelieur JP (2011) Synthesis of yellow and green pigments of zeolite LTA structure: Identification of their chromophores. Micropor Mesopor Mat 141:214–221

Goff F, Janik CJ (2000) Geothermal systems. *Encyclopaedia of Volcanoes*, pp 817–834

Goldberg RN, Parker VB (1985) Thermodynamics of solution of $SO_2$(g) in water and of aqueous sulfur dioxide solutions. J Res Nat Bur Stand 90(5):341–358

Goldfarb RJ, Baker T, Dube B, Groves DI, Hart CJR, Gosselin P (2005) Distribution, character, and genesis of gold deposits in metamorphic terranes. Econ Geol 100th Anniv Vol: 407–450

Goldhaber MB, Church SE, Doe BR, Aleinikoff JN, Brannon JC, Podosek FA, Mosier EL, Taylor CD, Gent CA (1995) Lead and sulfur isotope investigation of Paleozoic sedimentary rocks from the southern midcontinent of the United States; implications for paleohydrology and ore genesis of the Southeast Missouri lead belts. Econ Geol 90:1875–1910

Goldstein RH (2003) Petrographic analysis of fluid inclusions. In: Samson, I., Anderson, A. and Marshall, D. (eds), Fluid inclusions – analysis and interpretation. Miner Assoc Canada Short Course Ser 32:9–53

Goldstein RH, Reynolds TJ (1994) Systematics of fluid inclusions in diagenetic minerals. Soc Sedimentary Geol Short Course Ser 31:199 p.

Goodman JF, Robson P (1963) Decomposition of inorganic peroxyacids in aqueous alkali. J Chem Soc 1963:2871–2875

Goslar J, Lijewski S, Hoffmann SK, Jankowska A, Kowalak S (2009) Structure and dynamics of $S_3^-$ radicals in ultramarine-type pigment based on zeolite: Electron spin resonance and electron echo studies. J Chem Phys 130:204504

Graham UM, Ohmoto H (1994) Experimental study of formation mechanisms of hydrothermal pyrite. Geochim Cosmochim Acta 58:2187–2202

Grente I, Plyasunov AV, Runde WH, Konings RJM, Moore EE, Gaona X, Rao L, Grambow B, Smith AL (2020) Second update of the chemical thermodynamics of uranium, neptunium, plutonium, americium and technetium. OECD Nuclear Energy Agency Data Bank, Eds., OECD Publications, Paris, France

Grinenko VA, Grinenko LH, Zagriazhskaya (1969) Kinetic isotopic effect during high-temperature reduction of sulfate. Geokhimia 4:484–491 (in Russian).

Grishina S, Dubessy J, Kontorovich A, Pironon J (1992) Inclusions in salt beds resulting from thermal metamorphism by dolerite sills (eastern Siberia, Russia). Eur J Mineral 4:1187–1202

Cruywagen JJ (1999) Protonation, oligomerization, and condensation reactions of vanadate(V), molybdate(VI), and tungstate(VI). Adv Inorg Chem 49:127–182

Gu Y, Gammons CH, Bloom MS (1994) A one-term extrapolation method for estimating equilibrium constants of aqueous reactions at elevated temperatures. Geochim Cosmochim Acta 58:3545–3560

Guan Q, Mei Y, Liu W, Brugger J (2023) Different metal coordination in sub- and super-critical fluids: Do molybdenum(IV) chloride complexes contribute to mass transfer in magmatic systems? Geochim Cosmochim Acta 354:240–251

Guan Q, Mei Y, Etschmann B, Testemale D, Louvel M, Brugger J (2022) Yttrium complexation and hydration in chloride-rich hydrothermal fluids: A combined ab initio molecular dynamics and in situ X-ray absorption spectroscopy study. Geochim Cosmochim Acta 281:168–189

Guan Q, Mei Y, Etschmann B, Louvel M, Testemale D, Bastrakov E, Brugger J (2020) Yttrium speciation in sulfate-rich hydrothermal ore-forming fluids. Geochim Cosmochim Acta 325:278–295

Guillong M, Heinrich CA (2007) LA-ICP-MS analysis of inclusions: improved ablation and detection. 19th Biennial Conference *European Current Research on Fluid Inclusions, ECROFI-XIX*, Bern, p. 82.

Guillong M, Latkoczy C, Seo JH, Günther D, Heinrich CA (2008a) Determination of sulfur in fluid inclusions by laser ablation ICPMS. J Anal Atom Spectr 23:1581–1589

Guillong M, Meier DL, Allan MM, Heinrich CA, Yardley BWD (2008b) SILLS: A MATLAB-based program for the reduction of laser ablation ICP-MS data of homogeneous materials and inclusions. In: Sylvester, P. (Ed.), Laser Ablation ICP-MS in the Earth Sciences: Current Practices and Outstanding Issues. Miner Assoc Canada Short Course Ser 40:328–333

Gunn G (2014) *Critical Metals Handbook.* Wiley

Günther D, Audétat A, Frischknecht R, Heinrich CA (1998) Quantitative analysis of major, minor and trace elements in fluid inclusions using laser ablation inductively coupled plasma mass spectrometry. J Anal Atom Spectroscopy 13:263–270

Guo H, Audétat A, Dolejš D (2018) Solubility of gold in oxidized, sulfur-bearing fluids at 500–850 °C and 200–230 MPa: A synthetic fluid inclusion study. Geochim Cosmochim Acta 222:655–670

Haas JR, Shock EL, Sassani DC (1995) Rare earth elements in hydrothermal systems: Estimates of standard molal thermodynamic properties of aqueous complexes of rare earth elements at high pressures and temperatures. Geochim Cosmochim Acta 59:4329–4350

Hadley JH, Gordy W (1975) Nuclear coupling of $^{33}S$ and the nature of free radicals in irradiated crystals of cysteine hydrochloride and N-acetyl methionine. Proc Nat Acad Sci USA 72(9):3486–3490

Häkkinen H (2012) The gold-sulfur interface at the nanoscale. Nat Chem 4:443–455

Hamer WJ (1934) The ionization constant and heat of ionization of the bisulfate ion from electromotive force measurements. J Amer Chem Soc 56:860–864

Hamilton EMcC (1991) The chemistry of low valence sulfur compounds in the sulfur-water system. Unpublished Master's thesis, University of Minnesota, 155 p.

Hamisi J, Etschmann B, Tomkins A, Pitcairn I, Pintér Z, Wlodek A, Morrissey L, Micklethwaite S, Trcera N, Mills S, Brugger J (2023) Complex sulfur speciation in scapolite – Implications for the role of scapolite as a redox and fluid chemistry buffer in crustal fluids. Gondwana Res 121:418–435

Han L, Pan JY, Ni P, Chen H (2023) Cassiterite deposition induced by cooling of a single-phase magmatic fluid: Evidence from SEM-CL and fluid inclusion LA-ICP-MS analysis. Geochim Cosmochim Acta 342:108–127

Hanley JJ (2005) The aqueous geochemistry of the platinum-group elements (PGE) in surficial, low-T hydrothermal and high-T magmatic-hydrothermal environments. In: Exploration for Platinum Group Elements Deposits (ed. J. E. Mungall). Min Assoc Canada Short Course 35:35–56

Hannington M, Harðardóttir V, Garbe-Schönberg D, Brown KL (2016) Gold enrichment in active geothermal systems by accumulating colloidal suspensions. Nat Geosci 9:299–302

Harlaux M, Borovinskaya O, Frick DA, Tabersky D, Gschwind S, Richard A, Günther D, Mercadier J (2015) Capabilities of sequential and quasi-simultaneous LA-ICPMS for the multi-element analysis of small quantity of liquids (pl to nl): insights from fluid inclusion analysis. J Anal Atom Spect 30:1945–1969

Hartler N, Libert J, Teder A (1967) Rate of sulfur dissolution in aqueous sodium sulfide. Ind Eng Chem Process Des Dev 6(4):398–406

Hayashi KI, Ohmoto H (1991) Solubility of gold in NaCl- and $H_2S$-bearing aqueous solutions at 250-350°C. Geochim Cosmochim Acta 55:2111–2126

Hayashi K, Sugaki A, Kitakaze A (1990) Solubility of sphalerite in aqueous sulfide solutions at temperatures between 25 and 240 °C. Geochim Cosmochim Acta 54:715–725

Hayon E, Treinin A, Wilf J (1972) Electronic spectra, photochemistry, and autoxidation mechanism of the sulfide-bisulfite-pyrosulfite systems. $SO_2^-$, $SO_3^-$, $SO_4^-$, $SO_5^-$ radicals. J Amer Chem Soc 94(1):47–57

He M, Liu X, Lu X, Zhang C, Wang R (2016) Structures and acidity constants of silver−sulfide complexes in hydrothermal fluids: A first-principles molecular dynamics study. J Phys Chem A 120:8435−8443

He M, Liu X, Lu X, Zhang C, Wang R (2017) Structure, acidity, and metal complexing properties of oxythioarsenites in hydrothermal solutions. Chem Geol 471:131–140

He D-Y, Qiu K, Simon AC, Pokrovski GS, Yu H, Connolly JAD, Li S, Turner S, Wang Q-F, Yang M-F, Deng J (2024) Mantle oxidation by sulfur drives the formation of giant gold deposits in subduction zones. Proc Nat Acad Sci USA 121(52):e2404731121

Hedenquist JW (1995) The ascent of magmatic fluid: Discharge versus mineralization, *in* Thompson JFH, ed., Magmas, fluids, and ore deposits, 23, Mineralogical Association of Canada Shortcourse Series: 263-289

Hedenquist JW, Lowenstern JB (1994) The role of magmas in the formation of hydrothermal ore deposits. Nature 370:519–527

Hedenquist JW, Simmons SF, Giggenbach WF, Eldridgem CS (1993) White Island, New Zealand, volcanic-hydrothermal system represents the geochemical environment of high-sulfidation Cu and Au ore deposition. Geology 21:731–734

Hedenquist JW, Aoki M, Shinohara H (1994) Flux of volatiles and ore-forming metals from the magmatic-hydrothermal system of Satsuma Iwojima volcano. Geology 22:585–588

Hedenquist JW, Arribas Jr A, Gonzalez-Urien E (2000) Exploration for epithermal gold deposits. SEG Reviews 31:245–277

Heinrich CA (2005) The physical and chemical evolution of low-salinity magmatic fluids at the porphyry to epithermal transition: A thermodynamic study. Mineral Dep 39:864–889

Heinrich CA (2006) From fluid inclusion microanalysis to large-scale hydrothermal mass transfer in the Earth's interior. J Miner Petrol Sci 101:110–117

Heinrich CA, Ryan CG, Mernagh TP, Eadington PJ (1992) Segregation of ore metals between magmatic brine and vapor: a fluid inclusions study using PIXE microanalysis. Econ Geol 87:1566–1583

Heinrich CA, Günther D, Audétat A, Ulrich T, Frischknecht R (1999) Metal fractionation between magmatic brine and vapour, and the link between porphyry-style and epithermal Cu-Au deposits. Geology 27:755–758

Heinrich CA, Pettke T, Halter WE, Aigner-Torres M, Audétat A, Günther D, Hattendorf B, Bleiner D, Guillong M, Horn I (2003) Quantitative multi-element analysis of minerals, fluid and melt inclusions by laser-ablation inductively-coupled-plasma mass-spectrometry. Geochim Cosmochim Acta 67:3473–3497

Heinrich CA, Connoly JAD (2022) Physical transport of magmatic sulfides promotes copper enrichment in hydrothermal ore fluids. Geology 50:1101−1105

Helgeson HC (1969) Thermodynamic properties of hydrothermal systems at elevated temperatures and pressures. Am J Sci 267:729−804

Helgeson HC, Garrels RM (1968) Hydrothermal transport and deposition of gold. Econ Geol 63:622−635

Helgeson HC, Kirkham DH (1974) Theoretical prediction of the thermodynamic behavior of aqueous electrolytes at high pressures and temperatures: II. Debye-Hückel parameters for activity coefficients and relative partial molal properties. Amer J Sci 274:1199−1261

Helgeson HC, Kirkham DH, Flowers GC (1981) Theoretical prediction of the thermodynamic behavior of aqueous electrolytes at high pressures and temperatures: IV. Calculation of activity coefficients, osmotic coefficients and apparent molal and relative partial molal properties to 600° C and 5 kb. Amer J Sci 281:1249−1516

Helz GR, Tossell JA, Charnock JM, Pattrick RAD, Vaughan DJ, Garner CD (1995) Oligomerization in As(III) sulfide solutions – theoretical constraints and spectroscopic evidence. Geochim Cosmochim Acta 59:4591–4604

Helz GR (2021) Dissolved molybdenum asymptotes in sulfidic waters. Geochem Persp Let 19:23–26

Hemley JJ (1953) A study of lead sulfide solubility and its relation to ore deposition. Econ Geol 48:113–138

Hemley JJ, Cygan GL, Fein JB, Robinson GR, Dangelo WM (1992) Hydrothermal ore-forming processes in the light of studies in rock-buffered systems. 1. Iron-copper-zinc-lead sulfide solubility relations. Econ Geol 87:1–22

Herlinger AW, Long II TV (1969) An investigation of the structure of the disulfite ion in aqueous solution using Raman and infrared spectroscopies. Inorg Chem 8:2661–2665

Hettmann K, Wenzel T, Marks M, Markl G (2012) The sulfur speciation in S-bearing minerals: New constraints by a combination of electron microprobe analysis and DFT calculations with special reference to sodalite-group minerals. Am Miner 97:1653–1661

Hill PG (1990) A unified fundamental equation for the thermodynamic properties of $H_2O$. J Phys Chem Ref Data 19:1233–1274

Hnedkovsky L, Wood RH (1997) Apparent molar heat capacities of aqueous solutions of $CH_4$, $CO_2$, $H_2S$, and $NH_3$ at temperatures from 304 K to 704 K and pressures to 35 MPa. J Chem Thermodyn 29:731–747

Hnedkovsky L, Wood RH, Balashov VN (2005) Electrical conductances of aqueous $Na_2SO_4$, $H_2SO_4$, and their mixtures: limiting equivalent ion conductances, dissociation constants, and speciation to 673 K and 28 MPa. J Phys Chem B 109:9034−9046

Hnedkovsky L, Wood RH, Majer V (1996) Volumes of aqueous solutions of $CH_4$, $CO_2$, $H_2S$, and $NH_3$ at temperatures from 298.15 K to 705 K and pressures to 35 MPa. J Chem Thermodyn 28:125–142

Holden WM, Seidler GT, Cheah S (2018) Sulfur speciation in biochars by very high resolution benchtop $K_\alpha$ X-ray emission spectroscopy. J Phys Chem A 122:5153−5161

Holland HD (1959) Some applications of thermochemical data to problems of ore deposits. I. Stability relations among the oxides, sulfides, sulfates and carbonates of ore and gangue metals. Econ Geol 54:184–233

Holland HD (1965) Some applications of thermochemical data to problems of ore deposits. II. Mineral assemblages and the composition of ore forming fluids. Econ Geol 60:1101–1166

Holland TJB, Powell R (2011) An improved and extended internally consistent thermodynamic dataset for phases of petrological interest, involving a new equation of state for solids. J Metamorph Geol 29:333–383

Holwell DA, Adeyemi Z, Ward LA, Smith DJ, Graham SD, McDonald I, Smith JW (2017) Low temperature alteration of magmatic Ni-Cu-PGE sulfides as a source for hydrothermal Ni and PGE ores: A quantitative approach using automated mineralogy. Ore Geol Rev 91:718–740

Hormer DA, Connick RE (1986) Equilibrium quotient for the isomerization of bisulfite ion from $HSO_3^-$ to $SO_3H^-$. Inorg Chem 25:2414–2417

Hu M, Chou I-M, Wang R, Shang L, Chen C (2022) High solubility of gold in $H_2S$-$H_2O$ ± NaCl fluids at 100–200 MPa and 600–800 °C: A synthetic fluid inclusion study. Geochim Cosmochim Acta 330:116–130

Huang F, Sverjensky DA (2019) Extended Deep Earth Water model for predicting major element mantle metasomatism. Geochim Cosmochim Acta 254:192–230

Huang W, Ni P, Zhou J, Shui T, Ding J, Zhu R, Cai Y, Fan M. (2021) Discovery of disulfane ($H_2S_2$) in fluid inclusions in rubies from Yuanjiang, China, and its implications. Crystals 11(11), 1305

Hüfner S (1995) *Photoelectron Spectroscopy: Principles and Applications*. Springer Verlag

Hurai V, Černušák I, Randive K. (2019) Raman spectroscopic study of polysulfanes ($H_2S_n$) in natural fluid inclusions. Chem Geol 508:15–29

Iacomo-Marziano G, Le Vaillant M, Godel BM, Barnes SJ, Arbaret L (2022) The critical role of magma degassing in sulphide melt mobility and metal enrichment. Nat Comm 13:2359

Igumnov SA (1976) Sulfur isotope exchange between sulfide and sulfate in hydrothermal solutions. Geokhimia 4:497–503 (in Russian)

Ivanova GF, Lavkina NI, Nesterova LA, Zhudikova AP, Khodakovsky IL (1975) Equilibrium in the $MoO_3$-$H_2O$ system at 25-300 °C. Geochem Intl 12(1):163–176

Jacquemet N, Guillaume D, Zwick A, Pokrovski GS (2014) In situ Raman spectroscopy identification of the $S_3^-$ ion in S-rich hydrothermal fluids from synthetic fluid inclusions. Amer Miner 99:1109–1118

Jalilehvand F (2006) Sulfur: not a ''silent'' element any more. Chem Soc Rev 35:1256–1268

James RH, Elderfield H, Palmer MR (1995) The chemistry of hydrothermal fluids from the Broken Spur site, 29°N Mid-Atlantic Ridge. Geochim Cosmochim Acta 59:651–659

James-Smith J, Cauzid J, Testemale D, Liu W, ; Hazemann J-L, Proux O, Etschmann B, Philippot P, Banks D, Williams P, Brugger J (2010) Arsenic speciation in fluid inclusions using micro-beam X-ray absorption spectroscopy. Am Mineral 95(7):921–932

Jha MK, Kumari A, Panda R, Kumar JR, Yoo K, Lee JY (2016) Review on hydrometallurgical recovery of rare earth metals. Hydrometallurgy 165:2–26

Jiang L, Lyu J, Shao Z (2017) Sulfur metabolism of *hydrogenovibrio thermophilus* strain S5 and its adaptation to deep-sea hydrothermal vent environment. Front Microbiol 8:2513

Jiang L, Xin Y, Chou I-M, Chen Y (2017) Raman spectroscopic measurements of ν1 band of hydrogen sulfide over a wide range of temperature and density in fused-silica optical cells. J Raman Spectroscopy 49(2):343–350

Johnson J, Anderson F, Parkhurst DL (2000) Database thermo.com, V8.R6.230, Rev 1.11. Lawrence Livermore National Laboratory, Livermore, California

Johnson JW, Oelkers EH, Helgeson HC (1992) SUPCRT92: A software package for calculating the standard molal thermodynamic properties of minerals, gases, aqueous species, and reactions from 1 to 5000 bar and 0 to 1000°C. Computers & Geosci 18:899–947; updated version based on a series of subsequent papers reporting HKF parameters for most common ions and aqueous complexes is available at http://geopig.asu.edu/index.html#

Joly Y (2022) FDMNES user's guide, http://fdmnes.neel.cnrs.fr/

Jorgensen BB (2021) Sulfur biochemical cycle of marine sediments. Geochem Persp 10(2):145–307

Kaasalainen H, Stefánsson A (2011a) Chemical analysis of sulfur species in geothermal waters. Talanta 85:1897–1903

Kaasalainen H, Stefánsson A (2011b) Sulfur speciation in natural hydrothermal waters, Iceland. Geochim Cosmochim Acta 75:2777–2791

Kaasalainen H, Stefánsson A (2012) The chemistry of trace elements in surface geothermal waters and steam, Iceland. Chem Geol 330:60–85

Kalintsev A, Migdisov A, Xu H, Roback R, Brugger J (2019) Uranyl speciation in sulfate-bearing hydrothermal solutions up to 250 °C. Geochim Cosmochim Acta 267:75–91

Kamyshny Jr. A (2009) Solubility of cyclooctasulfur in pure water and sea water at different temperatures. Geochim Cosmochim Acta 73:6022–6028

Kamyshny Jr. A, Goifman A, Rizkov D, Lev O (2003a) Formation of carbonyl sulfide by the reaction of carbon monoxide and inorganic polysulfides. Environ Sci Technol 37:1865–1872

Kamyshny Jr. A., Goifman D., Rizkov D. and Lev O. (2003b) Kinetics of disproportionation of inorganic polysulfides in undersaturated aqueous solutions at environmentally relevant conditions. Aquat Geochem 9:291–304

Kamyshny Jr. A, Goifman A, Gun J, Rizkov D, Lev O (2004) Equilibrium distribution of polysulfide ions in aqueous solutions at 25°C: a new approach for the study of polysulfides equilibria. Environ Sci Technol 38:6633–6644

Kamyshny Jr. A, Ekeltchik I, Gun J, Lev O (2006) Method for the determination of inorganic polysulfide distribution in aquatic systems. Anal Chem 78:2631–2639

Kamyshny Jr. A, Gun J, Rizkov D, Voitsekovski T, Lev O (2007) Equilibrium distribution of polysulfide ions in aqueous solutions at different temperatures by rapid single phase derivatization. Environ Sci Technol 41:2395–2400

Kamyshny Jr A, Zilberbrand M, Ekeltchik I, Voitsekovski T, Gun J, Lev O (2008) Speciation of polysulfides and zerovalent sulfur in sulfide-rich water wells in Southern and Central Israel. Aquatic Geochem 14:171–192

Kamyshny Jr A, Druschel G, Mansaray ZF, Farquhar J (2014) Multiple sulfur isotopes fractionations associated with abiotic sulfur transformations in Yellowstone National Park geothermal springs. Geochem Trans 15:7

Kanitpanyacharoen W, Boudreau AE (2013) Sulfide-associated mineral assemblages in the Bushveld Complex, South Africa: platinum-group element enrichment by vapor refining by chloride–carbonate fluids. Miner. Deposita 48:193–210

Kas JJ, Vila FD, Pemmaraju D, Tan TS, Vimolchalao S, Rehr JJ (2021) Calculations of X-ray spectroscopies using FEFF10 and Corvus. arXiv:2106.13334 [cond-mat.mtrl-sci]

Kavčič M, Dousse J-C, Szlachetko J, Cao W (2007) Chemical effects in the $K_\beta$ X-ray emission spectra of sulfur. Nucl Instr Meth Phys Res B260:642–646

Keller NS, Stefánsson A, Sigfússon B (2014a) Arsenic speciation in natural sulfidic geothermal waters. Geochim Cosmochim Acta 142:15–26

Keller NS, Stefánsson A, Sigfússon B (2014b) Determination of arsenic speciation in sulfidic waters by ion chromatography hydride-generation atomic fluorescence spectroscopy (IC-HG-AFS). Talanta 128:466–472

Kerr A, Leitch AM (2005) Self-destructive sulfide segregation systems and the formation of high-grade magmatic ore deposits. Econ Geol 100:311–332

Kim Y, Konecke B, Fiege A, Simon A, Becker U (2017) An ab-initio study of the energetics and geometry of sulphide, sulphite, and sulphate incorporation into apatite: The thermodynamic basis for using this system as an oxybarometer. Am Mineral 102(8):1646–1656

Kiseeva ES, Fonseca ROC, Beyer C, Li Y (2025) Sulfur and sulfides in the Earth's mantle. This Book

Kishima N (1989) A thermodynamic study on the pyrite-pyrrhotite-magnetite-water system at 300–500°C with relevance to the fugacity/concentraton quotient of aqueous $H_2S$. Geochim Cosmochim Acta 53:2143–2155

Kleine BI, Gunnarsson-Robin J, Kamunya KM, Ono S, Stefánsson A (2021) Source controls on sulfur abundance and isotope fractionation in hydrothermal fluids in the Olkaria geothermal field, Kenya. Chem Geol 582:120446

Knox J (1906) Zur Kenntnis der ionenbildung des schwefels und der komplexionen des quecksilber. Zeit Elektrochem 12:477–481 (in German)

Kokh MA, Lopez M, Gisquet P, Lanzanova A, Candaudap F, Besson P, Pokrovski GS (2016) Combined effect of carbon dioxide and sulfur on vapor-liquid partitioning of metals in hydrothermal systems. Geochim Cosmochim Acta 187:311–333

Kokh MA, Akinfiev NN, Pokrovski GS, Salvi S, Guillaume D (2017) The role of carbon dioxide in the transport and fractionation of metals by geological fluids. Geochim Cosmochim Acta 197:433–466

Kokh MA, Assayag N, Mounic S, Cartigny P, Gurenko A, Pokrovski GS (2020) Multiple sulfur isotope fractionation in hydrothermal systems in the presence of radical ions and molecular sulfur. Geochim Cosmochim Acta 285:100–128

Kokh MA, Wilke M, Klemme S, Pokrovski GS (2024) The role of trisulfur radical ion in the transport of molybdenum by hydrothermal fluids. Abstracts of EMC2 4th European Mineralogical Conference, Dublin, p. 247

Kolonin GR, Laptev YV (1982) Study of process of dissolution of α-$Bi_2O_3$ (bismite) and complexation of bismuth in hydrothermal solutions. Geokhimiya 11:1621–1631 (in Russian)

Konecke BA, Fiege A, Simon AC, Parat F, Stechern A (2017) Co-variability of $S^{6+}$, $S^{4+}$, and $S^{2-}$ in apatite as a function of oxidation state: Implications for a new oxybarometer. Am Mineral 102(3):548–557

Korsakova NV, Kokina TA, Sushchevskaya TM, Varshal GM (1991) Potentiometric determination of sulfide sulfur in solutions in inclusions. Geochem Intl 28(8):42–51

Kouzmanov K, Bailly L, Ramboz C, Rouer O, Beny JM (2002) Morphology, origin and infrared microthermometry of fluid inclusions in pyrite from the Radka epithermal copper deposit, Srednogorie zone, Bulgaria. Miner Dep 37:599–613

Kouzmanov K, Pettke T, Heinrich CA (2010) Direct analysis of ore-precipitating fluids: combined IR microscopy and LA-ICP-MS study of fluid inclusions in opaque ore minerals. Econ Geol 105:351–373

Kouzmanov K, Pokrovski GS (2012) Hydrothermal controls on metal distribution in Cu(-Au-Mo) porphyry systems. In: Geology and Genesis of Geology and Genesis of Major Copper Deposits and Districts of the World: A Tribute to Richard H. Sillitoe (eds. J.W. Hedenquist, M. Harris, and F. Camus). Soc Econ Geol Spec Publ 16:573–618

Kozintseva TN (1964) Solubility of hydrogen sulfide in water at elevated temperatures. Geokhimia 8:758–765 (in Russian)

Krupp RE (1988) Solubility of stibnite in hydrogen sulfide solutions, speciation, and equilibrium constants, from 25 to 350 °C. Geochim Cosmochim Acta 52:3005–3015

Krupp RE (1990a) Comment on "As( III) and Sb( III) sulfide complexes: An evaluation of stoichiometry and stability from existing experimental data" by N. F. Spycher and M. H. Reed. Geochim Cosmochim Acta 54:3239–3240

Krupp RE (1990b) Response to the reply by N. F. Spycher and M. H. Reed. Geochim Cosmochim Acta 54:3245

Kryukov PA, Starostina LI (1978) The first ionization constant of hydrogen sulphide at temperatures to 150°C. Geokhimia 1:84–87 (in Russian)

Kryukov PA, Starostina LI, Tarasenko SY, Primanchuk MP (1974) Second constant of hydrogen sulfide ionization at temperatures up to 150 °C. Geochem. Intl. 688–698 (translated from Geokhimia, no7:1003–1013 in Russian)

Kudrin AV (1989) Behavior of Mo in aqueous NaCl and KCl solutions at 300–450 °C. Geochem Intl 26(8):87–99

Kulik DA, Wagner T, Dmytrieva SV, Kosakowski G, Hingerl FF, Chudnenko KV, Berner U (2013) GEM-Selektor geochemical modeling package: revised algorithm and GEMS3K numerical kernel for coupled simulation codes. Comput Geosci 17:1–24

Küster FW, Heberlein E (1905) Beitrage zur Kenntnis der Polysulficfe I. Zeit Anorg Chem 43:53–84 (in German)

Lai F, Zou S, Xu D (2022) Silver complexation in chloride- and sulfur-rich hydrothermal fluids: Insight from ab initio molecular simulations. Chem Geol 589:120684

Large RR, Bull SW, Maslennikov VV (2011) A carbonaceous sedimentary sourcerock model for Carlin-type and orogenic gold deposits. Econ Geol 106:331–358

Lasaga AC (1998) *Kinetic Theory in Earth Sciences*. Princeton University Press. Princeton, NJ, 811p.

Laskar C (2022) Impact du soufre sur le transport des platinoïdes par les fluides hydrothermaux. PhD thesis, University of Toulouse. https://theses.fr/2022TOU30068

Laskar C, Bazarkina EF, Kokh MA, Hazemann J-L, Vuilleumier R, Desmaele E, Pokrovski GS (2022) Stability and structure of platinum sulfide complexes in hydrothermal fluids. Geochim Cosmochim Acta 336:407–422

Laskar C, Bazarkina EF, Pokrovski GS (2025) The role of hydrosulfide complexes in palladium transport and fractionation from platinum by hydrothermal fluids. Geochim Cosmochim Acta (in press) https://doi.org/10.1016/j.gca.2024.12.019

Latimer WM (1952) *Oxidation Potentials*. Prentice-Hall, Englewood Cliffs, 392 p.

Le Vaillant M, Barnes SJ, Fiorentini ML, Santaguida F, Törmänen T (2016a) Effects of hydrous alteration on the distribution of base metals and platinum group elements within the Kevitsa magmatic nickel sulphide deposit. Ore Geol Rev 72:128–148

Le Vaillant M, Saleem A, Barnes SJ, Fiorentini ML, Miller J, Beresford S, Perring C (2016b) Hydrothermal remobilisation around a deformed and remobilised komatiite-hosted Ni-Cu-(PGE) deposit, Sarah's Find, Agnew Wiluna greenstone belt, Yilgarn Craton, Western Australia. Miner. Deposita 51:369–388

Leach DV, Sangster DF, Kelley KD, Large RR, Garven G, Allen CR, Gutzmer J, Walters S (2005). Sediment-hosted lead-zinc deposits: A global perspective. Econ Geology 100$^{th}$ Anniv Vol: 561–607

Leavitt WD, Bradley AS, Santos AA, Pereira IAC, Johnston DT (2015) Sulfur isotope effects of dissimilatory sulfite reductase. Front Microbiol 6:1–20

Ledé B, Demortier A, Gobeltz-Hautecoeur N, Lelieur J-P, Picquenard E, Duhayon C (2007) Observation of the $\nu_3$ Raman band of $S_3^-$ inserted into sodalite cages. J Raman Spectr 38:1461–1468

Lee JI, Mather AE (1977) Solubility of hydrogen sulfide in water. Ber Bunsenges Phys Chem 81:1020–1023

Leman L, Orgel L, Ghadiri R. (2004) Carbonyl sulfide–mediated prebiotic formation of peptides. Science 306:283–286.

Lemire RJ, Berner U, Musikas C, Palmer DA, Taylor P, Tochiyama O (2013) *Chemical Thermodynamics of Iron, Part 1*. NEA No. 6355, OECD

Lerchbaumer L, Audétat A (2012) High Cu concentrations in vapor-type fluid inclusions: An artifact? Geochim Cosmochim Acta 88:255–274

Lewis GN, Randall M (1918) Equilibrium in the reaction between water and sulfur at the boiling point of sulfur. J Amer Chem Soc 1918:362–367

Lewis GN, Randall M, Bichowsky FR (1918) Preliminary study of reversible reactions of sulfur compounds. J Amer Chem Soc 1918:356–362

Li L, Li Z, Zhong R, Du Z, Luan Z, Xi S, Zhang X (2023) Direct $H_2S$, $HS^-$ and pH measurements of high-temperature hydrothermal vent fluids with in situ Raman spectroscopy. Geophys Res Lett 50:e2023GL103195

Li N, Derrey IT, Holtz F, Horn I, Weyer S, Xi W (2021) Molybdenum solubility and partitioning in $H_2O$-$CO_2$-NaCl fluids at 600 °C and 200 MPa. Chem Geol 583:120438

Li W, Audétat A, Zhang J (2015) The role of evaporites in the formation of magnetite–apatite deposits along the Middle and Lower Yangtze River, China: Evidence from LA-ICP-MS analysis of fluid inclusions. Ore Geol Rev 67:264–278

Licht S, Manassen J (1987) The second dissociation constant of $H_2S$. J Electrochem Soc 134:918–921

Licht S, Davis J (1997) Disproportionation of aqueous sulfur and sulfide: Kinetics of polysulfide decomposition. J Phys Chem B 101(14):2540–2545

Licht S, Hodes G, Manassen J (1986) Numerical analysis of aqueous polysulfide solutions and its application to cadmium chalcogenide/polysulfide photoelectrochemical solar cells. Inorg Chem 25:2486–2489

Licht S, Forouzan F, Longo K (1990) Differential densometric analysis of equilibria in highly concentrated media: Determination of the aqueous second acid dissociation constant of $H_2S$. Anal Chem 62:1356–1360

Liebing P, Kühling M, Swanson C, Feneberg M, Hilfert L, Goldhahn R, Chivers T, Edelmann FT (2019) Catenated and spirocyclic polychalcogenides from potassium carbonate and elemental chalcogens. Chem Comm 55:14965–14967

Lietzke MH, Stoughton RW, Young TF (1961) The bisulfate acid constant from 25 to 225° as computed from solubility data. J Phys Chem 65:2247–2249

Lin J-F, Alp EE, Goncharov AF (2014) Raman and nuclear resonant spectroscopy. In: *Treatise on Geochemistry*, , 2$^{nd}$ ed. Elsevier, chap. 15.11:195–211

Littlejohn D, Walton SA, Chang S-G (1992) A Raman study of the isomers and dimer of hydrogen sulfite ion. Applied Spectr 46:848–851.

Liu W, McPhail DC (2005) Thermodynamic properties of copper chloride complexes and copper transport in magmatic-hydrothermal solutions. Chem Geol 221:21–39

Liu W, Brugger J, McPhail DC, Spiccia L (2002) A spectrophotometric study of aqueous copper(I)–chloride complexes in LiCl solutions between 100 °C and 250 °C. Geochim Cosmochim Acta 66:3615–3633

Liu W, Etschmann B, Brugger J, Spiccia L, Foran G, McInnes B (2006) UV–Vis spectrophotometric and XAFS studies of ferric chloride complexes in hyper-saline LiCl solutions at 25–90 °C. Chem Geol 231:326–349

Liu W, Migdisov A, Williams-Jones A (2012). The stability of aqueous nickel(II) chloride complexes in hydrothermal solutions: results of UV–Visible spectroscopic experiments. Geochim Cosmochim Acta 94:276–290

Liu W, Etschmann B, Mei Y, Guan Q, Testemale D, Brugger J (2020) The role of sulfur in molybdenum transport in hydrothermal fluids: Insight from in situ synchrotron XAS experiments and molecular dynamics simulations. Geochim Cosmochim Acta 290:162–179

Liu X, Lu X, Wang R, Zhou H, Xu S (2011) Speciation of gold in hydrosulphide-rich ore-forming fluids: Insights from first-principles molecular dynamics simulations. Geochim. Cosmochim. Acta 75:185–194
Liu X, Lu X, Wang R, Zhou H (2012) Silver speciation in chloride-containing hydrothermal solutions from first principles molecular dynamics simulations. Chem Geol 294–295:103–112
Liu X, Sprik M, Cheng J (2013a) Hydration, acidity and metal complexing of polysulfide species: A first principles molecular dynamics study. Chem Phys Lett 563:9–14
Liu X, Cheng J, Sprik M, Lu X (2013b) Solution structures and acidity constants of molybdic acid. J Phys Chem Lett 4:2926–2930
Lodders K, Fegley B (2025) Sulfur in the giant planets, their moons, and extrasolar gas giant planets. This Book
Lohmayer R, Reitmaier GMS, Bura-Nakic E, Planer-Friedrich B (2015) Ion-pair chromatography coupled to inductively coupled plasma mass spectrometry (IPC-ICP-MS) as a method for thiomolybdate speciation in natural waters. Anal Chem 87:3388–3395
Long DA (1977) *Raman Spectroscopy*. McGraw-Hill, 276 p.
Lorand J-P, Luguet A, Alard O (2008) Platinum-group elements: a new set of key tracers for the Earth's interior. Elements 4:247–252
Loucks RR, Mavrogenes JA (1999) Gold solubility in supercritical hydrothermal brines measured in synthetic fluid inclusions. Science 284:2159–2163
Louvel M, Bordage A, Da Silva-Cadoux C, Testemale D, Lahera E, Del Net W, Geaymond O, Dubessy J, Argoud R, Hazemann J-L (2015) A high-pressure high-temperature setup for in situ Raman spectroscopy of supercritical fluids. J Mol Liq 205:54–60
Luguet A, Pearson DG, Nowell GM, Dreher ST, Coggon JA, Spetsius ZV, Parman SW (2008) Enriched Pt-Re-Os isotope systematics in plume lavas explained by metasomatic sulfides. Science 319:453–456
Luo Y, Millero FJ (2007) Stability constants for the formation of lead chloride complexes as a function of temperature and ionic strength. Geochim Cosmochim Acta 71:326–334
Luther GWI, Glazer BT, Hohmann L, Popp J, Taillefert M, Rozan TF, Brendel PJ, Theberge SM, Nuzzio DB (2001) Sulfur speciation monitored in situ with solid state gold amalgam voltammetric microelectrodes: polysulfides as a special case in sediments, microbial mats and hydrothermal vent waters. J Environ Monit 3:61–66
Manning CE (1994) The solubility of quartz in $H_2O$ in the lower crust and upper mantle. Geochim Cosmochim Acta 58:4831–4839
Markússon SH, Stefánsson A (2011) Geothermal surface alteration of basalts, Krýsuvík Iceland—Alteration mineralogy, water chemistry and the effects of acid supply on the alteration process. J Volc Geotherm Res 206:46–59
Maronny G (1959) Constantes de dissociation de l'hydrogen surfuré. Electrochim Acta 1:58–69 (in French)
Márquez-Zavalía MF, Heinrich CA (2016) Low-salinity magmatic fluids in a volcanic-hosted epithermal carbonate–base-metal–gold vein system: Alto de la Blenda, Farallón Negro, Argentina. Min. Deposita 51:873–902
Marsala A, Wagner T, Wälle M (2013) Late-metamorphic veins record deep ingression of meteoric water: a LA-ICPMS fluid inclusion study from the fold-and-thrust belt of the Rhenish Massif, Germany. Chem Geol 351:134–53
Marsden JO, House CI (2009) *The Chemistry of Gold Extraction.* 2$^{nd}$ ed. Society for Mining, Metallurgy, and Exploration Inc
Marshall WL, Jones EV (1966) Second dissociation constant of sulfuric acid from 25 to 350° evaluated from solubilities of calcium sulfate in sulfuric acid solutions. J Phys Chem 70:4028–4040
Marshall WM, Franck EU (1981) Ion product of water substance, 0–1000°C, 1–10,000 bars, new international formulation and its background. J Phys Chem Ref Data 10:295–304
Matsushima Y, Okuwaki A (1988) The second dissociation constant of sulfuric acid at elevated temperatures from potentiometric measurements. Bull Chem Soc Jpn 61:3344–3346
Mavrogenes JA, Bodnar RJ (1994) Hydrogen movement into and out of fluid inclusions in quartz: experimental evidence and geologic implications. Geochim Cosmochim Acta 58:141–148
Mavrogenes JA, Berry AJ, Newvill M, Sutton SR (2002) Copper speciation in vapor-phase fluid inclusions from Mole Granite, Australia. Am Mineral 87:1360–1364
May PM, Batka D, Hefter G, Königsberger E, Rowland D (2018) Goodbye to $S^{2-}$ in aqueous solution. Chem Commun 54:1980–1983
McCleskey RB, Nordstrom DK, Susong DD, Ball JW, Taylor HE (2010) Source and fate of inorganic solutes in the Gibbon River, Yellowstone National Park, Wyoming, USA. II. Trace element chemistry. J Volcanol Geotherm Res 196:139–155
McCleskey RB, Roth DA, Nordstrom DK, Hurwitz S, Holloway JM, Bliznik PA, Ball JW, Repert DA, Hunt AG (2022) Water-chemistry and isotope data for selected springs, geysers, streams, and rivers in Yellowstone national park, Wyoming. U.S. Geological Survey data release
Mei Y, Sherman DM, Liu W, Brugger J (2013a). Ab initio molecular dynamics simulation and free energy exploration of copper(I) complexation by chloride and bisulfide in hydrothermal fluids. Geochim Cosmochim Acta 102:45–64

Mei Y, Sherman DM, Liu W, Brugger J (2013b) Complexation of gold in $S_3^-$-rich hydrothermal fluids: Evidence from ab-initio molecular dynamics simulations. Chem Geol 347:34–42

Mei Y, Liu W, Sherman DM, Brugger J (2014) Metal complexation and ion hydration in low density hydrothermal fluids: ab initio molecular dynamics simulation of Cu(I) and Au(I) in chloride solutions (25–1000 °C, 1–5000 bar). Geochim Cosmochim Acta 131:196–212

Mei Y, Sherman DM, Liu W, Etschmann B, Testemale D, Brugger J (2015a). Zinc complexation in chloride-rich hydrothermal fluids (25–600 °C): a thermodynamic model derived from ab initio molecular dynamics. Geochim Cosmochim Acta 150:265–284

Mei Y, Etschmann B, Liu W, Sherman DM, Barnes SJ, Fiorentini ML, Seward TM, Testemale D, Brugger J (2015b). Palladium complexation in chloride- and bisulfide-rich fluids: insights from ab initio molecular dynamics simulations and X-ray absorption spectroscopy. Geochim Cosmochim Acta 161:128–145

Mei Y, Etschmann B, Liu W, Sherman DM, Testemale D, Brugger J (2016) Speciation and thermodynamic properties of zinc in sulfur-rich hydrothermal fluids: insights from ab initio molecular dynamics simulations and X-ray absorption spectroscopy. Geochim Cosmochim Acta 179:32–52

Mei Y, Liu W, Brugger J, Guan Q (2020) Gold solubility in alkaline and ammonia-rich hydrothermal fluids: Insights from ab initio molecular dynamics simulations. Geochim Cosmochim Acta 291:62–78

Mel HC, Hugus ZZ Jr, Latimer WM (1956) The thermodynamics of thiosulfate ion. J Amer Chem Soc 78:1822–1826

Melent'yev BN, Ivanenko VV, Pamfilova LA (1969) Solubility of some ore-forming sulfides under hydrothermal conditions. Geochem Intl 7:416–460

Ménez B, Philippot P, Bonnin-Mosbah M, Simonovici A, Gibert F (2002) Analysis of individual fluid inclusions using Synchrotron X-Ray Fluorescence microprobe: progress toward calibration for trace elements. Geochim Cosmochim Acta 66:561–576

Mernagh TP, Mavrogenes J (2019) Significance of high temperature fluids and melts in the Grasberg porphyry copper-gold deposit. Chem Geol 508:210–224

Meshoulam A, Amrani, Shurki A (2023) The initiation stage of thermochemical sulfate reduction: An isotopic and computational study. J Anal Appl Pyrol 172:106011

Mesmer RE, Marshall WL, Palmer DA, Simonson JM, Holmes HF (1988) Thermodynamics of aqueous association and ionization reactions at high temperatures and pressures. J Sol Chem 17:699–718

Meyer B (1976) Elemental sulfur. Chem Rev 76:367–388

Meyer B, Ospina M, Peter LB (1980) Raman spectrometric determination of oxysulfur anions in aqueous systems. Anal Chim Acta 117:301–311

Meyer B, Ward K, Koshlap K, Peter L (1983) Second dissociation constant of hydrogen sulfide. Inorg Chem 22:2345–2346

Mi J-X, Pan Y (2018) Halogen-rich minerals: crystal chemistry and geological significances. In: Harlov D, Aranovich L (eds) *The Role of Halogens in Terrestrial and Extraterrestrial Geochemical Processes*. Springer Geochemistry, Springer, Cham, pp. 123–184

Migdisov AA, Bychkov AY (1998) The behaviour of metals and sulphur during the formation of hydrothermal mercury-antimony-arsenic mineralization, Uzon caldera, Kamchatka, Russia. J Volcanol Geotherm Res 84:153–171

Migdisov AA, Williams-Jones AE, Lakshtanov LZ, Alekhin YV (2002) Estimates of the second dissociation constant of $H_2S$ from the surface sulfidation of crystalline sulfur. Geochim Cosmochim Acta 66:1713–1725

Migdisov AA, Zezin D, Williams-Jones AE (2011) An experimental study of cobalt (II) complexation in $Cl^-$ and $H_2S$-bearing hydrothermal solutions. Geochim Cosmochim Acta 75:4065–4079

Migdisov AA, Bychkov AY, Williams-Jones AE, van Hinsberg VJ (2014) A predictive model for the transport of copper by HCl-bearing water vapour in ore-forming magmatic-hydrothermal systems: implications for copper porphyry ore formation. Geochim Cosmochim Acta 129:33–53

Migdisov A, Williams-Jones AE, Brugger J, Caporuscio FA (2016) Hydrothermal transport, deposition, and fractionation of the REE: Experimental data and thermodynamic calculations. Chem Geol 439:13–42

Mikhlin Y (2020) X-ray photoelectron spectroscopy in mineral processing studies. Appl Sci 10:5138

Mikucki EJ (1998) Hydrothermal transport and depositional processes in Archean lode-gold systems: A review. Ore Geol Rev 13:307–321

Miluska J et al. (2012) Zero-valent sulphur is a key intermediate in marine methane oxidation. Nature 491:541–546.

Minubayeva Z, Seward TM (2010) Molybdic acid ionisation under hydrothermal conditions to 300 °C. Geochim Cosmochim Acta 74:4365–4374

Miron GD, Wagner T, Wälle M, Heinrich CA (2013) Major and trace-element composition and pressure-temperature evolution of rock-buffered fluids in low-grade accretionary-wedge metasediments, Central Alps. Contrib Miner Petrol 165:981–1008

Miron GD, Kulik DA, Dmytrieva SV, Wagner T (2015): GEMSFITS: Code package for optimization of geochemical model parameters and inverse modeling. Applied Geochem 55:28–45

Mißbach H, Duda J-P, van den Kerkhof AM, Lüders V, Pack A, Reitner J, Thielet V (2021) Ingredients for microbial life preserved in 3.5 billion-year-old fluid inclusions. Nat Commun 12:1101

Molleman E, Dreisinger D (2002) The treatment of copper–gold ores by ammonium thiosulfate leaching. Hydrometallurgy 66:1–21

Monnin C (1999) A thermodynamic model for the solubility of barite and celestite in electrolyte solutions and seawater to 200 °C and to 1 kbar. Chem Geol 153:187–209

Moretti R, Baker DR (2008) Modeling the interplay of $fO_2$ and $fS_2$ along the FeS-silicate melt equilibrium. Chem Geol 256: 286–298

Morey GW, Hesselgesser JM (1951) The solubility of some minerals in superheated steam at high temperatures. Econ Geol 46:821–835

Mountain BW, Seward TM (1999) The hydrosulphide/sulphide complexes of copper(I): experimental determination of stoichiometry and stability at 22 °C and reassessment of high temperature data. Geochim Cosmochim Acta 63:11–29

Mountain BW, Seward TM (2003) Hydrosulfide/sulfide complexes of copper(I): Experimental confirmation of the stoichiometry and stability of $Cu(HS)_2^-$ to elevated temperatures. Geochim Cosmochim Acta 67:3005–3014.

Mudd GM, Jowitt SM (2014) A detailed assessment of global nickel resource trends and endowments. Econ Geol 109:1813–1841

Mukhin L (1974) Evolution of organic compounds in volcanic regions. Nature 251:50–51

Mungall JE, ed. (2005) Exploration for platinum-group elements deposits. Min Assoc Canada Short Course 35:1–494

Mungall JE, Naldrett AJ (2008). Ore deposits of the platinum-group elements. Elements 4:253–258

Mungall et al (2020) Transport of metals and sulphur in magmas by flotation of sulphide melt on vapour bubbles. Nat Geosci 8:216–219

Murowchick JB, Barnes HL (1986) Marcasite precipitation from hydrothermal solutions. Geochim Cosmochim Acta 50:2615–2629

Murray RC, Cubicciotti D (1983) Thermodynamics of aqueous sulfur species to 300°C and potential-pH diagrams. J Electrochem Soc 130:866–869

Naumov GB, Ryzhenko BN, Khodakovsky IL (1971) *Handbook of Thermodynamic Data*. Atomizdat, Moscow (in Russian). English translation is available from U.S. Department of Commerce, Washington D.C., 1974, PB-226

Naumov VB, Dorofeev VA, Mironova OF (2009) Principal physico-chemical parameters of natural mineralizing fluids. Geokhimia 8:825–851 (in Russian)

Neuville DR, de Ligny D, Henderson GS (2014) Advances in Raman spectroscopy applied to Earth and material sciences. Rev Miner Geochem 78:509–541

Ni H, Keppler H (2012). In situ Raman spectroscopic study of sulfur speciation in oxidized magmatic-hydrothermal fluids. Am Mineral 97:1348–1353

Nikolaeva NM, Gusyakova ZM (1989) Thermodynamic characteristics of the dissociation reaction of thiosulfuric acid. Izvest Sibir Otdel AN SSSR 2:15–18, 1989 (In Russian)

Nikolaeva NM, Erenburg AM, Antipina VA (1972) About the temperature dependence of the standard potentials of halogenide complexes of gold. Izv Sibirskogo Otdeleniya AN SSSR, No 9, Ser Khim Nauk Vip 4:126–128 (in Russian)

Niskanen J, Sahle CJ, Juurinen I, Koskelo J, Lehtola S, Verbeni R, Müller H, Hakala M, Huotari S (2015) Protonation dynamics and hydrogen bonding in aqueous sulfuric acid. J Phys Chem B 119:11732−11739

Nordstrom DK, McCleskey BR, Ball JW (2009) Sulfur geochemistry of hydrothermal waters in Yellowstone national park: IV acid-sulfate waters. Applied Geochem 24:191–207

Nordstrom DK, Majzlan J, Königsberger E (2014) Thermodynamic properties for arsenic minerals and aqueous species. Rev Mineral Geochem 79:217–255

Norman DI, Sawkins FJ (1987) Analysis of volatiles in fluid inclusions by mass spectrometry. Chem Geol 61:1–10.

Nriagu JO (1971) Studies in the system PbS-NaCl-$H_2S$-$H_2O$: Stability of lead(II) thiocomplexes at 90 °C. Chem Geol 8:299–310

Oduro H, Harms B, Sintim HO, Kaufman AJ, Cody G, Farquhar J (2011) Evidence of magnetic isotope effects during thermochemical sulfate reduction. Proc Nat Acad Sci USA 108:17635–17638

Oelkers EH, Sherman DM, Ragnarstottir KV, Collins C (1998) An EXAFS spectroscopic study of aqueous antimony(III)-chloride complexation at temperatures from 25 to 250 °C. Chem. Geol. 151: 21–27

Oelkers EH, Benezeth P, Pokrovski GS (2009) Thermodynamic databases for water-rock interaction. Rev Miner Geochem 70:1–46

Ogryzlo SP (1935) Hydrothermal experiments with gold. Econ Geol 30:400–424

Ohmoto H, Goldhaber MB (1997) Sulfur and Carbon isotopes. in: *Geochemistry of Hydrothermal Ore Deposits*, 3rd ed., HL Barnes (ed.), Wiley, pp 517–596

Ohmoto H, Lasaga AC (1982) Kinetics of reactions between aqueous sulfates and sulfides in hydrothermal systems. Geochim Cosmochim Acta 46:1727–1745

OlsenNJ, Mountain BW, Seward TM (2018) Antimony(III) sulfide complexes in aqueous solutions at 30 °C: A solubility and XAS study. Chem Geol 476:233−247

OlsenNJ, Mountain BW, Seward TM (2019) Antimony(III) speciation in hydrosulfide solutions from 70 to 400 °C and up to 300 bar. ACS Earth Space Chem 3:1058−1072
Oppenheimer C (2010) Ultraviolet sensing of volcanic eruptions Elements 6:87–92
Orr WL (1974) Changes in sulfur content and isotopic ratios of sulfur during petroleum maturation. Study of the big Horn Basin Paleozoic oils. Am Assoc Petrol Geol Bull 58:2295–2318
Oscarson JL, Izatt RM, Brown PR, Pawlak Z, Gillespie SE, Christensen JJ (1988) Thermodynamic quantities for the interaction of $SO_4^{2-}$ with $H^+$ and $Na^+$ in aqueous solution from 150 to 320°C. J Sol Chem 17(9):841–863
Ovchinnikov LN, Kozlov YD, Rafal'sky RP (1982) The solubility of stibnite in chloride solutions at elevated temperatures. Geochem. Int. 19(5):56–63
Page FM (1953) The dissociation constants of thiosulphuric acid. J Chem Soc 1953:1719–1724
Palmer DA, Bénézeth P, Simonson JM (2004) The solubility of copper oxides around the water/steam cycle. Power Plant Chem 6:81–88
Pan P, Wood SA (1994) Solubility of Pt and Pd sulfides and Au metal in aqueous bisulfide solutions: II. Results at 200° to 350°C and saturated vapor pressure. Miner Deposita 29:373–390
Pan Y, Nilges MJ (2014) Electron paramagnetic resonance spectroscopy: basic principles, experimental techniques and applications to earth and planetary sciences. Rev Min Geochem 78:655–690
Pan Y, Mi Y (2025) Sulfur in non-sulfide minerals: from sulfates, sulfites and thiocyanates to rock-forming minerals. This Book
Papp C, Steinrück H-P (2013) In situ high-resolution X-ray photoelectron spectroscopy – Fundamental insights in surface reactions. Surface Sci Rep 68:446–487
Parkhurst DL, Appelo CAJ (1999) User's guide to PHREEQC (version 2) - A computer program for speciation, batch-reaction, one-dimensional transport, and inverse geochemical calculations. U.S.G.S. Water-Res. Invest. Rep. 99-4259, Denver, Colorado, USA, 312 p.
Pascal TA, Pemmaraju CD, Prendergast D (2015) X-ray spectroscopy as a probe for lithium polysulfide radicals. Phys Chem Chem Phys 17:7743–7753
Pasteris JD, Bessac O (eds) (2020) Raman spectroscopy in earth and planetary sciences. Elements 16(2):87–122
Patai S, ed. (1977) *Chemistry of Cyanates and Their Derivatives*, vol. II, John Wiley, New York
Pearson RG (1963) Hard and soft acids and bases. J Amer Chem Soc 85:3533–3539
Pearson RG (1997) *Chemical Hardness*. Wiley
Peltzer ET., Zhang X, Walz PM, Luna M, Brewer PG (2016). In situ Raman measurement of $HS^-$ and $H_2S$ in sediment pore waters and use of the $HS^-$:$H_2S$ ratio as an indicator of pore water pH. Marine Chem 184:32–42
Pereda S, Thomsen K, Rasmussen P (2000) Vapor-liquid-solid equilibria of sulfur dioxide in aqueous electrolyte solutions. Chem Eng Sci 55:2663–2671
Perfetti E, Pokrovski GS, Ballerat-Busserolles K, Majer V, Gibert F (2008) Densities and heat capacities of aqueous arsenious and arsenic acid solutions to 350°C and 300 bar, and revised thermodynamic properties of $As(OH)_3^0$(aq), $AsO(OH)_3^0$(aq) and iron sulfarsenide minerals. Geochim Cosmochim Acta 72:713–731
Persaud A (2022) The aqueous chloride complexes of antimony(III) and lanthanum(III) from 25 °C up to 300 °C by Raman spectroscopy. PhD thesis, University of Guelph, Ontario, Canada
Pettke T, Oberli F, Audétat A, Guillong M, Simon AC, Hanley JJ, Klemm LM (2012) Recent developments in element concentration and isotope ratio analysis of individual fluid inclusions by laser ablation single and multiple collector ICP-MS. Ore Geol Rev 44:10–38
Phillips DJ, Phillips SL (2000) High temperature dissociation constants of $HS^-$ and the standard thermodynamic values for $S^{2-}$. J Chem Eng Data 45:981−987
Pin S, Huthwelker T, Brown MA, Vogel F (2013) Combined sulfur K-edge XANES-EXAFS study of the effect of protonation on the sulfate tetrahedron in solids and solutions. J. Phys. Chem. 117:8368−8376
Pitzer KS (1991) Ion interaction approach: theory and data correlation. In: Pitzer KS ed., *Activity Coefficients in Electrolyte Solutions*, 2nd edition, CRC Press, pp 75–155
Planer-Friedrich B, Scheinost AC (2011) Formation and structural characterization of thioantimony species and their natural occurrence in geothermal waters. Environ Sci Technol 45:6855−6863
Planer-Friedrich B, London J, McCleskey RB, Nordstrom DK, Wallschlager D (2007) Thioarsenates in geothermal waters of Yellowstone national park: Determination, preservation, and geochemical importance. Environ Sci Technol 41:5245−5251
Planer-Friedrich B, Forberg J, Lohmayer R, Kerl CF, Boeing F, Kaasalainen H, Stefánsson A (2020) Relative abundance of thiolated species of As, Mo, W, and Sb in hot springs of Yellowstone National Park and Iceland. Environ Sci Technol 54:4295−4304
Pluth MD, Tonzetich ZJ (2020) Hydrosulfide complexes of the transition elements: diverse roles in bioinorganic, cluster, coordination, and organometallic chemistry. Chem Soc Rev 49:4070–4134
Plyasunov AV (2020) An experimental study of the solubility and speciation of $MoO_3$(s) in hydrothermal fluids at temperatures up to 350°C – a discussion. Econ Geol 115:1871–1873

Plyasunov AV, Shock EL (2001) Correlation strategy for determining the parameters of the revised Helgeson-Kirkham-Flowers model for aqueous nonelectrolytes. Geochim Cosmochim Acta 65:3879−3900

Pokrovski GS (1996) Etude experimentale du comportement du germanium, du silicium et de l'arsenic et de la complexation de l'aluminium avec la silice dans les solutions naturelles. PhD Thesis, Université Paul-Sabatier, Toulouse, France

Pokrovski GS (2014) Use and misuse of chemical reactions and aqueous species distribution diagrams for interpreting metal transport and deposition in porphyry copper systems: Comment on Sun et al. (2013) "The link between reduced porphyry copper deposits and oxidized magmas, Geochim. Cosmochim. Acta 103, 263–275". Geochim Cosmochm Acta 126:635–638

Pokrovski GS (2025a) Thermodynamic modeling of hydrothermal ore deposit formation. Ore Geol Rev 178:106436

Pokrovski GS (2025b) Quantifying sulfur speciation in geological fluids at elevated temperatures and pressures. Extended abstracts of the SGA2025 Conference, Golden, Colorado (in press).

Pokrovski GS, Schott J (1998). Experimental study of the complexation of silicon and germanium with aqueous organic species: Implications for Ge and Si transport and the Ge/Si ratio in natural waters. Geochim Cosmochim Acta 62:3413–3428

Pokrovski GS, Dubrovinsky LS (2011) The $S_3^-$ ion is stable in geological fluids at elevated temperatures and pressures. Science 331:1052–1054

Pokrovski GS, Dubessy J (2015) Stability and abundance of the trisulfur radical ion $S_3^-$ in hydrothermal fluids. Earth Planet Sci Lett 411:298–309

Pokrovski GS, Schott J, Sergeyev AS (1995) Experimental determination of the stability constants of $NaSO_4^-$ and $NaB(OH)_4^0$ in hydrothermal solutions using a new sodium selective glass electrode. Implications for boron isotopic fractionation. Chem Geol 124:253–265

Pokrovski GS, Gout R, Schott J, Zotov AV, Harrichoury J-C (1996) Thermodynamic properties and stoichiometry of As(III) hydroxide complexes at hydrothermal conditions. Geochim Cosmochim Acta 60:737–749

Pokrovski GS, Bény J-M, Zotov AV (1999) Solubility and Raman spectroscopic study of As(III) speciation in organic compound-water solutions. A hydration approach for aqueous arsenic in complex solutions. J Solution Chem 28:1307–1327

Pokrovski GS, Zakirov IV, Roux J, Testemale D, Hazemann J-L, Bychkov AY, Golikova GV (2002a) Experimental study of arsenic speciation in vapor phase to 500°C: Implications for As transport and fractionation in low-density crustal fluids and volcanic gases. Geochim Cosmochim Acta 66:3453–3480

Pokrovski GS, Kara S, Roux J (2002b) Stability and solubility of arsenopyrite, FeAsS, in crustal fluids. Geochim Cosmochim Acta 66:2361–2378

Pokrovski GS, Roux J, Hazemann J-L, Testemale D (2005a) An X-ray absorption spectroscopy study of argutite solubility and germanium aqueous speciation in hydrothermal fluids to 500 °C and 400 bar. Chem Geol 217:127–145

Pokrovski GS, Roux J, Harrichoury J-C (2005b) Fluid density control on vapor-liquid partitioning of metals in hydrothermal systems. Geology 33:657–660

Pokrovski GS, Borisova AY, Roux J, Hazemann JL, Petdang A, Tella M, Testemale D (2006) Antimony speciation in saline hydrothermal fluids: A combined X-ray absorption fine structure spectroscopy and solubility study. Geochim Cosmochim Acta 70:4196–4214

Pokrovski GS, Borisova AY, Harrichoury J-C (2008a) The effect of sulfur on vapor-liquid fractionation of metals in hydrothermal systems. Earth Planet Sci Lett 266:345–362

Pokrovski GS, Roux J, Hazemann J-L, Borisova AY, Gonchar AA, Lemeshko MP (2008b) In situ X-ray absorption spectroscopy measurement of vapour-brine fractionation of antimony at hydrothermal conditions. Min Mag 72(2):667–681

Pokrovski GS, Tagirov BR, Schott J, Bazarkina EF, Hazemann J-L, Proux O (2009a) An in situ X-ray absorption spectroscopy study of gold-chloride complexing in hydrothermal fluids. Chem Geol 259:17–29

Pokrovski GS, Tagirov BR, Schott J, Hazemann J-L, Proux O (2009b) A new view on gold speciation in sulfur-bearing hydrothermal fluids from in situ X-ray absorption spectroscopy and quantum-chemical modeling. Geochim. Cosmochim. Acta, 73 2506–2527

Pokrovski GS, Borisova AY, Bychkov AY (2013a) Speciation and transport of metals and metalloids in geological vapors. Rev Miner Geochem 76:165–218

Pokrovski GS, Roux J, Ferlat G, Jonchiere R, Seitsonen AP, Vuilleumier R, Hazemann J (2013b) Siver in geological fluids from in situ X-ray absorption spectroscopy and first principles molecular dynamics. Geochim Cosmochim Acta 106:501–523

Pokrovski GS, Akinfiev NN, Borisova AY, Zotov AV, Kouzmanov K (2014) Gold speciation and transport in geological fluids: insights from experiments and physical-chemical modeling. In: Garofalo P, Ripley E (eds.), *Gold-Transporting Fluids in the Earth's Crust*. Geol Soc London Spec Publ 402:9–70

Pokrovski GS, Kokh MA, Guillaume D, Borisova AY, Gisquet P, Hazemann J-L, Lahera E, Del Net W, Proux O, Testemale D, Haigis V, Jonchière R, Seitsonen AP, Ferlat G, Vuilleumier R, Saitta AM, Boiron M-C, Dubessy J (2015) Sulfur radical species form gold deposits on Earth. Proc Nat Acad Sci USA 112(44):13484–13489

Pokrovski GS, Kokh MA, Proux O, Hazemann J-L, Bazarkina EF, Testemale D, Escoda C, Boiron M-C, Blanchard M, Aigouy T, Gouy S, de Parseval P, Thibaut M (2019) The nature and partitioning of invisible gold in the pyrite-fluid system. Ore Geol Rev 109:545–563

Pokrovski GS, Escoda C, Blanchard M, Testemale D, Hazemann J-L, Gouy S, Kokh MA, Boiron MC, de Parseval F, Aigouy T, Menjot L, de Parseval P, Proux O, Rovezzi M, Béziat D, Salvi S, Kouzmanov K, Bartsch T, Pöttgen R, Doert T (2021a) An arsenic-driven pump for invisible gold in hydrothermal systems. Geochem Persp Let 17:39–44

Pokrovski GS, Kokh MA, Desmaele E, Laskar C, Bazarkina EF, Borisova AY, Testemale D, Hazemann J-L, Vuilleumier R, Ferlat G, Saitta AM (2021b) The trisulfur radical ion $S_3^{\bullet-}$ controls platinum transport by hydrothermal fluids. Proc Nat Acad Sci USA 118:e2109768118

Pokrovski GS, Blanchard M, Poitrasson F, Saunier G (2021c) Mechanisms and rates of pyrite precipitation from hydrothermal fluids revealed by iron isotopes. Geochim Cosmochim Acta 304:281–304

Pokrovski GS, Sanchez-Valle C, Guillot S, Borisova AY, Muñoz M, Auzende A-L, Proux O, Roux J, Hazemann J-L, Testemale D, Shvarov YV (2022a) Redox dynamics of subduction revealed by arsenic in serpentinite. Geochem Persp Let 22:36–41

Pokrovski GS, Desmaele E, Laskar C, Bazarkina EF, Testemale D, Hazemann J-L, Vuilleumier R, Seitsonen AP, Ferlat G, Saitta AM (2022b) Gold speciation in hydrothermal fluids revealed by in situ high energy resolution X-ray absorption spectroscopy. Am Mineral 107:369–376

Pokrovski GS, Wilke M, Kokh MA (2024) Artificially high stability of the sodium hydromolybdate ion pair in aqueous solution. Comment on "Guan et al. (2023) Different metal coordination in sub- and super-critical fluids: Do molybdenum (IV) chloride complexes contribute to mass transfer in magmatic systems? Geochim Cosmochim Acta 354, 240–251". Geochim Cosmochim Acta 387:130–132

Pokrovski GS, Wilhelm F, Colin A, Kokh MA, Ledé B, Jacobs J, Garbarino G, Rogalev A (2025) Exploring sulfur speciation under high pressure using in situ X-ray absorption spectroscopy. Am Mineral, *submitted*

Popova MY, Khodakovsky IL, Ozerova NA (1975) Experimental determination of the thermodynamic properties hydroxo- and hydroxofluoride complexes of antimony at temperatures up to 200 °C. Geokhimiya 6:835–843

Prausnitz JM, Lichtenthaler RN, Gomes de Azeredo E. (1986) *Molecular Thermodynamics of Fluid-Phase Equilibria*. 2nd ed. Prentice Hall, New York

Prasetio R, Daud Y, Hutabarat J, Hendarmawan H (2020) Fluid geochemistry and isotope compositions to delineate physical processes in Wayang Windu geothermal reservoir, Indonesia. Geosci J 25:507–523

Pryor WA (1960) The kinetics of the disproportionation of sodium thiosulfate to sodium sulfide and sulfate. J Amer Chem Soc 82:4794–4797

Qiu Z, Fan H-R, Tomkins AG, Brugger J, Etschmann B, Liu X, Xing Y, Hu Y (2021) Insights into salty metamorphic fluid evolution from scapolite in the Trans-North China Orogen: Implication for ore genesis. Geochim Cosmochim Acta 293:256–276

Queizán M, Graña AM, Hermida-Ramón JM (2022) Achieving $S^{2-}$ in aqueous solution: An evaluation using first-principle molecular dynamics simulations. J Mol Liq 349:118109(1–7)

Quist AS, Marshall WL, Jolley HR (1965) Electrical conductances of aqueous solutions at high temperature and pressure. II. The conductances and ionization constants of sulfuric acid-water solutions from 0 to 800° and at pressures up to 4000 bars. J Phys Chem 69(8):2726–2735

Qureshi M, Nowak SH, Vogt LI, Cotelesage JJH, Dolgova NV, Sharifi S, Kroll T, Nordlund D, Alonso-Mori R, Weng T-C, Pickering IJ, George GN, Sokaras D (2021) Sulfur $K_\beta$ X-ray emission spectroscopy: Comparison with sulfur K-edge X-ray absorption spectroscopy for speciation of organosulfur compounds. Phys Chem Chem Phys 23:4500–4508

Rae AJ, Cooke DR, Brown KL (2011) The trace metal chemistry of deep geothermal water, Palinpinon geothermal field, Negros Island, Philippines: Implications for precious metal deposition in epithermal gold deposits. Econ Geol 106:1425–1446

Rankin AH, Ramsey MH, Coles B, Vanlangevelde F, Thomas CR (1992) The composition of hypersaline, iron-rich granitic fluids based on laser-ICP and synchrotron-XRF microprobe analysis of individual fluid inclusions in topaz, Mole Granite, eastern Australia. Geochim Cosmochim Acta 56:67–79

Rau H, Kutty TRN, Guedes de Carvalho JRF (1973) Thermodynamics of sulphur vapour. J Chem Thermodyn 5:833–844

Rauchenstein-Martinek K, Wagner T, Wälle M, Heinrich CA (2014) Gold concentrations in metamorphic fluids: a LA-ICPMS study of fluid inclusions from the Alpine orogenic belt. Chem Geol 385:70–83

Rauchenstein-Martinek K, Wagner T, Wälle M, Heinrich CA, Arlt T (2016) Chemical evolution of metamorphic fluids in the Central Alps, Switzerland: insight from LA-ICPMS analysis of fluid inclusions. Geofluids 16:877–908

Raulin K, Gobeltz N, Vezin H, Touati N, Ledé B, Moissette A (2011) Identification of the EPR signal of $S_2^-$ in green ultramarine pigments. Phys Chem Chem Phys 13:9253–9259

Readnour JM, Cobble JW (1969) Thermodynamic properties for the dissociation of bisulfate ion and the partial molal heat capacities of bisulfuric acid and sodium bisulfate over an extended temperature range. Inorg Chem 8:2174–2182

Reed M, Rusk B, Palandri J (2013) The Butte magmatic-hydrothermal system: one fluid yields all alteration and veins. Econ Geol 108:1376–1396

Rehkämper M, Halliday AN, Barfod D, Fitton JG, Dawson JB (1997) Platinum group element abundance patterns in different mantle environments. Science 278:1595–1598

Reimer J, Steele-MacInnis M, Wambach JM, Vogel F (2015) Ion association in hydrothermal sodium sulfate solutions studied by modulated FT-IR-Raman spectroscopy and molecular dynamics. J Phys Chem B 119:9847–9857

Reinen D, Lindner G-G (1999) The nature of the chalcogen colour centres in ultramarine-type solids. Chem Soc Rev 28:75–84

Reith F, Campbell SG, Ball AS, Pring A, Southam G (2014) Platinum in Earth surface environments. Earth Sci Rev 131:1–21

Rejmak P (2018) Structural, optical, and magnetic properties of ultramarine pigments: A DFT insight. J Phys Chem C 112:29338−29349

Rejmak P (2020) Computational refinement of the puzzling red tetrasulfur chromophore in ultramarine pigments. Phys Chem Chem Phys 22:22684–22698

Renders PJ, Seward TM (1989) The stability of hydrosulphido- and sulphido-complexes of Au(I) and Ag(I) at 25°C. Geochim Cosmochim Acta 53:245–253

Rhee J-S, Dasgupta PK (1985) The second dissociation constant of $SO_2 \bullet H_2O$. J Phys Chem 89:1799–1804

Ricci A., Kleine B.I., Fiebig J., Gunnarsson-Robin J., Kamunya K.M., Mountain B., Stefánsson A. (2022) Equilibrium and kinetic controls on molecular hydrogen abundance and hydrogen isotope fractionation in hydrothermal fluids. Earth Planet Sci Lett 579:117338

Richard A, Cathelineau M, Boiron MC, Mercadier J, Banks DA, Cuney M (2016) Metal-rich fluid inclusions provide new insights into unconformity-related U deposits (Athabasca basin and basement, Canada). Min Deposita 51:249–270

Rickard D, Luther III GW (2006) Metal sulfide complexes and clusters. Rev Miner Geochem 61:421–504

Rickard D, Luther III GW (2007) Chemistry of iron sulfides. Chem Rev 107:514–562

Ridley JR, Diamond LW (2000) Fluid chemistry of orogenic lode-gold deposits and implications for genetic models. Rev Econ Geol 13:141–162

Risberg ED, Eriksson L, Mink J, Pettersson LGM, Skripkin MY, Sandström M (2007) Sulfur X-ray absorption and vibrational spectroscopic study of sulphur dioxide, and sulfonate solutions and of the substituted sulfonate ions $X_3CSO_3^-$ (X = H, Cl, F). Inorg Chem 46:8332–8348

Robie RA, Hemingway BS (1995) Thermodynamic properties of minerals and related substances at 298.15 K and 1 bar ($10^5$ Pascals) pressure and at higher temperatures. US Geol Survey Bull 2131:1−461

Robinson BW (1973) Sulphur isotope equilibrium during sulphur hydrolysis at high temperatures. Earth Planet Sci Lett 18:443–450

Roedder E (1990) Fluid inclusion analysis - prologue and epilogue. Geochim Cosmochim Acta 54:495–507

Roedder E (ed) (1984) Fluid inclusions. Rev Mineral 12:1–644

Roedder E, Bodnar RJ (1980) Geologic pressure determinations from fluid inclusion studies. Ann Rev Earth Planet Sci 8:263–301

Rosasco GJ, Roedder E (1979) Application of a new Raman microprobe spectrometer to nondestructive analysis of sulfate and other ions in individual phases in fluid inclusions in minerals. Geochim Cosmochim Acta 43:1907–1915

Rudnick RL, Gao S (2014) Composition of the continental crust. In: *Treatise on Geochemistry*, 2nd edition, 4:1–51

Rudolph W (1996) Structure and dissociation constant of the hydrogen sulphate ion in aqueous solution over a broad temperature range: a Raman study. Zeit Phys Chem 194:73–95

Rumpf B, Maurer G (1992) Solubilities of hydrogen cyanide and sulfur dioxide in water at temperatures from 293.15 to 413.15 K and pressures up to 2.5 MPa. Fluid Phase Equil 81:241–260

Ryan CG, Cousens DR, Heinrich CA, Griffin WL, Sie SH, Mernagh TP (1991) Quantitative PIXE microanalysis of fluid inclusions based on a layered yield model. Nucl Instr Met Phys Res B54:229–297

Ryan CG, Etschmann BE, Vogt S, Maser J, Harland CL, van Achterbergh E, Legnini D (2005) Nuclear microprobe – synchrotron synergy: Towards integrated quantitative real-time elemental imaging using PIXE and SXRF. Nucl Instr Meth Phys Res B231:183–188

Ryzhenko BN (1964) Determination of the second dissociation constant of sulfuric acid, and precipitation of salts in the reciprocal system $Ca^{2+}$, $Ba^{2+}$ - $SO_4^{2-}$, $CO_3^{2-}$ under hydrothermal conditions. Geokhimia 1964(1):23–50 (English translation)

Ryzhenko BN (1981) *Equilibria in Hydrothermal Solutions*. Nauka, Moscow, 191 p. (in Russian)

Sadove G, Konecke BA, Fiege A, Simon AC (2019) Structurally bound $S^{2-}$, $S^{1-}$, $S^{4+}$, $S^{6+}$ in terrestrial apatite: The redox evolution of hydrothermal fluids at the Phillips mine, New York, U.S.A. Ore Geol Rev 107: 1084–1096

Sahle CJ, Mirone A, Niskanen J, Inkinen J, Krisch M, Huotari S (2015) Planning, performing and analyzing X-ray Raman scattering experiments. J Synchrotron Rad 22:400–409

Sakai H, Dickson FW (1978) Experimental determination of the rate and equilibrium fractionation factors of sulfur isotope exchange between sulfate and sulfide in slightly acid solutions at 300 °C and 1000 bars. Earth Planet Sci Lett 39:151–161

Sakashita M, Fujihisa H, Yamawaki H, Aoki K (2000) Molecular dissociation in deuterium sulfide under high pressure: Infrared and Raman study. J Phys Chem A 104:8838–8842

Sassani DC, Shock EL (1998) Solubility and transport of platinum-group elements in supercritical fluids: Summary and estimates of thermodynamic properties for ruthenium, rhodium, palladium, and platinum solids, aqueous ions, and complexes to 1000°C and 5 kbar. Geochim Cosmochim Acta 62:2643–2671

Saunier G, Pokrovski GS, Poitrasson F (2011) First experimental determination of iron isotope fractionation between hematite and aqueous solution at hydrothermal conditions. Geochim Cosmochim Acta 75:6629–6654

Scheuermann PP, Tutolo BM, Seyfried WE (2019) Anhydrite solubility in low-density hydrothermal fluids: Experimental measurements and thermodynamic calculations. Chem Geol 524: 184–195

Schiavi F, Bolfan-Casanova N, Buso R, Laumonier M, Laporte D, Medjoubi K, Venugopal S, Gomez-Ulla A, Cluzel N, Hardiagon M. (2020) Quantifying magmatic volatiles by Raman microtomography of glass inclusion-hosted bubbles. Geochem Persp Let 16:17–24

Schirra M, Laurent O, Zwyer T, Driesner T, Heinrich CA (2022) Fluid evolution at the Batu Hijau porphyry Cu-Au deposit, Indonesia: Hypogene sulfide precipitation from a single-phase aqueous magmatic fluid during chlorite–white mica alteration. Econ Geol 117:979–1012

Schmidt C (2009) Raman spectroscopic study of a $H_2O$+$Na_2SO_4$ solution at 21–600°C and 0.1 MPa to 1.1 GPa: Relative differential $\nu_1$-$SO_4^{2-}$ Raman scattering cross sections and evidence of the liquid-liquid transition. Geochim Cosmochim Acta 73:425–437

Schmidt C, Seward TM (2017) Raman spectroscopic quantification of sulfur species in aqueous fluids: Ratios of relative molar scattering factors of Raman bands of $H_2S$, $HS^-$, $SO_2$, $HSO_4^-$, $SO_4^{2-}$, $S_2O_3^{2-}$, $S_3^-$ and $H_2O$ at ambient conditions and information on changes with pressure and temperature. Chem Geol 467:64–75

Schmidt C, Jahn S (2024) Raman spectra of oxidized sulfur species in hydrothermal fluids. J Volcan Geotherm Res 454:108146

Schmidt K, Koschinsky A, Garbe-Schönberg D, de Carvalho LM, Seifert R (2007) Geochemistry of hydrothermal fluids from the ultramafic-hosted Logatchev hydrothermal field, 15°N on the Mid-Atlantic Ridge: Temporal and spatial investigation. Chem Geol 242:1–21

Scholten L, Schmidt C, Lecumberri-Sanchez P, Newville M, Lanzirotti A, Sirbescu M-LC, Steele-MacInnis M. (2019) Solubility and speciation of iron in hydrothermal fluids. Geochim Cosmochim Acta 252:126–143

Schoonen MAA, Barnes HL (1988) An approximation of the second dissociation constant for $H_2S$. Geochim Cosmochim Acta 52:649–654

Schoonen MA, Barnes HL (1991) Mechanisms of pyrite and marcasite formation from solution: III. Hydrothermal processes. Geochim Cosmochim Acta 55:3491–3504

Schulte MD, Rogers KL (2004) Thiols in hydrothermal solution: standard partial molal properties and their role in the organic geochemistry of hydrothermal environments. Geochim Cosmochim Acta 68:1087–1097

Schulte MD, Shock EL, Wood RH (2001) The temperature dependence of the standard-state thermodynamic properties of aqueous nonelectrolytes. Geochim Cosmochim Acta 65:3919–3930

Schwamm RJ, Lein M, Coles MP, Fitchett CM (2017) Bismuth(III) complex of the $[S_4]^{\bullet-}$ radical anion: dimer formation via pancake bonds. J Am Chem Soc 139:16490–16493

Schwarzenbach EM, Evans K (2025) Behavior of sulfur during subduction of oceanic lithosphere. This Book

Schwarzenbach G, Fischer A (1960) Die Acidität der Sulfane und die Zusammensetzung wässeriger Polysulfidlösungen. Helv Chim Acta 43:1365–1390

Schwarzenbach G, Widmer G (1966) Die Löslichkeit von Metalsulfiden. II. Silbersulfid. Helv Chim Acta 49:111–123.

Schwarzenbach G, Gübeli O, Züst H (1958) Thiokomplexe des Silbers und die Löslichkeit von Silbersulfid. Chimia 12:84–86

Selleck FT, Carmichael LT, Sage BH (1952) Phase behavior of the hydrogen sulfide-water system. Ind Eng Chem 44:2219–2226

Seo JH, Heinrich CA (2013) Selective copper diffusion into quartz-hosted vapor inclusions: Evidence from other host minerals, driving forces, and consequences for Cu-Au ore formation. Geochim Cosmochim Acta 113:60–69

Seo JH, Guillong M, Heinrich CA (2009) The role of sulfur in the formation of magmatic-hydrothermal copper-gold deposits. Earth Planet Sci Lett 282:323–328

Seo JH, Guillong M, Aerts M, Zajacz Z, Heinrich CA (2011) Microanalysis of S, Cl, and Br in fluid inclusions by LA–ICP-MS. Chem Geol 284:35–44

Seo JH, Guillong M, Heinrich CA (2012) Separation of molybdenum and copper in porphyry deposits: the roles of sulfur, redox, and pH in in ore mineral deposition at Bingham Canyon. Econ Geol 107:333–356

Seyfried WE, Pester NJ, Tutolo BM, Ding K (2015) The Lost City hydrothermal system: Constraints imposed by vent fluid chemistry and reaction path models on subseafloor heat and mass transfer processes. Geochim Cosmochim Acta 163:59–79

Seward TM (1973) Thio complexes of gold and the transport of gold in hydrothermal ore solutions. Geochim Cosmochim Acta 37:379–399

Seward TM (1976) The stability of chloride complexes of silver in hydrothermal solutions up to 350 °C. Geochim Cosmochim Acta 40:1329–1341

Seward TM (1981) Metal complex formation in aqueous solutions at elevated temperatures and pressures. In: Physics and Chemiswy of the Earth (eds. F. Wickman and D. Rickard), 13–14:113–129. Pergamon, Oxford

Seward TM (1984) The formation of lead (II) chloride complexes to 300 °C - a spectrophotometric study. Geochim Cosmochim Acta 48:121–134

Shang L, Williams-Jones AE, Wang X, Timofeev A, Hu R, Bi X (2020) An experimental study of the solubility and speciation of $MoO_3$(s) in hydrothermal fluids at temperatures up to 350°C. Econ Geol 115:661–669

Sharygin AV, Inglese A, Sedlbauer J, Wood RH (1997) Apparent molar heat capacities of aqueous solutions of phosphoric acid and sulfur dioxide from 303 to 623 K and a pressure of 28 MPa. J Sol Chem 26:183–197

Sheldrick WS (2013) Polychalcogenides. in: *Handbook of Chalcogen Chemistry: New Perspectives in Sulfur, Selenium and Tellurium*, eds. FA Devillanova and WW du Mont. RSC Publishing, 2nd ed., vol. 1, chap. 9.2:514–545

Shenberger DM, Barnes HL (1989) Solubility of gold in aqueous sulfide solutions from 150 to 350 °C. Geochim Cosmochim Acta 53:269–278

Sherman DM (2007) Complexation of $Cu^+$ in hydrothermal NaCl brines: ab initio molecular dynamics and energetics. Geochim Cosmochim Acta 71:714–722

Shock EL, Helgeson HC (1988) Calculation of the thermodynamic and transport properties of aqueous species at high pressures and temperatures: correlation algorithms for ionic aqueous species and equation of state predictions to 5 kb and 1000 °C. Geochim Cosmochim Acta 52:2009–2036

Shock EL, Helgeson HC (1990) Calculation of the thermodynamic and transport properties of aqueous species at high pressures and temperatures: Standard partial molal properties of organic aqueous species. Geochim Cosmochim Acta 54:915–945

Shock EL, Koretsky CM (1993) Metal-organic complexes in geochemical processes: Calculation of standard partial molal thermodynamic properties of aqueous acetate complexes at high pressures and temperatures. Geochim Cosmochim Acta 57:4899–4922

Shock EL, Helgeson HC, Sverjensky DA (1989) Calculation of the thermodynamic and transport properties of aqueous species at high pressures and temperatures: Standard partial molal properties of inorganic neutral species. Geochim. Cosmochim.Acta 53:2157–2183

Shock EL, Koretsky CM (1995) Metal-organic complexes in geochemical processes: Estimation of standard partial molal thermodynamic properties of aqueous complexes between metal cations and monovalent organic acid ligands at high pressures and temperatures. Geochim. Cosmochim. Acta 59:1497–1532

Shock EL, Oelkers EH, Johnson JW, Sverjensky DA, Helgeson HC (1992) Calculation of the thermodynamic and transport properties of aqueous species at high pressures and temperatures: Effective electrostatic radii, dissociation constants and the standard molal properties to 1000 °C and 5 kbar. J Chem Soc Faraday Trans 88:803–826

Shock EL, Sassani DC, Willis M, Sverjensky DA (1997) Inorganic species in geologic fluids: Correlations among standard molal thermodynamic properties of aqueous ions and hydroxide complexes. Geochim Cosmochim Acta 61:907–950

Shvarov YV (2008) HCh: New potentialities for the thermodynamic simulation of geochemical systems offered by windows. Geochem Intl 46:834–839

Shvarov YV (2015) A suite of programs, OptimA, OptimB, OptimC, and OptimS, compatible with the Unitherm database, for deriving the thermodynamic properties of aqueous species from solubility, potentiometry and spectroscopy measurements. Appl Geochem 55:17–27. Programs are available at http://www.geol.msu.ru/deps/geochems/soft/index_e.html

Sillitoe RH (2010) Porphyry copper systems. Econ Geol 105:3–41

Simmons SF, Brown KL, Tutolo BM (2016) Hydrothermal transport of Ag, Au, Cu, Pb, Te, Zn, and other metals and metalloids in New Zealand geothermal systems: spatial patterns, fluid-mineral equilibria, and implications for epithermal mineralization. Econ Geol 111:589–618

Simmons SF, Tutolo B, Barker SL, Goldfarb R, Robert F (2020) Hydrothermal gold deposition in epithermal, Carlin, and orogenic deposits geothermal resources. Soc Econ Geol Spec Pub 23:823–845

Simon AC, Pettke T (2009) Platinum solubility and partitioning in a felsic melt–vapor–brine assemblage. Geochim Cosmochim Acta 73:438–454

Simon AC, Wilke M (2025) The behavior of sulfur in silicate magmas. This Book

Simon AC, Pettke T, Candela PA, Piccoli PM, Heinrich CA (2004) Magnetite solubility and iron transport in magmatic-hydrothermal environments. Geochim Cosmochim Acta 68:4905–4914

Singh R (2002) C. V. Raman and the discovery of the Raman effect. Phys Persp 4:399–420

Skirrow RC, Walshe JL (2002) Reduced and oxidized Au-Cu-Bi iron oxide deposits of the Tennant Creek inlier, Australia: An integrated geologic and chemical model. Econ Geol 97:1167–1202

Smith RW, Popp CJ, Norman DI (1986) The dissociation of oxy-acids at elevated-temperatures. Geochim Cosmochim Acta 50:137–142

Smith RM, Martell AE, Motekaitis RJ (2004) *NIST Critically Selected Stability Constants of Metal Complexes Database*, version 8.0, NIST, Gaithersburg, USA

Song P, Rao W, Chivers T, Wang S-Y (2023) Applications of trisulfur radical ion $S_3^{\bullet-}$ in organic synthesis. Org Chem Front 10:3378–3400

Song Y, Liu Z, Mao H-K, Hemley RJ, Herschbach DR (2005) High-pressure vibrational spectroscopy of sulfur dioxide. J Chem Phys 122:174511

Sorby HC (1869) On the microscopical structure of some precious stones. Month Microscop J 1:220–224

Sośnicka M, Lüders V (2021) Phase transitions in natural C-O-H-N-S fluid inclusions - implications for gas mixtures and the behavior of solid $H_2S$ at low temperatures. Nat Comm 12:6975

Sośnicka M, Lüders V, Duschl F, Kraemer D, Laurent O, Niedermann S, Banks DA, Wilke F, Wohlgemuth-Ueberwasser C, Wiedenbeck M (2023) Metal budget and origin of aqueous brines depositing deep-seated Zn-Pb mineralization linked to hydrocarbon reservoirs, North German Basin. Miner Deposita 58:1143–1170

Spycher NF, Reed MH (1989) As(III) and Sb(III) sulfide complexes – an evaluation of stoichiometry and stability from existing experimental data. Geochim Cosmochim Acta 53:2185–2194

Spycher NF, Reed MH (1990a) Reply to comments by R. E. Krupp on "As(III) and Sb(III) sulfide complexes: An evaluation of stoichiometry and stability from existing experimental data". Geochim Cosmochim Acta 54:3241–3243

Spycher NF, Reed MH (1990b) Response to the response of R. E. Krupp. Geochim Cosmochim Acta 54:3246

Sretenskaya NG (1977) Dissociation of hydrogen sulfide under pressure. Geokhimia 1977(3):430–438

Stefánsson A (2017) Gas chemistry of Icelandic thermal fluids. J Volcanol Geotherm Res 346:81–94

Stefánsson A, Arnórsson S (2000) Feldspar saturation state in natural waters. Geochim Cosmochim Acta 64:2567–2584

Stefánsson A, Arnórsson S (2002) Gas pressures and redox reactions in geothermal systems in Iceland. Chem Geol 190:251–271

Stefánsson A, Seward TM (2003a) Experimental determination of the stability and stoichiometry of sulphide complexes of silver(I) in hydrothermal solutions to 400°C. Geochim Cosmochim Acta 67:1395–1413

Stefánsson A, Seward TM (2003b) The hydrolysis of gold(I) in aqueous solutions to 600°C and 1500 bar. Geochim Cosmochim Acta 67:1677–1688

Stefánsson A, Seward TM (2003c) Stability of chloridogold(I) complexes in aqueous solutions from 300 to 600°C and from 500 to 1800 bar. Geochim Cosmochim Acta 67:4559–4576

Stefánsson A, Seward TM (2004) Gold(I) complexing in aqueous sulphide solutions to 500°C at 500 bar. Geochim Cosmochim Acta 68:4121–4143

Stefánsson A, Seward TM (2008) A spectrophotometric study of iron(III) hydrolysis in aqueous solutions to 200 °C. Chem Geol 249:227–235

Stefánsson A, Driesner T, Bénézeth P (eds) (2013) *Thermodynamics of Geothermal Fluids*. Rev Miner Geochem 76:1–350, Miner. Soc. Amer. and Geochem. Soc.

Stefánsson A, Keller NS, Gunnarsson Robin J, Ono S (2015) Multiple sulfur isotope systematics of Icelandic geothermal fluids and the source and reactions of sulfur in volcanic geothermal systems at divergent plate boundaries. Geochim Cosmochim Acta 165:307–323

Stefánsson A, Keller NS, Gunnarsson-Robin J, Kaasalainen H, Björnsdóttir S, Pétursdóttir S, Jóhannesson H, Hreggvidsson GÓ (2016) Quantifying mixing, boiling, degassing, oxidation and reactivity of thermal waters at Vonarskard, Iceland. J Volcan Geotherm Res 309:53–62

Stephens HP, Cobble JW (1971) Thermodynamic properties of the aqueous sulfide and bisulfide ions and the second ionization constant of hydrogen sulfide over extended temperatures. Inorg Chem 10:619–625

Steudel R (2003) *Elemental Sulfur and Sulfur-Rich Compounds I, II.* Springer, Berlin

Steudel R (2020) The chemical sulfur cycle. In: *Environmental Technologies to Treat Sulfur Pollution: Principles and Engineering*, 2nd edn., PNL Lens (ed.), Chap.2:11–53

Steudel R, Chivers T (2019) The role of polysulfide dianions and radical anions in the chemical, physical and biological sciences, including sulfur-based batteries. Chem Soc Rev 48:3279–3319

Steudel R, Holdt G, Göbel T, Hazeu W (1987) Chromatographic separation of higher polythionates $[S_nO_6]^-$ ($n$ = 3–22) and their detection in cultures of *thiobacillus ferrooxidans*; molecular composition of bacterial sulfur secretions. Angew Chem Int Ed 26:151–153

Steudel R, Prenzel A (1989) Raman spectroscopic discovery of the hydrogenthiosulphate anion, $HSSO_3^-$, in solid $NH_4HS_2O_3$. Zeit Naturforsch B 44:1499–1502

Steudel R, Steudel Y (2009a) Sulfur dioxide and water: structures and energies of the hydrated species $SO_2•nH_2O$, $[HSO_3]^-•nH_2O$, $[SO_3H]^-•nH_2O$, and $H_2SO_3•nH_2O$ ($n$ = 0–8). Eur J Inorg Chem 2009:1393–1405

Steudel R, Steudel Y (2009b) Microsolvation of thiosulfuric acid and its tautomeric anions $[HSSO_3]^-$ and $[SSO_2(OH)]^-$ studied by B3LYP-PCM and G3X(MP2) calculations. J. Phys. Chem. A 113:9920–9933

Stöhr J (1992) NEXAFS Spectroscopy. Springer Series in Surface Sciences, vol. 25

Stumm W, Morgan JJ (1996) *Aquatic Chemistry: Chemical Equilibria and Rates in Natural Waters*. 3rd Edn., Wiley & Sons, NJ, USA, 1022 p.

Sue K, Uchida M, Adschiri T, Arai K (2004) Determination of sulfuric acid first dissociation constant to 400 °C and 32 MPa by potentiometric pH measurements. J Supercrit Fluids 31:295–299

Sugaki A, Scott SD, Hayashi K, Kitakaze A (1987) $Ag_2S$ solubility in sulfide solutions up to 250°C. Geochem J 21:291–305

Suleimenov OM, Krupp RE (1994) Solubility of hydrogen sulfide in pure water and in NaCl solutions, from 20 to 320°C and at saturation pressures. Geochim Cosmochim Acta 58:2433–2444

Suleimenov OM, Seward TM (1997) A spectrophotometric study of hydrogen sulphide ionization in aqueous solutions to 350°C. Geochim Cosmochim Acta 61:5187–5198

Suleimenov OM, Ha T-K (1998) Ab initio calculations of the thermochemical properties of polysulphanes ($H_2S_n$). Chem Phys Lett 290:451–457

Sullivan NA, Zajacz Z, Brenan JM, Hinde JC, Tsay A, Yin Y (2022a) The solubility of gold and palladium in magmatic brines: Implications for PGE enrichment in mafic-ultramafic and porphyry environments. Geochim Cosmochim Acta 316:230–252

Sullivan NA, Zajacz Z, Brenan JM, Tsay A (2022b) The solubility of platinum in magmatic brines: Insights into the mobility of PGE in ore-forming environments. Geochim Cosmochim Acta 316:253–272

Sullivan PD (1968) Hyperfine splitting from naturally occurring sulfur-33 in electron paramagnetic resonance spectra. J Am Chem Soc 90(14):3618–3622

Sulpizi M, Sprik M (2008) Acidity constants from vertical energy gaps: density functional theory based molecular dynamics implementation. Phys Chem Chem Phys 10:5238–5249

Syverson DD, Borrok DM, Seyfried WE (2013) Experimental determination of equilibrium Fe isotopic fractionation between pyrite and dissolved Fe under hydrothermal conditions. Geochim Cosmochim Acta 122:170–183

Syverson DD, Ono S, Shanks WC, Seyfried WE Jr (2015) Multiple sulfur isotope fractionation and mass transfer processes during pyrite precipitation and recrystallization: An experimental study at 300 and 350 °C. Geochim Cosmochim Acta 165:418–434

Sverjensky DA (2025) Sulfur speciation matters. Nat Geosci 18:3–4

Sverjensky DA, Shock EL, Helgeson HC (1997) Prediction of the thermodynamic properties of aqueous metal complexes to 1000 °C and 5 kb. Geochim Cosmochim Acta 61:1359–1412

Sverjensky DA, Harrison B, Azzolini D (2014) Water in the deep Earth: The dielectric constant and the solubilities of quartz and corundum to 60 kb and 1200 °C. Geochim Cosmochim Acta 129:125–145

Svoronos PDN, Bruno TJ (2002) Carbonyl sulfide: a review of its chemistry and properties. Ind Eng Chem Res 41:5321–5336

Sweeton FH, Baes SF Jr (1970) The solubility of magnetite and hydrolysis of ferrous ion in aqueous solutions at elevated temperatures. J Chem Thermodyn 2:479–500

Tagirov BR, Schott J (2001) Aluminum speciation in crustal fluids revisited. Geochim Cosmochim Acta 65:3965–3992.

Tagirov BR, Seward TM (2010) Hydrosulfide/sulfide complexes of zinc to 250°C and the thermodynamic properties of sphalerite. Chem Geol 269:301–311

Tagirov BR, Zotov AV, Akinfiev NN (1997) Experimental study of the dissociation of HCl from 350 to 500 °C and from 500 to 2500 bar. Thermodynamic properties of $HCl^0$ (aq). Geochim Cosmochim Acta 61:4267–4280

Tagirov BR, Salvi S, Schott J, Baranova NN (2005) Experimental study of gold-hydrosulphide complexing in aqueous solutions at 350-500°C, 500 and 1000 bars using mineral buffers. Geochim Cosmochim Acta 69:2119–2132

Tagirov BR, Baranova NN, Zotov AV, Schott J, Bannykh LN (2006) Experimental determination of the stabilities $Au_2S$(cr) at 25°C and $Au(HS)_2^-$ at 25–250°C. Geochim Cosmochim Acta 70:3689–3701

Tagirov BR, Suleimenov OM, Seward TM (2007) Zinc complexation in aqueous sulfide solutions: determination of the stoichiometry and stability of complexes via $ZnS_{(cr)}$ solubility measurements at 100 °C and 150 bars. Geochim Cosmochim Acta 71:4942–4953

Tagirov BR, Baranova NN, Zotov AV, Akinfiev NN, Polotnyanko NA., Shikina ND, Shvarov YV, Bastrakov EV (2013) The speciation and transport of palladium in hydrothermal fluids: experimental modeling and thermodynamic constraints. Geochim Cosmochim Acta 117:348–373

Tagirov BR, Baranova NN, Bychkova YV (2015) Thermodynamic properties of platinum chloride complexes in aqueous solutions: Derivation of consistent parameters from literature data and experiments on $Pt_{(cr)}$ solubility at 400–475 °C and 1 kbar. Geochem Intl 53(4):327–340

Tagirov BR, Trigub AL, Filimonova ON, Kvashnina KO, Nickolsky MS, Lafuerza S, Chareev DA (2019a) Gold transport in hydrothermal chloride-bearing fluids: Insights from in situ X-ray absorption spectroscopy and ab initio molecular dynamics. ACS Earth Space Chem 3:240−261

Tagirov BR, Filimonova ON, Trigub AL, Akinfiev NN, Nickolsky MS, Kvashnina KO, Chareev DA, Zotov AV (2019b) Platinum transport in chloride-bearing fluids and melts: insights from in situ X-ray absorption spectroscopy and thermodynamic modeling. Geochim Cosmochim Acta 254:86–101

Taillefert M, Rozan TF (eds.) (2002) *Electrochemical Methods for the Environmental Analysis of Trace Elements Biogeochemistry*. American Chemical Society, Washington, DC

Takano B (1987) Correlation of volcanic activity with sulfur oxyanion speciation in a crater lake. Science 235:1633–1635

Tang YH, Han CM, Bao ZK, Huang YY, He W, Hua W (2005) Analysis of apatite crystals and their fluid inclusions by synchrotron radiation X-ray fluorescence microprobe. Spectrochim Acta B60:439–446

Tanger JC, Helgeson CH (1988) Calculation of the thermodynamic and transport properties of aqueous species at high pressures and temperatures; revised equations of state for the standard partial molal properties of ions and electrolytes. Amer J Sci 288(1):19–98

Tantardini C, Oganov AR (2021) Thermochemical electronegativities of the elements. Nat Comm 12:2087

Tauson VL, Goettlicher J, Sapozhnikov AN., Mangold S, Lustenberg EE (2012) Sulphur speciation in lazurite-type minerals $(Na,Ca)_8[Al_6Si_6O_{24}](SO_4,S)_2$ and their annealing products: a comparative XPS and XAS study. Eur J Mineral 24:133–152

Tella M, Pokrovski GS (2009) Antimony(III) complexing with O-bearing organic ligands in aqueous solution: An X-ray absorption fine structure spectroscopy and solubility study. Geochim Cosmochim Acta 73(2):268–290

Tella M, Pokrovski GS (2012) Stability and structure of pentavalent antimony complexes with aqueous organic ligands. Chem Geol 292–293:57–68

Tello E, Verma MP, Tovar R (2000) Origin of acidity in the Los Humeros, Mexico, geothermal reservoir. Proc World Geothermal Congres 2000: 2956–2967

Teo BK (1986) EXAFS: *Basic Principles and Data Analysis*. Inorganic Chemistry Concepts, vol 9, Springer

Testemale D, Brugger J, Liu W, Etschmann B, Hazemann J-L (2009) In situ X-ray absorption study of iron(II) speciation in brines up to supercritical conditions. Chem Geol 264:295–310

Testemale D, Pokrovski GS, Lahera E, Prat A, Kieffer I, Delnet W, Proux O, Louvel M, Sanchez-Valle C, Hazemann J-L (2024) In situ X-ray absorption spectroscopy using the FAME autoclave: a window into fluid-mineral-melt interactions in the Earth's crust. High Pressure Res 104:277–293

Toland WG (1960) Oxidation of organic compounds with aqueous sulfate. J Amer Chem Soc 82:1911–1916

Tomkins AG (2010) Windows of metamorphic sulfur liberation in the crust: Implications for gold deposit genesis. Geochim Cosmochim Acta 74:3246–3259

Tooth B, Brugger J, Ciobanu CL, Liu W (2008) Modeling of gold scavenging by bismuth melts coexisting with hydrothermal fluids. Geology 36:815–818

Tooth B, Ciobanu CL, Green L, O'Neill B, Brugger J (2011) Bi-melt formation and gold scavenging from hydrothermal fluids: an experimental study. Geochim Cosmochim Acta 75:5423–5443

Tooth B, Etschmann B, Pokrovski GS, Testemale D, Hazemann J-L, Grundler PV, Brugger J (2013) Bismuth speciation in hydrothermal fluids: an X-ray absorption spectroscopy and solubility study. Geochim Cosmochim Acta 101:156–172

Tossell JA (2012) Calculation of the properties of the $S_3^-$ radical anion and its complexes with $Cu^+$ in aqueous solution. Geochim Cosmochim Acta 95:79–92

Tossell JA, Zimmermann MD (2008) Calculation of the structures, stabilities and vibrational spectra of arsenites, thioarsenites and thioarsenates in aqueous solution. Geochim Cosmochim Acta 72:5232–5242

Tremaine PR, LeBlanc JC (1980) The solubility of magnetite and the hydrolysis and oxidation of $Fe^{2+}$ in water to 300 °C. J Soln Chem 9:415–442

Trigub AL, Tagirov BR, Kvashnina KO, Lafuerza S, Filimonova ON, Nickolsky MS (2017) Experimental determination of gold speciation in sulfide-rich hydrothermal fluids under a wide range of redox conditions. Chem Geol 471:52–64

Trofimov ND, Tagirov BR, Akinfiev NN, Reukov VL, Nickolsky MS, Nikolaeva IY, Tarnopolskaya ME, Afanasyev AA (2023) Chalcocite $Cu_2S$ solubility in aqueous sulfide and chloride fluids. Thermodynamic properties of copper(I) aqueous species and copper transport in hydrothermal systems. Chem Geol 625:121413

Truche L, Bazarkina EF, Barre G, Thomassot E, Berger G, Dubessy J, Robert P (2014) The role of $S_3^-$ ion in thermochemical sulphate reduction: geological and geochemical implications. Earth Planet Sci Lett 396:190–200

Tsonopoulos C, Coulson DM, Inman LB (1976) Ionization constants of water pollutants. J Chem Eng Data 21:190–193

Turner DR, Whitefield M, Dickson AG (1981) The equilibrium speciation of dissolved components in freshwater and seawater at 25°C and 1 atm pressure. Geochim Cosmochim Acta 45:855–881

Ueno Y, Johnson MS, Daniel SO, Eskebjerg C, Pandey A, Yoshida N. (2009) Geological sulfur isotopes indicate elevated OCS in the Archean atmosphere, solving faint young sun paradox. Proc Nat Acad Sci USA 106(35): 14784–14789

Ulrich T, Mavrogenes J. (2008) An experimental study of the solubility of molybdenum in $H_2O$ and KCl-$H_2O$ solutions from 500 °C to 800 °C, and 150 to 300 MPa. Geochim Cosmochim Acta 72:2316–2330

Ullrich MK, Pope JG, Seward TM, Wilsona N, Planer-Friedrich B (2013) Sulfur redox chemistry governs diurnal antimony and arsenic cycles at Champagne Pool, Waiotapu, New Zealand. J Volcanol Geotherm Res 262:164–177

Uyama F, Chiba H, Kusakabe M, Sakai H (1985) Sulfur isotope exchange reaction in the aqueous system: thiosulfate–sulfide–sulfate at hydrothermal temperature. Geochem J 19:301–315

Vairavamurthy A, Manowitz B, Luther GW III, Jeon Y (1993) Oxidation state of sulfur in thiosulfate and implications for anaerobic energy metabolism. Geochim Cosmochim Acta 57:1619–1623

Vallance J, Galdos R, Balboa M, Berna B, Cabrera O, Baya C, Van De Vyver C, Viveen W, Béziat D, Salvi S, Brusset S, Baby P, Pokrovski GS (2024) Combined effect of organic carbon and arsenic on the formation of sediment-hosted gold deposits: A case study of the Shahuindo epithermal deposit, Peru. Econ Geol 119:85–112

Van den Kerkhof A, Hein UF (2001) Fluid inclusion petrography. Lithos 55:27–47

Vanko DA, Sutton SR, Rivers ML, Bodnar RJ (1993) Major-element ratios in synthetic fluid inclusions by synchrotron X-ray fluorescence microprobe. Chem Geol 109:125–134

Vanko DA, Bonnin-Mosbah M, Philippot P, Roedder E, Sutton SR (2001) Fluid inclusions in quartz from oceanic hydrothermal specimens and the Bingham, Utah porphyry-Cu deposit: a study with PIXE and SXRF. Chem Geol 173:227–238

Vassileva E, Becker A, Broekaert JAC (2001) Determination of arsenic and selenium species in groundwater and soil extracts by ion chromatography coupled to inductively coupled plasma mass spectrometry. Anal Chim Acta 441, 135–146

Vlassopoulos D, Wood SA (1990) Gold speciation in natural waters: I. Solubility and hydrolysis reactions of gold in aqueous solution. Geochim Cosmochim Acta 54:3–12

Voge HH (1939) Exchange reactions with radiosulfur. J Am Chem Soc 61:1032–1035

Voge HH, Libby WF (1937) Exchange reactions with radiosulfur. J Am Chem Soc 59:2474

Von Damm KL, Bischoff JL (1987) Chemistry of hydrothermal solutions from the southern Juan de Fuca Ridge. J Geophys Res 92:11334–11346

Wagman DD, Evans WH, Parker VB, Halow I, Bailey SM, Schumm RH (1968) Selected values of chemical thermodynamic properties. Natl Bur Stand US Government Office, Washington D.C., Tech Note 270-3

Wagman DD, Evans WH, Parker VB, Schumm RH, Halow I, Bailey SM, Churney KL, Nuttall RL (1982) The NBS tables of chemical thermodynamic properties. Selected values for inorganic and C1 and C2 organic substances in SI units. J Phys Chem Ref Data 11, suppl. 2

Wallace PJ (2005) Volatiles in subduction zone magmas: concentrations and fluxes based on melt inclusion and volcanic gas data. J Volcanol Geotherm Res 140:217–240

Wallace PJ, Edmonds M (2011) The sulfur budget in magmas: evidence from melt inclusions, submarine glasses, and volcanic gas emissions. Rev Mineral Geochem 73:215–246

Wallace RM (1966) Determination of the second dissociation constant of sulphuric acid by Donnan membrane equilibrium. J Phys Chem 70(12):3922–3927

Walter BF, Steele-MacInnis M, Markl G (2017) Sulfate brines in fluid inclusions of hydrothermal veins: Compositional determinations in the system $H_2O$-Na-Ca-Cl-$SO_4$. Geochim Cosmochim Acta 209:184–203

Walter BF, Burisch M, Fusswinkel T, Marks MAW, Steele-MacInnis M, Wälle M, Apukhtina O, Markl G (2018) Multi-reservoir fluid mixing processes in rift-related hydrothermal veins, Schwarzwald, SW-Germany. J Geochem Explor 186:158–186

Walter BF, Kortenbruck P, Scharrer M, Zeitvogel C, Wälle M, Mertz-Kraus R, Markl G (2019) Chemical evolution of ore-forming brines-Basement leaching, metal provenance, and the redox link between barren and ore-bearing hydrothermal veins. A case study from the Schwarzwald mining district in SW-Germany. Chem Geol 506:126–148

Walter BF, Jensen JL, Coutinho P, Laurent O, Markl G, Steele-MacInnis M (2020a) Formation of hydrothermal fluorite-hematite veins by mixing of continental basement brine and redbed-derived fluid: Schwarzwald mining district, SW-Germany. J Geochem Explor 212:106512

Walter BF, Steele-MacInnis M, Giebel RJ, Marks MAW, Markl G (2020b) Complex carbonate-sulfate brines in fluid inclusions from carbonatites: estimating compositions in the system $H_2O$-Na-K-$CO_3$-$SO_4$-Cl. Geochim Cosmochim Acta 277:224–242

Walter BF, Giebel RJ, Steele-MacInnis M, Marks MAW, Kolb J, Markl G (2021) Fluids associated with carbonatitic magmatism: A critical review and implications for carbonatite magma ascent. Earth Sci Rev 215:103509

Wan Q, Wang X, Hu W, Wan Y, Chou I-M (2024) Reaction pathway, mechanism and kinetics of thermochemical sulfate reduction: insights from in situ Raman spectroscopic observations at elevated temperatures and pressures. Geochim Cosmochim Acta 381:25–42

Wang C, Fu L, Yang S, Zheng H, Wang T, Gao J, Su M, Yang J, Wu G, Zhang W, Zhang Z, Li G, Zhang DH, Jiang L, Yang X (2022) Infrared spectroscopy of stepwise hydration motifs of sulfur dioxide. J Phys Chem Lett 13:5654−5659

Wang F, Tessier A (2009) Zero-valent sulfur and metal speciation in sediment porewaters of freshwater lakes. Environ Sci Technol 43;7252–7257

Wan Y, Wang X, Chou I-M, Li X (2021) Role of sulfate in the transport and enrichment of REE in hydrothermal systems. Earth Planet Sci Lett 569:117068

Wan Y, Chou I-M, Wang X, Wang R, Li X (2023) Hydrothermal sulfate-booming promotes REE transport and mineralization. Geology 51(5):449–453

Warr LN (2021) IMA-CNMNC approved mineral symbols. Mineral Mag 85: 291–320

Weston RE, Schwartz HA (1972) *Chemical Kinetics*. Prentice-Hall, Englewood Cliffs, NJ

Wilhelm F, Garbarino G, Jacobs J, Vitoux H., Steinmann R., Guillou F., Snigirev A, Snigireva I, Voisin P, Braithwaite D, Aoki D, Brison J-P, Kantor I, Lyatun I, Rogalev A. (2016) High pressure XANES and XMCD in the tender X-ray energy range. High Pressure Res 36(3):445–457

Wilke M, Klimm K, Kohn SC (2011) Spectroscopic studies of sulfur speciation in synthetic and natural glasses. Rev Miner Geochem 73:41–78

Wilkin RT, Wallschlager D, Ford RG (2003) Speciation of arsenic in sulfidic waters. Geochem Trans 4:1–7

Wilkinson JJ (2001) Fluid inclusions in hydrothermal ore deposits. Lithos 55:229–272

Wilkinson JJ (2010) A review of fluid inclusion constraints on mineralization in the Irish ore field and implications for the genesis of sedimentary-hosted Zn-Pb deposits. Econ Geol 105:417–442

Wilkinson JJ, Rankin AH, Mulshaw SC, Nolan J, Ramsey MH (1994) Laser ablation-ICP-AES for the determination of metals in fluid inclusions - An application to the study of magmatic ore fluids. Geochim Cosmochim Acta 58:1133–1146

Wilkinson JJ, Stoffell B, Wilkinson CC, Jeffries TE, Appold MS (2009) Anomalously metal-rich fluids form hydrothermal ore deposits. Science 323:764–767

Williams PJ, Barton MD, Johnson DA, Fontboté L, de Haller A, Mark G, Oliver NHS, Marschik R (2005) Iron oxide copper-gold deposits: geology, space-time distribution, and possible models of origin. Econ Geol 100$^{th}$ Anniv Vol, pp 371–405

Williamson MA, Rimstidt JD (1992) Correlation between structure and thermodynamic properties of aqueous sulfur species. Geochim Cosmochim Acta 56:3867–3880

Wood SA (1990) The aqueous geochemistry of the rare-earth elements and yttrium: 2. Theoretical predictions of speciation in hydrothermal solutions to 350°C at saturation water vapor pressure. *Chem Geol* 88:99–125

Wood SA (1991) Experimental determination of the hydrolysis constants of $Pt^{2+}$ and $Pd^{2+}$ at 25°C from the solubility of Pt and Pd in aqueous hydroxide solutions. Geochim Cosmochim Acta 55:1759–1767

Wood SA (2002) The aqueous geochemistry of the platinum-group elements with applications to ore deposits. In: *The Geology, Geochemistry, Mineralogy and Mineral Beneficiation of Platinum-Group Elements*. Canadian Institute of Mining, Metallurgy and Petroleum, spec vol 54:211–249

Wood SA, Samson IM (1998) Solubility of ore minerals and complexation of ore metals in hydrothermal solutions. Rev Econ Geol 10:33–80

Wood SA, Crerar DA, Borcsik MP (1987) Solubility of the assemblage pyrite-pyrrhotite-magnetite-sphalerite-galena-gold-stibine-bismuthinite-argentite-molybdenite in $H_2O$-NaCl-$CO_2$ solutions from 200° to 350 °C. Econ Geol 82:1864–1887

Wood SA, Mountain BW, Pan P (1992) The aqueous geochemistry of platinum, palladium and gold – recent experimental constraints and a reevaluation of theoretical predictions. Can Mineral 30:955–982

Wood SA, Pan P, Zhang Y, Mucci A (1994) The solubility of Pt and Pd sulfides and Au in bisulfide solutions I. Results at 25°–90 °C and 1 bar pressure. Miner Deposita 29:309–317

Wright RH, Maas O (1932) The electrical conductivity of aqueous solutions of hydrogen sulphide and the state of the dissolved gas. Can J Research 6:588–595

Xiang T, Johnston KP, Wofford WT, Gloyna EF (1996) Spectroscopic measurements of pH in aqueous sulfuric acid and ammonia from sub- to supercritical conditions. Ind Eng Chem Res 35:4788–4795

Xiao Z, Gammons CH, Williams-Jones AE (1998) Experimental study of copper(I) chloride complexing in hydrothermal solutions at 40 to 300 °C and saturated water vapor pressure. Geochim Cosmochim Acta 62:2949–2964

Xing Y, Brugger J, Tomkins A, Shvarov Y (2019) Arsenic evolution as a tool for understanding formation of pyritic gold ores. Geology 47:335–338

Xing Y, Scheuermann P, Seyfried WE (2021) Experimental study on Fe solubility in vapor-rich hydrothermal fluids at 400–500 °C, 215–510 bar: Implication for Fe mobility in seafloor vent systems. Geochim Cosmochim Acta 314:209–222.

Xu Y, Schoonen MAA., Nordstrom DK, Cunningham KM, Ball JW (1998) Sulfur geochemistry of hydrothermal waters in Yellowstone National Park: I. The origin of thiosulfate in hot spring waters. Geochim Cosmochim Acta 62:3729–3743

Xu Y, Schoonen MAA, Nordstrom DK, Cunningham KM, Ball JW (2000) Sulfur geochemistry of hydrothermal waters in Yellowstone National Park, Wyoming, USA. II. Formation and decomposition of thiosulfate and polythionates in cinder pool. J Volcanol Geotherm Res 97:407–423

Yam VWW (2010) Behind platinum's sparkle. Nat Chem 2:790

Yardley BWD (1983) Quartz veins and devolatilization during metamorphism. J Geol Soc 140: 657–663

Yardley BWD (2005) Metal concentrations in crustal fluids and their relationship to ore formation. Econ Geol 100$^{th}$ Anniv Spec Vol 100:613–632

Yin Y, Zajacz Z (2018) The solubility of silver in magmatic fluids: Implications for silver transfer to the magmatic-hydrothermal ore-forming environment. Geochim Cosmochim Acta 238:235–251

Yuan S, Chou I-M, Burruss RC, Wang X, Li J (2013) Disproportionation and thermochemical sulfate reduction reactions in S-$H_2O$-$CH_4$ and S-$D_2O$-$CH_4$ systems from 200 to 340°C at elevated pressures. Geochim. Cosmochim. Acta 118:263–275

Zajacz Z, Seo JH, Candela PA, Piccoli PM, Heinrich CA, Guillong M (2010) Alkali metals control the release of gold from volatile-rich magmas. Earth Planet Sci Lett 297:50–56

Zajacz Z, Seo JH, Candela PA, Piccoli PM, Tossell JA (2011) The solubility of copper in high-temperature magmatic vapors: a quest for the significance of various chloride and sulfide complexes. Geochim Cosmochim Acta 75:2811–2827

Zakaznova-Herzog VP, Seward TM, Suleimenov OM (2006a) Arsenous acid ionisation in aqueous solutions from 25 to 300 °C. Geochim Cosmochim Acta 70:1928–1938.

Zakaznova-Herzog VP, Seward TM (2006b) Antimonous acid protonation/deprotonation equilibria in hydrothermal solutions to 300 °C. Geochim Cosmochim Acta 70:2298–2310

Zakaznova-Herzog VP, Seward TM (2012) A spectrophotometric study of the formation and deprotonation of thioarsenite species in aqueous solution at 22 °C. Geochim Cosmochim Acta 83:48–60

Zezin DY, Migdisov AA, Williams-Jones AE (2011) *PVTx* properties of $H_2O$–$H_2S$ fluid mixtures at elevated temperature and pressure based on new experimental data. Geochim Cosmochim Acta 75:5483–5495

Zhang JZ, Millero FJ (1991) The rate of sulfite oxidation in seawater. Geochim Cosmochim. Acta 55:677–685

Zhang L, Audétat A, Dolejš D (2012) Solubility of molybdenite ($MoS_2$) in aqueous fluids at 600–800 °C, 200 MPa: a synthetic fluid inclusion study. Geochim. Cosmochim. Acta 77:175–185

Zhang N, Königsberger E, Duan S, Lin K, Yi H, Zeng D, Zhao Z, Hefter G (2019) Nature of monomeric molybdenum(VI) cations in acid solutions using theoretical calculations and Raman spectra. J Phys Chem B 123:3304–3311

Zhang Z, Ewing GE (2002) Infrared spectroscopy of $SO_2$ aqueous solutions. Spectrochim Acta A 58:2105–2113

Zimmer K. Zhang YL, Lu P, Chen YY, Zhang GR, Dalkilic M, Zhu C (2016) SUPCRTBL: A revised and extended thermodynamic dataset and software package of SUPCRT92. Comput Geosci 90:97–111

Zhong R, Brugger J, Tomkins AG, Chen Y, Li W (2015) Fate of gold and base metals during metamorphic devolatilization of a pelite. Geochim Cosmochim Acta 171:338–352

Zhu X, Nordstrom DK, McCleskey RB, Wang R (2016) Ionic molal conductivities, activity coefficients, and dissociation constants of $HAsO_4^{2-}$ and $H_2AsO_4^-$ from 5 to 90 °C and ionic strengths from 0.001 up to 3 mol kg$^{-1}$ and applications in natural systems. Chem Geol 441:177–190

Zopfi J, Ferdelman T, Fossing H (2004). Distribution and fate of sulfur intermediates–sulfite, tetrathionate, thiosulfate, and elemental sulfur–in marine sediments. In: *Sulfur Biogeochemistry - Past and Present*. Geol Soc Amer 379:97–116

Zotov AV, Baranova NN, Dar'ina TG, Bannykh LN (1991) The solubility of gold in aqueous chloride fluids at 350-500°C and 500-1500 atm: thermodynamic properties of $AuCl_2^{-}{}_{,aq}$ up to 750 °C and 5000 atm. Geochem Intern 28(2):63–71

Zotov AV, Kudrin AV, Levin KA, Shikina ND, Var'yash LN (1995) Experimental studies of the solubility and complexing of selected ore elements (Au, Ag, Cu, Mo, As, Sb, Hg) in aqueous solutions. In: *Fluids in the Crust - Equilibrium and Transport Properties* (eds. KI Shmulovich, BWD Yardley, GG Gonchar). Chapman & Hall, London, pp. 95–138

Zotov AV, Shikina ND, Akinfiev NN (2003) Thermodynamic properties of the Sb(III) hydroxide complex $Sb(OH)_3$(aq) at hydrothermal conditions. Geochim Cosmochim Acta 67:1821–1836

Zotov AV, Kuzmin NN, Reukov VL, Tagirov BR (2018) Stability of $AuCl_2^-$ from 25 to 1000 °C at pressures to 5000 bar and consequences for hydrothermal gold mobilization. Minerals 8:286

Zotov AV, Diagileva DR, Koroleva LA (2020) Silver solubility in supercritical fluids in a wide range of NaCl concentration (0.6–50 wt%) – Experimental and thermodynamic description. ACS Earth Space Chem 4:2403–2413

## Figures

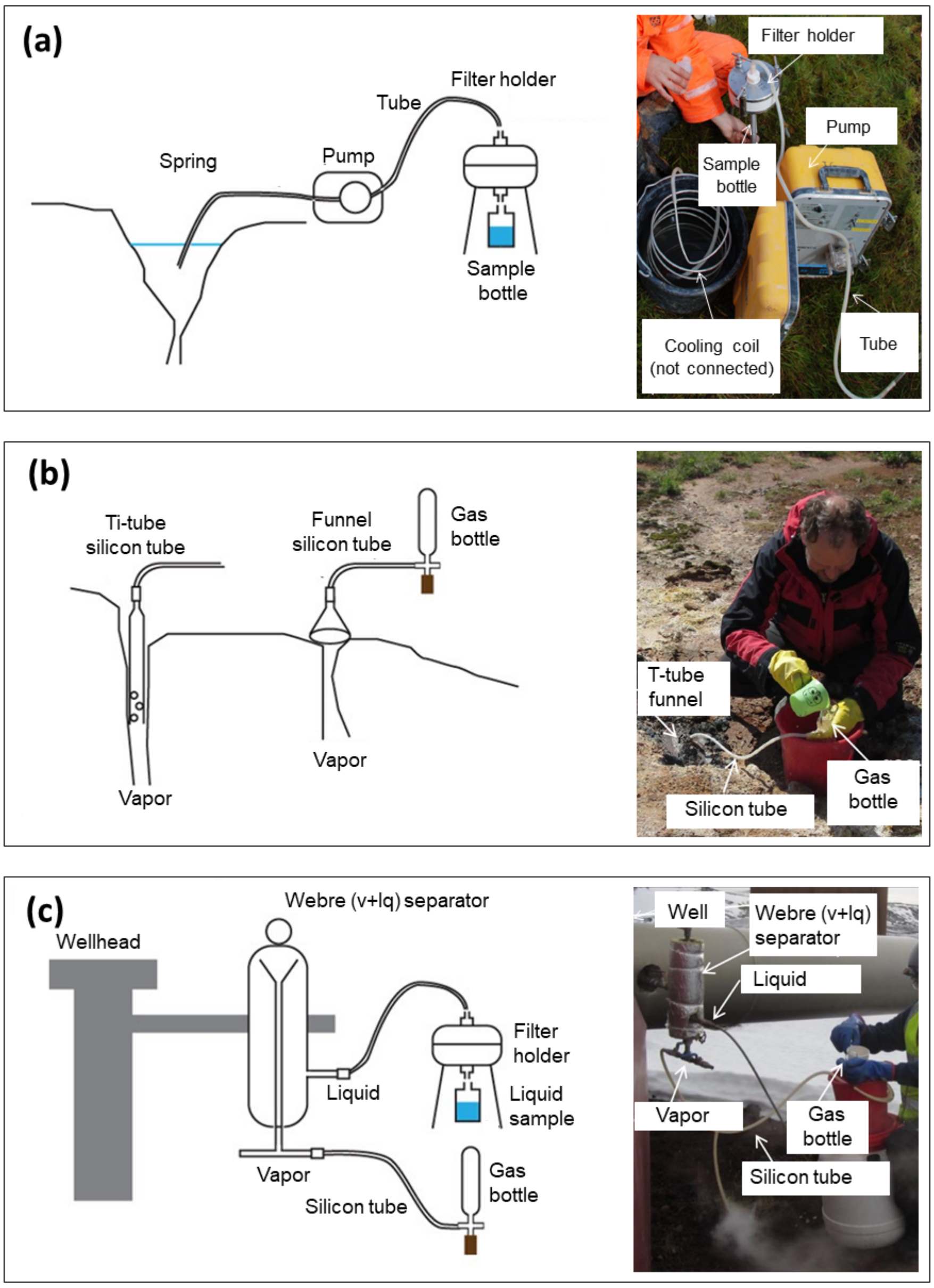

**Fig. 4.1** Sampling methods for hot springs (liquid water) (a), fumaroles (vapor) (b) and two-phase (liquid and vapor) well discharges (c). The hot springs are sampled using pump and a filter holder containing filter with <0.45 µm pore size. The fumarole samples are collected by placing a funnel over or inserting a titanium tube into the vapor stream, connecting the outflow by silicone tube to a gas bottle. For the two-phase well discharges the liquid and vapor phases are separated using a Webre separator and the two phases sampled as for hot springs and fumaroles. For further details see Arnórsson et al. (2006).

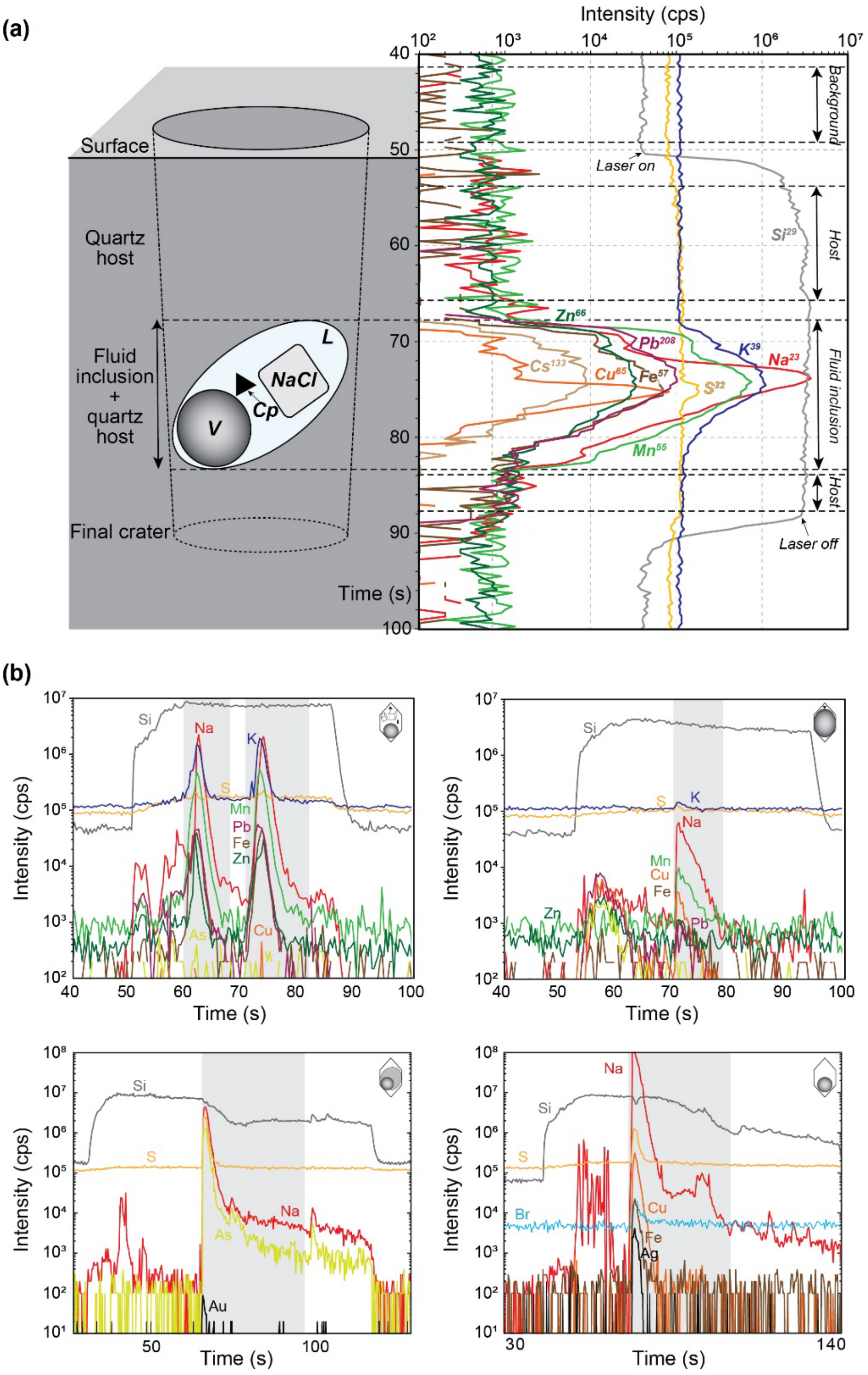

**Fig. 4.2** (a) Schematic representation of a quartz-hosted aqueous saline fluid inclusion, the ablation pit, and the resulting analytical signal for selected elements. The parts of the spectrum corresponding to the background signal, the signal from the host quartz, and the fluid inclusion signal are indicated. The analysis is stopped after the ablation of the entire inclusion. Note a shift of the signal for some elements, due to ablation of solids such as halite (NaCl) or chalcopyrite (Cp, $CuFeS_2$) inside the inclusion cavity. Judging from the overlap in time between the maxima of the S, Cu and Fe intensities, the sulfur signal is mainly associated with the ablation of chalcopyrite present in the inclusion as daughter crystals. (b) Representative LA-ICPMS elemental signals of single brine (upper left), vapor (upper right), aqueous-carbonic (lower left) and aqueous (lower right) fluid inclusions. Brine and vapor are magmatic-hydrothermal fluids from the Morococha polymetallic district, central Peru (data from Catchpole et al. 2015); aqueous-carbonic and aqueous – metamorphic fluids from the Alpine orogenic belt (from Rauchenstein-Martinek et al. 2016).

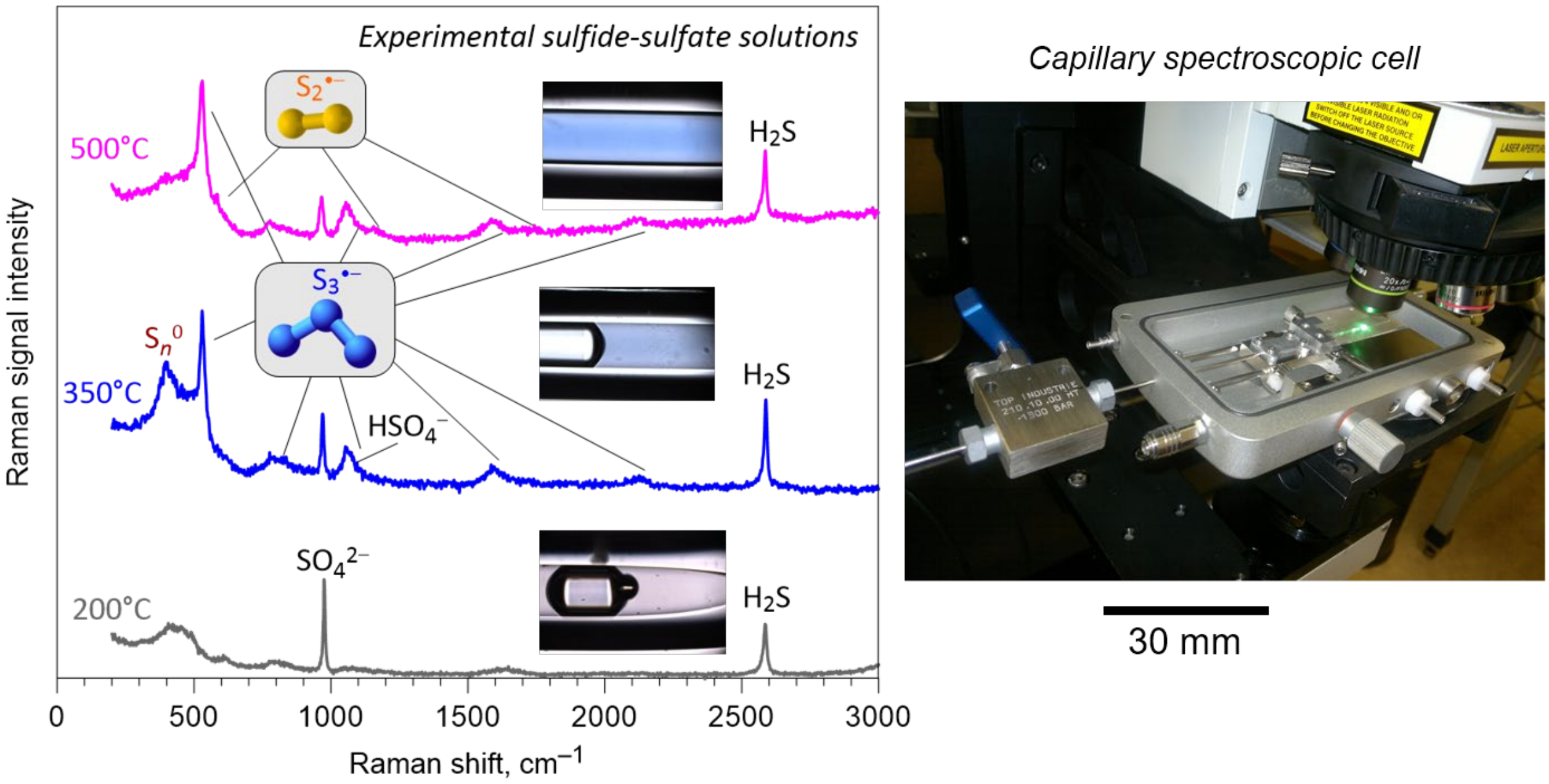

**Fig. 4.3** Example of Raman spectra of an aqueous S-bearing fluid initially containing sulfide ($H_2S$) and sulfate ($SO_4^{2-}$) as a function of temperature from $P_{sat}$ to 1 kbar bar, measured using the capillary cell technique shown on the right (see Pokrovski and Dubessy 2015 for details). Raman peaks of the major sulfur aqueous species are indicated. The photos of the capillary cell at each corresponding temperature show color changes in the fluid with changing temperature, reflecting the growing formation of the blue chromophore $S_3^{\bullet -}$ ion at the expense of sulfate and sulfide that are the only species detected at $T$< 200°C. Chain-like $S_n^0$ polymers also form at moderate temperatures (250–350°C), but disappear at the expense of the radical ions at higher temperatures. At the highest $T$, the $S_2^{\bullet -}$ disulfur radical ion is also clearly detected. Quenching results in immediate disapearence of the blue color and the radical ion spectral signature, demonstrating that the radical ions can only be studied using in situ methods. The spectral pattern and colors are fully reversible in cooling-heating cycles, and are stable at a given temperature within at least the experiment duration (few days).

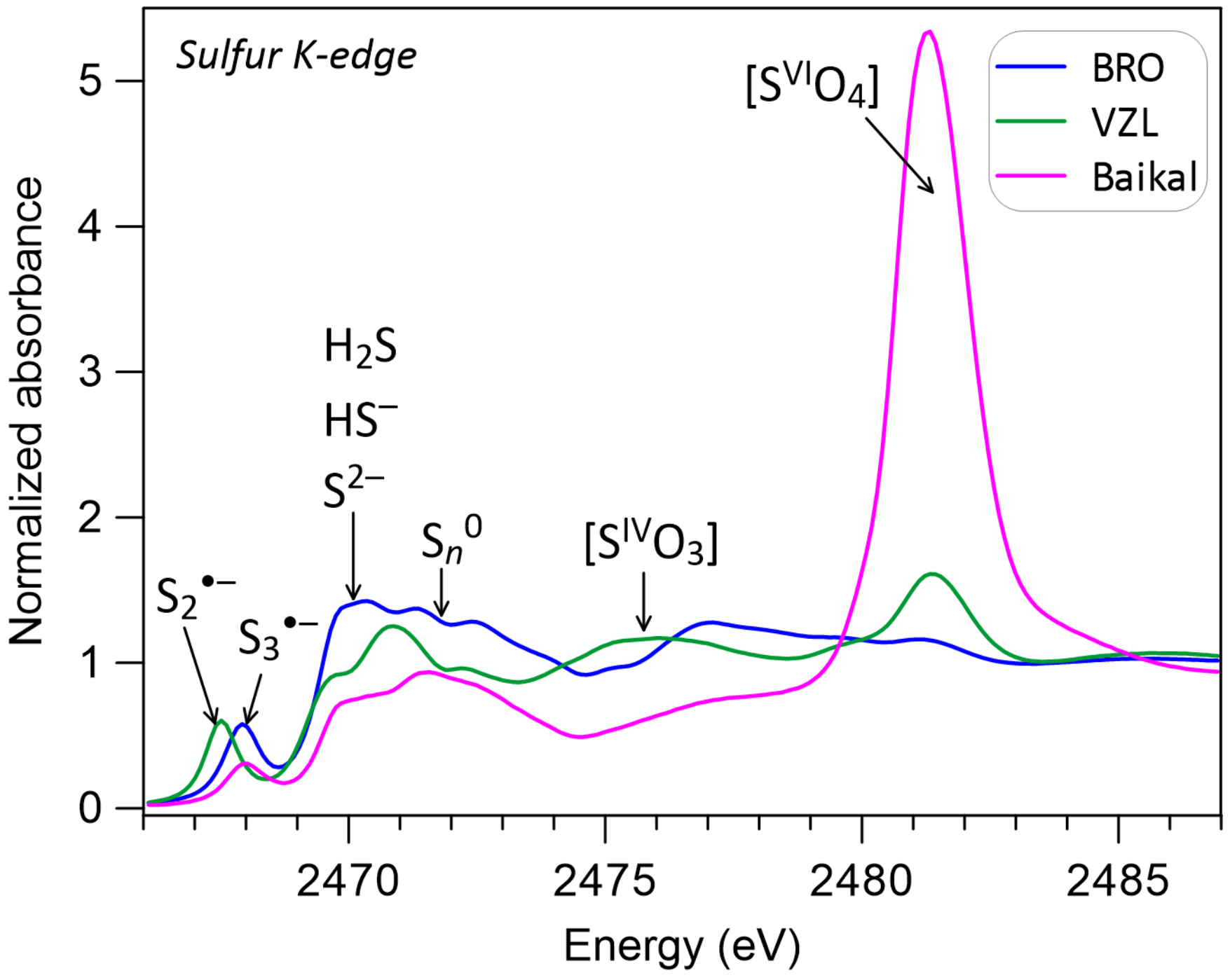

**Fig. 4.4** Normalized S K-edge XANES spectra of three representative ultramarine minerals recorded at ID12 beamline (ESRF) at ambient conditions. BRO and VZL = synthetic blue and green ultramarines, respectively (courtesy of B. Ledé); Baikal = natural blue sodalite from the lake Baikal metamorphic complex, Siberia (coutesy of A. Bychkov). The typical energy positions and shapes of the sulfur major formal oxidation states are indicated, together with the radical ions $S_3^{\bullet-}$ and $S_2^{\bullet-}$ that can be distinguished using their respective pre-edge peak positions (spectral data from Pokrovski et al. 2025).

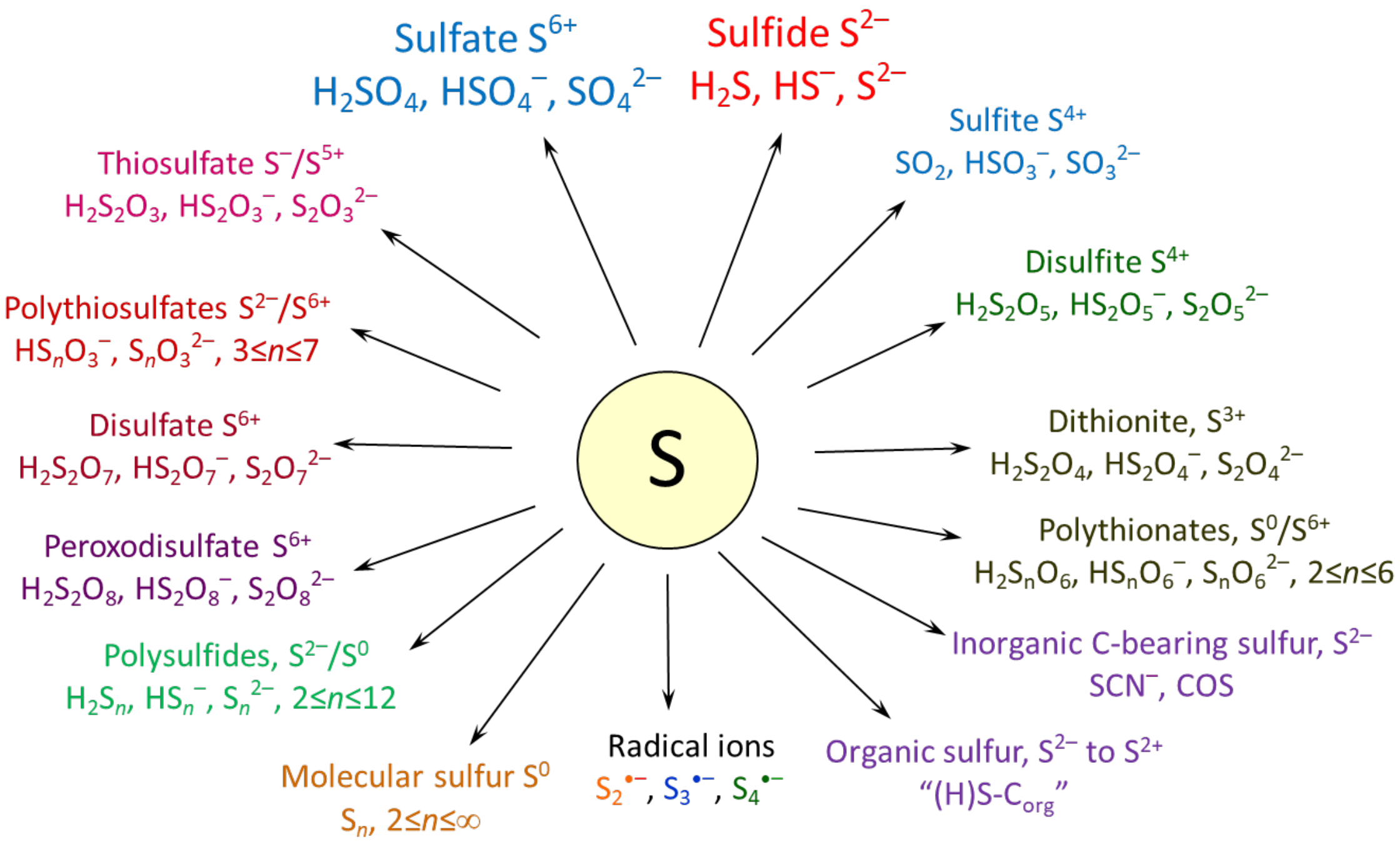

**Fig. 4.5** Chemical forms of sulfur known in aqueous solution, solid or gas phases at near-ambient conditions (modified and extended from Pokrovski and Dubessy 2015). Formal redox state of sulfur is indicated (where appropriate).

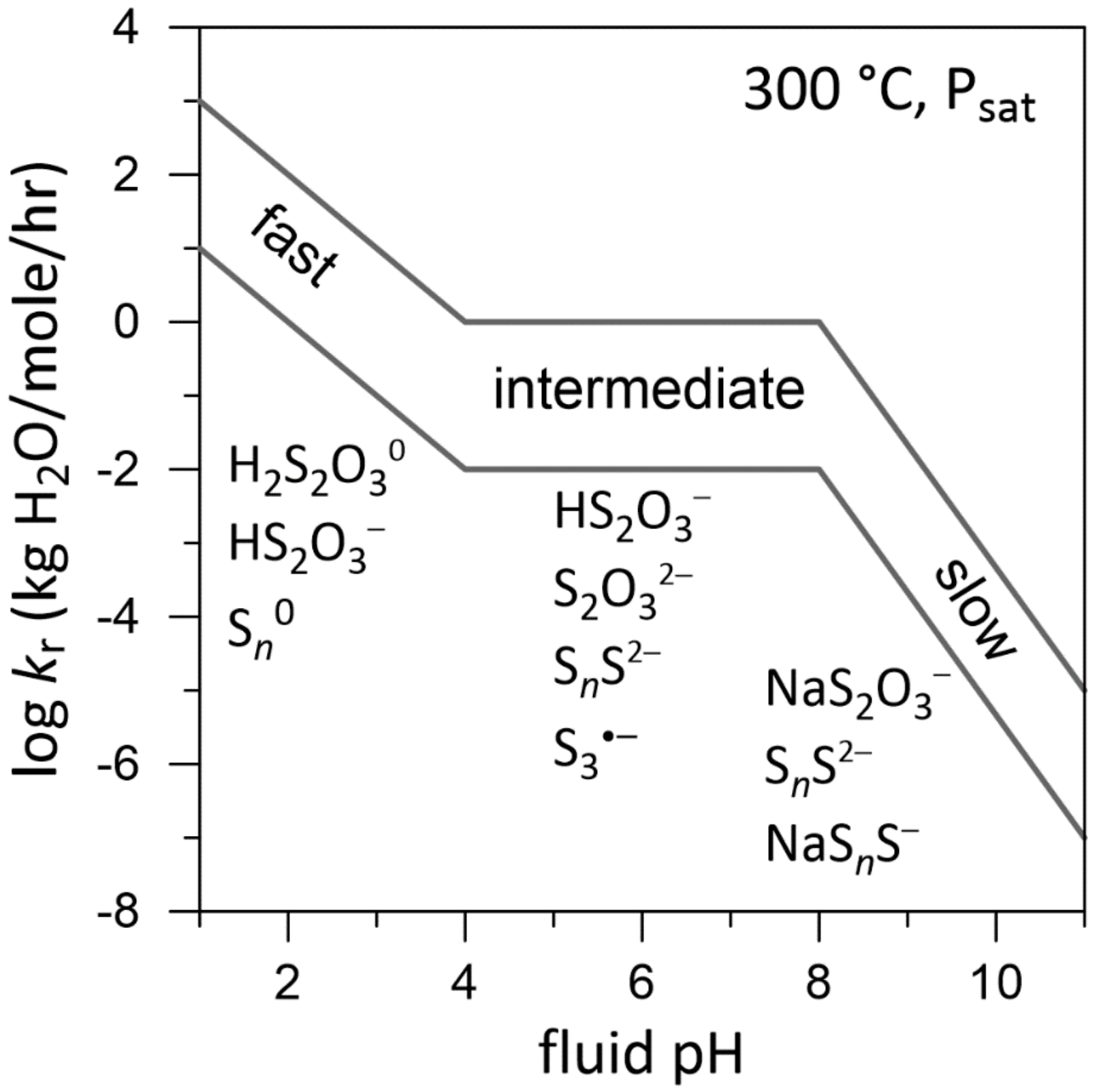

**Fig. 4.6** Typical range of experimental and predicted rate constants ($k_r$) of the reaction between sulfide and sulfate as a function of fluid pH at 300°C and saturated vapor pressure, highlighting three major trends in the constant depending on the pH range. Intermediate-valence sulfur species that might act as a rate-controlling species are indicated for each pH range (*as per* Ohmoto and Lasaga 1982; Chu et al. 2004; Pokrovski and Dubessy 2015).

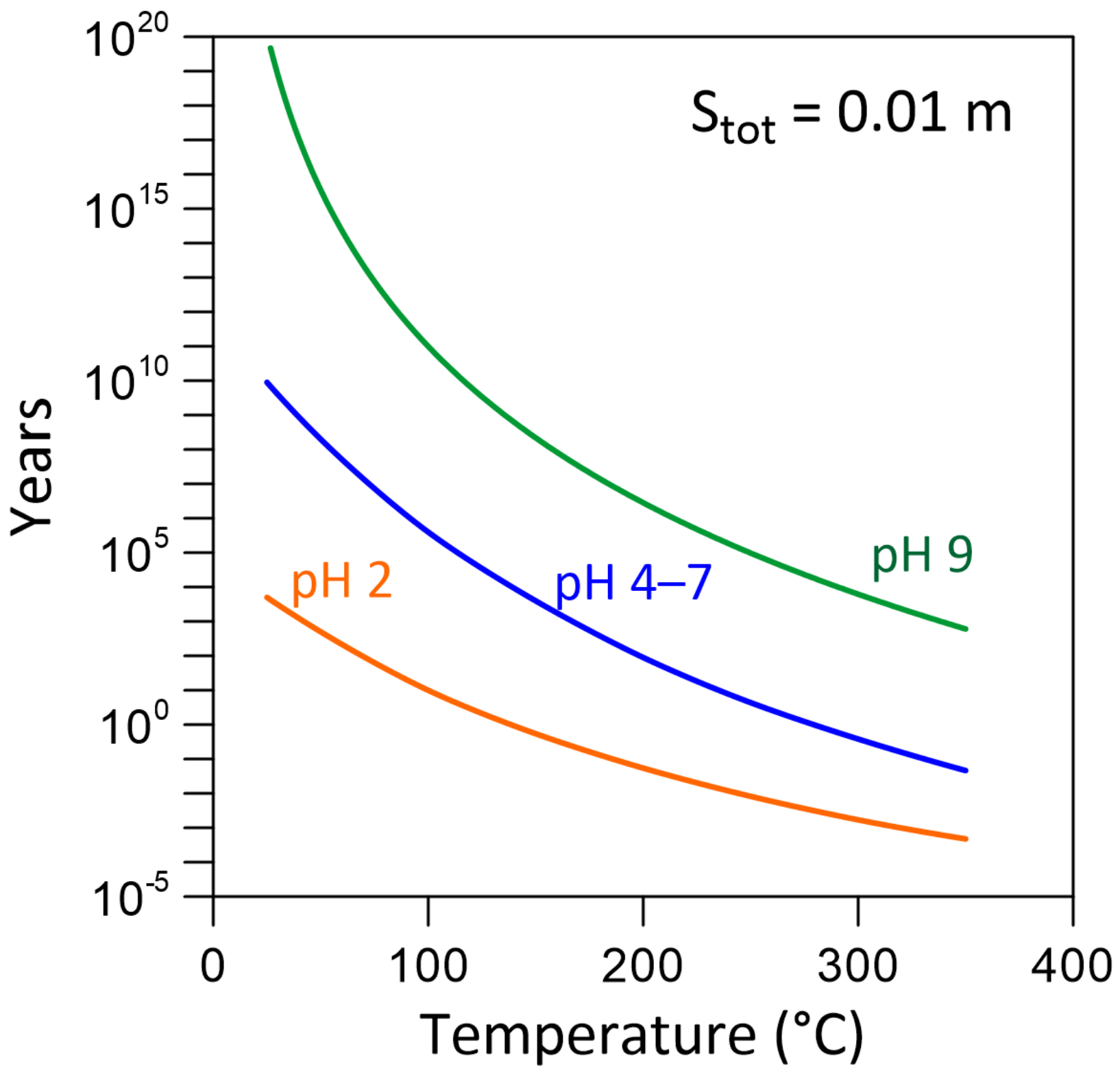

**Fig. 4.7** Time (in years) necessary to reach 90% of equilibrium for the sulfate-sulfide pair in aqueous solution at a total sulfur concentration of 0.01 mol/kg, as a function of temperature at the indicated pH values, according to the thiosulfate kinetic model of Ohmoto and Lasaga (1982).

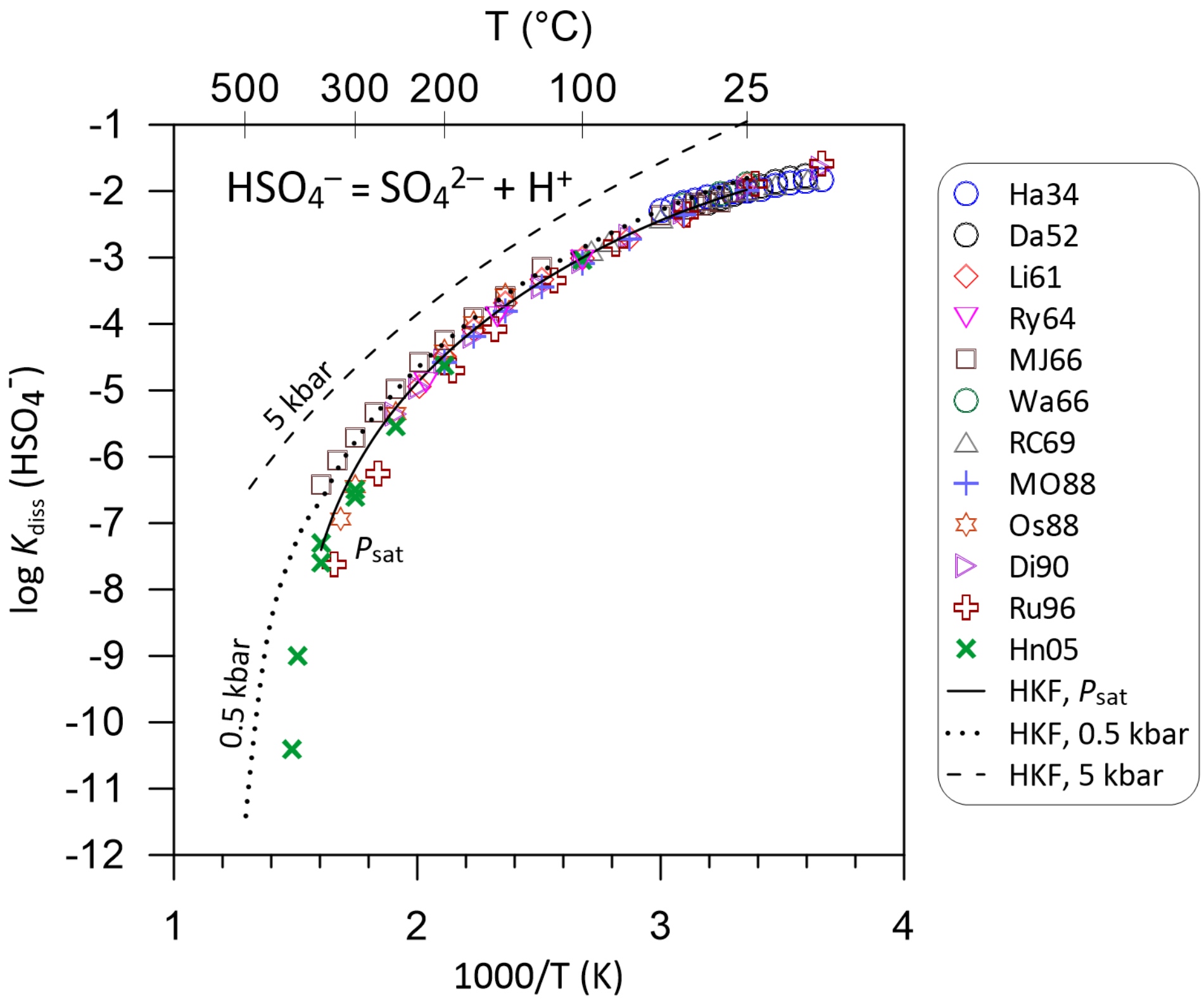

**Fig. 4.8** Dissociation constant of hydrogen sulfate in aqueous solution as a function of the reciprocal of absolute temperature according to the available experimental data and thermodynamic predictions by HKF model (Shock and Helgeson, 1988). Ha34 = Hamer (1934), Da52 = Davis et al. (1952), Li61 = Lietzke et al. (1961), Ry64 = Ryzhenko (1964), MJ66 = Marshall and Jones (1966), Wa66 = Wallace (1966), RC69 = Readnour and Cobble (1969), MO88 = Matsushima and Okuwaki (1988), Os88 = Oscarson et al. (1988), Di90 = Dickson et al. (1990), Ru96 = Rudolph (1996), Hnedkovsky et al. (2005). Note that the linear van't Hoff equation (4.9) would fail to account for the curvature of log *K vs.* 1/*T*; more extended polynomial equations (e.g., 4.11 and 4.12) were used in some studies to describe the data to 300–350°C (not shown), but cannot account for the effect of pressure. In contrast, the HKF model, based on pre-1970 measurements, accurately reproduces the whole set of data including more recent ones and accounts for the pressure effect.

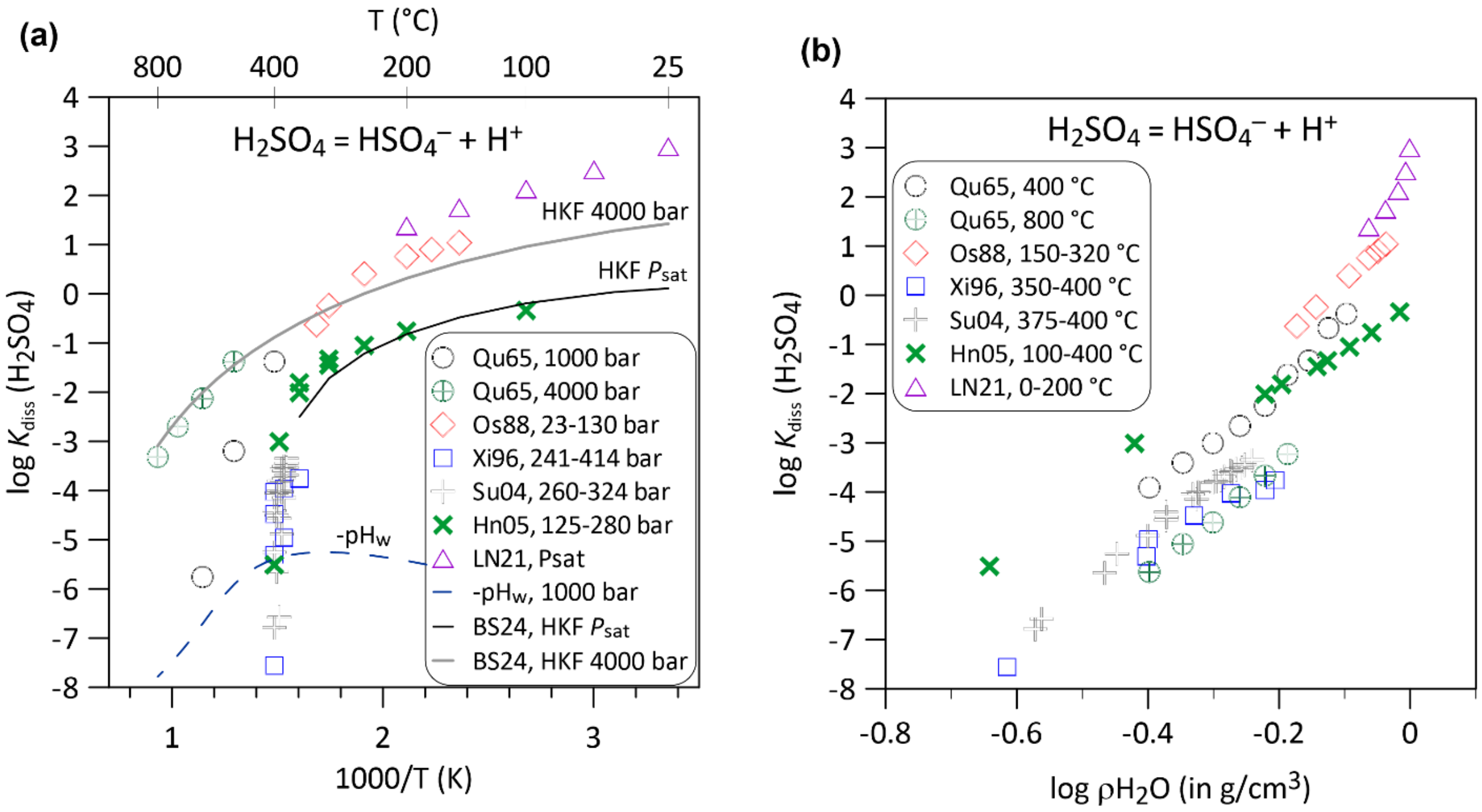

**Fig. 4.9** Dissociation constant of sulfuric acid in aqueous solution as a function of the reciprocal of absolute temperature (a) and water density (b), according to available experimental data. Qu65 = Quist et al. (1965), Os88 = Oscarson et al. (1988), Xi96 = Xiang et al. (1996), Su04 = Sue et al. (2004), Hn05 = Hnedkovsky et al. (2005), LN21 = LLNL database (thermo.dat, Johnson et al. 2000), BS24 = Baudry and Sverjensky (2024), which is an HKF fit of Qu65; $-pH_w$ = log $a(H^+)$ of pure water. The dissociation constants are strongly *P*-dependent in the supercritical region. While the density model allows for significant linearization of $K_{diss}$ trends, the values from different sources yet remain rather discrepant, spanning over 2 to 4 log units at high (>300 °C) and low (<100 °C) temperatures, respectively.

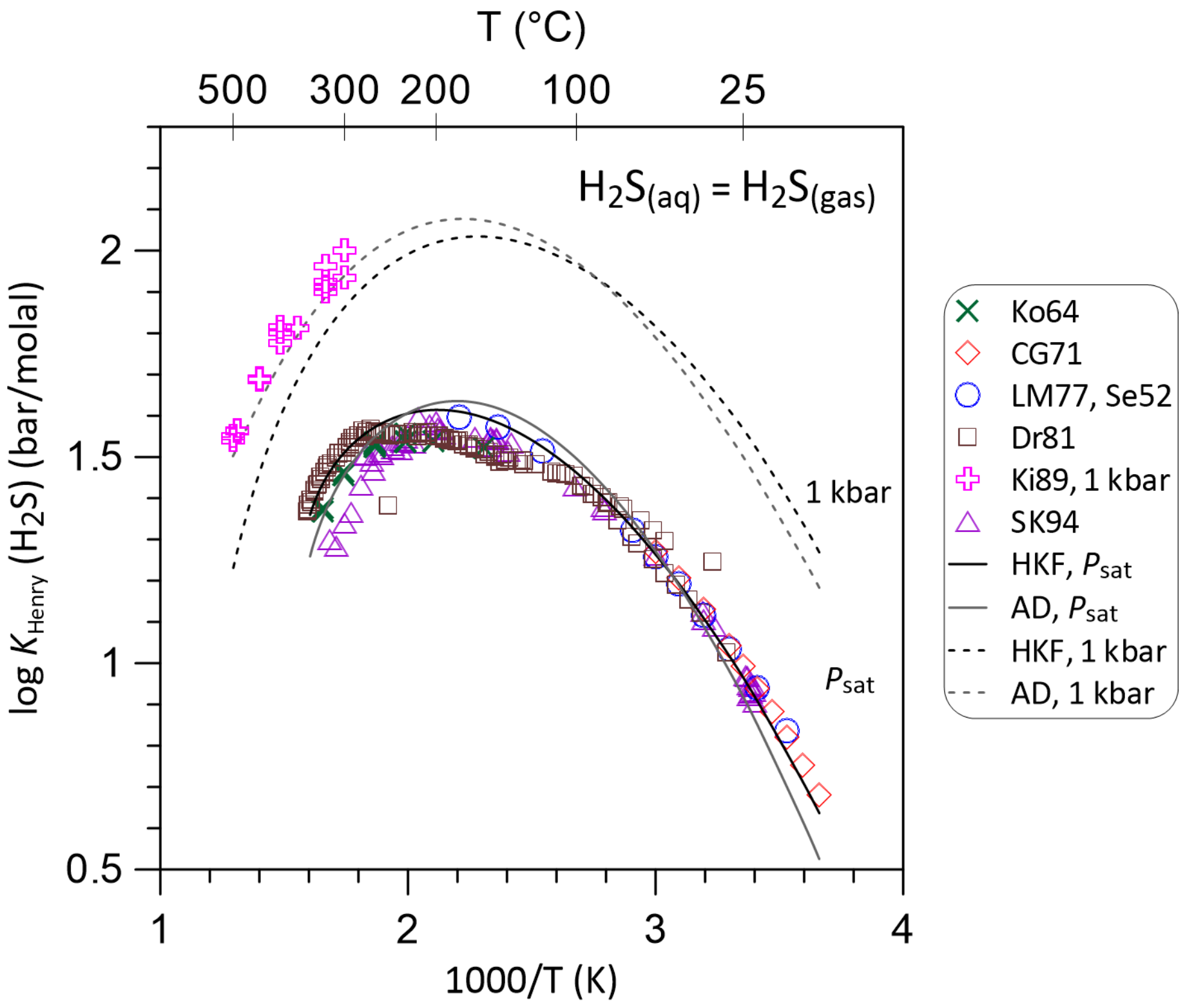

**Fig. 4.10** Henry's constant of $H_2S$ solubility in pure water as a function of the reciprocal of absolute temperature, according to major available experimental data sets and thermodynamic model predictions at saturated vapor pressure ($P_{sat}$) and at 1 kbar. Ko64 = Kozintseva (1964), CG71 = Clarke and Glew (1971), LM77 = Lee and Mather (1977) including Selleck et al. (1952) data, Dr81 = Drummond (1981), Ki89 = Kishima (1989), SK94 = Suleimenov and Krupp (1994), HKF = HKF model updated by Schulte et al. (2001); AD = Akinfiev and Diamond (2003) model. Note that the AD model provides a slightly better match of the available experimental data than the HKF model, in particular at elevated temperatures and pressures. Note that the differences between the models grow with increasing *T* and *P* and may be significant outside the *T*-*P* range covered so far by experiments.

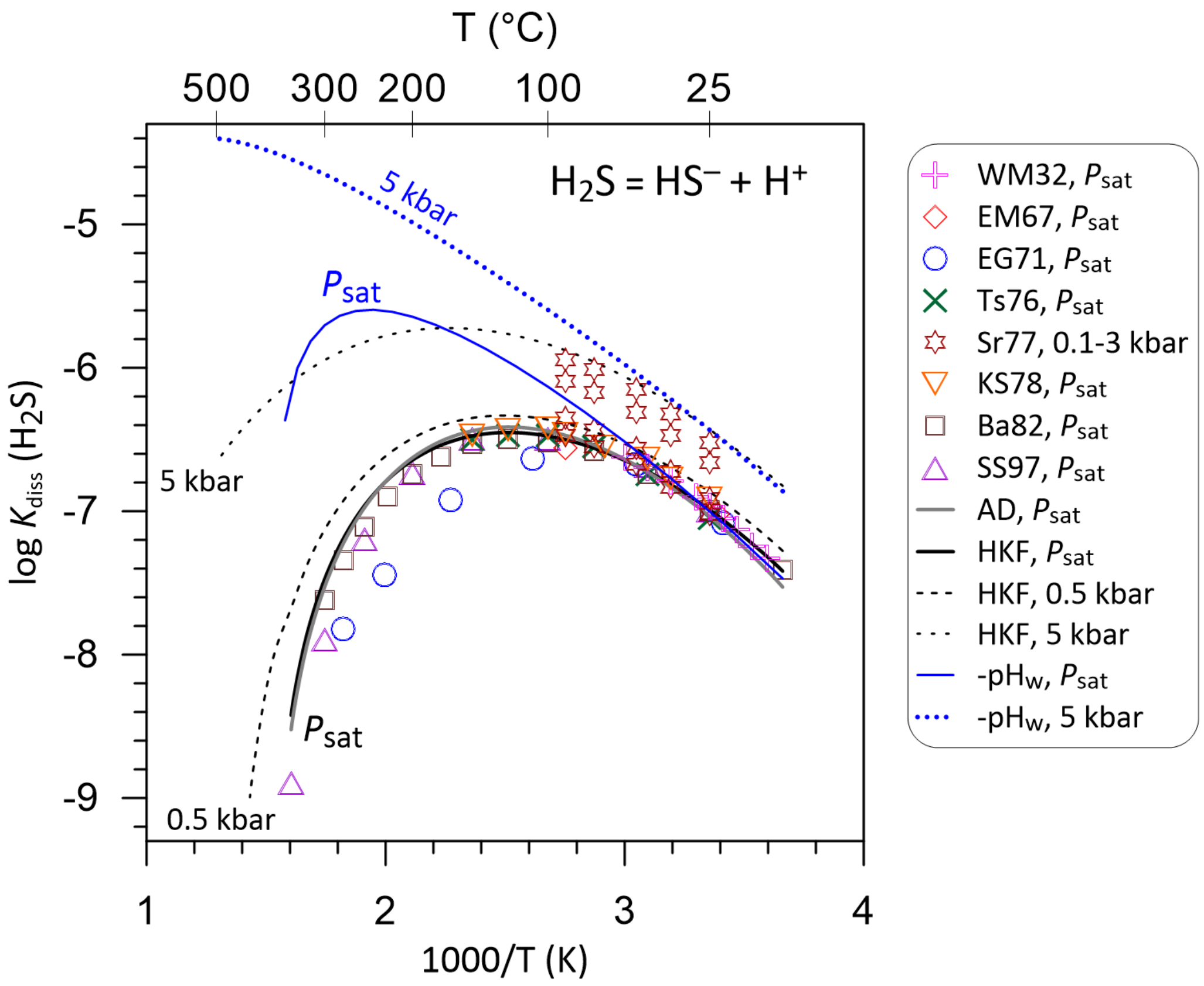

**Fig. 4.11** First dissociation constant of hydrogen sulfide in aqueous solution as a function of the reciprocal of absolute temperature, according to the available experimental data and thermodynamic predictions by the HKF model for $HS^-$ (Shock and Helgeson 1988) and $H_2S$ (updated by Schulte et al. 2001) and the AD model for $H_2S$ (Akinfiev and Diamond 2003). WM32 = Wright and Maass (1932), EM67 = Ellis and Millestone (1967), EG71 = Ellis and Giggenbach (1971), Ts76 = Tsonopoulos et al. (1976), Sr77 = Sretenskaya (1977), KS78 = Kryukov and Starostina (1978), Barbero et al. (1982), SS97 = Suleimenov and Seward (1997). Note that both models give almost undistinguishable results (curves overlap at all pressures), by accurately reproducing most of experimental data acquired from 0 to 350 °C at $P_{sat}$, and yielding very similar predictions of the pressure effect, in good agreement with rare $P$-dependent experimental data (to at least 90 °C and 3 kbar, Sretenskaya 1977). For comparison are also plotted the negative pH values of the neutrality point of pure water ($pH_w$) at $P_{sat}$ and 5 kbar (blue curves), see text.

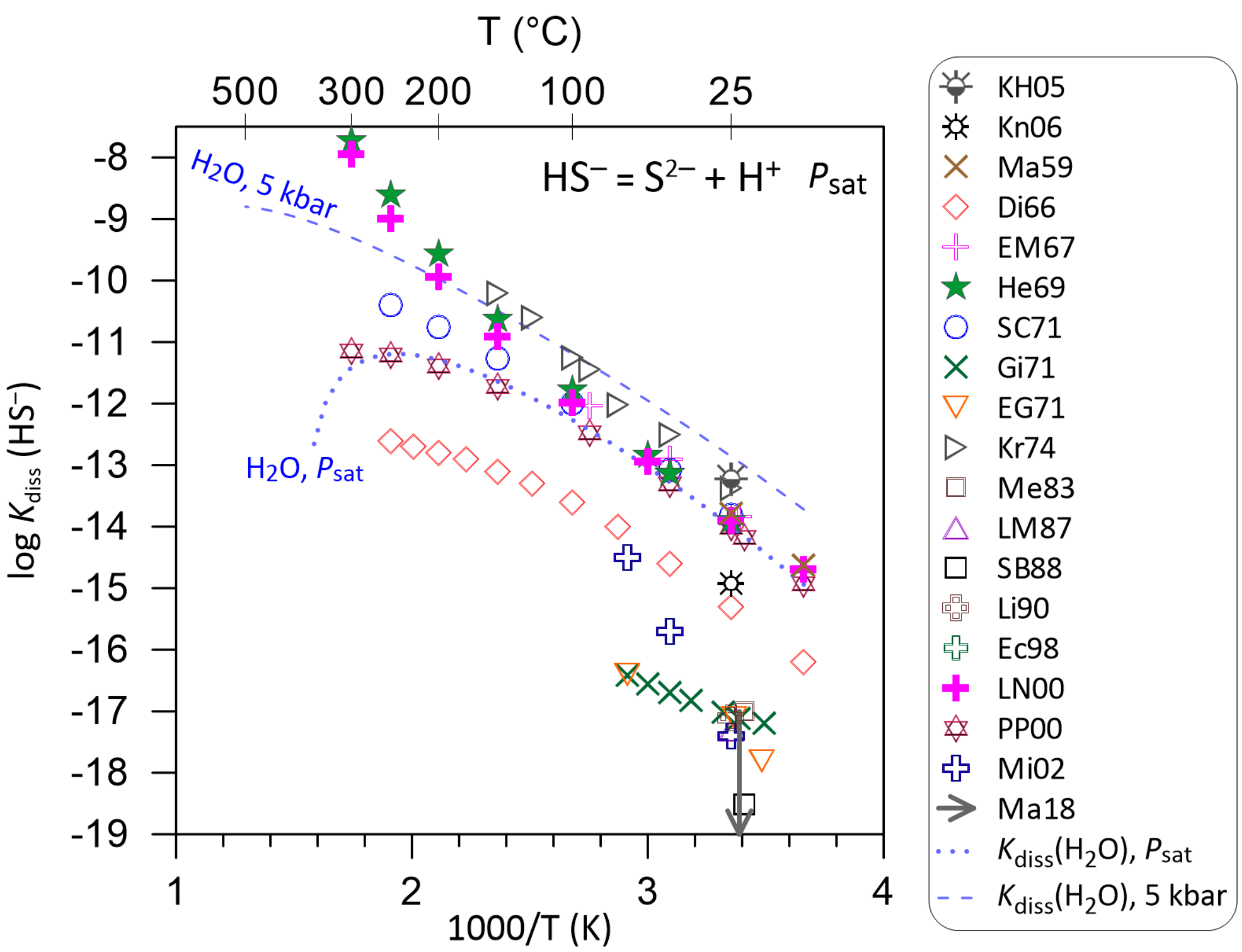

**Fig. 4.12** Second dissociation constant of hydrogen sulfide in aqueous solution as a function of the reciprocal of absolute temperature, according to the major available experimental data and compilations (all at saturated vapor pressure, $P_{sat}$): KH05 = Küster and Heberlein (1905), Kn06 = Knox (1906), Ma59 = Maronny (1959), Di66 = Dickson (1966), He69 = Helgeson (1969), EM67 = Ellis and Milestone (1967), SC71 = Stephens and Cobble (1971), Gi71 = Giggenbach (1971a), EG71 = Ellis and Giggenbach (1971), Kr74 = Kryukov et al. (1974), Me83 = Meyer et al. (1983), LM87 = Licht and Manassen (1987), SB88 = Schoonen and Barnes (1988), Li90 = Licht et al. (1990), Ec98 = Eckert (1998), LN00 = LLNL database (Johnson et al. 2000), PP00 = Phillips and Phillips (2000), Mi02 = Migdisov et al. (2002), Ma18 = May et al. (2018) and references therein. The arrow symbol of Ma18 indicates the most plausible range of maximum values of $K_{diss}(HS^-)$ at ambient conditions according to those authors. For comparison, water dissociation constant ($K_{diss}$ ($H_2O$), $H_2O = H^+ + OH^-$) is also plotted at $P_{sat}$ and 5 kbar pressure (blue curves).

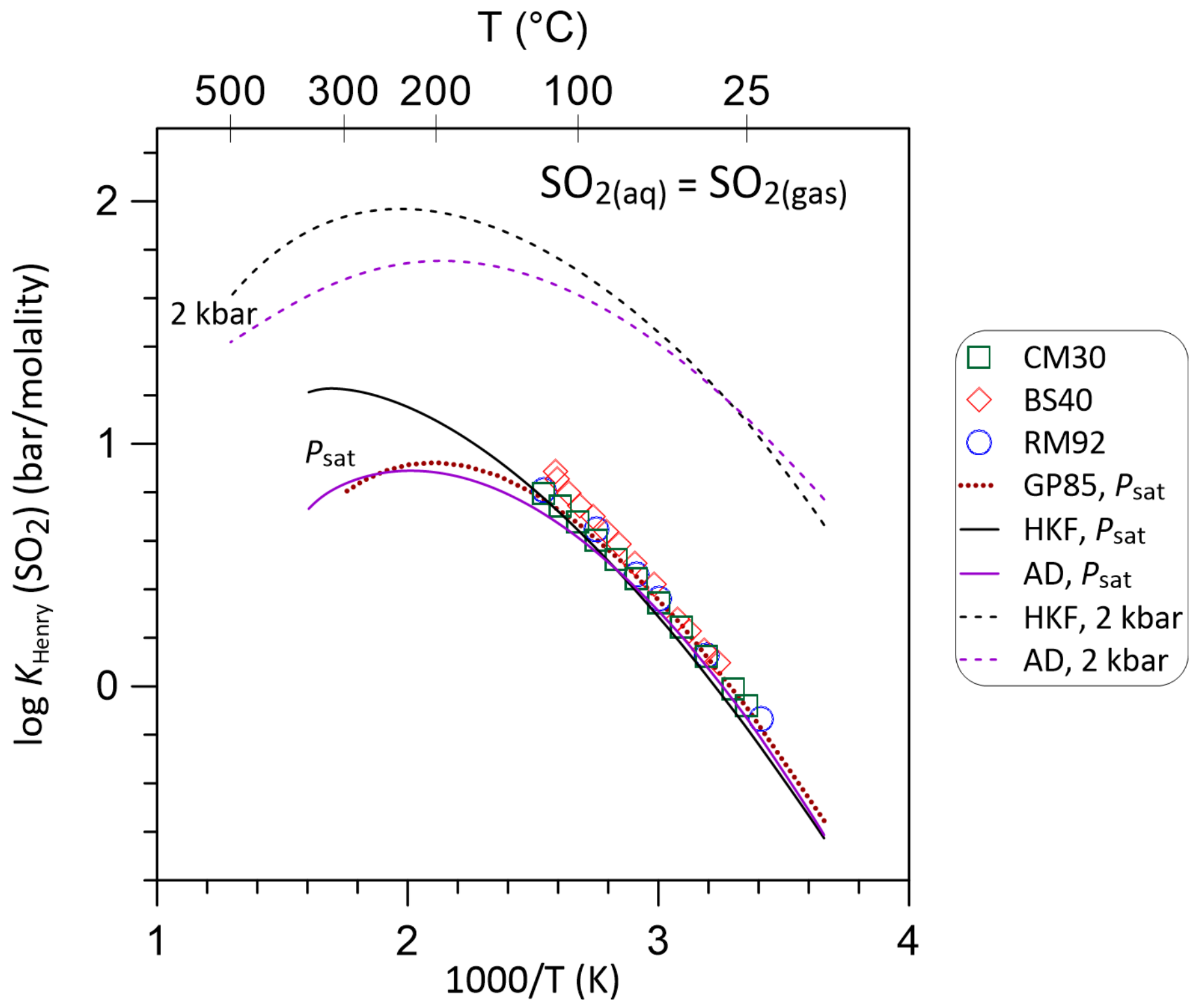

**Fig. 4.13** Henry's constant of $SO_2$ solubility in pure water as a function of the reciprocal of absolute temperature, according to selected available high-*T* experimental data sets at saturated vapor pressure ($P_{sat}$) and thermodynamic model predictions at $P_{sat}$ and 2 kbar. CM30 = Campbell and Maass (1930), BS40 = Beuschlein and Simenson (1940), RM92 = Rumpf and Maurer (1992), GP85 = Goldberg and Parker (1985) based on critical evaluation of extensive previous data (validity 0–150 °C), HKF = HKF model updated by Schulte et al. (2001); AD = Akinfiev and Diamond (2003) model.

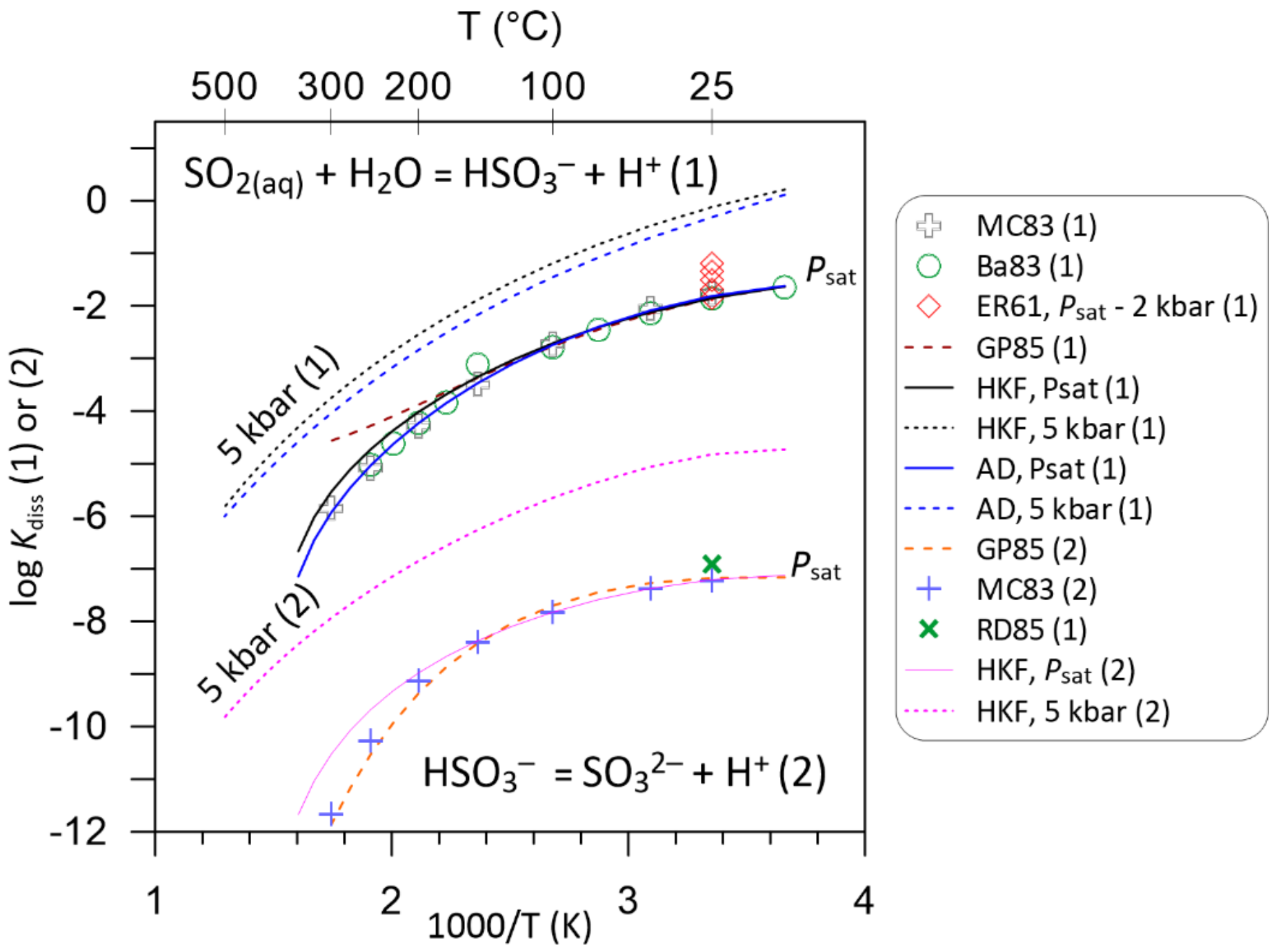

**Fig. 4.14** First (1) and second (2) dissociation constants of sulfur dioxide in aqueous solution as a function of the reciprocal of absolute temperature, according to the major available experimental or compilation datasets and thermodynamic models (at saturated vapor pressure, $P_{sat}$, unless indicated): MC85 = Murray and Cubicciotti (1983), Ba83 = Barbero et al. (1983), ER61 = Ellis and Anderson (1961), GP85 = Goldberg and Parker (1985), RD85 = Rhee and Dasgupta (1985), HKF model = Schulte et al. (2001), AD model = Akinfiev and Diamond (2003). The bisulfite ion, $HSO_3^-$, includes both isomers, $(HS)O_3^-$ and $(HO)SO_2^-$.

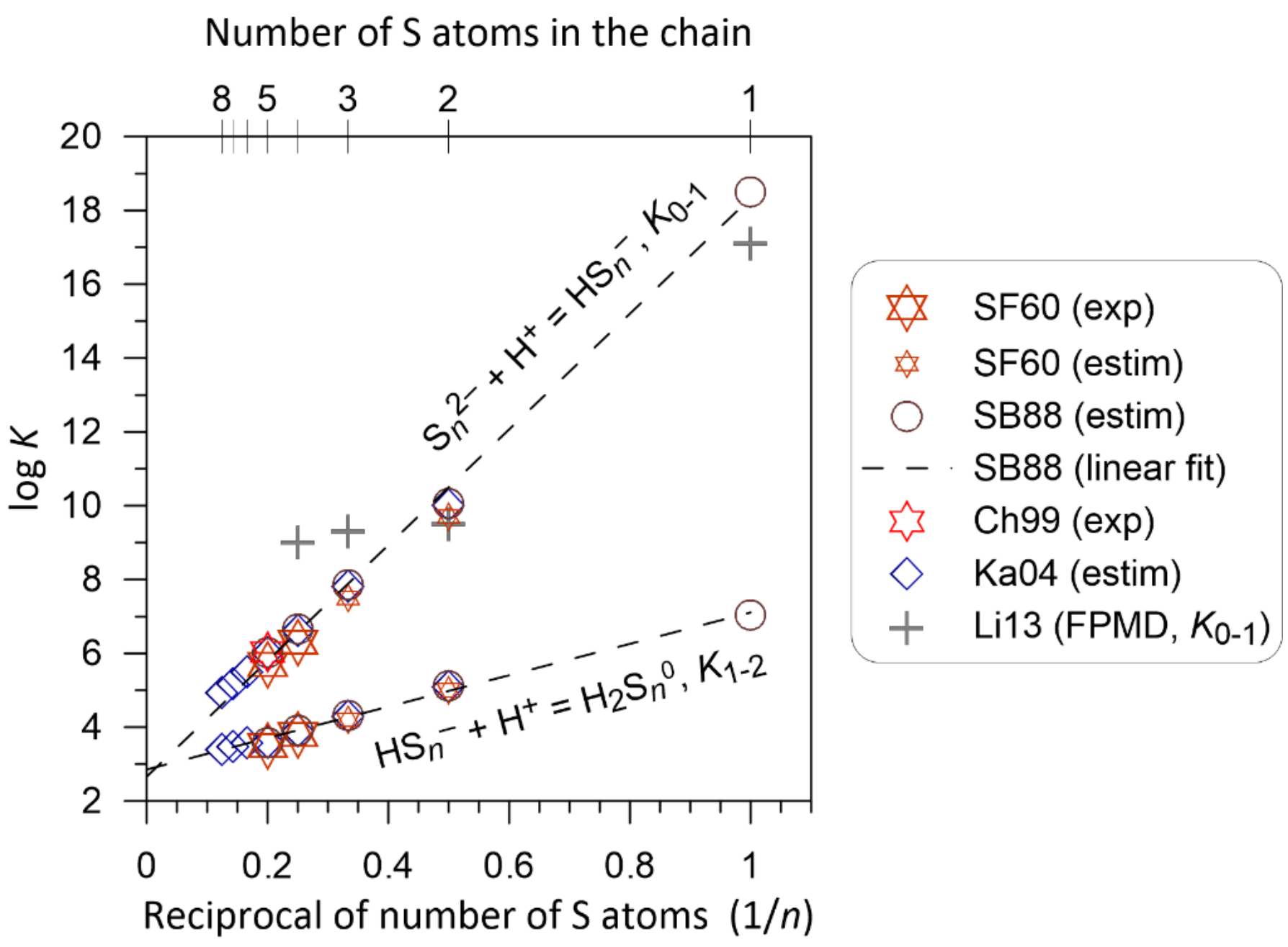

**Fig. 4.15** First ($K_{0\text{-}1}$) and second ($K_{1\text{-}2}$) protonation constants of polysulfide dianions at ambient conditions as a function of the reciprocal of the number of sulfur atoms in the polysulfide chain (1/*n*) according to available data: SF60 = Schwarzenbach and Fischer (1960), both experiment and estimation, SB88 = Schoonen and Barnes (1988), Ch99 = Chadwell et al. (1999), Ka04 = Kamyshny et al. (2004), Li13 = Liu et al. (2013a). exp = experiment, estim = estimation, FPMD = first principles molecular dynamics calculations.

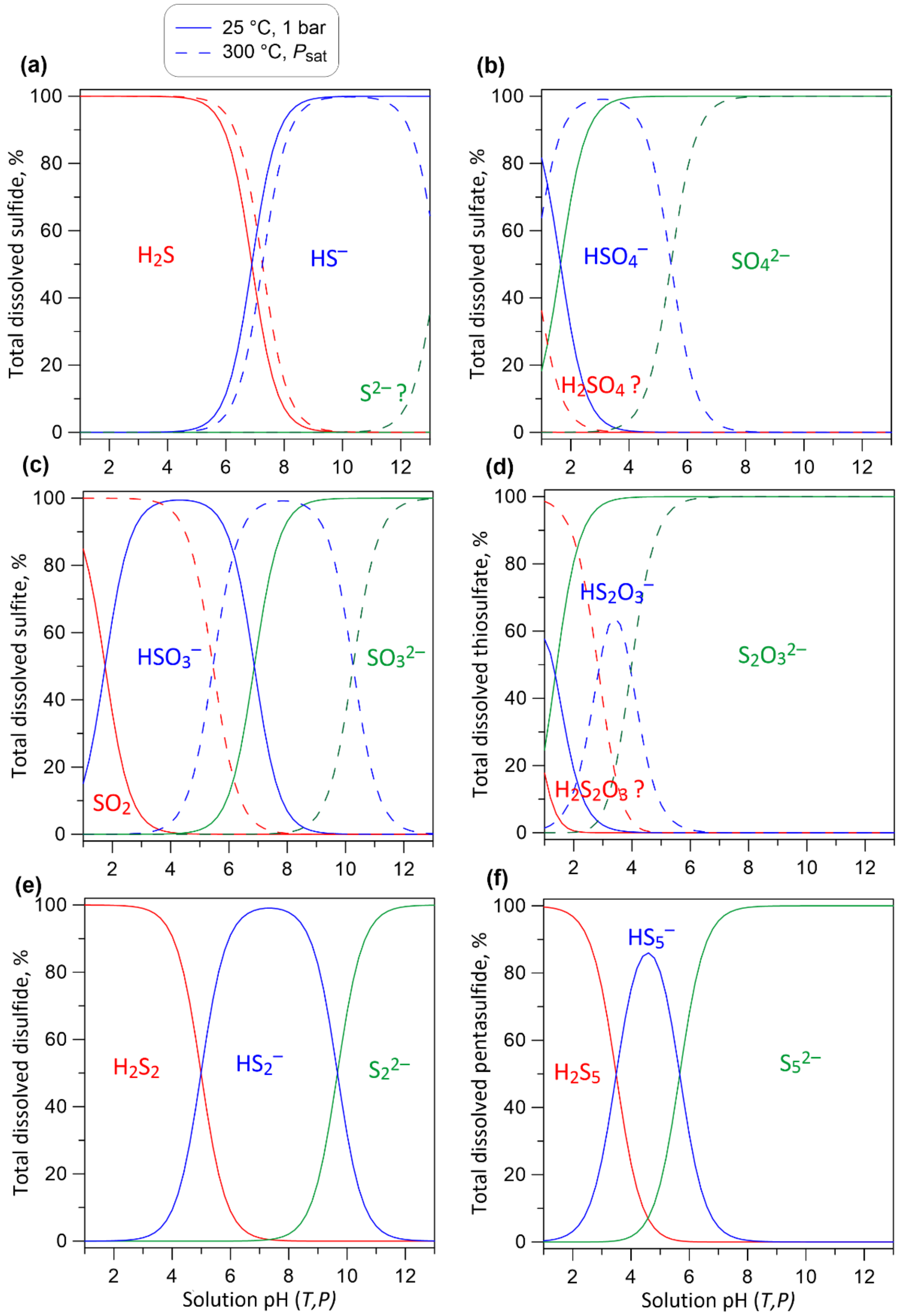

**Fig. 4.16** Distribution of the indicated individual sulfur forms as a function of pH at ionic strength of 0.1 and 25 °C, 1 bar (solid curves) and 300 °C, $P_{sat}$ (dashed curves), calculated using the thermodynamic data adopted in this study as overviewed in the text. Data for the species with question marks yet remain uncertain.

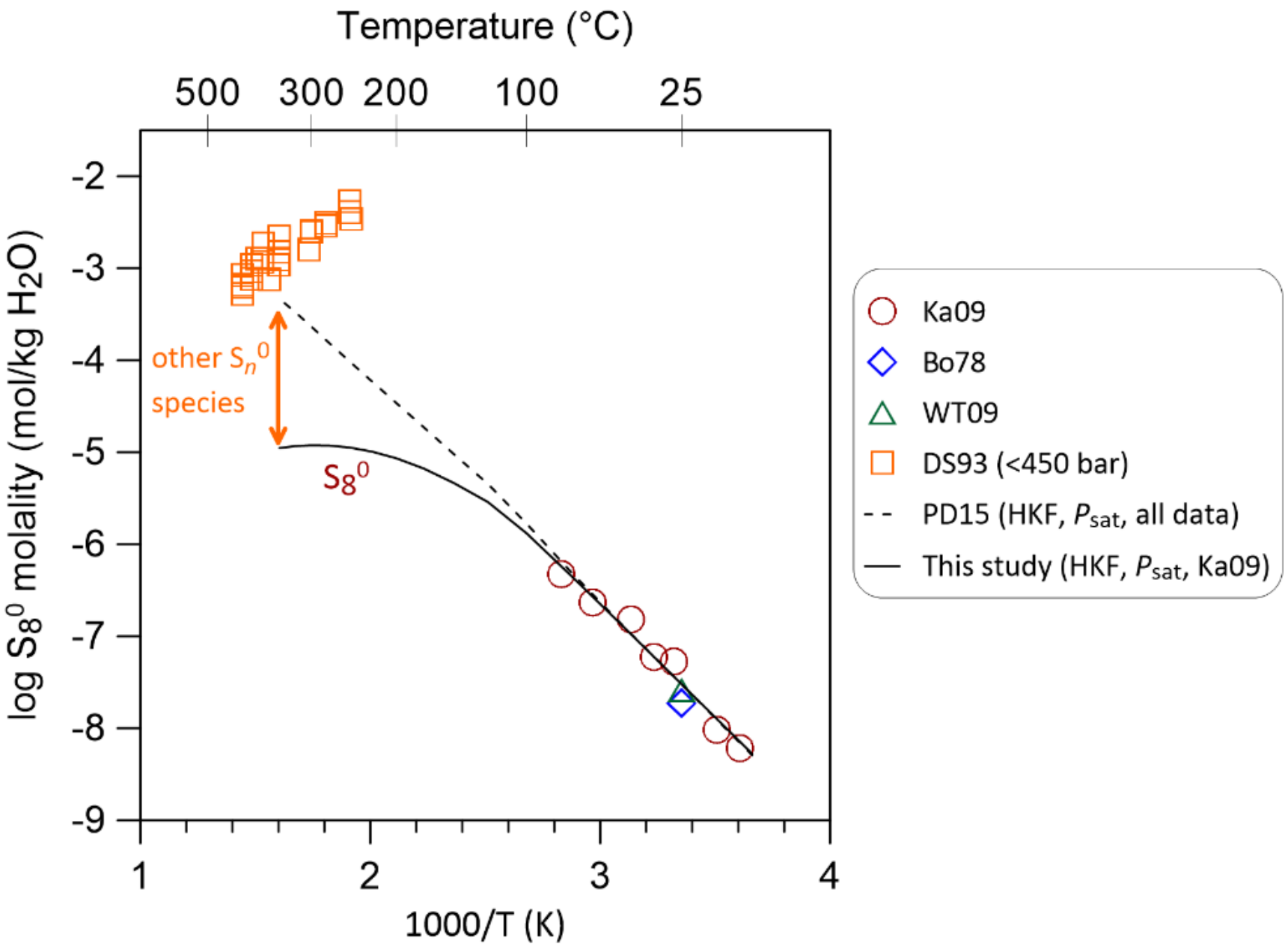

**Fig. 4.17** Dissolved molecular sulfur concentrations in water in equilibrium with solid α-allotrope of sulfur (<120 °C) and molten sulfur at higher temperatures, according to available experimental data and thermodynamic models. Ka09 = Kamishny (2009), Bo78 = Boulëgue (1978b), WT09 = Wand and Tessier (2009); DS93 = Dadze and Sorokin (1993); PD15 = Pokrovski and Dubessy (2015). The discrepancy between the two HKF model predictions at elevated temperatures (see Table 4.5 for parameters) indicates the formation of molecular sulfur species other than $S_8^0$ in hydrothermal fluids (indicated by the vertical orange arrow).

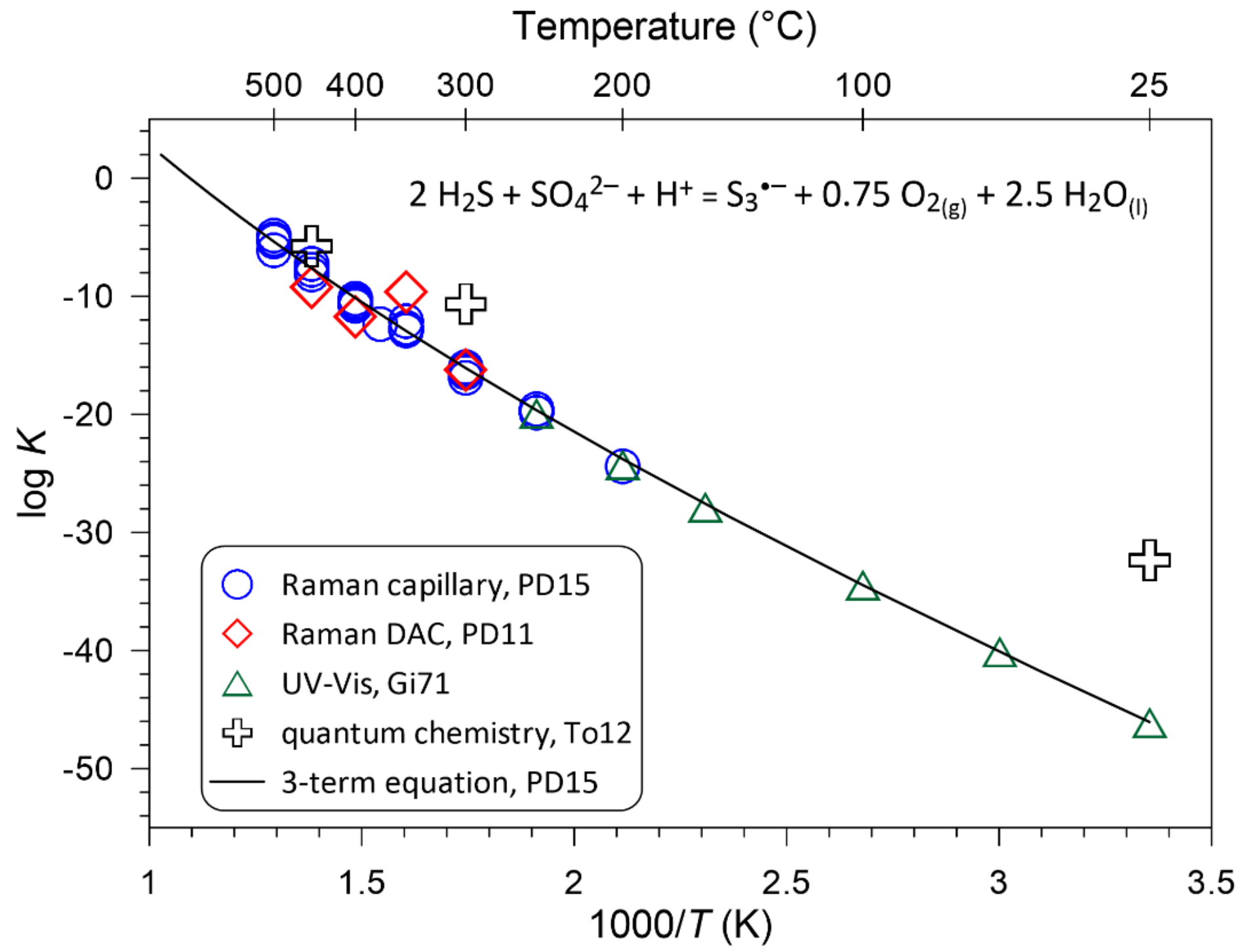

**Fig. 4.18** Decimal logarithm of the equilibrium constant of reaction (4.44) versus the reciprocal of absolute temperature (in Kelvin). Symbols show the data derived in the indicated studies: PD11 = Pokrovski and Dubrovinsky (2011), PD15 = Pokrovski and Dubessy (2015), Gi71 = Giggenbach (1971b), To12 = Tossell (2012). Solid curve is the 3-term equation from PD15: $\log K_{44} = -123.3(\pm 20.5) - 10585(\pm 1380)/T + 45.57(\pm 6.50)\times \log T$. The corresponding set of HKF-model parameters is available in PD15 consistent with those of $H_2S_{(aq)}$ and $SO_4^{2-}$ from the SUPCRCT database. Note that this set is attached to the thermodynamic properties of $H_2S_{(aq)}$ (Schulte et al. 2001) that were used in its deviation (see Pokrovski and Dubessy 2015 for in-depth discussion of errors related to the use of other models and datasources).

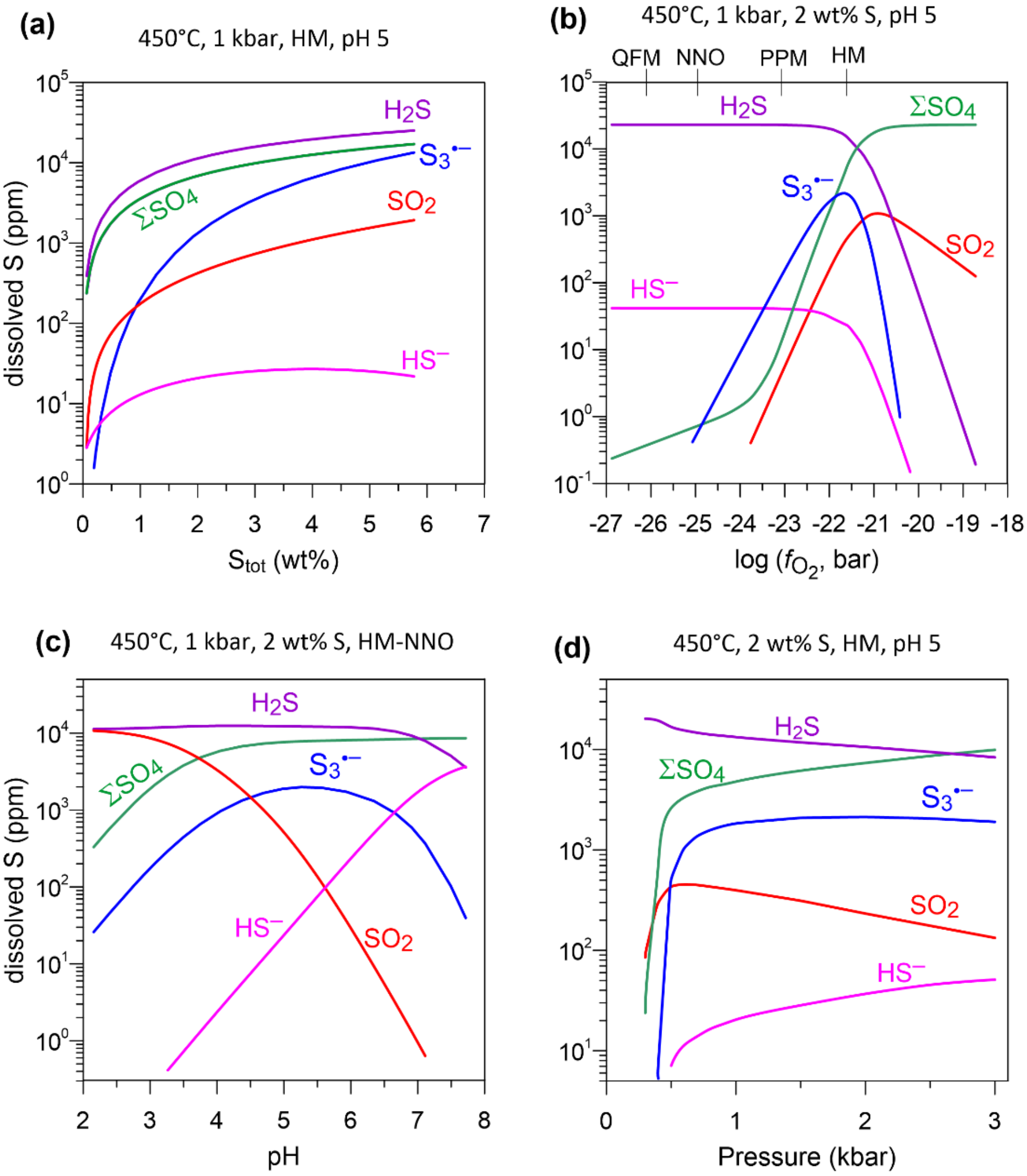

**Fig. 4.19** Concentrations of $S_3^{\bullet-}$ and other major sulfur forms (expressed as ppm of S) calculated in model hydrothermal fluids of 10 wt% salinity (NaCl+KCl) equilibrated with the quarz-muscovite-potassic feldspar assemblage, at 450°C and indicated conditions as a function of (a) total S content, (b) oxygen fugacity, (c) pH, and (d) pressure. Curves show concentrations of each indicated species ($\Sigma SO_4$ stands for the sum of $SO_4^{2-}$, $HSO_4^-$, and their $K^+$ and $Na^+$ ion pairs). QFM = quartz-fayalite-magnetite, NNO = nickel-nickel oxide, PPM = pyrite-pyrrhotite-magnetite, and HM = hematite-magnetite (modified after Pokrovski and Dubessy 2015).

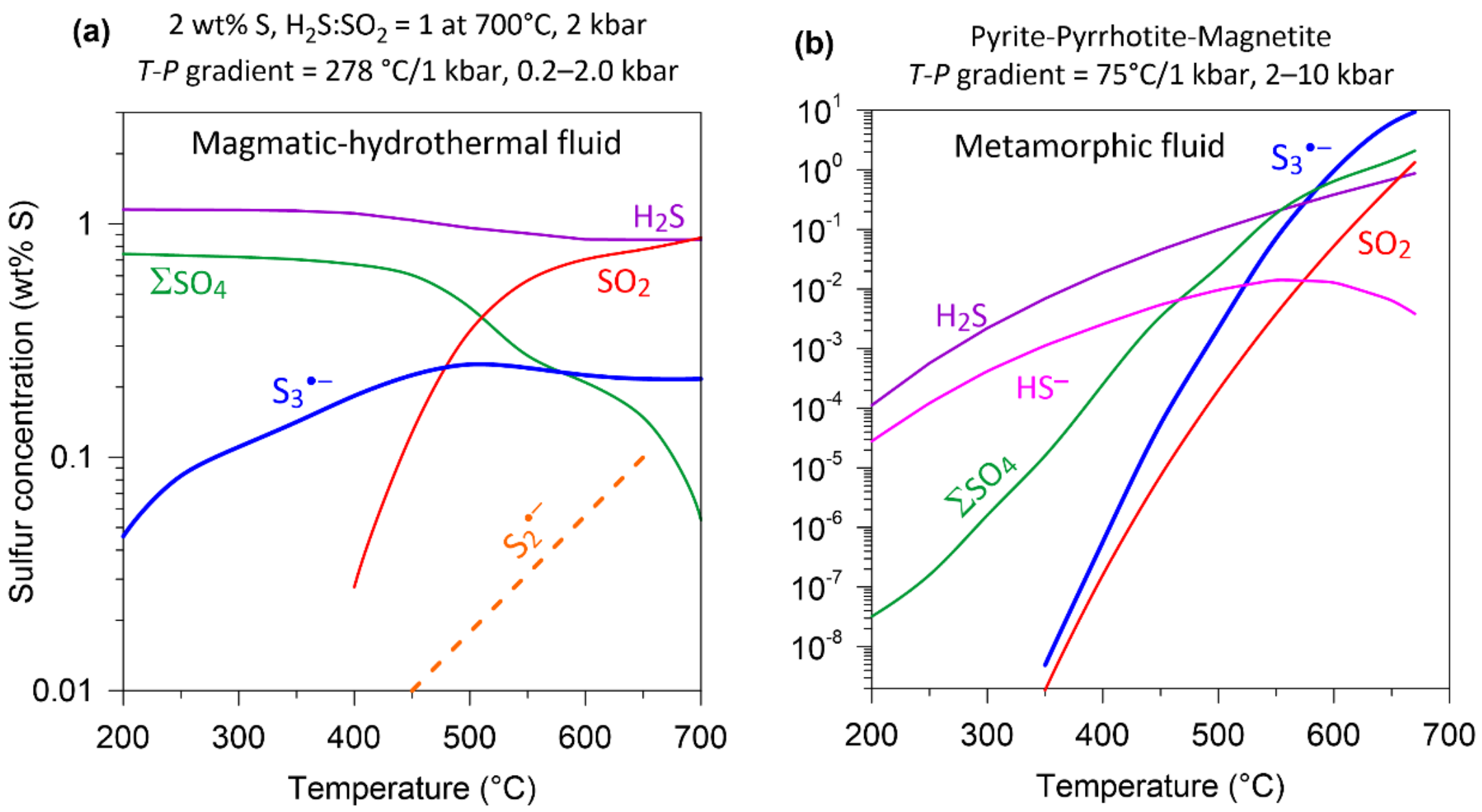

**Fig. 4.20** Concentrations of the major forms of sulfur, including the radical ions, in two typical types of fluids as a function of temperature, (a) magmatic–hydrothermal fluid and (b) metamorphic fluid both in equilibrium with the quartz-muscovite-potassic feldspar assemblage. The sulfur species concentrations were predicted using the stability constants of sulfur forms discussed in this chapter. The concentration of $S_2^{\bullet-}$ is tentative (see text). The $S_3^{\bullet-}$, and potentially $S_2^{\bullet-}$, ions may represent a significant part of dissolved sulfur content over a wide *T*–*P* range.

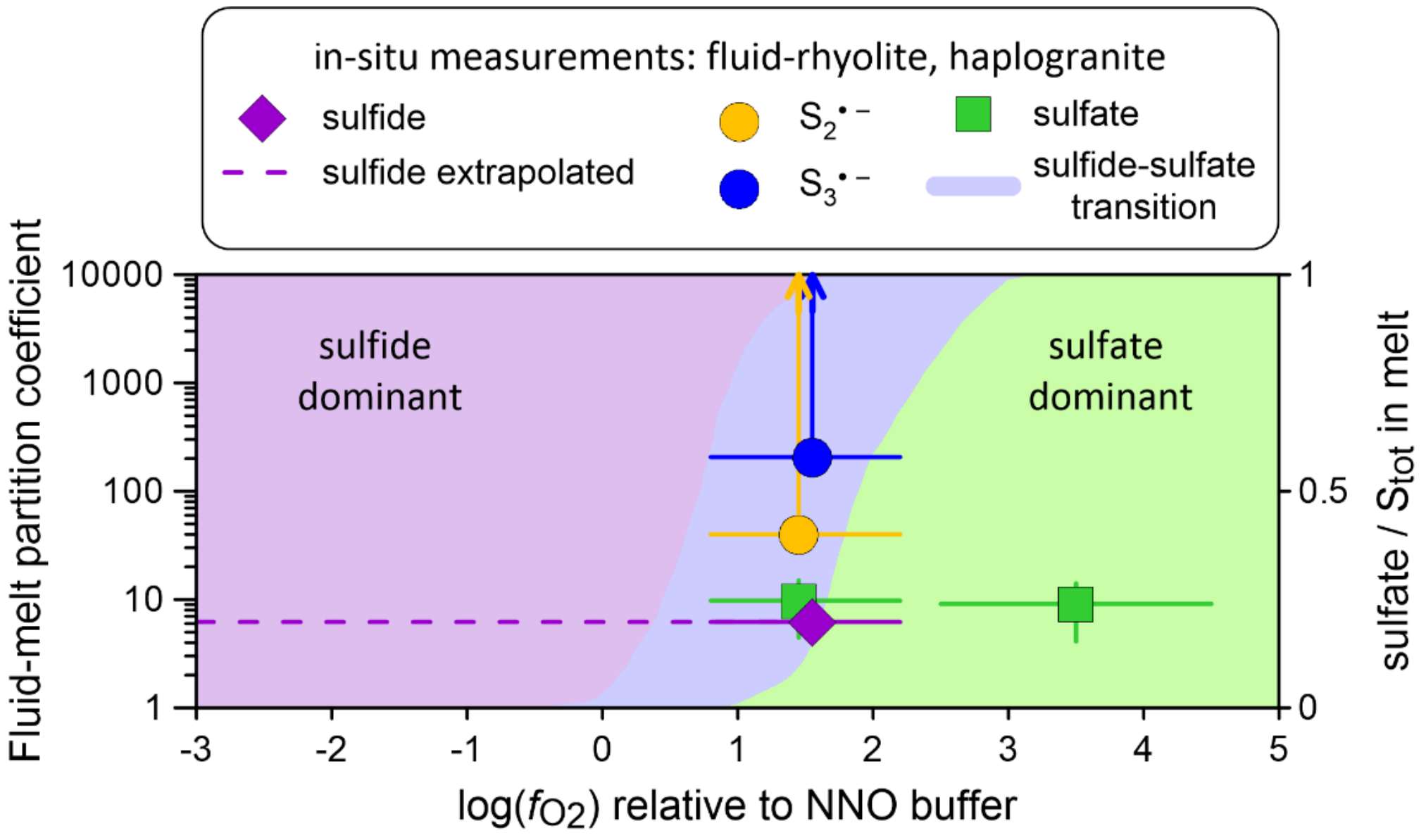

**Fig. 4.21** In situ fluid/melt partition coefficients of sulfate, sulfide, $S_3^{\bullet -}$ and $S_2^{\bullet -}$, derived from in situ measurements for felsic melts at 700 °C and 3–15 kbar as a function of $f_{O2}$. Error bars are 1 s.d.; only upper vertical error bars for the radical ions could be estimated (shown by arrows), because their concentrations in the melt were close to the detection limit. Blue area outlines the sulfate-sulfide boundary in water saturated melt (estimated according to Moretti and Baker 2008); horizontal dashed line shows partition coefficients of sulfide extrapolated to the more reduced conditions. Note that a partition coefficient of an individual sulfur species is independent of redox conditions (modified after Colin et al. 2020).

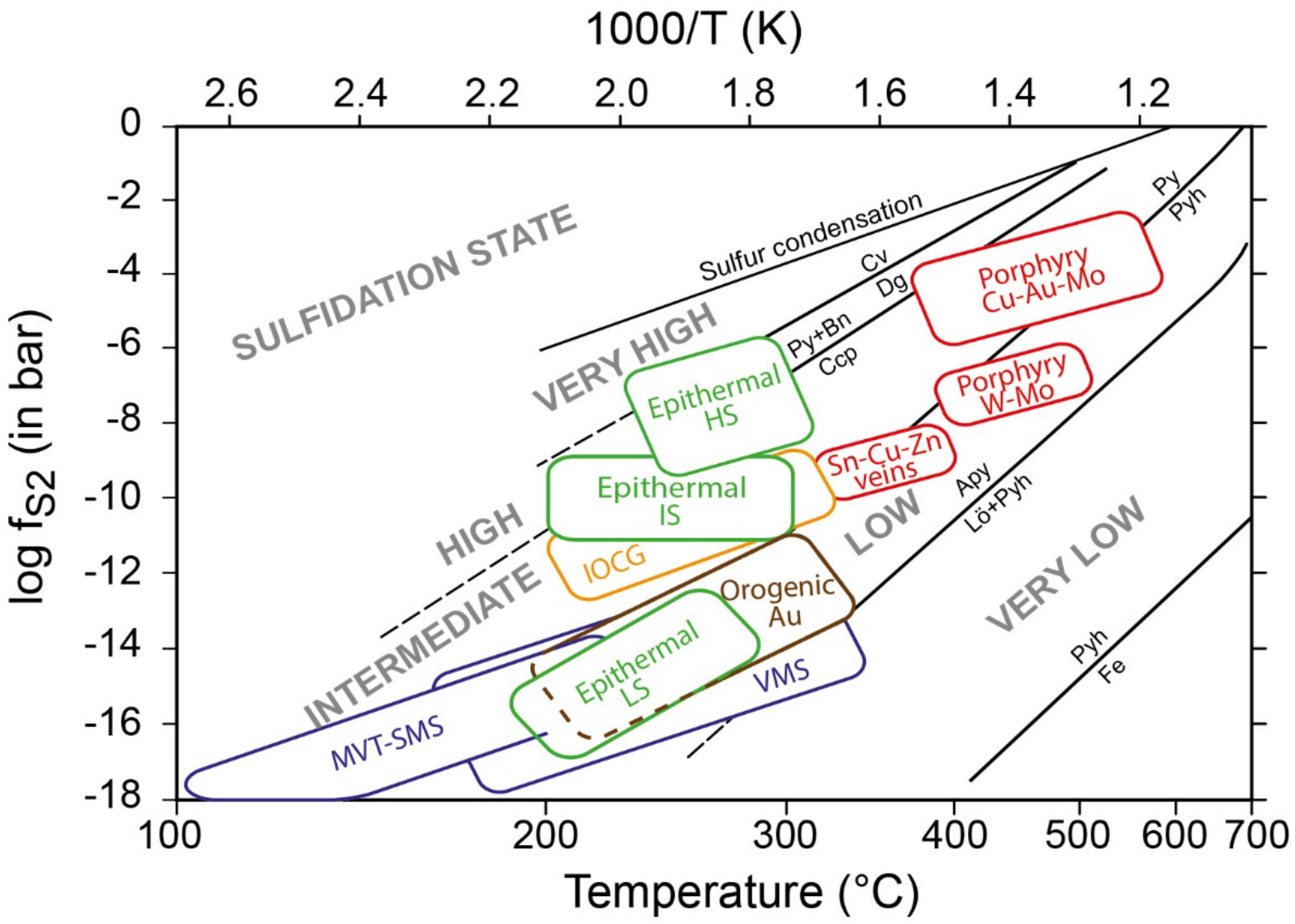

**Fig. 4.22** Sulfidation states of hydrothermal fluids, in sulfur fugacity-temperature space, based on natural sulfide mineral assemblages. Lines represent the indicated mineral equilibria estimated at 1 bar pressure (from Einaudi et al. 2003). Different types of hydrothermal ore deposits are shown (colors for visualization aids only; modified from Fontboté et al. 2017). Abbreviations: IOCG = iron oxide copper gold; MVT = Mississippi Valley type, SMS = sedimentary-hosted massive sulfide; VMS = volcanic-hosted massive sulfide. Mineral abbreviations (according to Warr 2021): Apy = arsenopyrite; Bn = bornite; Ccp = chalcopyrite; Cv = covellite; Dg = digenite; Fe = iron; Lö = löllingite; Py = pyrite; Pyh = pyrrhotite.

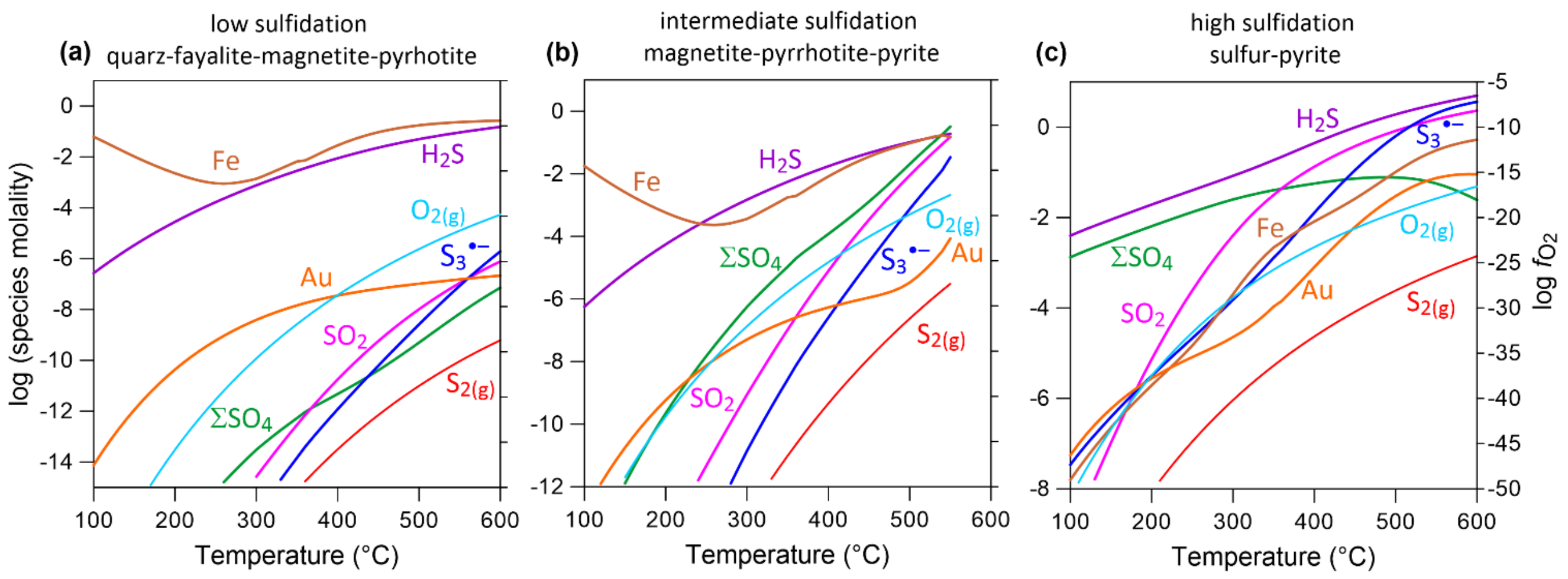

**Fig. 4.23** Sulfur speciation in the fluid phase at different sulfidation states shown in Fig. 22. Concentrations of aqueous sulfur species in hydrothermal saline fluid (KCl+NaCl, 10 wt% NaCl eq.) are shown as a function of temperature, along a typical pressure gradient (from 10 bar at 100 °C to 1000 bar at 600 °C) of hydrothermal ore deposit formation, in equilibrium with (a) quartz-faylite-magnetite-pyrrhotite, (b) pyrite-pyrrhotite-magnetite, and (c) liquid sulfur-pyrite, which are relevant to low, intermediate, and high sulfidation states, respectively. In (a) and (b), the fluid acidity is buffered by the quartz-muscovite-potassic feldspar assemblage (pH ~5–6), whereas in (c) the acidity is imposed by protonation reactions of sulfate and sulfite forming by dissolusion of sulfur (pH ~2–4). Sulfur and oxygen fugacity values are also plotted as $S_{2(g)}$ and $O_{2(g)}$ (in bars) to the left and right Y axes, respectively. Solubility of Fe (as $Fe^{2+}$, $FeCl^{+}$ and $FeCl_2^0$) and Au (dominantly $Au(HS)_2^-$ and $Au(HS)S_3^-$) is also shown. In (b) the curves are drawn to 550 °C, above which pyrite is unstable at the chosen pressure conditions (<1 kbar).

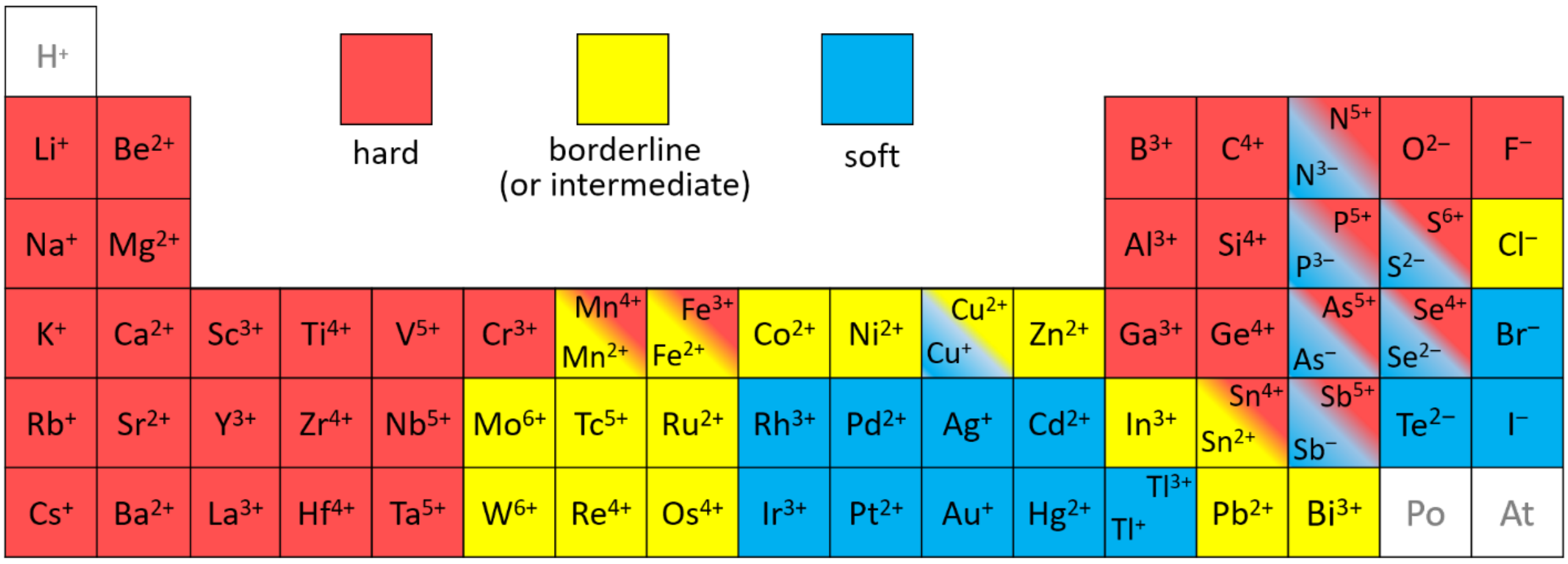

**Fig. 4.24** Hard-soft classification of chemical elements for their naturally representative oxidation states (modified from Pokrovski 2025a).

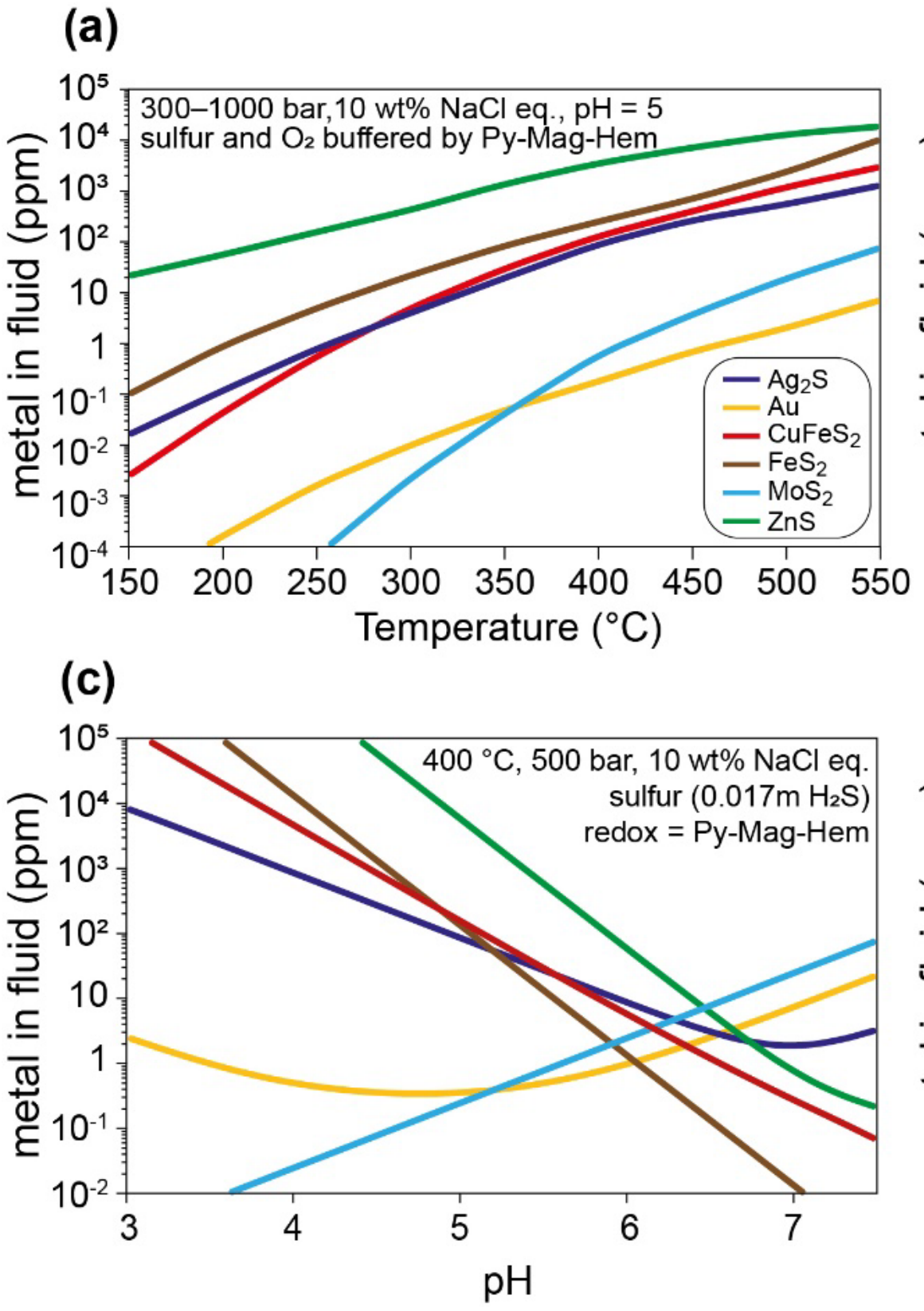

**Fig. 4.25** Solubility of metals in hydrothermal fluids in equilibrium (i.e. saturated) with major sulfide minerals as a function of four parameters (a) temperature, (b) fluid salinity, (c) fluid pH, and (d) $H_2S$ concentration. In each case, the other physical-chemical parameters of the fluid are fixed. Galena solubility, not shown, is slightly lower than that of sphalerite. Abbreviations: hem = hematite, py = pyrite, mag = magnetite. Gold solubility in (a) increases with temperature mainly due to the growing abundance of hydrosulfide (and trisulfur ion) ligands due to enhanced solubility of the Py-Mag-Hem assemblage with temperature. Note that predictions for Mo are tentative and given for illustration purposes only, in light of poor constraints on the stability of its chloride, sulfide and alkali complexes (see section 4.4.10). Modified from Kouzmanov and Pokrovski (2012) and Fontboté et al. (2017).

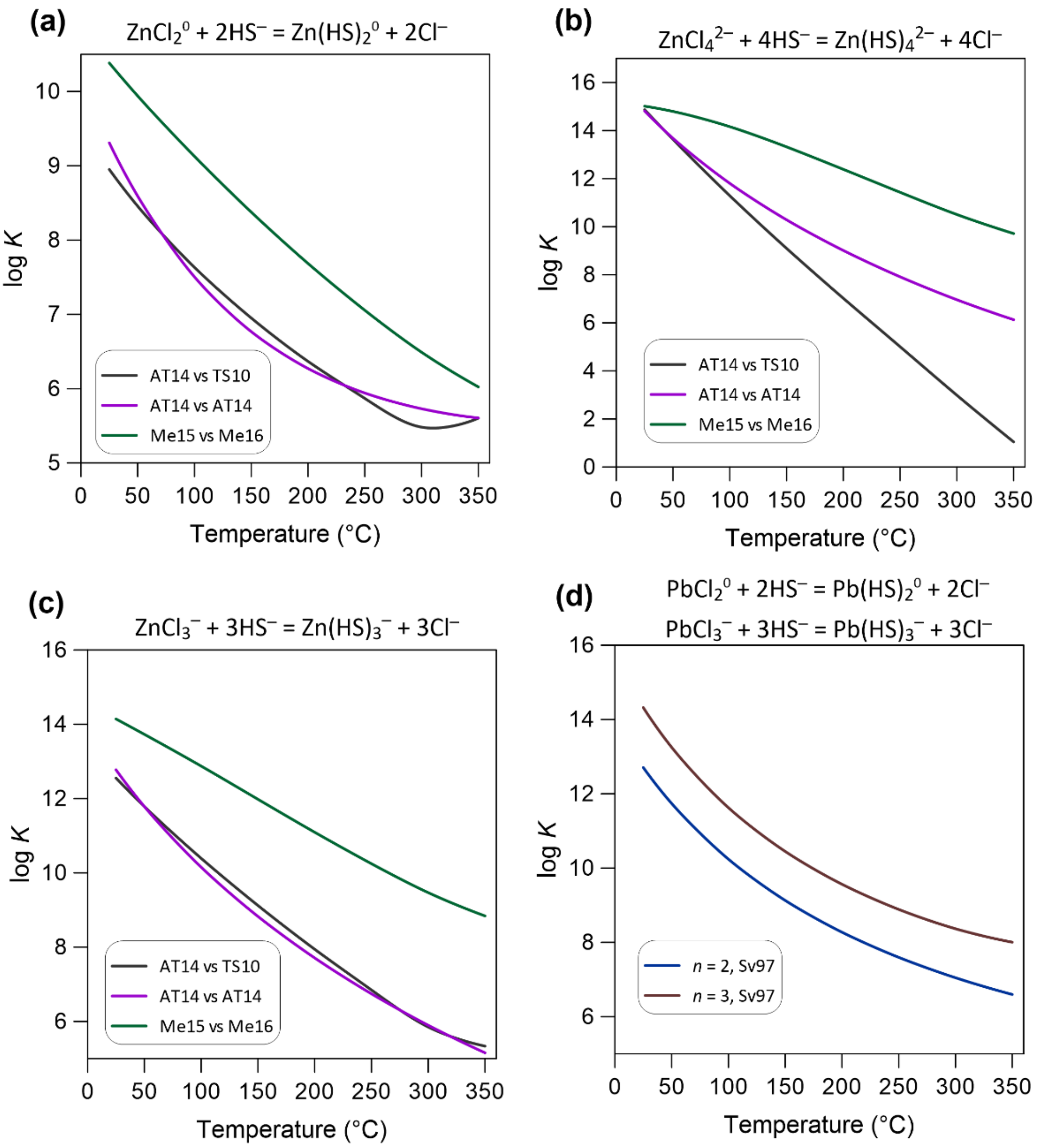

**Fig. 4.26** Decimal logarithms of the constants of the $Cl^-$ vs $HS^-$ ligand exchange reactions (4.68) and (4.69) indicated in the figure for the major Zn and Pb complexes as a function of temperature (at $P_{sat}$), according to the available model equations discussed in the text. Sv97 = Sverjensky et al. (1997) from the SUPCRT92 database, TS10 = Tagirov and Seward (2010), AT14 = Akinfiev and Tagirov (2014), Me15 = Mei et al. (2015a), Me16 = Mei et al. (2016). First reference in the legend corresponds to the datasource for the chloride complex versus the second reference for its sulfide counterpart.

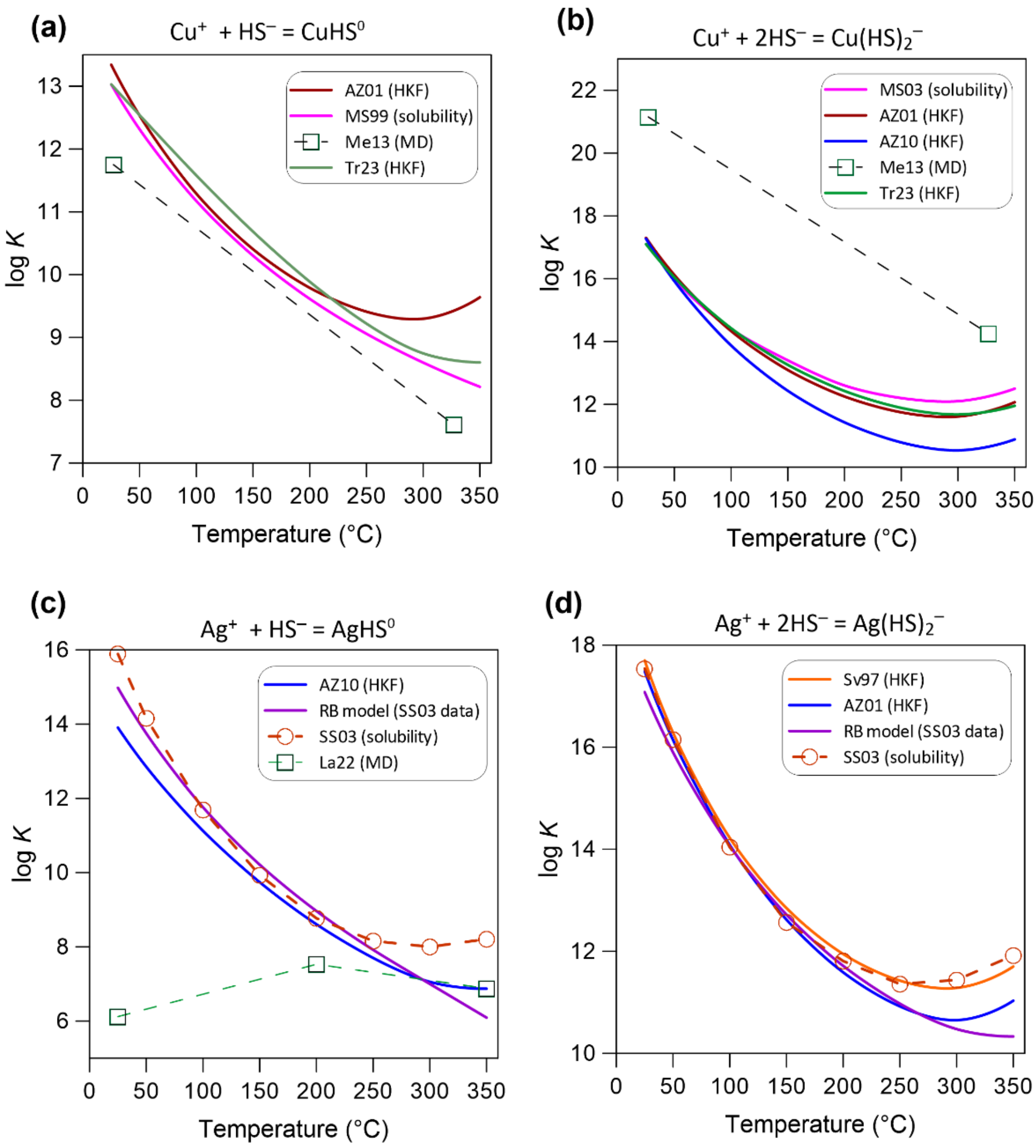

**Fig. 4.27** Decimal logarithms of the reactions constants of formation of the major $Cu^{I}$ (a, b) and $Ag^{I}$ (c, d) hydrosulfide complexes, according to the thermodynamic data sources discussed in the text. Sv97 = Sverjensky et al. (1997) from the SUPCRT92 database, AZ01 = Akinfiev and Zotov (2001), AZ10 = Akinfiev and Zotov (2010), MS03 = Mountain and Seward (2003), SS03 = Stefansson and Seward (2003a), Me13 = Mei et al. (2013), La22 = Lai et al. (2022), Tr23 = Trofimov et al. (2023). RB model = SS03 data were regressed, in this study for comparison, using the RB model and thermodynamic properties of the $Ag^+$ ion from AZ01 ($AgHS^0 = Ag^+ + HS^-$, $pK_{298} = 14.98$, A = –0.194; $Ag(HS)_2^- = Ag^+ + 2HS^-$, $pK_{298} = 17.08$, A = 0.39).

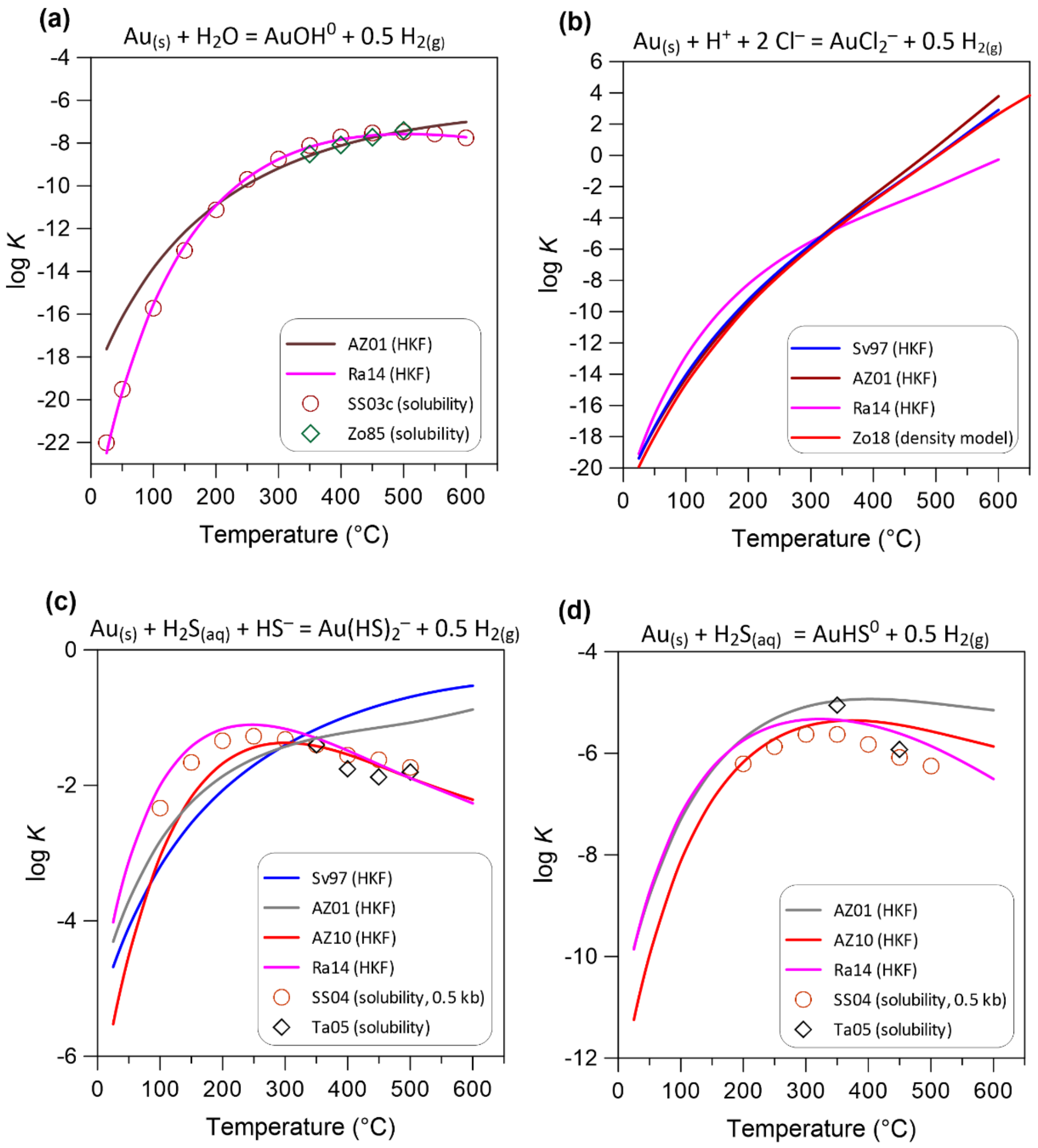

**Fig. 4.28** Decimal logarithms of the solubility constants at 1 kbar (unless indicated) of the key Au aqueous species, $AuOH^0$ (a), $AuCl_2^-$ (b), $Au(HS)_2^-$ (c), and $AuHS^0$ (d), according to the major indicated thermodynamic datasources, discussed in the text: Sv97 = Sverjensky et al. (1997) from the SUPCRT92 database, AZ01 = Akinfiev and Zotov (2001), AZ10 = Akinfiev and Zotov (2010), Ra14 = Rauchenstein-Martinek et al. (2014), SS03b = Stefánsson and Seward (2003b), SS04 = Stefánsson and Seward (2004), Ta05 = Tagirov et al. (2005), Zo18 = Zotov et al. (2018).

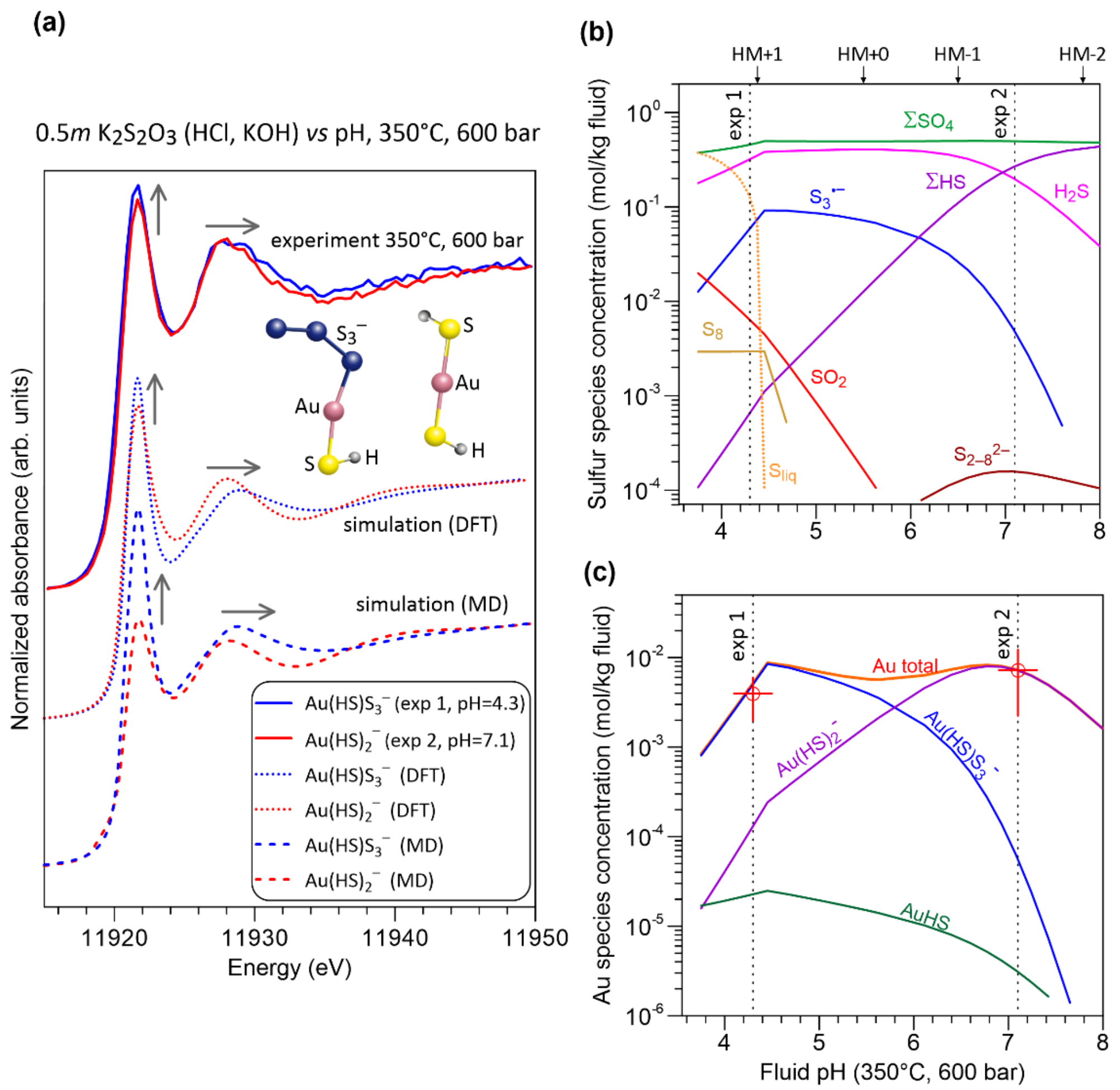

**Fig. 4.29** Gold-trisulfur ion complexes in model hydrothermal fluids revealed by a combination of in situ spectroscopy and solubility measurements (modified from Pokrovski et al. 2022b). (a) Comparison between measured and calculated Au $L_3$-edge HERFD-XANES spectra of the two major aqueous Au species with sulfur ligands. Measured spectra (solid curves) are from the two experiments in acidic and slightly alkaline thiosulfate solutions (pH of the neutrality point of water is 5.5 at 350 °C and 600 bar). The solution detailed sulfur (b) and gold (c) speciation is calculated, using the thermodynamic datasources discussed in the text, as a function of pH and corresponding redox (indicated in $\log_{10}f_{O_2}$ relative to the conventional hematite-magnetite buffer, HM). Simulated XANES spectra in (a) are obtained by quantum-chemistry calculations (FDMNES program) using the species structures shown in ball-and-stick style (Au = pink, S = yellow, $S_3$ = blue, H = grey) from static DFT (dotted curves) and dynamic MD (dashed curves) simulations. Grey arrows in (a) indicate differences in the experimental spectra, also apparent in the corresponding calculated spectra, supporting the predicted change in Au speciation from $Au(HS)_2^-$ to $Au(HS)S_3^-$.

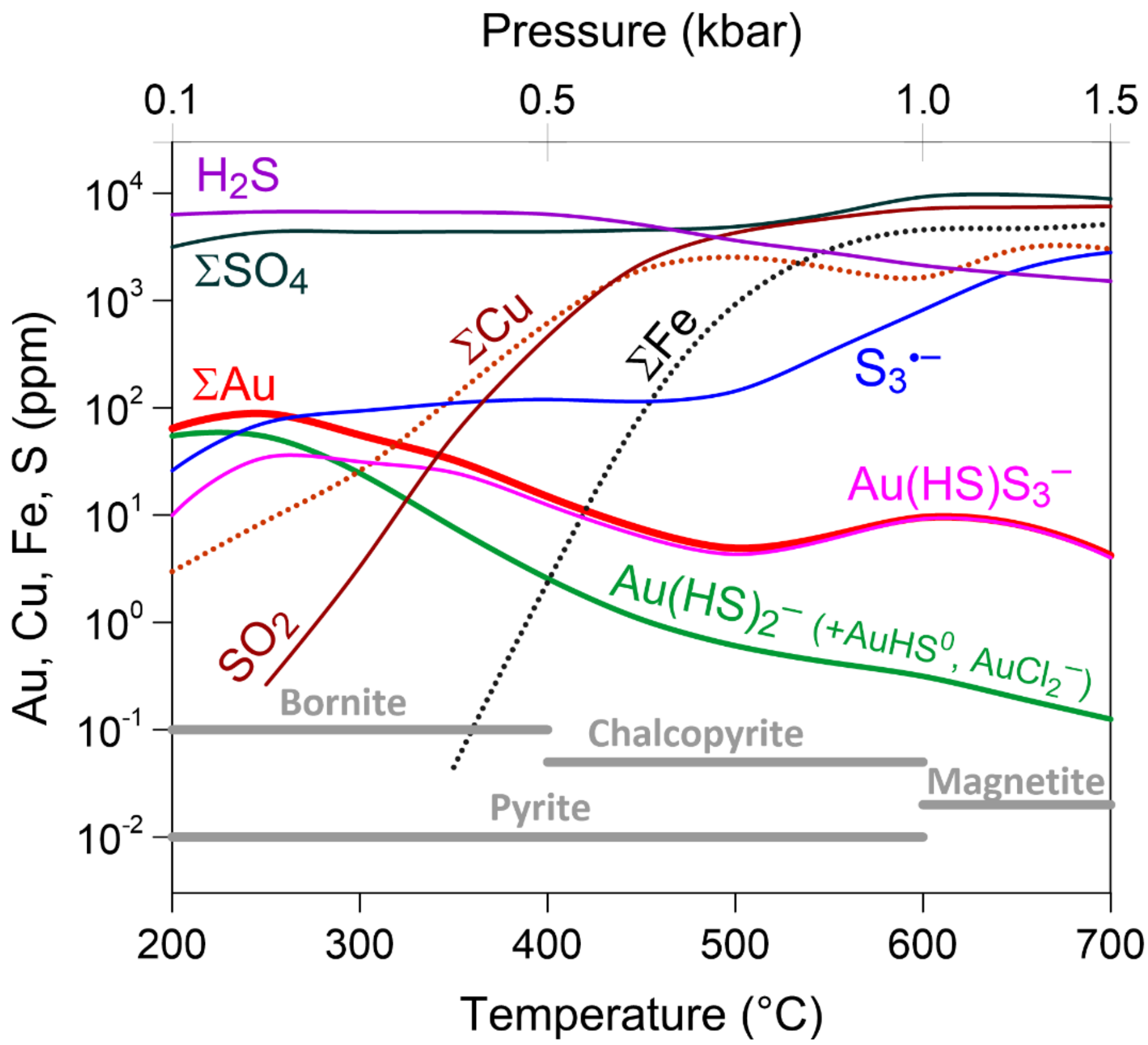

**Fig. 4.30** Gold, copper, iron and sulfur solubility and speciation in the hydrothermal fluid phase during its evolution in a generic context of porphyry-epithermal Cu-Au-Mo deposits (modified from Pokrovski et al. 2015). An initial magmatic fluid has the following typical composition as inferred from fluid inclusions and geochemical studies: 10 wt% NaCl equivalent, 2 wt% S, $H_2S$:$SO_2$ molal ratio = 1, 0.75 wt% Fe as $FeCl_2$, 0.30 wt% as CuCl. This fluid degasses from magma at 700 °C and 1.5 kbar and cools and decompresses in the liquid phase in equilibrium with native Au and alkali aluminosilicate rocks (quartz-muscovite-potassic feldspar assemblage, QMK, pH of 5–6 at all temperatures); this is a common scenario of fluid evolution in a porphyry-epithermal setting (see Kouzmanov and Pokrovski 2012 for details). Curves show the concentrations of the major indicated sulfur forms (in ppm S), Au total solubility (ΣAu, red curve) which is the sum of the major di-hydrosulfide and subordinate mono-hydrosulfide and di-chloride complexes (green curve) and the dominant $Au(HS)S_3^-$ complex (pink curve), and total Fe and Cu solubility (ΣFe and ΣCu, as dominantly chloride complexes, Table 4.6). Gray horizontal bars show the temperature range of stability of the indicated minerals that precipitate on fluid cooling.

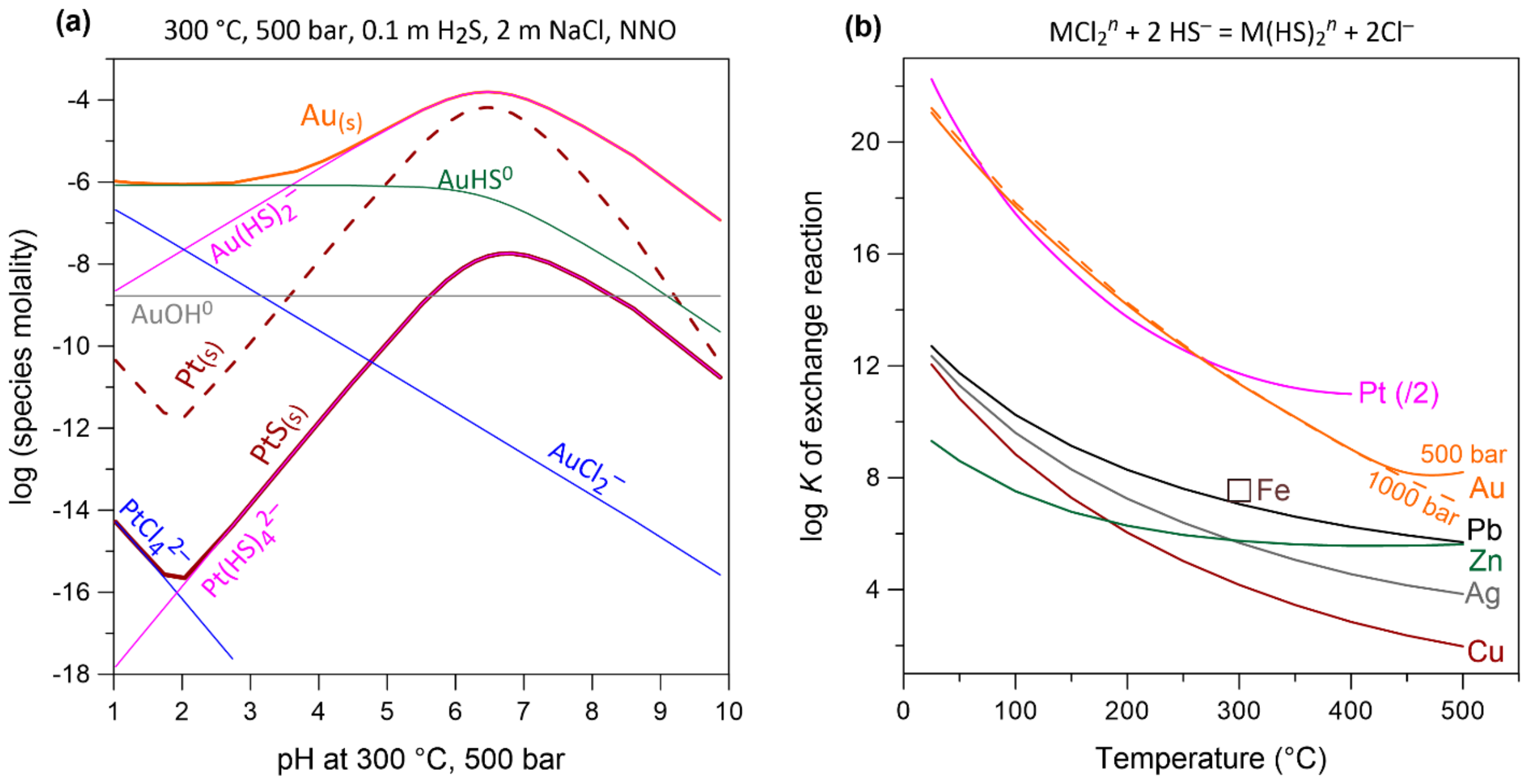

**Fig. 4.31** (a) Comparison of the solubility of $PtS_{(s)}$, and native Pt and Au in an aqueous fluid of 2 m NaCl and 0.1 m $H_2S$ at the oxygen fugacity of the conventional nickel-nickel oxide buffer (NNO, log $f_{O2}$ = –33.9) as a function of pH at 300 °C and 500 bar. The thermodynamic data of all Au species, $PtCl_4^{2-}$, and $Pt(HS)_4^{2-}$ are, respectively, from Pokrovski et al. (2014), Tagirov et al. (2019), and Laskar et al. (2022). Adopted and modified from Laskar et al. (2022). (b) Comparison of the equilibrium constant values of the indicated $Cl^-$ vs $HS^-$ ligand exchange reaction between di-chloride and di-hydrosulfide complexes of $Cu^I$, $Ag^I$, $Zn^{II}$, $Pb^{II}$, $Fe^{II}$, $Au^I$, and $Pt^{II}$, as a function of temperature and at 500 bar pressure for all metals (and at 1 kbar for Au). Note that the $Pt^{II}$ reaction constant between tetrachloride and tetrahydrosulfide complexes (reaction 4.80) was scaled to 2 exchanging ligands in the reaction to allow more direct comparisons with the other metals. The following datasources were used, as discussed in the text: Cu and Ag - Akinfiev and Zotov (2001, 2010); Au - Pokrovski et al. (2014) and Zotov et al. (2018); Zn - Akinfiev and Tagirov (2014); Pb - Sverjensky et al. (1997), Fe - Laskar (2022), Pt - Laskar et al. (2022).

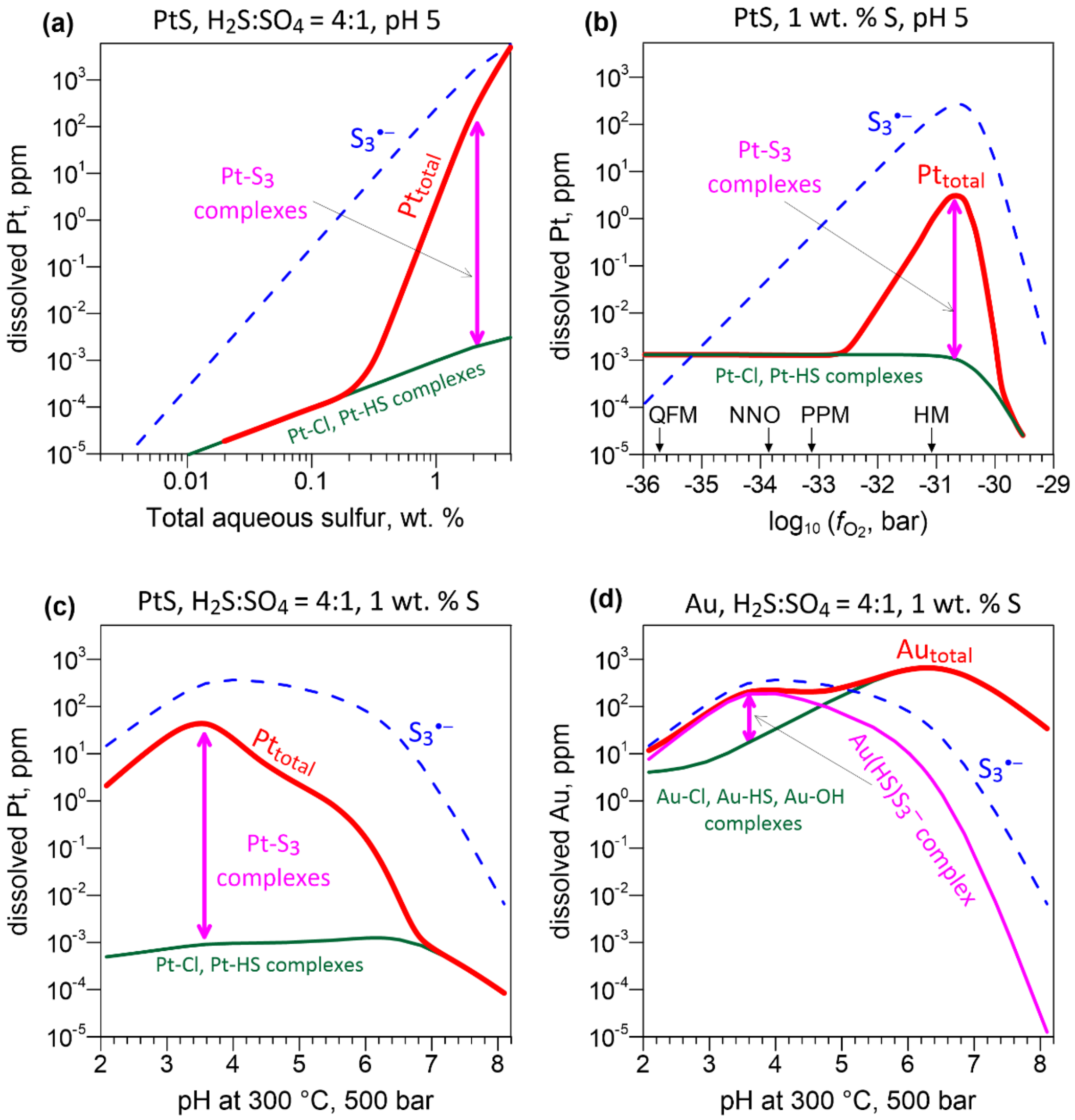

**Fig. 4.32** The effect of the trisulfur ion on platinum solubility in hydrothermal fluids. Platinum dissolved concentration in equilibrium with $PtS_{(s)}$ in hydrothermal fluid at 300°C and 500 bar and its comparison with gold as a function of (a) total dissolved S concentration at pH ~5 and $f_{O2}$ buffered by the sulfide-sulfate equilibrium with $(H_2S)_{tot}$:$(SO_4)_{tot}$ molal ratio of 4:1 corresponding to the maximum of $S_3^{\bullet-}$ abundance (see reaction 4.45); (b) oxygen fugacity corresponding to common redox buffers, QFM – quartz-fayalite-magnetite, NNO – nickel-nickel oxide, PPM – pyrite-pyrrhotite-magnetite, and HM – hematite-magnetite; (c) fluid acidity, pH = –log $a(H^+)$ at $f_{O2}$ buffered by sulfide-sulfate equilibrium ($H_2S$:$SO_4$ = 4:1); (d) in equilibrium with gold metal as a function of pH at the conditions identical to (c). The curves denote the concentrations of $S_3^{\bullet-}$ (blue, in ppm of S, same scale as for Pt and Au), sum of known Pt/Au-Cl, Pt/Au-HS, Pt-$SO_4$ and Au-OH species (green) and the total Pt or Au solubility (red) including $Pt(HS)_2(S_3)_2^{2-}$ and $Pt(HS)_3(H_2O)(S_3)_2^-$ or $Au(HS)S_3^-$. The difference between the two latter curves (vertical pink arrows) corresponds to the contribution of Pt-$S_3$ or Au-$S_3$ types of complexes. The thermodynamic properties of Au and Pt species and other fluid constituents are from datasources of Tables 4.5 and 4.6.

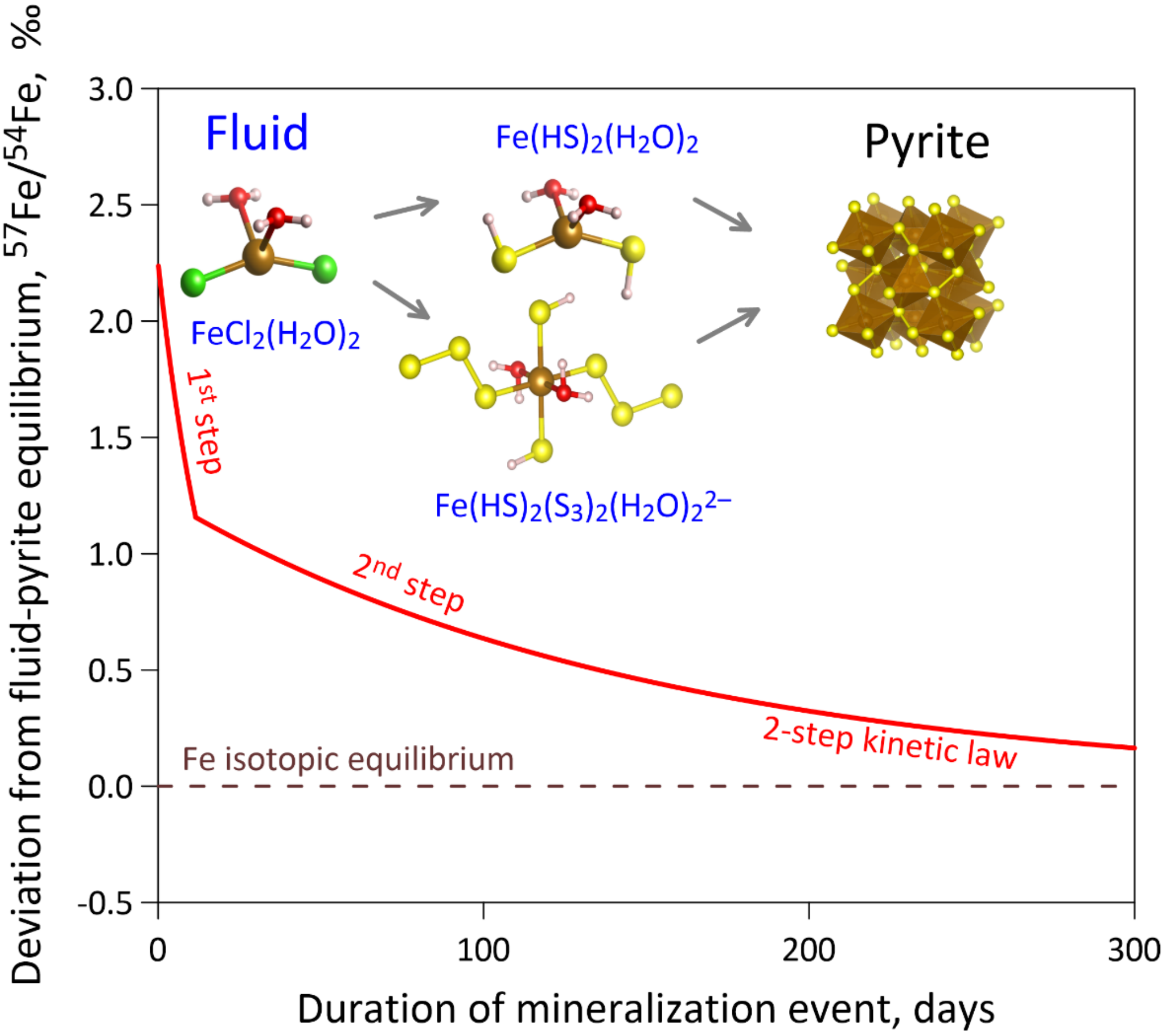

**Fig. 4.33** Two-step mechanism of pyrite formation and equilibration between the mineral and hydrothermal fluid controlled by [$Fe^{II}$-HS-$S_3$]-type aqueous species, as revealed using Fe isotopes (Pokrovski et al. 2021c). The first, both short and fast (<10 days), step of pyrite precipitation occurs via aqueous $Fe^{II}$ (poly)sulfide species precursors exemplified here, forming in reaction between the dominant Fe-bearing complex, $FeCl_2^0$, and hydrosulfide and polysulfide anions. The second, much slower and longer (100s to 1000 days), step of Fe isotope equilibration with the fluid occurs through slower pyrite recrystallization and re-equilibration with the fluid and requires about 1 year to reach 90% of Fe isotope equilibrium even at such high temperatures as 300–450 °C. The kinetic law established may be used for estimating, using Fe isotope signatures, the dynamic of relatively short (<year-scale) mineralizing events in ancient hydrothermal metal sulfide deposits that cannot be assessed using traditional dating approaches.

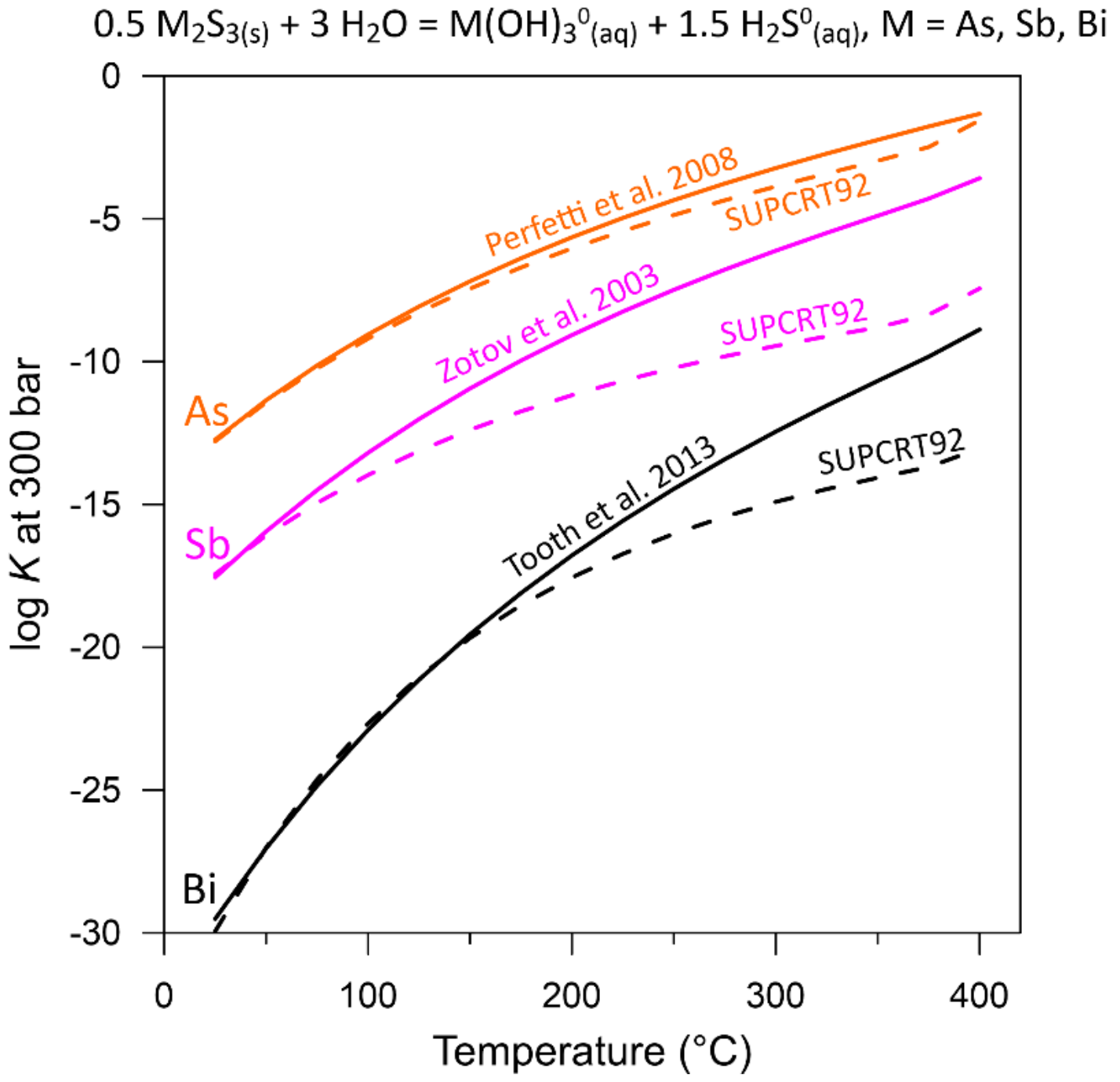

**Fig. 4.34** Comparison of the equilibrium constants of the dissolution reaction of the major As, Sb and Bi sulfide minerals (orpiment, stibnite and bithmutite) with formation of aqueous trihydroxide complexes as a function of temperature at 300 bar, according to the available thermodynamic data sources discussed in the text.

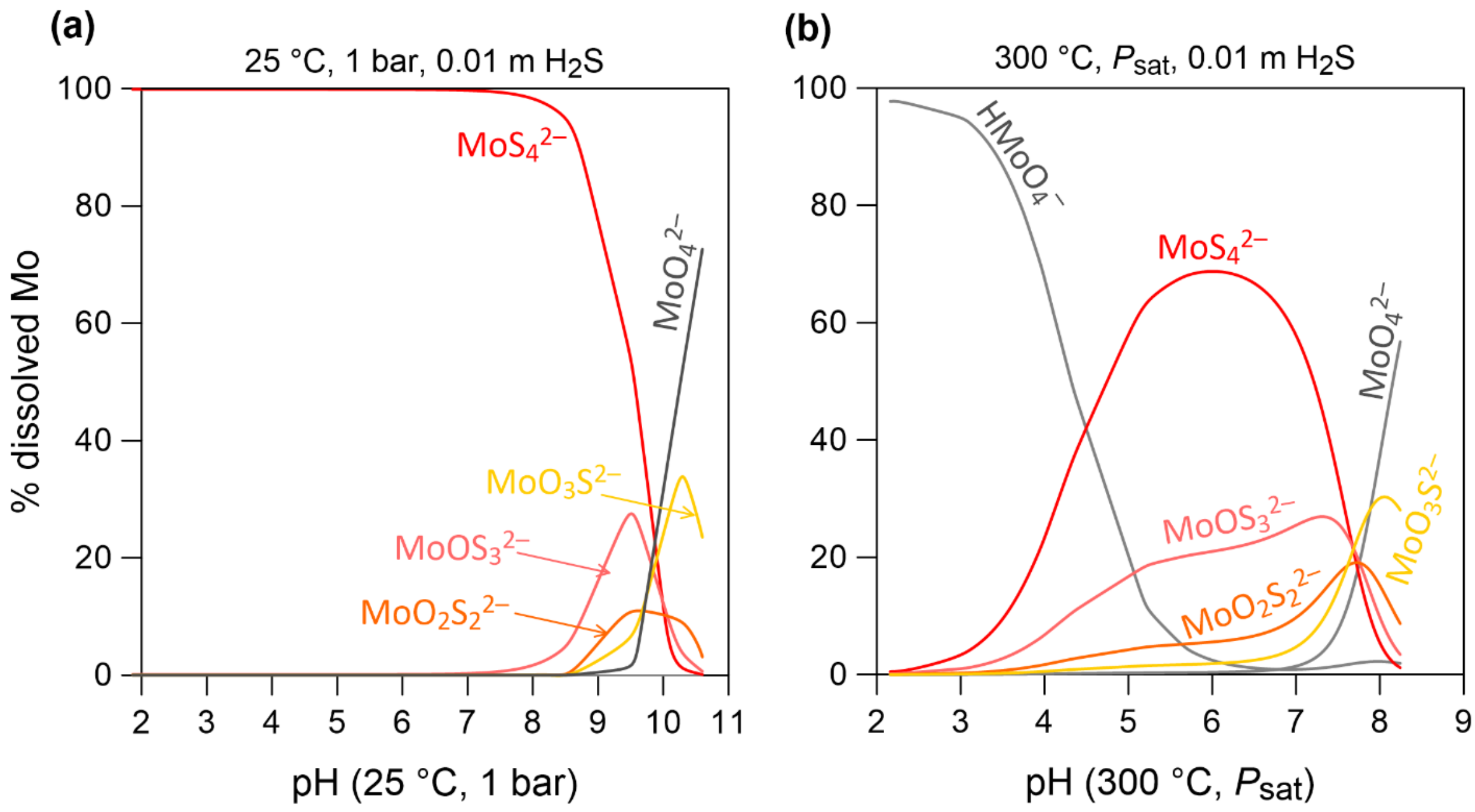

**Fig. 4.35** Speciation of $Mo^{VI}$ in aqueous fluid at 25 °C, 1 bar (a) and 300 °C, $P_{sat}$ (b) as a function of pH with 0.01 m total $H_2S$ and 0.15 m total salt concentration (NaCl+HCl+NaOH). The fractions of $H_2MoO_4^0$ and $HMoO_4^-$ in (a) and of $H_2MoO_4^0$ in (b) are <1% (not shown).

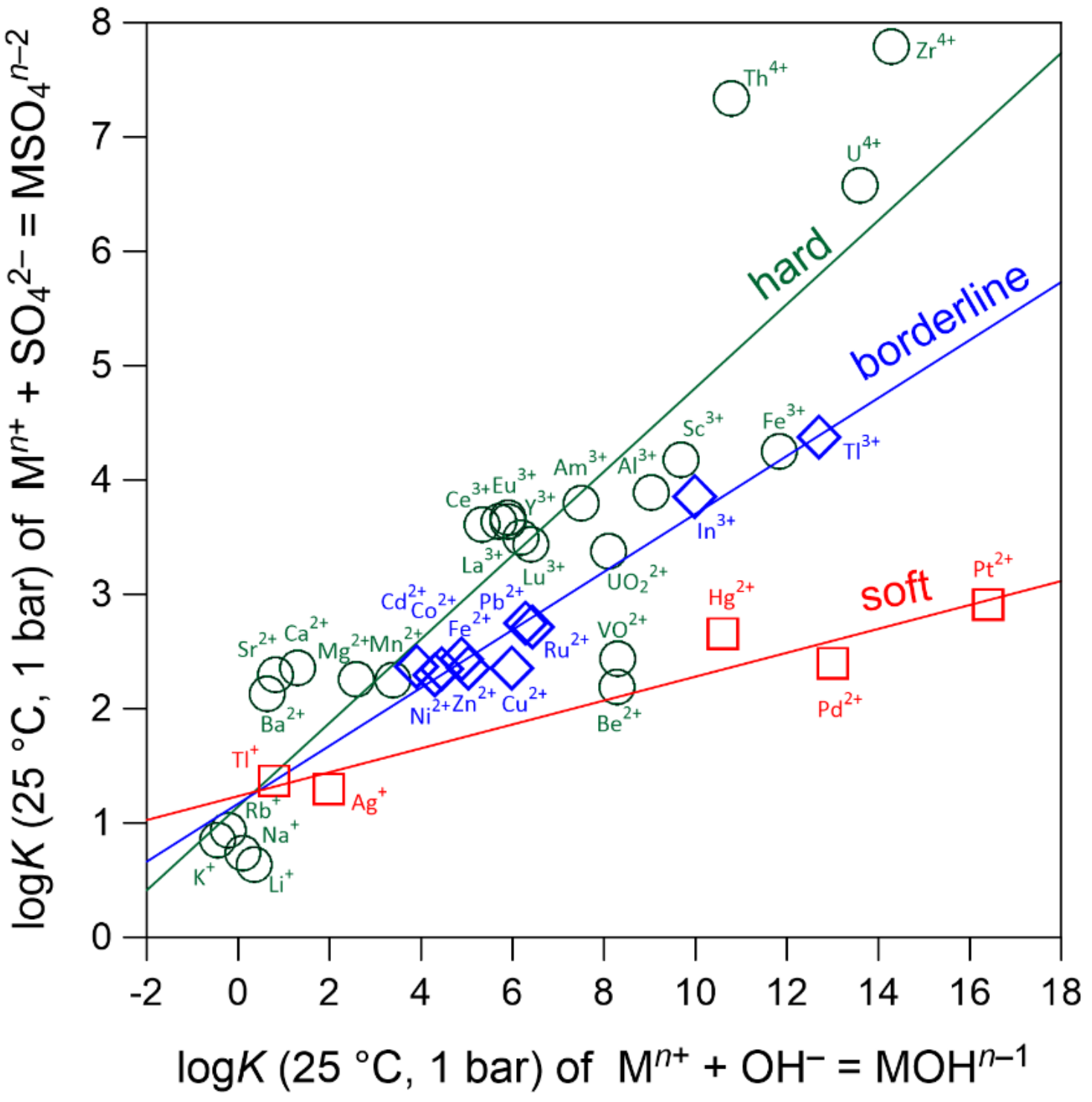

**Fig. 4.36** Equilibrium thermodynamic formation constants for the 1:1 metal sulfate complexes plotted versus their corresponding first hydrolysis constants, at 25 °C and 1 bar and zero ionic strength, for the three groups of cations – hard, intermediate and soft, shown in green, blue and red, respectively. The straight lines are least-square linear regressions of data for each corresponding metal group. Data are from compilations of Baes and Mesmer (1976), Turner et al. (1981), Johnson et al. (1992), Smith et al. (2004).

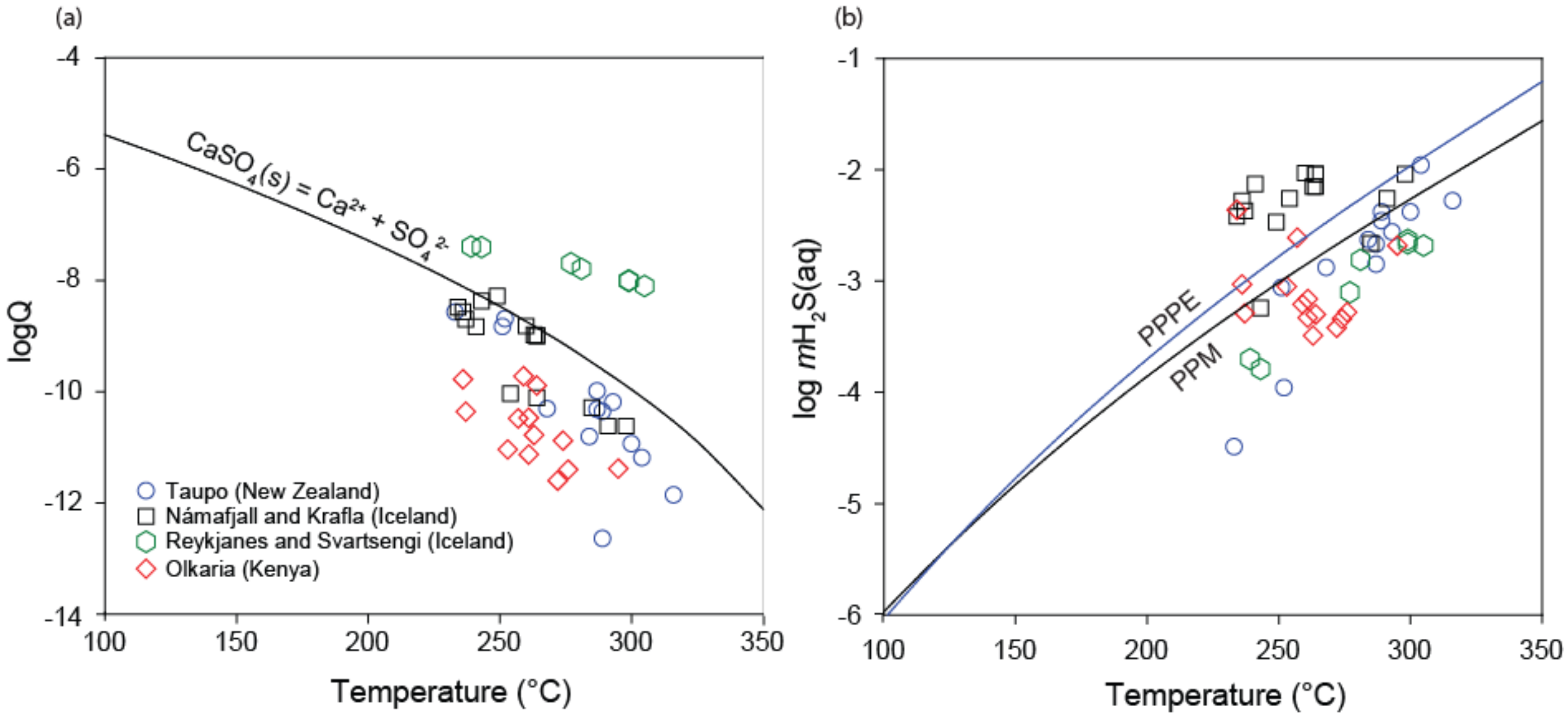

**Fig. 4.37** Mineral saturation state for anhydrite (a) and dissolved sulfide buffers (b) in reservoir hydrothermal fluids. The curves represent the equilibrium solubility of anhydrite and PPM (pyrite, pyrrhotite and magnetite) and PPPE (pyrite, pyrrhotite, prehnite and epidote) mineral buffers calculated with the aid of the SUPCRT92 program (Johnson et al. 1992). The symbols indicate reaction quotients for hydrothermal reservoir fluids for geothermal system sourced by seawater in basalts (Reykjanes and Svarstengi) and sourced by meteoric water in basalts (Námafjall and Krafla), trachyte (Olkaria) and andesite (Taupo). Data on natural hydrothermal fluids are from Stefánsson et al. (2016), Kleine et al. (2021) and Ricci et al. (2022).

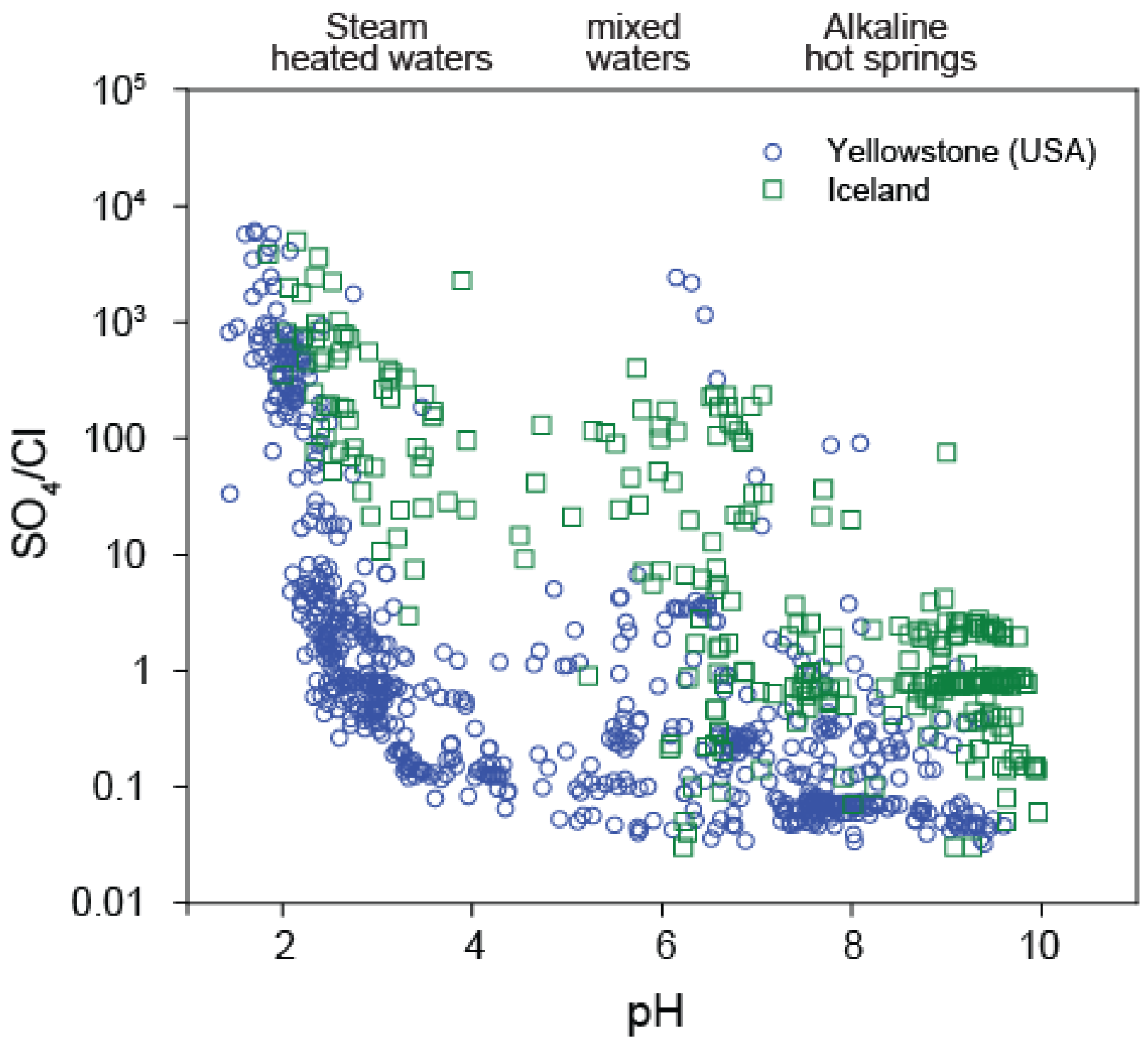

**Fig. 4.38** The relationship between the sulfate/chloride mole ratio and pH for hydrothermal waters at surface. Included are data from Yellowstone (USA) (McCleskey et al. 2022) and various geothermal fields in Iceland (Kaasalanen and Stefánsson 2012; Markússon and Stefánsson 2011; Björke et al. 2015; Stefánsson et al. 2016). Alkaline hot springs represent boiled reservoir hydrothermal water at surface and are characterized by similar $SO_4^{2-}/Cl^-$ ratios as the reservoir fluid and pH >8. Steam heated waters are produced upon condensation at surface of $H_2S$-enriched vapor formed upon boiling of reservoir hydrothermal fluids, followed by oxidation to form waters enriched in sulfuric acid resulting in the low pH and elevated $SO_4^{2-}/Cl^-$ ratios. Waters with mildly acid to mildly alkaline pH values can often be mixtures of hydrothermal water and cold or non-thermal water.

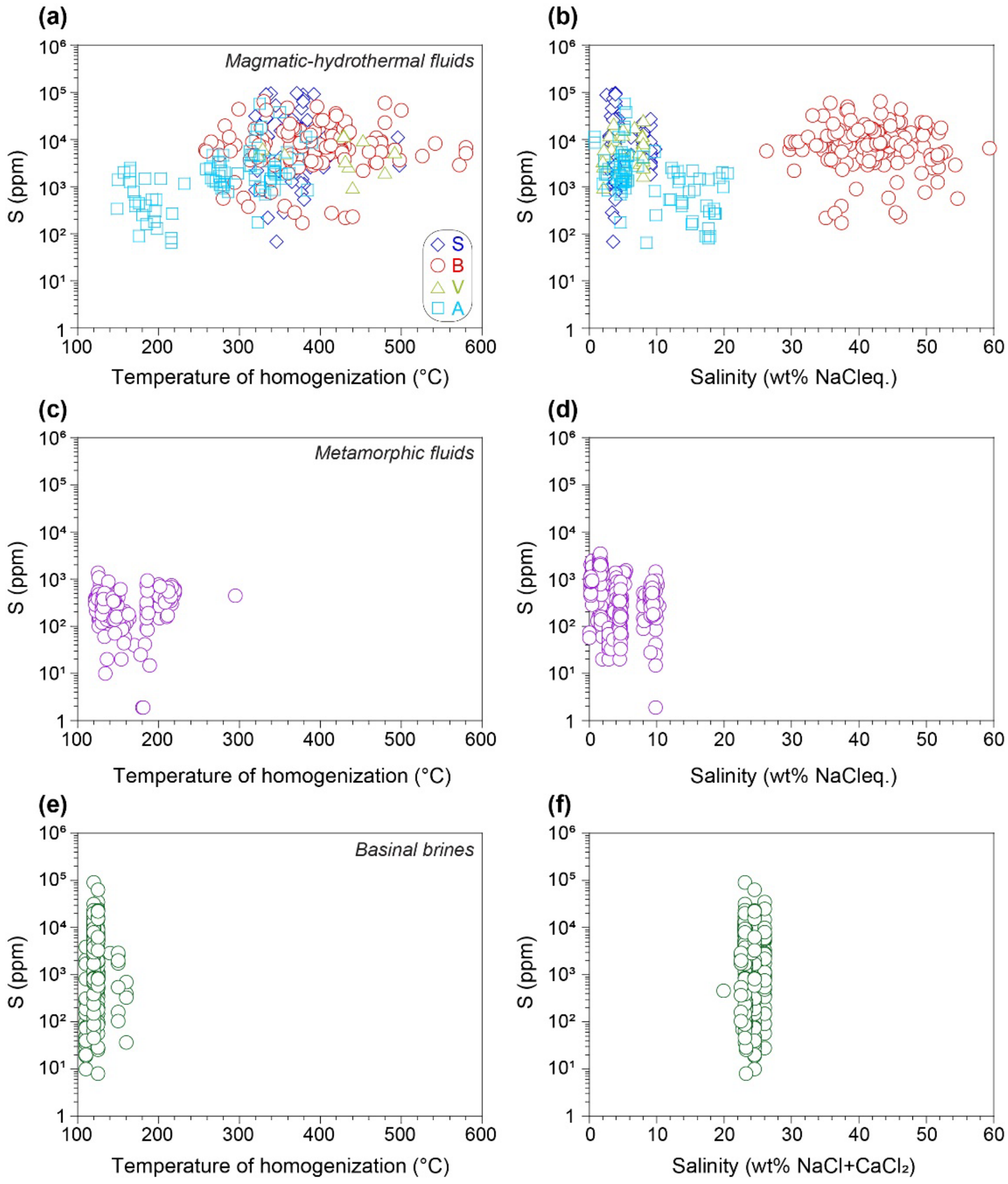

**Fig. 4.39** Concentration of sulfur (in ppm) in various fluid inclusion types from magmatic-hydrothermal systems (a, b), metamorphic fluids (c, d) and basinal brines (e, f), as a function of temperature of homogenization (in °C) and apparent fluid salinity (in wt% NaCl eq.). See Table 4.2 and Supplementary Table 4.S1 for data sources. Abbreviations for fluid inclusion types in magmatic-hydrothermal systems: A – aqueous; B – hypersaline aqueous (brine); S – supercritical fluid; V – vapor.

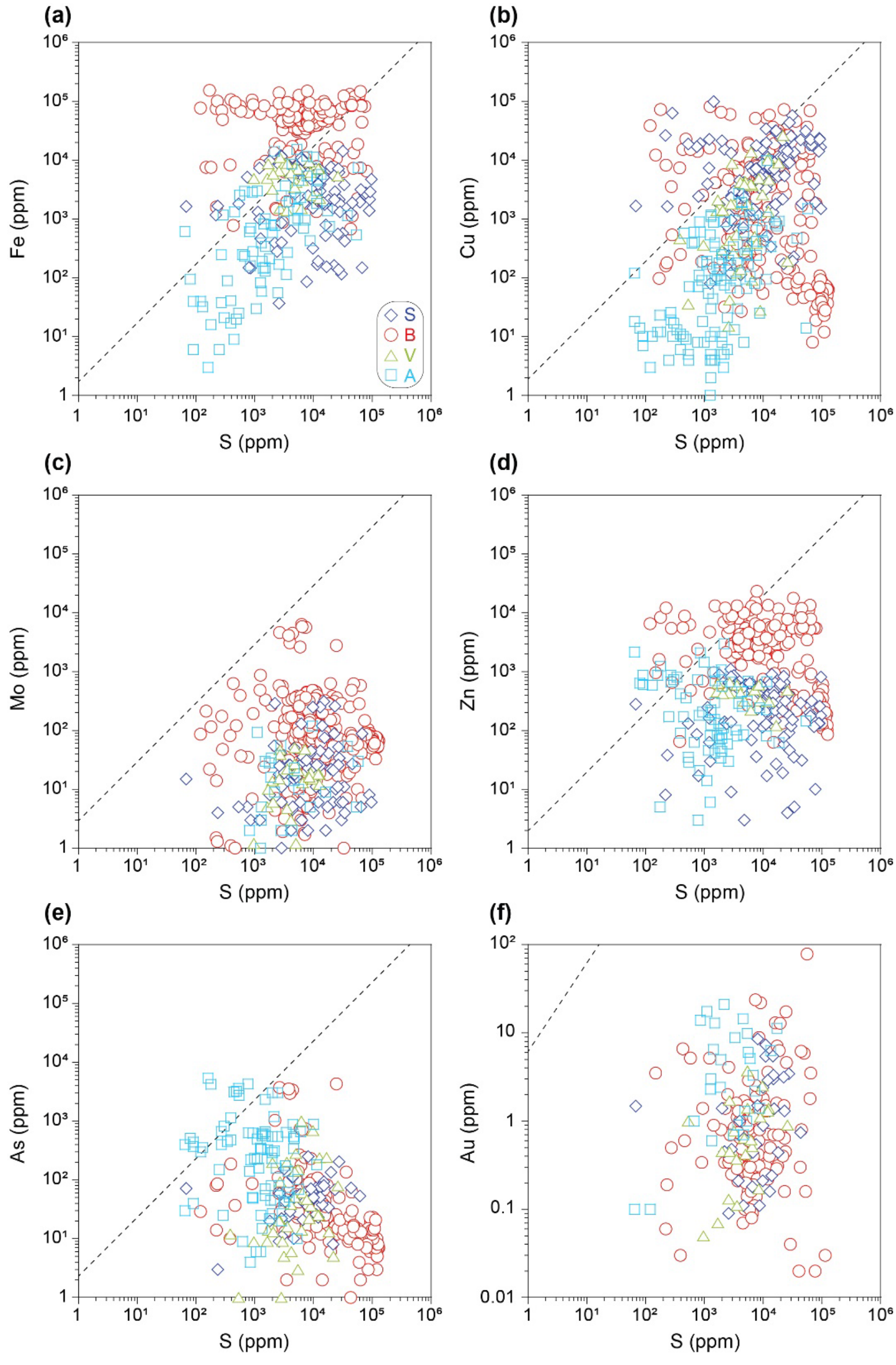

**Fig. 4.40** Concentration of selected metals in fluid inclusions from magmatic-hydrothermal systems as a function of sulfur concentration (in ppm). See Tables 4.2 and S1 for data sources. Abbreviations for fluid inclusion types in magmatic-hydrothermal systems: A – aqueous; B – hypersaline aqueous (brine); S – supercritical; V – vapor. Dashed lines indicate the 1:1 atomic concentration ratio of metal to sulfur.

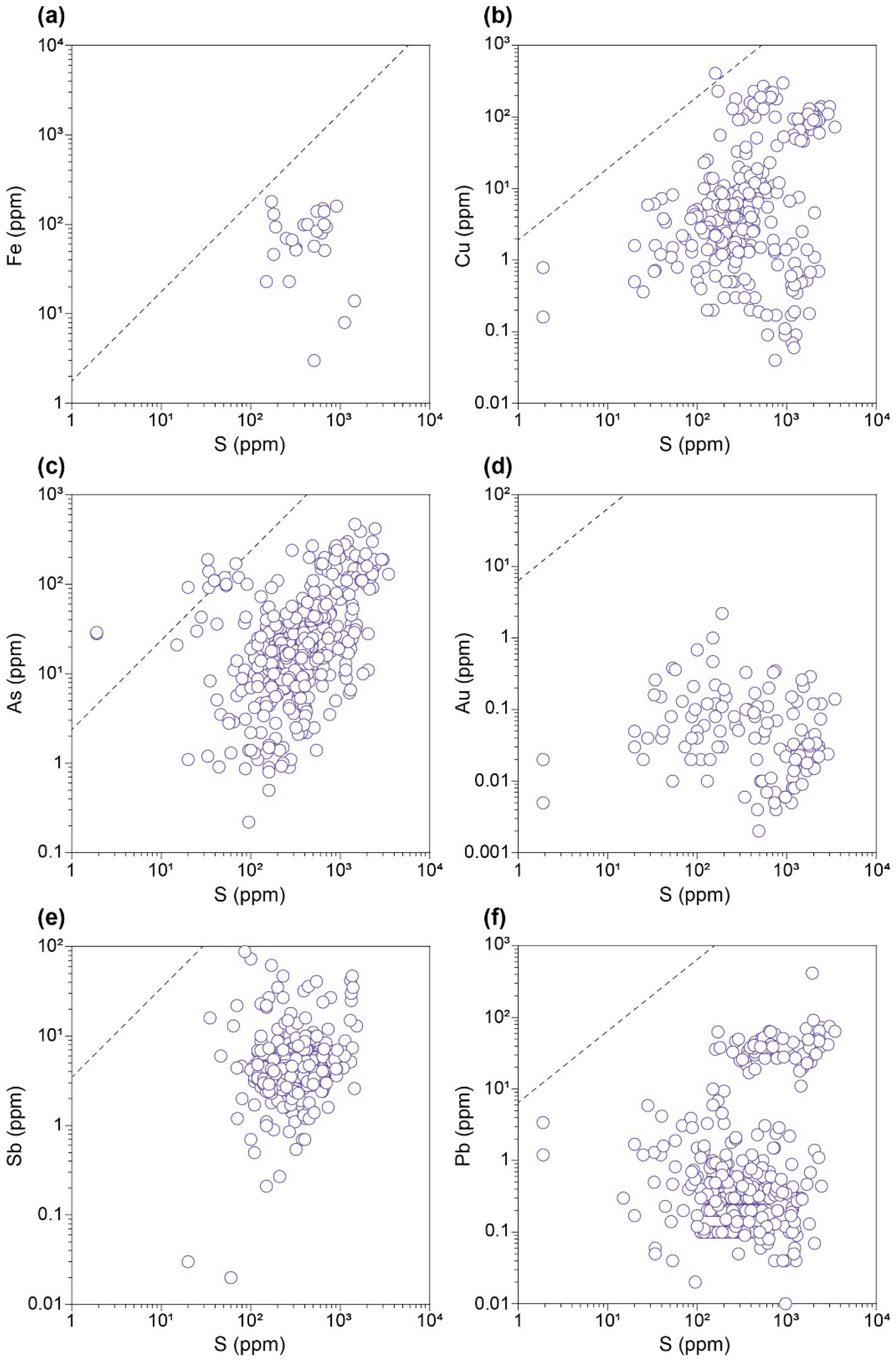

**Fig. 4.41** Concentration of metals in fluid inclusions from metamorphic environments as a function of sulfur concentration (in ppm). See the Fig. 4.40 caption and Table 4.2 for details and data sources.

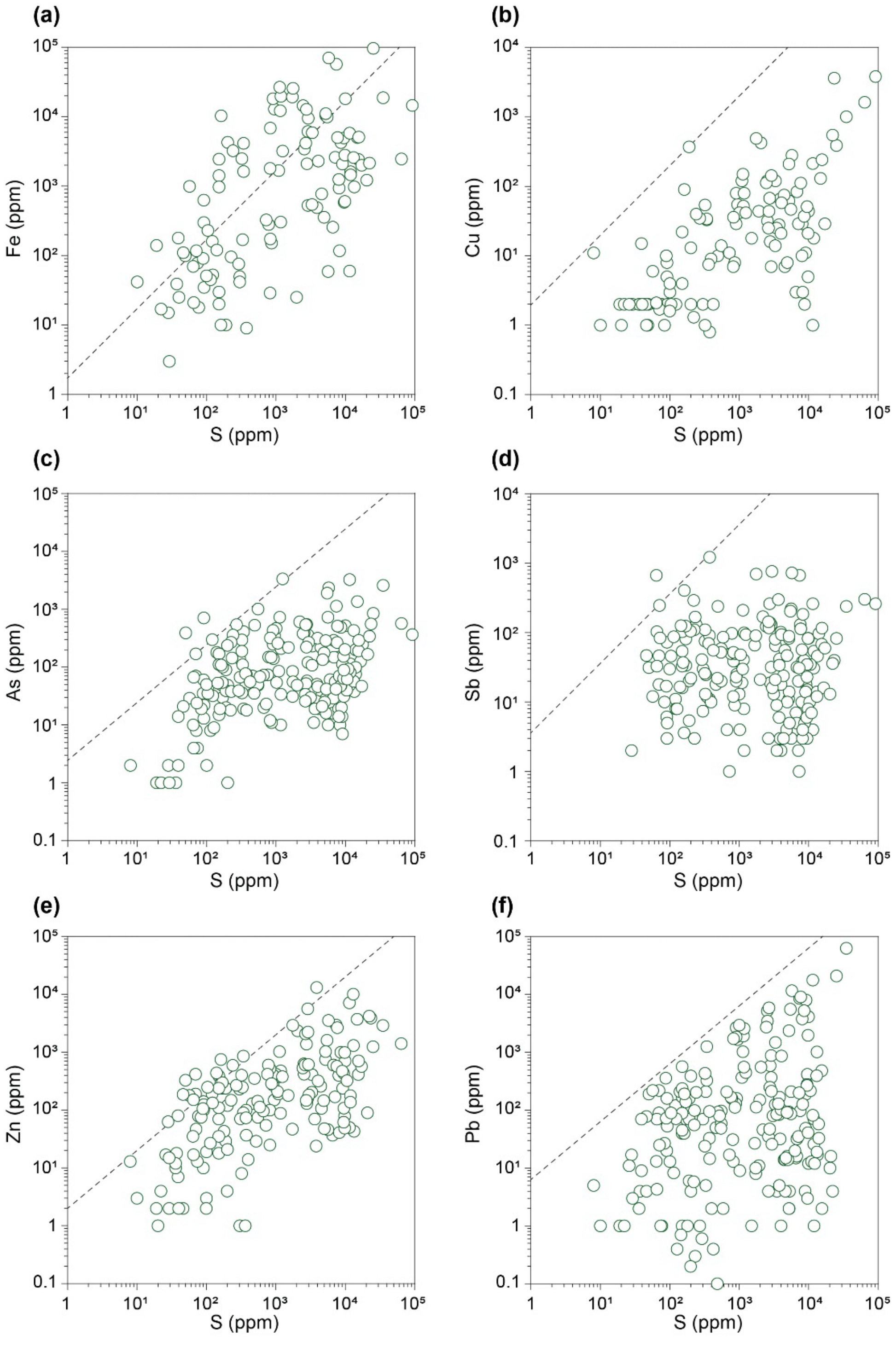

**Fig. 4.42** Concentration of metals in fluid inclusions entrapping basinal brine-type fluids as a function of sulfur concentration (in ppm). See Fig. 4.40 and Table 4.2 for details and data sources.

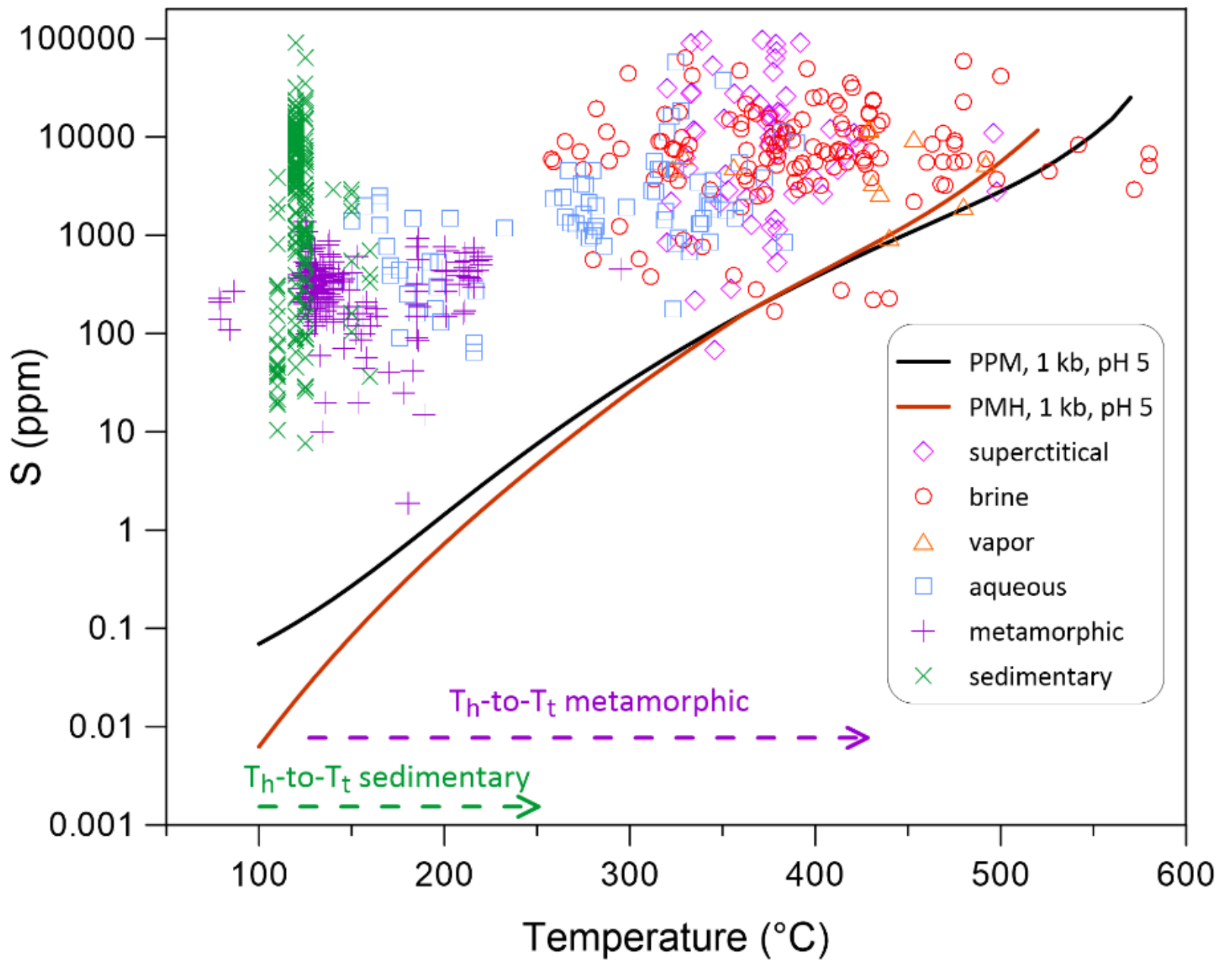

**Fig. 4.43** Comparison between sulfur concentrations from natural fluid inclusions from major types of fluids indicated in the legend and predicted S concentrations in equilibrium with the common $f_{S2}$ and $f_{O2}$ iron oxide-sulfide mineral buffers, pyrite-pyrrhotite-magnetite (PPM) and pyrite-magnetite-hematite (PMH), calculated as a function of temperature at 1 kbar pressure for a typical 10 wt% NaCl eq. fluid whose pH is buffered by equilibruim with a common silicate rock (quartz-muscovite-feldspar, pH 4.5–5.5 at all temperatures). Note that fluid inclusion homogenization temperatures ($T_h$) must be shifted by ~100–300 °C to the right, to account for true entrapment temperatures ($T_t$) for metamorphic and sedimentary fluids (dashed horizontal arrows).

# Tables

**Table 4.1** Concentrations of sulfur aqueous forms (in mmol/kg S equivalent) in modern natural hydrothermal fluids.

| Location | t °C | $pH_T$ | Cl | $SO_4$ | $S_2O_3$ | $SO_3$ | $S_nO_6$ | $H_2S$ |
|---|---|---|---|---|---|---|---|---|
| | | | *Reservoir hydrothermal fluids* | | | | | |
| Ohaaki, New Zealand[a] | 268 | 6.61 | 19.9 | 0.25 | – | – | – | 1.60 |
| Rotokawa, New Zealand[a] | 300 | 6.16 | 30.4 | 0.01 | – | – | – | 4.28 |
| Reykjanes, Iceland[a] | 299 | 4.88 | 498 | 0.76 | – | – | – | 2.38 |
| Krafla, Iceland[a] | 258 | 6.89 | 0.24 | 0.09 | – | – | – | 5.58 |
| Kizildere, Turkey[a] | 243 | 6.86 | 2.8 | 8.49 | – | – | – | 0.13 |
| Olkaria, Kenya[b] | 295 | 7.19 | 10.0 | 0.05 | – | – | – | 2.81 |
| Momotombo, Nicaragua[c] | 250 | 5.61 | 75.9 | 0.31 | – | – | – | 0.86 |
| Los Humeros, Mexico[d] | 300 | 6.63 | 1.8 | 1.67 | – | – | – | 2.49 |
| Wayang Windu, Indonesia[e] | 290 | 4.88 | 410 | 0.12 | – | – | – | 2.62 |
| Fusime, Japan[c] | 272 | 4.37 | 563 | 0.24 | – | – | – | 3.46 |
| Palinpinon, Philippines[f] | 240 | 5.50 | 202 | 0.14 | – | – | – | 2.53 |
| | | | *Surface springs and ponds* | | | | | |
| Geysir, Iceland[g] | 82 | 8.85 | 3.72 | 1.02 | 0.089 | 0.003 | – | 0.038 |
| Laugarvatn, Iceland[g] | 100 | 8.63 | 1.01 | 0.37 | 0.022 | 0.002 | – | 0.27 |
| Krýsuvík, Iceland[g] | 98 | 2.20 | 0.037 | 52.7 | 0.017 | – | – | 1.16 |
| Cinder Pool, YNP, USA[h] | 92 | 4.22 | 16.9 | 1.00 | 0.038 | – | 0.008 | 0.047 |
| Azure Spring, YNP, USA[h] | 84 | 8.70 | 8.62 | 0.43 | 0.055 | – | – | 0.011 |
| Acid Inkpot, Yellowstone USA[h] | 76 | 2.92 | 0.28 | 33.4 | 0.001 | – | <0.001 | 0.24 |
| Champagne Pool, New Zealand[i] | 76 | 5.5 | 1.36 | 0.68 | 0.37 | – | | 0.16 |
| Champagne Pool, New Zealand[i] | 22 | 2.5 | 1.07 | 1.12 | 0.003 | – | | 0.003 |
| | | | *Fumaroles* | | | | | |
| Nisyros, Greece[a] | 97-102 | – | – | – | – | – | – | 90-340 |
| Krafla, Iceland[j] | 100 | – | – | – | – | – | – | 46.3 |
| Norris Canyon, YNP, USA[k] | 93 | – | – | – | – | – | – | 18.9 |
| By Inkpot, YNP, USA[k] | 92 | – | – | – | – | – | – | 42.6 |
| | | | *Hydrothermal vent fluids* | | | | | |
| Lost City[l] | 94 | 12.1 | 545–570 | <4.2 | – | – | – | 0.1 |
| Logatchev[l] | 346–352 | 3.3–3.9 | 511–515 | <2.3 | – | – | – | 1.1–1.4 |
| Juan de Fuca[m] | 224–285 | 3.2 | 896–1090 | <1.0 | – | – | – | 3.0–4.4 |
| Rainbow[l] | 360–365 | 2.8–3.0 | 750–796 | – | – | – | – | 1.2–1.4 |
| Borken Spur[l] | 356–360 | – | 469 | – | – | – | – | 8.5–11.0 |

‘–’ means not available; $pH_T$ is at the fluid temperature; pH is measured at lower temperature and then slided to higher temperatures applying a mass balance equation for $H^+$; this procedure is often called conservation of alkalinity or charge balance.

[a] Ricci et al. (2022); [b] Kleine et al. (2021); [c] Stefánsson and Arnórsson (2000); [d] Tello et al. (2000); [e] Prasetio et al. (2020); [f] Rae et al. (2011); Kaasalainen and Stefánsson (2011a,b); [h] Xu et al. (1998); [i] Ulrich et al. (2013); [j] Stefánsson (2017); [k] Chiodini et al. (2012); [l] Charlou et al. (2010) and references therein; [m] Von Damm and Bischoff (1987).

**Table 4.2** Typical ranges of temperature, salinity, sulfur and selected metal concentrations (range and median value) in major types of fluids from magmatic-hydrothermal systems, metamorphic terrains and basinal continental brines, based on analyses of individual fluid inclusions by microthermometry and LA-ICPMS (please use the Table 4.2 native xlsx file for production).

| Hydrothermal fluid type | Environment | Fluid inclusion types | | Temperature of hom. (°C) | Salinity (wt% NaCl eq.) | Density (g/cm$^3$) | S (ppm) | Fe (ppm) | Cu (ppm) | Mo (ppm) | Au (ppm) | Ag (ppm) | As (ppm) | Sb (ppm) | Zn (ppm) | Pb (ppm) |
|---|---|---|---|---|---|---|---|---|---|---|---|---|---|---|---|---|
| Magmatic-hydrothermal* | Deep barren and deep central and peripheral Cu-Au (Mo) ore zones in porphyry copper deposits; porphyry-associated skarns | S 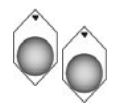 | *Supercritical inclusions* (S) are intermediate-density aqueous inclusions containing equal proportions of liquid and vapor; many contain small opaque daughter crystals of constant relative size. | 300 - 650 | 2 - 16 | 0.4 - 0.7 | 70 - 96000<br>10500 | 10 - 47000<br>3900 | 5 - 90000<br>3800 | <1 - 1200<br>50 | <1 - 20<br>1.6 | <1 - 600<br>20 | 3 - 2000<br>210 | 1- 30<br>10 | 2 - 6500<br>600 | <1 - 4500<br>330 |
| | Stockwork mineralization at moderate to shallow depth in central parts of Cu-Au (Mo) ore zones and upper parts of Cu-rich periphery; porphyry-associated skarns; IOCG; Sn-W deposits. | V & B 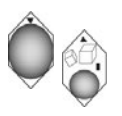<br>"boiling" FIA 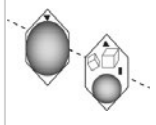 | *Vapor-rich inclusions (V)* can be high- (60-90% vapor) or low-density (>90% vapor) and commonly contain small opaque daughter crystals. *Hypersaline liquid inclusions (brines - B)* always contain a vapor bubble, a saline aqueous liquid and halite, additionally polyphase inclusions contain a few other transparent and opaque daughter phases. In many cases vapor-rich and hypersaline liquid inclusions form boiling assemblages. | 260 - 650, rarely up to 800 | *vapor:*<br>0.2 - 20 | *vapor:*<br>*0.01 - 0.4* | 385 - 26000<br>5000 | 2 - 31000<br>6000 | <1 - 33000<br>4300 | 1 - 1025<br>20 | 0.05 - 11<br>1.9 | <1 - 600<br>45 | 2 - 1200<br>175 | 1 - 150<br>35 | <1 - 6800<br>660 | <1 - 4000<br>340 |
| | | | | | *hypersaline liquid:*<br>26 - 75 | *hypersaline liquid:*<br>*0.8 - >1* | 120 - 120000<br>8800 | 15 - 221000<br>51000 | 10 - 86000<br>3300 | 1 - 9300<br>70 | 0.03 - 90<br>0.7 | <1 - 184<br>95 | 2 - 6600<br>230 | 1 - 1200<br>95 | 90 - 40000<br>3200 | 5 - 60000<br>3000 |
| | Late-stage porphyry veining (quartz-sericite-pyrite); IOCG; Sn-W deposits; base and percious metal mineralization in skarns, carbonate-replacement and epithermal ore bodies. | V & A 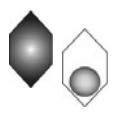<br>"boiling" FIA 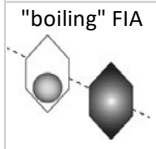 | *Aqueous (A) low- to intermediate-salinity inclusions* with predominantly liquid water and smaller vapor bubbles ranging from 20 to 40 vol % may occur together or not with *low-density vapor-rich (V) inclusions* (>90 % vapor). Aqueous inclusions are usually free of daughter crystals, however some may contain accidentally trapped sericite crystals. Coexistence of aqueous and low-density vapor inclusions is indicative of fluid boiling at low pressure. | 150 - 450 | *vapor:*<br><0.5 | *vapor:*<br>*<0.02* | no data available | | | | | | | | | |
| | | | | | *aqueous:*<br>0.5 - 22 | *aqueous:*<br>*0.1 - 0.6* | 100 - 57000<br>1750 | 6 - 33600<br>915 | <1 - 9700<br>85 | <1 - 270<br>15 | 0.08 - 54<br>2.1 | <1 - 250<br>10 | 2 - 5300<br>330 | <1 - 950<br>30 | <1 - 9200<br>215 | <1 - 5600<br>30 |
| Metamorphic** | Metamorphic fluids trapped in late-metamorphic quartz veins and Alpine fissure veins hosted by metamorphic rocks formed from subgreenschist- to amphi-bolite-facies conditions. | A & AC 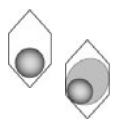 | *Aqueous (A) inclusions* are common for veins in subgreenschist- and greenschist-facies rocks, and *aqueous-carbonic (AC)inclusions* in amphibolite-facies rocks. Volatile content ($CO_2$, sulfur) increases systematically with the metamorphic grade. | 80 - 300 | 0.4 - 10.3 | 0.1 - 0.5 | 2 - 3400<br>290 | 3 - 180<br>85 | <0.1 - 410<br>5 | no data<br>available | <0.1 - 2.2<br>0.05 | <0.1 - 2.5<br>0.5 | <1 - 470<br>15 | <0.1 - 95<br>5 | <0.1 - 200<br>10 | <0.1 - 420<br>0.5 |
| Basinal brines*** | Basement-hosted Pb-Zn deposits formed by complex fluid processe at the interface between sedimentary cover rocks and crystalline basement. | B & A 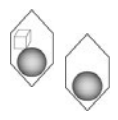 | *High- (B), moderate- to low-salinity(A) fluid inclusions*, sulfate and/or $CO_2$-bearing. Complex composition within the Na-Ca-Cl-$SO_4$-$HCO_3$- system, produced by mixing of fluids with different origin. | 100 - 180 | 2 - 26 | 0.1 - 0.8 | 10 - 91600<br>1200 | 3 - 97100<br>730 | <1 - 3900<br>15 | <0.1 - 950<br>50 | no data<br>available | <0.1 - 20<br>0.5 | <1 - 4500<br>100 | <0.1 - 8400<br>40 | <1 - 13300<br>200 | <1 - 63000<br>90 |

*Data from: Kouzmanov and Pokrovski (2012) and references therein; Audétat (2015); Catchpole et al. (2015); Li et al. (2015); Marquez-Zavalia and Heinrich (2016); Chang et al. (2018); Berni et al. (2020); Schirra et al. (2022); Han et al. (2023).

**Data from: Marsala et al. (2013); Miron et al. (2013); Rauchenstein-Martinek et al. (2014); Rauchenstein-Martinek et al. (2016)

***Data from: Fusswinkel et al. (2013); Burisch et al. (2016); Walter et al. (2018); Walter et al. (2019).

**Table 4.3** Major band Raman spectral frequencies and characteristic X-ray absorption energy positions at ambient conditions of the key aqueous inorganic species of sulfur that may be used for in situ detection of S species in fluid inclusions and hydrothermal fluids in laboratory experiments.

| **Sulfur species (aqueous, unless indicated)** | **Raman shift (at max intensity and at ambient *T, P*) wavenumber, $cm^{-1}$** | **Nature of Raman band** | **Integrated intensity relative to $SO_4^{2-}$ (=1)** | **S K-edge energy (at max intensity, main edge, unless indicated), eV** |
|---|---|---|---|---|
| $SO_4^{2-}$ | 980 | SO ss | 1 | 2481.5 |
| $HSO_4^-$ = $(HO)SO_3^-$ | 1050 | SO ss | 1 | 2181.5±0.5 |
| $H_2SO_4$ = $(HO)_2SO_2$ | 970, 1140, 1370, 1220 | $SO_2$ ss,as, $SO_4$ as | – | – |
| $SO_2$ | 1150 | SO ss | 0.5 | 2473.0[c]; 2478.0 |
| $(HS)O_3^-$ | 1050, 2520 | SO ss, SH ss | 0.6 | 2477.5 |
| $SO_2(OH)^-$ | 1020 | SO s | 0.7 | 2476.3 |
| $SO_3^{2-}$ | 930, 965 | SO as,ss | 0.5 | 2477.2 |
| $S_2O_3^{2-}$ | 450<br>672, 1002, 1125 | SS s<br>SO b,s,as | 1<br>0.3–0.5 | 2470.7<br>2479.7 |
| $HS_2O_3^-{}_{(s)}$ | 403, 642, 1027, 2486 | SS s, SO b,s, SH s | – | – |
| $COS_{(g)}$ | 857 | SO s | – | 2471.0[d] |
| $SCN^-$ | 745, 2068 | SC ss, CN ss | – | – |
| $S_{8(s,l,aq)}$ | 150, 215, 470 | SS b,b,s | – | 2471.5 |
| $S_n^0$, $S_n^{2-}$ | 400, 450 | SS ss,as | – | 2470–2473 |
| $H_2S$ | 2590 $_{(aq)}$, 2610 $_{(g)}$ | SH ss | 1–3 | 2469 |
| $HS^-$ | 2570 | SH ss | 1–3 | 2469 |
| $H_2S_2$ | 2500 | SH s | – | – |
| $S_3^{\bullet-}$ | 535, 1070 | SS ss ($\nu_1$, $2\nu_1$) | 10[a], 30–1000[b] | 2467.9[c] |
| $S_2^{\bullet-}$ | 590, 1180 | SS ss ($\nu_1$, $2\nu_1$) | 10[a] | 2467.5[c] |

Relative integrated intensities are indicative; they vary depending on laser frequency, Raman setup, fluid color, and temperature and pressure; see Meyer et al. (1980), Burke (2001), Risberg et al. (2007), Ni and Keppler (2012), Pokrovski and Dubessy (2015), Huang et al. (2021); Schmidt and Seward (2017), Eldridge et al. (2018), Schmidt and Jahn (2024) for details and more references. S K-edge energy positions are corrected relative to sulfate, set to 2481.5 eV in this study. Raman band types: s = stretch with no indication of particular mode; ss = symmetric stretch; as = asymmetric stretch, b = bend; $\nu_n$ = particular vibrational mode as per molecular geometry; [a] = for 458 and 473 nm laser; [b] = for 515 and 532 nm laser; [c] = pre-edge; [d] = tentative estimation; '–' = not available.

**Table 4.4** Sulfur species/minerals identified in natural fluid inclusions by Raman spectroscopy.

| Species | Geological setting, rock type | Host mineral/inclusion type, major composition | Associated species | Representative references |
|---|---|---|---|---|
| $H_2S$ ($HS^-$) | magmatic, porphyry-epithermal, sedimentary-hosted, metamorphic, shallow-mantle settings | large range of fluid compositions trapped in quartz, topaz, calcite, fluorite, garnet, pyroxene | $S_8$, $SO_4^{2-}$ | Dubessy et al. 1984, 1989, 1992; Bény et al. 1982; Burke 2001; Frezzotti et al. 2012 and references therein |
| $SO_4^{2-}$, $HSO_4^-$ | metamorphic, magmatic, volcanic settings | quartz | $H_2S$ | Rosasco and Roedder 1979; Dubessy et al. 1983, 1992; Boiron et al. 1999; Borisova et al. 2014 |
| $SO_2$ | magmatic, carbonatites in mantle xenoliths | $CO_2$-rich inclusions in pyroxene | $CO_2$ | Frezzotti et al. 2002 |
| $S_8^0$ | evaporites, oil-field sediments, metamorphic rocks | quartz, fluorite | $H_2S$, $SO_4^{2-}$ | Bény et al. 1982; Giuliani et al. 2003; Barré et al. 2017 |
| COS | metamorphic rocks | quartz | $S_8$, $H_2S$ | Grishina et al. 1992; Giuliani et al. 2003 |
| $H_2S_n$ | metamorphic quartzite, corundum-bearing marble | $H_2S$-$CH_4$-$CO_2$ quartz and ruby hosted | $S_8$, $H_2S$, COS | Hurai et al. 2019; Huang et al. 2021 |
| $S_3^{\bullet-}$, $S_n^{2-}$ | Carnian evaporites, French Alps | quartz, fluorite, saline (~25 wt% NaCl eq.), heated to 300°C | $H_2S$, $S_8$, $SO_4^{2-}$ | Barré et al. 2017 |
| $\alpha,\beta,\gamma$-$H_2S_{(s)}$ | organic-rich carbonates hosting fluorite, Southern Permian basin, Germany | fluorite, gas-rich, $H_2S$-$CO_2$-$N_2$-$CH_4$, cooled to -190°C | $H_2S$-clatrates, $CO_2$, $CH_4$ | Sośnicka and Lüders 2021 |
| Sulfide minerals | magmatic, sedimentary, metamorphic | e.g., pyrite, pyrrhotite, chalcopyrite hosted in quartz | – | Frezzotti et al. 2012; Mernagh and Mavrogenes 2019; Schiavi et al. 2020 |
| Sulfate minerals | magmatic, sedimentary, metamorphic | e.g., anhydrite, gypsum, barite hosted in quartz | – | Frezzotti et al. 2012 |

"–" = not reported.

**Table 4.5** Recommended sources of thermodynamic data for major sulfur aqueous and gaseous species and mineral phases considered in this work at hydrothermal conditions.

| Chemical species [a] | Data source |
|---|---|
| **Aqueous species** | |
| $H_2O$, $H^+$, $OH^-$, $Cl^-$, $Na^+$, $NaCl^0$, $NaOH^0$, $KCl^0$, $K^+$, $KOH^0$, $HSO_4^-$, $SO_4^{2-}$, $KSO_4^-$, $KHSO_4^0$, $HS^-$, $S_2O_3^{2-}$, $HSO_3^-$, $SO_3^{2-}$ | Johnson et al. 1992 (SUPCRT);<br>Zimmer et al. 2016 (SUPCRTBL)[b] |
| $H_2S^0$, $SO_2^0$, $H_2^0$, $O_2^0$ | <1–2 kbar; $\rho_{H2O}$<0.45 g/cm$^3$: Akinfiev and Diamond 2003;<br>≥2 kbar, $\rho_{H2O}$≥0.45 g/cm$^3$: Schulte et al. 2001 |
| $HCl^0$ | <1–2 kbar: Akinfiev and Diamond 2003;<br>≥2 kbar: Tagirov et al. 1997 |
| $NaSO_4^-$ | Pokrovski et al. 1995 |
| $S_{(2-8)}^{2-}$ | Barré et al. 2017 |
| $NaHS^0$, $KHS^0$, $NaHSO_4^0$ | Pokrovski and Dubessy 2015[c] |
| $S_3^{\bullet-}$ | Pokrovski and Dubessy 2015 |
| $S_2^{\bullet-}$ | Pokrovski et al. 2019[d] |
| $S_8^0$ (possibly including other $S_n^0$) | Pokrovski and Dubessy 2015[e] |
| $S_8^0$ | this study[f] |
| **Pure solids and liquids** | |
| $S_{(s)}$, $S_{(l)}$, $K_2SO_{4(s)}$, $Na_2SO_{4(s)}$ | Chase 1998 (JANAF database) |
| gold, pyrite, pyrrhotite, magnetite, hematite, quartz, muscovite, microcline, sanidine, andalusite, albite, chalcopyrite, bornite, argentite, galena, molybdenite | Robie and Hemingway 1995 |
| **Gases** | |
| $O_2$, $H_2$, $H_2S$, $SO_2$ | Robie and Hemingway 1995 |
| $S_{(2-8)}$ | Chase 1998 |
| $H_2S_{(2-5)}$ | Suleimenov and Ha 1998 |
| **Activity coefficient models for aqueous solution** | |
| $\log \gamma_i = -A z_i^2 \sqrt{I}/(1+B \mathring{a}_i \sqrt{I}) + \Gamma_\gamma$, for charged species<br>$\log \gamma_\iota = \Gamma_\gamma + b_i I$, for neutral species | Helgeson et al. 1974, 1981[g] |

[a] Thermodynamic properties of $H^+$ are equal to 0 at all *T* and *P*; the standard states for the solid phases and $H_2O$ are unit activity for the pure phase at all *T* and *P*; for gases - unit fugacity of the ideal gas at 1 bar and all *T*; for aqueous species, the standard state convention corresponds to unit activity coefficient for a hypothetical one molal solution whose behavior is ideal. Note that other thermodynamic datasets for the $OH^-$ ion are available from the Unitherm database (Shvarov 2008) and the DEW model (Sverjensky et al. 2014), which yield differences of a few to 10s kJ/mol for the Gibbs energy value of $OH^-$ in particular near the water isochore of 0.4±0.05 g/cm$^3$.

[b] The updated version of the original SUPCRT database (2007) based on a series of subsequent papers reporting HKF parameters for most ions and aqueous complexes is available on line at http://geopig.asu.edu/index.html#. These data were transferred and complemented with additional aqueous species and minerals in a more recent SUPCRTBL database available on line at https://models.earth.indiana.edu/supcrtbl.php. This database takes into account the expansivity and compressibility of the major mineral phases from Holland and Powell (2011). Note that for a given aqueous species the thermodynamic coefficients and the *T-P* allowed ranges of calculations are identical in both databases.

[c] Formation constant [cation + ligand = ion pair] at any T and P is assumed to be equal, respectively, to those of $NaCl^0$, $KCl^0$ and $KHSO_4^0$ from the SUPCRT database, for consistency.

[d] The apparent molal Gibbs energy values ($G°_{T,P}$) of $S_2^{\bullet-}$ at 450 °C, 700 bar (16±5 kJ/mol) and 500 °C, 700–1400 bar (19±9 kJ/mol) were derived from the $S_2^{\bullet-}$ concentrations estimated from Raman spectra in thiosulfate and sulfide-sulfate solutions reported by Pokrovski and Dubessy (2015). These values correspond, respectively, to stability constant values, $\log_{10}K$, of –16.2±0.5 and –18.4±3.7, for the reaction: 13 $H_2S(aq)$ + 3 $SO_4^{2-}$ = 8 $S_2^{\bullet-}$ + $2H^+$ + 12 $H_2O(liq)$.

[e] HKF parameters were obtained by regression of both Kamyshny (2009) low-temperature and Dadze and Sorokin (1993) high-temperature solubility data to optimize $c_1$ and $c_2$, and using HKF parameter correlations from Sverjensky et al. (2014) and Shvarov (2015): $\Delta_f G°_{298}$, = 10.26 kcal/mol, $S°_{298}$ = 64.87 cal/mol K, $C_p°_{298}$ = 43.40 cal/mol K, $V°_{298}$ = 124.08 cm$^3$/mol, $a_1$×10 = 25.6 cal/mol bar, $a_2$×10$^{-2}$ = 20.2 cal/mol, $a_3$ = -11.7 cal K/mol bar, $a_4$×10$^{-4}$ = -3.6 cal K/mol, $c_1$ = 84.8 cal/mol K, $c_2$×10$^{-4}$= -20.0 cal K/mol, $\omega$×10$^{-5}$ = 0.08 cal/mol.

[f] HKF parameters were obtained based on Kamyshny (2009) sulfur solubility data to 80 °C only and using the HKF correlations (see also Fig. 4.17): $\Delta_f G°_{298}$, = 10.26 kcal/mol, $S°_{298}$ = 64.87 cal/mol K, $C_p°_{298}$ = 43.40 cal/mol K, $V°_{298}$ = 124.08 cm$^3$/mol, $a_1$×10 = 25.6 cal/mol bar, $a_2$×10$^{-2}$ = 20.2 cal/mol, $a_3$ = -12.0 cal K/mol bar, $a_4$×10$^{-4}$ = -3.6 cal K/mol, $c_1$ = 31.6 cal/mol K, $c_2$×10$^{-4}$= 5.8 cal K/mol, $\omega$×10$^{-5}$ = 0 cal/mol.

[g] A and B are the Debye-Hückel electrostatic parameters; I is the effective molal ionic strength ($I = 0.5 \sum z_i^2 m_i$); $z_i$ and $\mathring{a}_i$ are the ionic charge and the distance of the closest approach for i$^{th}$ species, respectively; $\Gamma_\gamma$ is the mole fraction to molality conversion factor, $\Gamma_\gamma$

= log(1+0.018m*), where m* is the sum of the molalities of all solute species. We adopted a value for $\mathring{a}_i$ of 4.5 Å for all charged species. For neutral species, $b_i$ is the empirical Setchenov coefficient, which may be assumed, in the first approximation, to be close to zero for all neutral species, which yields activity coefficients close to one.

**Table 4.6** Chemical speciation of major gangue and ore metals in saline hydrothermal fluids ($T$ = 150–500 °C, $P$ <2 kbar, fluid density >0.35 g/cm$^3$,) and their major solubility-controlling solid phases at conditions relevant to magmatic-hydrothermal deposit formation.

| Element | Solubility-controlling solid phases | Major species in aqueous fluid [1] | Uncertainty in log units[2] | Key recommended references [3] |
|---|---|---|---|---|
| As | arsenopyrite, arsenian pyrite | $As(OH)_3^0$ | <0.5 | Pokrovski et al. 2002a,b; Perfetti et al. 2008; Xing et al. 2019 |
| Sb | stibnite, Cu-Fe-Sn sulfosalts | $Sb(OH)_3^0$, $Sb(OH)_2Cl^0$, $HSb_2S_4^-$, $Sb_2S_4^{2-}$ | 1 | Zotov et al. 2003; Pokrovski et al. 2006; Olsen et al. 2019 |
| Zn | sphalerite | $ZnCl_2^0$, $ZnCl_3^-$, $ZnCl_4^{2-}$ $Zn(HS)_2^0$, $Zn(HS)_3^-$ | 1 | Sverjensky et al. 1997; Tagirov and Seward 2010; Akinfiev and Tagirov 2014 |
| Pb | galena | $PbCl_2^0$, $PbCl_3^-$, $PbCl_4^{2-}$ $Pb(HS)_2^0$, $Pb(HS)_3^-$ | 1 | Sverjensky et al. 1997 |
| Fe | pyrite, magnetite, hematite, pyrrhotite | $Fe^{II}Cl_2^0$ *$Fe^{II}Cl_4^{2-}$* *$Fe^{III}Cl_4^-$* | 0.5 *(>2)* *(>3)* | Sverjensky et al. 1997; Testemale et al. 2009; Saunier et al. 2011; Gammons and Allin 2022 |
| Cu | chalcocite, chalcopyrite, bornite, enargite | *$CuCl^0$*, $CuCl_2^-$, *$CuCl_3^{2-}$* *$CuHS^0$*, $Cu(HS)_2^-$ | 0.5 | Akinfiev and Zotov 2001; Liu and McPhail 2005; Trofimov et al. 2023 |
| Ag | argentite, sulfosalts | $AgCl_2^-$ *$AgCl_3^{2-}$* | 0.5 | Pokrovski et al. 2013b |
| Au | native gold, invisible gold in arsenian pyrite and arsenopyrite as $AuAs_nS_{6-n}$ | $Au(HS)_2^-$, $Au(HS)S_3^-$ $AuHS^0$, $AuCl_2^-$ | 0.5-1.0 1.5 | Stefánsson and Seward 2003c, 2004; Pokrovski et al. 2014, 2015; Zotov et al. 2018 |
| Mo | molybdenite | $HMoO_4^-$, $MoO_4^{2-}$, $MoS_4^{2-}$ *oxysulfides, oxychlorides, alkali ion pairs, polysulfides* | >1 *(>3)* | Dadze et al. 2018a; Liu et al. 2020 |
| Pt | cooperite, sperrylite; chemically bound in pyrite, pyrrhotite, chalcopyrite | $Pt(HS)_4^{2-}$, $Pt(HS)_2(S_3)_2^{2-}$, $PtCl_4^{2-}$ *$Pt(HS)_2^0$* | >1 | Kokh et al. 2017; Tagirov et al. 2019b; Filimonova et al. 2021; Pokrovski et al. 2021b; Laskar et al. 2022 |
| Pd | braggite; chemically bound in pyrite, pyrrhotite, chalcopyrite | $Pd(HS)_4^{2-}$, $Pd(HS)_2(S_3)_2^{2-}$, $PdCl_3^-$, $PdCl_4^{2-}$ *$Pd(HS)_2^0$* | >2 | Bazarkina et al. 2014; Laskar 2022; Laskar et al. 2025 |
| Si | quartz, silicate minerals | $Si(OH)_4^0$ *$Si_2O(OH)_6^0$* | <0.5 | Sverjensky et al. 2014 |
| Al | muscovite, boehmite, corundum | $Al(OH)_3^0$, $Al(OH)_4^-$, $(Na,K)Al(OH)_4^0$ | <0.5 | Tagirov and Schott 2001; Sverjensky et al. 2014 |
| REE | monazite, bastnäsite, fluocerite | $REECl^{2+}$, $REECl_2^+$, $REE(SO_4)_2^-$, $REEF^+$, $REE(OH)_3^0$ | >1 | Migdisov et al. 2016 |
| Ca, Ba | calcite, gypsum, barite | $(Ca,Ba)^{2+}$, $(Ca,Ba)Cl^+$, $(Ca,Ba)Cl_2^0$ | <1 | Sverjensky et al. 1997 |

[1] Species shown *in italic* are either subordinate in typical hydrothermal fluids or unconstrained.
[2] Rough estimate, in terms of the variation of metal dissolved concentration in equilibrium with its solubility-controlling solid phase for typical porphyry-epithermal fluid compositions at a given $T$, $P$, pH and sulfur and oxygen fugacity, and using the range of published stability constants for the corresponding aqueous species from different studies. Values in brackets are for the case where the unconstrained species (shown in italic) are included in the calculations.
[3] Major recent experimental or theoretical studies explicitly reporting thermodynamic data, recommended in this article, for aqueous complexes and mineral phases, which can be used for mineral solubility predictions (numerous older references can be found therein).